\documentclass[12pt,preprint]{emulateapj}

\slugcomment{Astrophysical Journal}

\shorttitle{AKARI IRC 2.5--5 $\mu$m spectroscopy of LIRGs}
\shortauthors{Imanishi et al.}

\begin{document}

\title{AKARI IRC infrared 2.5--5 $\mu$m spectroscopy of a large sample of 
luminous infrared galaxies}


\author{Masatoshi Imanishi\altaffilmark{1}}
\affil{National Astronomical Observatory, 2-21-1, Osawa, Mitaka, Tokyo
181-8588, Japan}
\email{masa.imanishi@nao.ac.jp}

\author{Takao Nakagawa, Mai Shirahata}
\affil{Institute of Space and Astronautical Science, Japan Aerospace
Exploration Agency, 3-1-1 Yoshinodai, Sagamihara, Kanagawa 252-5210,
Japan}

\author{Yoichi Ohyama}
\affil{Institute of Astronomy and Astrophysics
Academia Sinica, P.O. Box 23-141, Taipei 10617, Taiwan, R.O.C.}

\and

\author{Takashi Onaka}
\affil{
Department of Astronomy, Graduate School of Science, University of
Tokyo, Bunkyo-ku, Tokyo 113-0033, Japan}

\altaffiltext{1}{Department of Astronomy, School of Science, Graduate
University for Advanced Studies (SOKENDAI), Mitaka, Tokyo 181-8588}

\begin{abstract}
We present the results of our systematic infrared 2.5--5 $\mu$m 
spectroscopy of 60 luminous infrared galaxies
(LIRGs) with infrared luminosities L$_{\rm IR} $=$
10^{11-12}$L$_{\odot}$, and 54 ultraluminous infrared galaxies 
(ULIRGs) with L$_{\rm IR}$ $\geq$ 10$^{12}$L$_{\odot}$, using AKARI IRC. 
AKARI IRC slit-less spectroscopy allows us to probe the full range of emission
from these galaxies, including spatially extended components. 
The 3.3 $\mu$m polycyclic aromatic hydrocarbon (PAH) emission features, 
hydrogen recombination emission lines, and various absorption features
are detected and used to investigate the properties of these galaxies.
Because of the relatively small effect of dust extinction in the infrared 
range, quantitative discussion of these dusty galaxy populations is possible. 
For sources with clearly detectable Br$\beta$ (2.63 $\mu$m) and
Br$\alpha$ (4.05 $\mu$m) emission lines, the flux ratios are found to
be similar to that predicted by case B theory. 
Starburst luminosities are estimated from both 3.3 $\mu$m PAH
and Br$\alpha$ emission, which roughly agree with each other.
In addition to the detected starburst activity, a significant fraction
of the observed sources display signatures of obscured AGNs, such
as low PAH equivalent widths, large optical depths of  dust absorption
features, and red continuum emission. 
The energetic importance of optically elusive buried AGNs in optically
non-Seyfert galaxies tends to increase with increasing galaxy infrared
luminosity, from LIRGs to ULIRGs.  
\end{abstract}


\section{Introduction}

In luminous infrared galaxies (LIRGs) with infrared luminosities of 
L$_{\rm IR}$ = 10$^{11-12}$L$_{\odot}$ \citep{sam96} and 
ultraluminous infrared galaxies (ULIRGs) with L$_{\rm IR}$ $\geq$
10$^{12}$L$_{\odot}$ \citep{san88a} 
\footnote{
In this paper, galaxies with L$_{\rm IR}$ =
10$^{11-12}$L$_{\odot}$ are called LIRGs, while   
those with L$_{\rm IR}$ $\geq$ 10$^{12}$L$_{\odot}$ 
are referred to as ULIRGs.
},
infrared emission is the dominant component of bolometric
luminosities.   
This means that 
(1) (U)LIRGs possess very luminous energy sources 
with L $>$10$^{11}$L$_{\odot}$,  
(2) the energy sources are hidden by dust, which absorbs most of the 
primary energetic radiation, and
(3) the heated dust radiates this energy as infrared dust thermal 
emission.   
The energy sources can be starburst activity (energy generation
by the nuclear fusion reactions occurring inside rapidly formed stars)
and/or AGN activity (in which the release of gravitational energy by a
spatially compact, mass-accreting, super-massive black hole (SMBH)
produces strong radiative energy).  
Because (U)LIRGs become an important population with increasing redshift,
in terms of cosmic infrared radiation density
\citep{lef05,cap07,mag09}, understanding the physical nature of (U)LIRGs
is important in clarifying the AGN-starburst connections in the
dust-obscured galaxy population of the early universe.  

Given that the bulk of the primary radiation from the energy sources is 
absorbed by dust, and so is not directly visible in (U)LIRGs, investigating
their energy sources is difficult, particularly when spatially compact
AGNs are surrounded by large amounts of dust along virtually every
line of sight, and become {\it buried}.   
In an AGN surrounded by a toroidal (torus-shaped) dusty medium,
the so-called "narrow-line regions" (NLRs), which are photo-ionized by the
ionizing radiation of the AGN, should be well developed along the axis 
of the torus above the torus scale height.
Because the optical emission line flux ratios of NLRs in AGNs are different
from those of clouds photo-ionized by stars in starbursts, this type of AGN 
is optically classified as Seyfert, and is thus distinguishable from a
normal starburst galaxy by optical spectroscopy
\citep{vei87,kau03,kew06}.  
However, a buried AGN is not readily detectable by optical
spectroscopy, because well-developed NLRs are lacking.
 
Infrared 2.5--5 $\mu$m (rest-frame) spectroscopy is a powerful tool for 
detecting these optically elusive \citep{mai03} buried AGNs, because the
effect of dust extinction is relatively small \citep{nis09}.
More importantly, starburst and AGN activity can be distinguished, based
on the infrared spectral shapes of galaxies. 
First, strong, large-equivalent-width emission of polycyclic aromatic
hydrocarbons (PAH) is usually seen at rest-frame 3.3 $\mu$m in a normal
starburst galaxy, whereas a pure AGN exhibits a PAH-free continuum, 
originating in AGN-heated, larger-sized hot dust \citep{moo86,imd00}. 
Second, in a normal starburst, the stellar energy sources and dust are
spatially well mixed \citep{pux91,mcl93,for01}, while in a buried AGN,
the energy source (i.e., a compact mass-accreting SMBH) is more centrally
concentrated than the surrounding dust
\citep{soi00,sie04,ima07a,ima08}. Thus, the optical depths of dust 
absorption features found at 2.5--5 $\mu$m cannot exceed certain
thresholds in a normal starburst with mixed dust/source geometry, but
can be arbitrarily large in a buried AGN with centrally concentrated
energy source geometry \citep{im03,idm06}.

Using these properties, infrared 2.8--4.2 $\mu$m ($L$-band) and 
4.5--5.0 $\mu$m ($M$-band) spectroscopy has previously been applied to many
galaxies, using infrared spectrographs attached to large ground-based 
telescopes,  
to examine the nature  of dust-obscured energy sources
\citep{idm06,ris06,ima06,san08,ris10}.  
However, the sample has been limited, primarily to ULIRGs with L$_{\rm IR}$
$\geq$ 10$^{12}$L$_{\odot}$.
Infrared dust emission from nearby ULIRGs is usually dominated by 
a spatially compact component \citep{soi00}, so that ground-based 
{\it slit} spectroscopy with a width of less than few arcsec can probe most 
of the emission. 
On the other hand, in nearby LIRGs with 
L$_{\rm IR}$ = 10$^{11-12}$L$_{\odot}$,  
spatially extended infrared dust emission of up to several arcsec
becomes important \citep{soi01}, and thus ground-based slit spectroscopy
could miss a significant fraction of the infrared radiation.  
{\it Slit-less} spectroscopy using the IRC infrared spectrograph on board the
AKARI infrared satellite \citep{ona07,mur07} is best suited for studying the
origins of infrared emission from such spatially extended LIRGs.
By covering both LIRGs and ULIRGs, we can better investigate 
AGN-starburst connections as a function of galaxy infrared luminosity.
In addition to its slit-less spectroscopic capability, AKARI IRC has a 
wide continuous wavelength coverage of 2.5--5 $\mu$m, unhindered by
Earth's atmosphere, thus enabling us to investigate the 
2.63 $\mu$m Br$\beta$ emission lines and 4.26 $\mu$m CO$_{2}$ absorption 
features of nearby sources, which are totally inaccessible from the ground. 
Furthermore, the strengths of the very broad 3.1 $\mu$m H$_{2}$O ice
absorption features \citep{gib04} and 4.67 $\mu$m CO absorption
features, as well as continuum slopes, are better estimated using AKARI IRC
spectra with wider wavelength coverage, rather than ground-based spectra with  
limited wavelength coverage.

In this paper, we present the results of systematic AKARI IRC 2.5--5
$\mu$m slit-less spectroscopy of LIRGs. 
Many ULIRGs have also been observed, to augment the 
ULIRG sample previously studied with AKARI IRC \citep{ima08}. 
Throughout this paper, $H_{0}$ $=$ 75 km s$^{-1}$ Mpc$^{-1}$,
$\Omega_{\rm M}$ = 0.3, and $\Omega_{\rm \Lambda}$ = 0.7 are
adopted to be consistent with our previous publications. 

\section{Targets}

LIRGs with L$_{\rm IR}$ = 10$^{11.1-12}$L$_{\odot}$ are selected from
the Bright Galaxy Sample (BGS) \citep{soi87,san95} and the revised BGS
\citep{san03}. 
ULIRGs with L$_{\rm IR}$ $\geq$ 10$^{12}$L$_{\odot}$ are selected from
the IRAS 1 Jy sample, compiled by \citet{kim98}.
These LIRG and ULIRG samples have IRAS 60 $\mu$m fluxes $\geq$ 5.24 Jy
and $\geq$ 1 Jy, respectively. 
Because of AKARI's sun-synchronous polar orbit, which follows
the boundary between night and day \citep{mur07}, objects with high (low)
ecliptic latitudes have high (low) visibilities. 
It is difficult to observe a statistically complete sample with AKARI,
because the probability of observing objects 
with low ecliptic latitudes is small. 
Moreover, due to the orbit, any particular object is observable
only twice a year, and the position angle of the AKARI IRC is fixed to 
within $<$1$^{\circ}$ and is not arbitrarily adjustable (excluding sources 
at the north and south ecliptic poles). 
For sources with multiple nuclei that happen to be aligned in the direction 
of spectral dispersion, the spectra of these nuclei overlap, and reliable
extraction of individual nuclear spectra is not possible.   
For these reasons, our sample consists of a statistically significant
number of LIRGs and ULIRGs, but is not a complete sample. 
However, the observed sources are limited only by their sky positions
and the elongation of multiple nuclei, and hence there should be no 
selection bias regarding the physical nature of the objects.  
These LIRGs and ULIRGs are used primarily for statistical discussion.
Tables 1 and 2 summarize the pertinent information for the observed LIRGs
and ULIRGs, respectively.

In addition to these unbiased samples, several additional
interesting LIRGs and ULIRGs are also observed. 
These sources display strong absorption features at 5--20 $\mu$m and/or 
luminous buried AGN signatures from previous observations at other
wavelengths.  
The observed sources are NGC 4418
\citep{roc91,dud97,spo01,ima04,lah07,ima10b}, NGC 1377
\citep{rou06,ima06}, IRAS 00183$-$7111 \citep{spo04}, IRAS 06035$-$7102 
\citep{spo02,dar07,far09}, IRAS 20100$-$4156 \citep{fra03,lah07}, 
IRAS 20551$-$4250 \citep{fra03,ris06,nar09}, and IRAS 23128$-$5919 
\citep{spo02,fra03,far09}.  
It is of particular interest to confirm whether the AKARI IRC infrared
spectral shapes of these galaxies are also indicative of luminous buried AGNs, 
and/or to carry out a detailed investigation of the absorption features found 
in the 2.5--5 $\mu$m wavelength range of the AKARI IRC. 
Finally, Mrk 231 is a galaxy optically classified as Seyfert 1, and 
is listed in the IRAS 1 Jy ULIRG sample \citep{kim98,vei99}.
Our basic sample selection of ULIRGs excludes optically Seyfert 1 
galaxies, because our primary scientific aim is to study obscured AGNs,
and the presence of unobscured AGNs is obvious in optical Seyfert 1 galaxies.
However, despite its optical Seyfert 1 classification, the infrared
spectrum of Mrk 231 shows absorption features \citep{lah07,dar07,bra08}, 
making it an enigmatic object. 
Obtaining a high-quality AKARI IRC 2.5--5 $\mu$m spectrum should help
to determine the true nature of Mrk 231. 
Thus, Mrk 231 is placed in the category of "additional interesting
sources", and is included in our observations.  
The observed properties of these sources are summarized in Tables 1
and 2, according to their infrared luminosities. 

\section{Observations and Data Analysis}

Infrared 2.5--5 $\mu$m spectroscopy of the LIRGs and ULIRGs was performed with
the IRC infrared spectrograph \citep{ona07} on board the AKARI infrared
satellite \citep{mur07}.
All data were collected as part of the mission program called "AGNUL". 
Tables 3 and 4 summarize the observation logs for the LIRGs and ULIRGs,
respectively.  
The NG grism mode was used for our observations. 
In this mode, the entire 2.5--5.0 $\mu$m wavelength range is 
simultaneously covered at an effective spectral resolution of R $\sim$
120 at 3.6 $\mu$m for point sources \citep{ona07}. 
Objects were located inside a 1 $\times$ 1 arcmin$^{2}$ window to
avoid spectral overlap from nearby sources \citep{ona07,ohy07}. 
The pixel scale of AKARI IRC is 1$\farcs$46 $\times$ 1$\farcs$46. 
One to five pointings were assigned to each source,
depending on its brightness and visibility.
The total net on-source exposure time for one pointing was $\sim$6 min. 
Because we used the IRC04 (phases 1 and 2; liquid-He cool holding
period, before 2007 August) and IRCZ4 (phase 3; post liquid-He warm
mission cooled by the onboard cryocooler, after 2008 June) observing
modes, one pointing consisted of eight or nine independent frames \citep{ona07}.
Hence, the effects of cosmic ray hits were essentially removable, even for
sources with only one pointing. 
For sources with more than one pointing, if independent datasets were
of similar quality, all of them were added together to obtain the final spectra.
However, if some of the datasets were substantially
inferior to others, only the better datasets were used to obtain the final
spectra, because the addition of poorer-quality data often degraded the 
quality of the final spectra. 
This was especially true when data for a particular object were 
collected during widely separated and dissimilar periods,
specifically during early and late phase 3 of AKARI, because the background 
noise gradually increased as phase 3 proceeded, and was
substantially higher during late phase 3 (because of
the temperature increase of the IRC).

Spectral analysis was performed in a standard manner, using the IDL
package prepared for the reduction of AKARI IRC spectra.
The actual software packages used for our data reduction were "IRC
Spectroscopy Toolkit for Phase 3 data Version 20090211" for phase 3
data and "IRC Spectroscopy Toolkit Version 20090211" for data collected 
during phases 1 and 2, both of which can be found at 
http://www.ir.isas.jaxa.jp/ASTRO-F/Observation/DataReduction/IRC/.  
Further details concerning these data analysis tools can be found in \citet{ohy07}.
Each frame was dark-subtracted, linearity-corrected, and 
flat-field-corrected. 
Many LIRGs display clear signals of spatially extended emission. 
We varied the aperture sizes for spectral extraction, depending
on the actual signal profile of each source. 
The background signal level was estimated from data points on both
sides, or in some cases only one side of the object position, in 
the direction perpendicular to the spectral dispersion
direction of the AKARI IRC, and was subtracted.  
Wavelength and flux calibrations were made with the data analysis
toolkits. 
According to \citet{ohy07}, the wavelength calibration accuracy is 
$\sim$1 pixel or $\sim$0.01 $\mu$m. 
The absolute flux calibration accuracy is $\sim$10\% at the central
wavelength of the spectra, and can be as large as $\sim$20\% at the edge
of the NG spectra (close to 2.5 $\mu$m and 5.0 $\mu$m). To reduce the 
scatter of the data points, appropriate binning of spectral elements was 
performed, particularly for faint sources.

\section{Results}

Figures 1 and 2 show the AKARI IRC infrared 2.5--5 $\mu$m spectra of
the LIRGs and ULIRGs, respectively. 
For the LIRGs VV 114, Arp 299 (IC 694 + NGC 3690), NGC6285/6, and NGC
7592, the spectra of both the double nuclei are of adequate quality, and
so the individual spectra are plotted separately.   
For ESO 244-G012, IC 2810, VV 250, and NGC 5256, only the spectra of the N,
NW, SE, and SW nuclei, respectively, are shown.  
For VV 705 and IRAS 17132+5313, the spectra of combined emission from the 
S + N and W + E nuclei, respectively, are extracted and displayed.   
In total, the spectra of 64 LIRG nuclei and 54 ULIRGs are shown, more than
tripling the number of available AKARI IRC 2.5--5 $\mu$m 
LIRG and ULIRG spectra \citep{ima08}.  

In Figures 1 and 2, 3.3 $\mu$m PAH emission features are detected 
at $\lambda_{\rm obs}$ $\sim$ (1 + $z$) $\times$ 3.29 $\mu$m in the
observed frame, for the majority of the observed sources 
\footnote{
A Pf$\delta$ ($\lambda_{\rm rest}$ $\sim$ 3.3 $\mu$m) emission line is present
at a similar wavelength, but is so weak ($\sim$10\% strength of
Br$\alpha$ for 10$^{4}$ K case B; Wynn-Williams 1984) that its
contribution is usually negligible in optically non-Seyfert (U)LIRGs
\citep{idm06,ima08}.
}. 
To estimate the strengths of the 3.3 $\mu$m PAH emission features of
LIRGs, caution must be exercised.
Unlike ULIRGs (the 2.5--5 $\mu$m emission of which is usually spatially compact),
many LIRGs exhibit spatially extended emission components. 
For such spatially extended sources, the profiles of the 3.3 $\mu$m PAH
emission features are broadened in comparison with intrinsic ones in
AKARI IRC slit-less spectra, because signals from spatially different
positions fall into different array positions along the spectral
dispersion directions. 
Accordingly, although the 3.3 $\mu$m PAH emission features and the 3.4
$\mu$m PAH sub-peaks \citep{tok91} should be resolvable at the AKARI IRC
spectral resolution of R $\sim$ 120 (as is the case for LIRGs dominated 
by spatially compact components, such as NGC 34 and NGC 7469), these
features are sometimes spectrally blended, particularly for spatially
extended LIRGs (e.g., UGC 2982 and NGC 3110). 
Thus, we let the width of the 3.3 $\mu$m PAH emission feature be a free
parameter. 
Nevertheless, for distant, compact ULIRGs with weak PAH emission, we assume 
the intrinsic 3.3 $\mu$m PAH profile (type A of Tokunaga et al. 1991) to
estimate the PAH strength or its upper limit.
We exclude the 3.4 $\mu$m PAH sub-peak from the 3.3
$\mu$m PAH emission strength by fitting the PAH emission after
removing the data points at the 3.4 $\mu$m shoulders, because the 3.3
$\mu$m PAH emission strength has been estimated and calibrated only for
the 3.3 $\mu$m main components in many earlier works
\citep{mou90,imd00,idm06}. 
We assume a single Gaussian component for the 3.3 $\mu$m PAH emission
feature. 
Tables 5 and 6 summarize the fluxes, luminosities, and rest-frame
equivalent widths (EW$_{\rm 3.3PAH}$) of the 3.3 $\mu$m PAH emission
features of the LIRGs and ULIRGs, respectively.

The power of AKARI IRC slit-less spectroscopy as a tool for probing all
the emission from a galaxy is highlighted in the LIRG NGC 7469. 
NGC 7469 has a Seyfert 1 nucleus at the center, with ring-shaped,
spatially extended (1--2 arcsec in radius) surrounding circum-nuclear
starburst activity \citep{soi03,gal05,dia07,reu10}.
In a ground-based slit spectrum with an aperture width $\sim$1 arcsec, 
only the nuclear emission was probed, and no 3.3 $\mu$m PAH emission was
detected \citep{iw04}. 
However, our AKARI IRC slit-less spectrum of NGC 7469 clearly recovers
the 3.3 $\mu$m PAH emission feature from the spatially extended starburst
ring (Fig. 1).

In many LIRGs and bright ULIRGs, Br$\alpha$ emission at 
$\lambda_{\rm rest}$ = 4.05 $\mu$m is clearly visible, primarily because
of their high S/N ratios in the continua.
For sources with particularly strong Br$\alpha$ emission, 
Br$\beta$ emission at $\lambda_{\rm rest}$ = 2.63 $\mu$m is often
discernible. 
Those emission lines above the linear continuum levels determined
from data points of the shorter and longer wavelength components are fit 
with single Gaussian profiles.
The central wavelength and normalization are chosen as free parameters 
for both the Br$\alpha$ and Br$\beta$ emission. 
The line width is taken to be a free parameter for the Br$\alpha$
lines, but given the faintness of the Br$\beta$ emission, we assume that
the line width of Br$\beta$ is the same as that of Br$\alpha$ in velocity.  
The estimated fluxes and luminosities of these Br emission lines are
listed in Tables 7 and 8, respectively, for LIRGs and ULIRGs for 
which the strengths of the Br$\alpha$ emission lines are estimated with
reasonable accuracy in the AKARI IRC R $\sim$ 120 spectra.
The calibration uncertainties of AKARI IRC spectra become large when
$\lambda_{\rm obs}$ $>$ 4.8 $\mu$m. 
Hence, the strengths of Br$\alpha$ emission redshifted into (or close to) this
wavelength could be quantitatively uncertain, and are not discussed.
The signatures of Pf$\beta$ lines at $\lambda_{\rm rest}$ = 4.65 $\mu$m are
recognizable in many LIRGs and ULIRGs, but the flux estimates could be
uncertain because Pf$\beta$ emission lines spectrally overlap 
with the 4.67 $\mu$m CO absorption features in AKARI IRC spectra with R
$\sim$ 120. 
Pf$\gamma$ emission at $\lambda_{\rm rest}$ = 3.74 $\mu$m is
intrinsically much weaker than Br$\alpha$ (Pf$\gamma$/Br$\alpha$ 
$\sim$ 0.15 for 10$^{4}$ K case B; Wynn-Williams 1984), and its signature is
clearly seen only in the LIRG Arp 193.  
Pf$\beta$ and Pf$\gamma$ emission lines are not discussed in detail.

Another prominent feature detected in the AKARI IRC 2.5--5 $\mu$m
spectra of LIRGs and ULIRGs is a broad H$_{2}$O ice absorption feature,
caused by ice-covered dust grains, centered at $\lambda_{\rm rest}$ =
3.05--3.1 $\mu$m and 
extending from $\lambda_{\rm rest}$ $\sim$ 2.75 $\mu$m to $\sim$3.55
$\mu$m. 
This H$_{2}$O ice absorption feature is detected in many of the observed
LIRGs and ULIRGs.  
Because of the 3.4 $\mu$m PAH sub-peaks, the continuum level at the
shorter wavelength side of the 3.3 $\mu$m PAH emission could be 
depressed in comparison to that on the longer wavelength side, 
even in the absence of the 3.1 $\mu$m H$_{2}$O ice absorption feature.  
Thus, it is important to use data points with $\lambda_{\rm rest}$ $>$ 3.5
$\mu$m as the longer wavelength side of the continuum level, to
determine the optical depth of the 3.1 $\mu$m H$_{2}$O ice absorption
feature ($\tau_{3.1}$).  
In AKARI IRC 2.5--5 $\mu$m spectra, because data points with wavelengths as short as
$\lambda_{\rm obs}$ = 2.5 $\mu$m are covered, the continuum level on the
shorter wavelength side of the 3.1 $\mu$m H$_{2}$O ice absorption
feature is well determined.  
We assume linear continuum levels determined from data points at 
$\lambda_{\rm rest}$ $<$ 2.75 $\mu$m and 
$\lambda_{\rm rest}$ $>$ 3.55 $\mu$m, and unaffected by other absorption and
emission lines.  
The adopted continuum levels are shown as dotted lines in Figures 1 and
2, for sources with clearly detectable 3.1 $\mu$m H$_{2}$O ice absorption
features. 
Table 9 summarizes the estimated $\tau_{3.1}$ for clearly detected
sources. 

Other absorption features are also clearly detected in a fraction of
the LIRGs and ULIRGs. 
Examples include the 3.4 $\mu$m bare carbonaceous dust, 4.26
$\mu$m CO$_{2}$, and 4.67 $\mu$m CO absorption features. 
For these absorption features, linear continua are determined from data
points of shorter and longer wavelength parts that are free 
of other obvious emission and absorption features.
Unlike the very broad 3.1 $\mu$m H$_{2}$O ice absorption feature, these
absorption features are relatively thinner, so that continuum
determination is less ambiguous. 
The observed optical depths of these absorption features ($\tau_{3.4}$, 
$\tau_{\rm CO2}$, and $\tau_{\rm CO}$) are summarized in Table 9 for
detected sources.

The CO absorption feature shows a relatively narrow profile around
$\lambda_{\rm rest}$ = 4.67 $\mu$m in a solid phase \citep{chi98}.
However, gas-phase CO displays many sharp absorption features 
(v = 1--0), extending from 4.58--4.66 $\mu$m (R-branch) and 4.675--4.73
$\mu$m (P-branch), when the rotational level is $<$10 \citep{mit89,mon01}.
In the AKARI IRC low-resolution (R $\sim$ 120) spectra, these sharp
absorption features are widened and overlap each other, resulting in 
two broad absorption peaks in the P- and R-branches \citep{spo04,ima08,ima09}.
When the XCN ($\lambda_{\rm rest}$ = 4.62 $\mu$m) absorption feature and
the Pf$\beta$ ($\lambda_{\rm rest}$ = 4.65 $\mu$m) emission line are
superimposed on the broad 4.67 $\mu$m CO absorption feature in a similar
wavelength range, it is not easy to disentangle these features in the AKARI
IRC low-resolution spectra. 
Furthermore, in AKARI IRC spectra, systematic uncertainty can become
large when $\lambda_{\rm obs}$ $>$ 4.8 $\mu$m. 
For these reasons, an estimate of $\tau_{\rm CO}$ is only
attempted when the CO absorption feature is clearly detected and
the redshifted CO absorption profile is below $\lambda_{\rm obs}$ $=$
4.8 $\mu$m.

Sources with detectable 4.26 $\mu$m CO$_{2}$ absorption features usually
display strong 3.1 $\mu$m H$_{2}$O ice absorption features as well
(Fig. 1, 2), as seen in the Galactic highly obscured sources
\citep{gib04}. 
CO$_{2}$ molecules cannot be efficiently formed in gas-phase reactions 
\citep{her86}, but can be through UV photolysis of dust grains covered
with an ice mantle \citep{dhe86}. 
Thus, the coincidence of CO$_{2}$ and H$_{2}$O ice absorption features
seems reasonable \citep{pon08}.

The LIRG and ULIRG spectra shown in Figures 1 and 2 exhibit a variety of 
continuum slopes, ranging from a nearly flat spectral shape to a red,
steeply rising continuum with increasing wavelength. 
One may argue that the infrared 2.5--5 $\mu$m continuum slope can also
be used as an energy diagnostic tool for galaxies, in that a very red
continuum is the signature of an obscured AGN \citep{ris06,san08,ris10}.
Thus, we estimate the continuum slope $\Gamma$ (F$_{\nu}$ $\propto$
$\lambda^{\Gamma}$), using data points unaffected by absorption
features (broad 3.1 $\mu$m H$_{2}$O ice, 3.4 $\mu$m bare
carbonaceous dust, 4.26 $\mu$m CO$_{2}$, and 4.67 $\mu$m CO) and 
emission features (3.3 $\mu$m and 3.4 $\mu$m PAH, and hydrogen
recombination lines, such as 4.05 $\mu$m Br$\alpha$, 2.63 $\mu$m
Br$\beta$, 4.65 $\mu$m Pf$\beta$, and 3.74$\mu$m Pf$\gamma$). 
The estimated $\Gamma$ values for the LIRGs and ULIRGs observed in this
paper are listed in Tables 5 and 6 (column 7), respectively.
In Table 10, we also estimate the $\Gamma$ values of ULIRGs previously 
observed with AKARI IRC by \citet{ima08}.
A small fraction of sources (e.g., IRAS 15250$+$3609 and 13305$-$1739)
exhibit a break in the continuum slope, in such a way that the continuum
emission is flat (blue) in the shorter wavelength component of the AKARI IRC
2.5--5 $\mu$m spectrum, but rises steeply (red) in the longer
wavelength component (Fig. 1, 2). 
This behavior could be explained if the longer wavelength components
are dominated by hot (T = 100--1000K) dust emission heated by
obscured energy sources, while weakly-obscured starburst-related
(stellar photospheric and/or starburst heated dust) emission contributes
to the shorter wavelength components of the AKARI IRC 2.5--5 $\mu$m
spectra.    
We estimate the $\Gamma$ values for these sources from the continuum 
slopes in the longer wavelength regions.

We note that our continuum slope $\Gamma$ is defined by 
F$_{\nu}$ $\propto$ $\lambda^{\Gamma}$, whereas it was defined by 
F$_{\lambda}$ $\propto$ $\lambda^{\Gamma}$ in earlier research
\citep{ris06,san08,ris10}. 
Thus, the values of our $\Gamma_{\rm ours}$ differ from the $\Gamma_{\rm prev}$
values reported in previous papers in accordance with $\Gamma_{\rm ours}$ =
$\Gamma_{\rm prev}$ + 2.  
The wider wavelength coverage of AKARI IRC (2.5--5 $\mu$m) is 
superior to the limited wavelength coverage of the ground-based 
$L$-(2.8--4.2 $\mu$m) and $M$-band (4.5--5.0 $\mu$m) spectra for 
obtaining precise estimates of the continuum slope. 
For sources with strong 3.1 $\mu$m ice absorption features, 
although the continuum level must be determined outside this broad
absorption feature ($\lambda_{\rm rest}$ = 2.8--3.5 $\mu$m), 
the task is not readily accomplished using ground-based spectra.
An example is the ULIRG IRAS 00188$-$0856. 
Our AKARI spectrum provides a reliable $\Gamma$ estimate by 
covering the continuum emission on both the shorter and longer
wavelength sides of the broad 3.1 $\mu$m H$_{2}$O ice absorption
feature. However, with the ground-based spectrum, the continuum 
slope was estimated using data points inside the broad 3.1 $\mu$m H$_{2}$O 
ice absorption feature \citep{ris10}.   

\section{Discussion}

\subsection{Modestly obscured starbursts}

Based on the recently estimated dust extinction curve \citep{nis08,nis09}, 
the amount of dust extinction at 3.3 $\mu$m is as small as $\sim$1/30 of
that in the optical $V$-band ($\lambda$ = 0.55 $\mu$m).
This indicates that the flux attenuation of the 3.3 $\mu$m PAH emission
is not significant (a factor of $<$2.5) if the dust extinction 
A$_{\rm V}$ is less than 30 mag.
For this reason, the observed 3.3 $\mu$m PAH emission luminosities 
roughly map out the intrinsic luminosities of modestly obscured 
(A$_{\rm V}$ $<$ 30 mag) PAH-emitting normal starburst activity.

The relationship L$_{\rm 3.3PAH}$/L$_{\rm IR}$ $\sim$
10$^{-3}$ has been used for this purpose \citep{mou90,ima02}. 
\citet{mou90} observed a limited number of starburst galaxies, and in
most of these, not all of the starburst regions were covered by their
apertures.  
The proportion L$_{\rm 3.3PAH}$/L$_{\rm IR}$ $\sim$ 10$^{-3}$ was derived
after aperture correction.
The aperture correction was based on the ratio of the infrared
aperture size to the optical extent of each galaxy, which could entail
an uncertainty.
\citet{ima02} supported the proportion L$_{\rm 3.3PAH}$/L$_{\rm IR}$ 
$\sim$ 10$^{-3}$ for starbursts on the basis of the rough agreement of
3.3$\mu$m-PAH-derived starburst luminosities (without a dust
extinction correction) and UV-derived starburst luminosities (with a dust extinction
correction) in three Seyfert galaxies.  
For the prototypical starburst galaxy M82, \citet{sat95} argued that 
the proportion L$_{\rm 3.3PAH}$/L$_{\rm IR}$ $\sim$ 0.8 $\times$
10$^{-3}$ holds, once the effects of dust extinction have been corrected.  
We now have AKARI IRC slit-less spectra for a large number of LIRGs, many
of which may be starburst-dominated, without strong AGN signatures in
the infrared range.
Because the 3.3 $\mu$m PAH emission from spatially extended
starburst regions of LIRGs should be covered in the AKARI IRC 
slit-less spectra, we can use observational data to directly test 
the validity of the proportion 
L$_{\rm 3.3PAH}$/L$_{\rm IR}$ $\sim$ 10$^{-3}$ 
for estimating intrinsic starburst luminosity, without 
introducing a potentially uncertain aperture correction. 

Figure 3 compares the 3.3 $\mu$m PAH emission luminosities
measured with ground-based slit spectroscopy and AKARI IRC slit-less
spectroscopy.
For ULIRGs (Fig. 3a), the two sets of measurements roughly agree, with some
scatter (partly originating from measurement errors for faint sources), as
expected from the physically compact nature of the infrared emission of 
ULIRGs in general ($<$300 pc; Soifer et al. 2000), and their relatively large
distances. 
The largest discrepancy among the 3.3 $\mu$m PAH luminosity measurements is
found in Mrk 231 (z = 0.042), one of the nearest ULIRGs in the sample,
as indicated in the figure. 
Mrk 231 exhibits spatially extended starburst activity \citep{kra97}, the
bulk of which can be missed with ground-based narrow-slit 
($<$1--2 arcsec wide) spectroscopy, because of the large apparent
extension resulting from its proximity.  
However, except for the nearest ULIRG, Figure 3a suggests that the
bulk of the 3.3 $\mu$m PAH emission from ULIRGs can be recovered 
effectively from ground-based narrow-slit spectra. 

Figure 3b compares the 3.3 $\mu$m PAH luminosity measurements
obtained from ground-based narrow slit spectroscopy and AKARI IRC
slit-less spectroscopy for LIRGs. 
The AKARI IRC measurements are systematically larger, which is
reasonable because the infrared emission from LIRGs is, in most cases, 
spatially extended (up to $\sim$kpc; Soifer et al. 2001), and
LIRGs are generally closer than ULIRGs, so that their apparent spatial
extensions are also larger.

Figure 4 compares the rest-frame equivalent widths of the 3.3 $\mu$m PAH
emission features (EW$_{\rm 3.3PAH}$) measured by ground-based slit
spectroscopy and AKARI IRC slit-less spectroscopy.
If the 3.3 $\mu$m PAH emission properties do not exhibit substantial
spatial variation, the EW$_{\rm 3.3PAH}$ values are less susceptible to
aperture effects.   
The two sets of measurements roughly agree, with some scatter, part of
which may originate from measurement errors for faint sources  
(Fig. 4a). The LIRGs plotted in Figure 4b are generally bright, and so 
we do not expect large measurement errors. 
In Figure 4b, many LIRGs display smaller EW$_{\rm 3.3PAH}$ values
in the AKARI IRC data.  
A possible explanation is that ground-based slit spectroscopy preferentially
probes strongly PAH-emitting, very active nuclear starburst regions,
while stellar photospheric continuum emission from inactive old stars
(which lack PAH-exciting UV photons) in the outer regions of galaxies
may contribute to the AKARI IRC slit-less spectra and reduces the
EW$_{\rm 3.3PAH}$ values.   
 
In Figure 5a, the observed ratios of 3.3 $\mu$m PAH to infrared luminosity 
(L$_{\rm 3.3PAH}$/L$_{\rm IR}$) for starburst-classified LIRGs (i.e., no
AGN signatures in the AKARI IRC 
2.5--5 $\mu$m spectra; see next subsection) are represented by open circles. 
The upper bound of the L$_{\rm 3.3PAH}$/L$_{\rm IR}$ distribution for
these starburst-classified LIRGs is indeed $\sim$10$^{-3}$.
We have now obtained direct evidence from AKARI IRC slit-less
spectroscopic observations that the relationship 
L$_{\rm 3.3PAH}$/L$_{\rm IR}$ = 10$^{-3}$ is appropriate for
quantitatively estimating the intrinsic power of modestly obscured
starburst activity. 

The L$_{\rm 3.3PAH}$/L$_{\rm IR}$ ratios of LIRGs and ULIRGs are
summarized in column 4 of Tables 5 and 6, respectively.   
Figure 5a shows a plot of the L$_{\rm 3.3PAH}$/L$_{\rm IR}$ ratios 
of the observed LIRGs and ULIRGs. 
As the figure indicates, the majority of starburst-classified LIRGs are
distributed between the 50\% and 100\% lines, suggesting that the bulk of
the infrared luminosities could be explained by PAH-probed
modestly obscured starburst activity. 
On the other hand, only a very
small fraction of the ULIRGs are above the 50\% line. 
At face value, PAH-probed modestly obscured
starbursts alone are insufficient to account for the large infrared
luminosities of most ULIRGs. 
We note that the quality of the AKARI IRC 2.5--5 $\mu$m spectra of most of 
the LIRGs is high, because their relative proximity permits high continuum
flux and high S/N ratios to be achieved. 
For these objects, the 3.4 $\mu$m PAH sub-peaks are separated from the
3.3 $\mu$m PAH main peaks, and the contributions from the 3.4 $\mu$m PAH
sub-peaks are removed from the final L$_{\rm 3.3PAH}$ values.
ULIRGs are generally fainter than LIRGs at infrared 2.5--5 $\mu$m,
simply because of the larger overall distances, and hence it is sometimes 
difficult to separate the 3.3 $\mu$m PAH peak and 3.4 $\mu$m PAH sub-peak. 
Thus, the 3.4 $\mu$m PAH sub-peaks may contribute
to the derived L$_{\rm 3.3PAH}$ values for some fraction of ULIRGs.
This possible ambiguity may even reduce the true 
L$_{\rm 3.3PAH}$/L$_{\rm IR}$ ratios for ULIRGs.
The implied trend of 3.3 $\mu$m PAH luminosity suppression in
ULIRGs is therefore robust.

\subsection{Possible origin of 3.3 $\mu$m PAH depression in ULIRGs}

There are several plausible scenarios for the generally small observed
L$_{\rm 3.3PAH}$/L$_{\rm IR}$ ratios of ULIRGs. 
(1) The 3.3 $\mu$m PAH emission is more highly flux-attenuated in ULIRGs,
because of the larger dust extinction, whereas the 8--1000 $\mu$m infrared
fluxes are not significantly reduced.  
(2) Powerful AGN activity produces large infrared luminosities, but
virtually no PAH emission in ULIRGs.
(3) PAH emission from ULIRG starbursts is intrinsically weak.
A normal starburst galaxy is observationally considered to be a
composite of spatially well-mixed HII regions, molecular gas, and
photo-dissociation regions \citep{pux91,mcl93,for01}.
If the starburst magnitudes per unit volume are larger in ULIRGs, 
more intense radiation fields could lead to greater destruction of 
3.3 $\mu$m PAH carriers than in LIRG starbursts \citep{smi07}.
Alternatively, under the effect of stronger radiation fields, a larger
fraction of stellar UV photons could be absorbed by dust inside HII regions
\citep{abe09}, 
producing stronger infrared dust continuum emission, instead of creating
photo-dissociation regions, which are the main source of 
3.3 $\mu$m PAH emission \citep{sel81}.     
A higher fraction of UV photons could also be absorbed by dust,
if the HII regions are dustier in ULIRGs than in LIRGs \citep{luh03}.
In the third scenario, the L$_{\rm 3.3PAH}$/L$_{\rm IR}$ ratios could
decrease, even if the total amount of dust extinction is similar.

Regarding scenario (3), if the PAH emission from starbursts in ULIRGs
is intrinsically more suppressed than in LIRGs, then
even starburst-dominated ULIRGs should exhibit weak 3.3
$\mu$m PAH emission, and the upper envelope of the EW$_{\rm 3.3PAH}$
distribution should decrease in ULIRGs, compared with LIRGs.  
Figure 5b shows a plot of the EW$_{\rm 3.3PAH}$ distribution as a function of 
L$_{\rm IR}$. 
There is no clear trend that the upper envelope systematically decreases
for higher L$_{\rm IR}$, and so at least for ULIRGs with minimum AGN
contributions, the 3.3 $\mu$m PAH emission is as strong as in
LIRG starbursts. 
Thus, scenario (3) is not strongly supported by our AKARI IRC
observational data.

In scenario (1), the EW$_{\rm 3.3PAH}$ distribution should be 
relatively insensitive to dust extinction. 
This is because the 3.3 $\mu$m PAH and continuum emission are
comparably flux-attenuated in a pure normal starburst galaxy with 
spatially well-mixed HII regions/molecular gas/photo dissociation regions.
Specifically, the dust extinction of starbursts only decreases the
L$_{\rm 3.3PAH}$ values, while the EW$_{\rm 3.3PAH}$ values remain
relatively unchanged \citep{idm06,ima08}. 
In scenario (2), the EW$_{\rm 3.3PAH}$ values diminish when
the contributions from AGNs to the observed 2.5--5 $\mu$m fluxes
become important.
In Figure 5b, many ULIRGs show small EW$_{\rm 3.3PAH}$ 
values, suggesting that scenario (2) occurs in at least a
fraction of the observed ULIRGs.

We note that stellar {\it photospheric} {\it continuum} emission could
also affect the EW$_{\rm 3.3PAH}$ values even for galaxies dominated by
normal starbursts, if its contribution to the AKARI IRC 2.5--5 $\mu$m
spectra is important and varies with different galaxies.
The stellar photospheric emission shows a decreasing continuum flux 
with increasing wavelength at $>$1.8 $\mu$m \citep{saw02}.
The 3.5 $\mu$m to 2.5 $\mu$m flux
ratio in F$_{\nu}$ is $\sim$0.75 ($\Gamma$ = $-$0.85; F$_{\nu}$
$\propto$ $\lambda^{\Gamma}$), which is insensitive to stellar ages, as
long as they are $>$10 Myr (Bruzual \& Charlot 1993; see also Figure 1
of Sawicki 2002).   
The stellar photospheric emission is bluer than starburst-heated dust
continuum emission ($\Gamma$ $\sim$ 0; see $\S$5.4), and so its
contribution becomes relatively less important at longer wavelengths.
Sources with blue observed continuum emission at 2.5--5 $\mu$m are
candidates largely contaminated by the stellar photospheric emission.    
Assuming that stellar photospheric and starburst-heated hot dust
continuum emission have $\Gamma$ = $-$0.85 and $\Gamma$ = 0,
respectively, the stellar photospheric contribution to the 3.5 $\mu$m
flux is estimated to be $<$30\%, if the observed 3.5 $\mu$m to 2.5
$\mu$m flux ratio in F$_{\nu}$ is $>$0.9.
Almost all sources meet this criterion (Figures 1 and 2), suggesting that
the stellar photospheric contributions are not so important to largely
decrease the EW$_{\rm 3.3PAH}$ values in the majority of the observed
LIRGs and ULIRGs.  
Furthermore, given that the observed continuum emission of ULIRGs is
generally redder than LIRGs ($\S$5.4), this possible stellar
photospheric contribution cannot explain the smaller EW$_{\rm 3.3PAH}$ 
trend in ULIRGs than LIRGs at all. 

\subsection{AGNs with weak starbursts}

To find sources that may possess luminous AGNs on the basis of
depressed EW$_{\rm 3.3PAH}$ values, we follow \citet{ima08}.
Sources with EW$_{\rm 3.3PAH}$ $\lesssim$ 40 nm are classified as
galaxies that contain luminous AGNs, and the AGN contributions to the
observed 2.5--5 $\mu$m fluxes are significant.  
Tables 5 and 6 (column 6) indicate whether or not AGN signatures are found. 

Based on this low EW$_{\rm 3.3PAH}$ method, 15 of 62 LIRG nuclei and 
15 of 48 ULIRG nuclei display AGN signatures, after excluding
additional interesting sources. 
Among the 15 LIRGs with detectable AGN signatures, six nuclei
(MCG-03-34-064, NGC 5135, NGC 7469, IC 5298, NGC 7592W, NGC 7674) are
optically classified as Seyferts.
For the ULIRGs, nine of the 15 AGN-detected sources are optical Seyferts. 
After removing these optically identified AGNs surrounded by torus-shaped
dusty media, nine LIRG nuclei and six ULIRG nuclei are possible
containers for optically elusive luminous buried AGNs.

\subsection{AGNs with strong starbursts}

The low EW$_{\rm 3.3PAH}$ method provides clear AGN signatures if the
observed 2.5--5 $\mu$m flux originates primarily in AGN-heated PAH-free
hot dust emission.
This can happen if (1) the surrounding starbursts are weak
and/or (2) the AGN emission is not heavily obscured. 
However, luminous AGN signatures could be overlooked by this method if the
AGN is dust-obscured (i.e., the AGN emission is flux-attenuated)
to the point that the AGN contribution to the observed 2.5--5 $\mu$m flux is 
insignificant compared with weakly obscured starbursts in the foreground.  
Such a highly obscured AGN may be distinguished by measuring the optical
depths of dust absorption features.
While stellar energy sources and dust are spatially well mixed in a
normal starburst (see $\S$1), in an obscured AGN, the energy source (i.e., a
mass-accreting compact SMBH) is more centrally
concentrated than dust.
In the former mixed dust/source geometry, there are upper limits for the
optical depths of the dust absorption features found at 2.5--5 $\mu$m
\citep{im03,idm06}, because the observed flux originates primarily from
weakly obscured, less flux-attenuated foreground emission (which has
weak dust absorption features), with a small contribution from
highly obscured, highly flux-attenuated emission from the obscured regions
(which should exhibit strong dust absorption features).
On the other hand, in a centrally concentrated energy source geometry, 
the optical depths of the dust absorption features can be
larger than the upper limits of the mixed dust/source geometry, because
dust extinction and flux attenuation can be approximated by a
foreground screen dust model. 
It has been determined that the optical depths of 3.1 $\mu$m ice-covered
dust absorption features with $\tau_{3.1}$ $>$ 0.3 and 3.4 $\mu$m
bare (ice mantle free) carbonaceous dust absorption features 
with $\tau_{3.4}$ $>$ 0.2 cannot be reproduced in the mixed
dust/source geometry, if the Galactic elemental abundance patterns and
dust compositions are assumed \citep{im03,idm06}. 
 
In Table 9 (column 6), sources with $\tau_{3.1}$ $>$ 0.3 and/or
$\tau_{3.4}$ $>$ 0.2 are indicated as possible hosts for
obscured AGNs with centrally concentrated energy source geometry.
Three LIRGs and 16 ULIRGs meet the criteria, of which one LIRG (IC 5298)
and two ULIRGs (IRAS 12072$-$0444 and 16156+0146) are optically
classified as Seyfert 2s. 
Thus, the remaining two LIRGs and 14 ULIRGs are buried AGN candidates. 
The number and fraction of sources with large $\tau_{3.1}$ and/or
$\tau_{3.4}$ is substantially larger for ULIRGs than for LIRGs, suggesting
that 2.5--5 $\mu$m ULIRG emission generally comes from more obscured
regions.

The continuum slope $\Gamma$ (F$_{\nu}$ $\propto$ $\lambda^{\Gamma}$)
can also be used to find candidates for obscured AGNs with 
coexisting strong starbursts in the foreground \citep{ris06,san08,ris10}.  
The logic is basically the same. 
The 2.5--5 $\mu$m continuum emission from a normal starburst with 
mixed dust/source geometry cannot be so red, while that from 
an obscured AGN can become very red \citep{ris06}. 
Figure 6a shows the distribution of $\Gamma$ for the LIRGs and ULIRGs
observed in this paper. 
In Figure 6b, the same distribution is shown with the ULIRGs studied by
\citet{ima08} included. 
A larger value of $\Gamma$ indicates a redder, rising continuum with increasing
wavelength.
In both plots, a larger fraction of the ULIRGs are distributed in the red, large
$\Gamma$ tails compared to the LIRGs, again suggesting that infrared 2.5--5 $\mu$m 
emission from ULIRGs is generally more obscured than those of LIRGs. 
After excluding the very red outliers, most of the LIRGs and ULIRGs 
have $\Gamma$ = $-$1 $\sim$ 1, with a median value of $\Gamma$ $\sim$ 0 
(Tables 5, 6, 10).
This $\Gamma$ value may be regarded as a typical value for weakly obscured
starburst/AGN.

If we interpret values of $\Gamma$ $>$ 1 as obscured AGN signatures, 
then only six LIRGs correspond to this case, of which two sources 
(MCG-03-34-064, NGC 7674) are optical Seyferts. 
11/34 (32\%) of optically non-Seyfert ULIRGs and 9/14 (64\%) of
optically Seyfert ULIRGs in the IRAS 1 Jy sample have large $\Gamma$
values ($>$1), indicative of obscured AGNs.
Among the additional interesting sources, red continuum emission is seen in
four sources (IRAS 00183$-$7111, 06035$-$7102, 20100$-$4156,
20551$-$4250).
Tables 5 and 6 (column 8) summarize the detection or non-detection of AGN
signatures, based on $\Gamma$ values. 

In previous research \citep{ris06,san08,ris10}, 
$\Gamma$ is defined by F$_{\lambda}$ $\propto$ $\lambda^{\Gamma}$, and 
$\Gamma_{\rm prev}$ = $-$0.5 is assumed to be a typical value for
a weakly obscured starburst/AGN.
Because $\Gamma_{\rm ours}$, as defined in our paper (F$_{\nu}$ $\propto$
$\lambda^{\Gamma}$), is related to $\Gamma_{\rm prev}$ by 
$\Gamma_{\rm ours}$ = $\Gamma_{\rm prev}$ + 2,
the typical $\Gamma_{\rm ours}$ values of 0 ($-$1 $\sim$ 1)
correspond to $\Gamma_{\rm prev}$ = $-$2 ($-$3 $\sim$ $-$1), which is
smaller (bluer) than the previously assumed values for weakly obscured
starbursts/AGNs.

Figure 6c compares the $\Gamma$ and EW$_{\rm 3.3PAH}$ values.
Red continuum sources with large $\Gamma$ values are preferentially
distributed in the low EW$_{\rm 3.3PAH}$ regions, usually occupied by
AGN-important galaxies, demonstrating that large $\Gamma$ values can be 
a good criterion for finding obscured AGNs.
However, there are many sources that have low EW$_{\rm 3.3PAH}$ values,
and yet exhibit blue continuum emission (small $\Gamma$).
Obviously, weakly obscured AGNs cannot be detected by large 
$\Gamma$ values. 

To relate $\Gamma$ value to obscuration, we estimate the 
dust extinction toward the observed 2.5--5 $\mu$m continuum emission
regions, based on the observed optical depths of the 3.1 $\mu$m
ice-covered dust ($\tau_{3.1}$) and 3.4 $\mu$m bare carbonaceous dust
absorption features ($\tau_{3.4}$).  
The $\tau_{3.1}$ value reflects the column density of dust grains covered
with an ice mantle deep inside molecular clouds, and is roughly proportional
to the dust extinction ($\tau_{3.1}$/A$_{\rm V}$ = 0.06; Tanaka et al. 1990;
Smith et al. 1993; Murakawa et al. 2000).
The $\tau_{3.4}$ value can be used to estimate the column density of
bare, ice-mantle-free dust grains in the diffuse interstellar medium
(outside molecular clouds), and is correlated with the dust extinction
($\tau_{3.4}$/A$_{\rm V}$ = 0.004--0.007; Pendleton et al. 1994;
Imanishi et al. 1996; Rawlings et al. 2003).  
Because the 3.4 $\mu$m dust absorption feature is absent from ice-covered
dust grains \citep{men01}, we can estimate the total column densities of
dust, including both ice-covered and ice-mantle-free, from the
combination of $\tau_{3.1}$ and $\tau_{3.4}$ values. 
Assuming that A$_{\rm V}$ = $\tau_{3.1}$/0.06 + $\tau_{3.4}$/0.006, the 
continuum slope $\Gamma$ and dust extinction towards the 2.5--5 $\mu$m
continuum emission regions (A$_{\rm V}$) are compared in Figure 6d.
For sources with undetectable $\tau_{3.1}$ and $\tau_{3.4}$, we
tentatively adopt A$_{\rm V}$ = 0 mag.
It can be seen that all the sources with A$_{\rm V}$ $>$ 40 mag show
red continua ($\Gamma$ $>$ 1), supporting the hypothesis that large 
$\Gamma$ values are caused by dust obscuration.

\subsection{Combination of energy diagnostic methods}

Here, we summarize the fraction of sources for which AGN signatures are
observed using at least one of the methods previously discussed  
(low EW$_{\rm 3.3PAH}$, large $\tau_{3.1}$
and/or $\tau_{3.4}$, large $\Gamma$), after excluding the additional
interesting sources.
The low EW$_{\rm 3.3PAH}$ method is relatively sensitive to
weakly obscured AGNs, while both the large $\tau$ and large $\Gamma$
techniques can selectively detect highly obscured AGNs.
Thus, these methods complement each other for the purpose of
finding AGNs.

AGN signatures are found in 17/62 (27\%) LIRG nuclei. 
The AGN detection rate is 6/8 (75\%) for optically Seyfert LIRG
nuclei, and 11/54 (20\%) for optically non-Seyfert LIRG nuclei.

For ULIRGs, the fraction of AGN-detected sources is 29/48 (60\%) for all
nuclei, 10/14 (71\%) for nuclei classified as optically Seyfert, and 
19/34 (56\%) for nuclei classified as optically non-Seyfert. 
The detection rate for buried AGNs in galaxies classified as optically 
non-Seyfert is higher in ULIRGs (19/34; 56\%) than in LIRGs (11/54; 20\%). 
For optically identified AGNs (optical Seyferts) of LIRGs and ULIRGs
combined, our AKARI IRC 2.5--5 $\mu$m spectroscopic methods 
recover AGN signatures in 16/22 (73\%).  
A small fraction of optically identified AGNs does not exhibit clear AGN
signatures according to our methods (see also Nardini et al. 2010), possibly
because of a relatively small covering of dust around a central AGN.
For such an AGN, the narrow line regions develop so well that optical AGN
detection becomes easy, while AGN-heated hot dust emission (on which 
our infrared AGN diagnostic is based) is not so strong.

For the additional interesting sources, AGN signatures are detected in 7/8
(88\%) by at least one method.
Because these sources are selected for their strong dust absorption features and/or 
AGN signatures at other wavelengths, the high AGN detection rate is
exactly what we would expect.

\subsection{Hydrogen recombination emission lines}

In many LIRGs and a small fraction of bright ULIRGs with high S/N
ratios, strong Br$\alpha$ ($\lambda_{\rm rest}$ = 4.05 $\mu$m) emission
lines are clearly seen, particularly for PAH-strong, starburst-important
galaxies.  
Br$\beta$ ($\lambda_{\rm rest}$ = 2.63 $\mu$m) emission lines are often
recognizable in sources with very strong 
Br$\alpha$ emission with larger equivalent widths. 
Although an AGN can also produce Br$\alpha$ and Br$\beta$ emission, 
their equivalent widths are expected to be low, because of dilution by
strong, AGN-heated hot dust continuum emission. 
Thus, Br$\alpha$ and Br$\beta$ emission from strong Br emission sources
is assumed to originate mostly from starburst activity.
Both lines are relatively strong, and their dust extinction effects are much
smaller than the widely used hydrogen recombination lines, such as
Ly$\alpha$ ($\lambda_{\rm rest}$ = 0.12 $\mu$m) in the UV range, or  
H$\alpha$ ($\lambda_{\rm rest}$ = 0.66 $\mu$m) and 
H$\beta$ ($\lambda_{\rm rest}$ = 0.49 $\mu$m) in the optical range. 
Thus, Br$\alpha$ and Br$\beta$ emission lines could provide a good tracer
for probing the properties of starbursts in the dusty LIRG and ULIRG
populations, by substantially reducing the uncertainties of dust
extinction corrections.

Br$\alpha$ ($\lambda_{\rm rest}$ = 4.05 $\mu$m) is
observable from the ground in the longest wavelength component of the $L$-band
(2.8--4.2 $\mu$m) Earth's atmospheric window for only the nearest
sources with small redshifts.
However, for such nearby sources, Br$\beta$ ($\lambda_{\rm rest}$ = 2.63
$\mu$m) emission is not covered by the ground-based $L$-band. 
Even if the Br$\alpha$ line is covered, it is difficult to measure
with ground-based spectroscopy because of the considerable amount of
Earth's atmospheric background noise in the 4.05--4.2 $\mu$m range.   
If a galaxy is redshifted, the Br$\beta$ emission line enters the
$L$-band atmospheric window. However, for such redshifted sources,
the Br$\alpha$ emission is shifted beyond the $L$-band's longest wavelength
range.
Consequently, in ground-based spectroscopy, it is impossible 
to observe both the Br$\alpha$ and Br$\beta$ emission lines. 
The simultaneous coverage of 2.5--5 $\mu$m provided by AKARI IRC is
quite unique, in that it enables the comparison of Br$\alpha$ and
Br$\beta$ emission lines. 

For sources in which both Br$\alpha$ and Br$\beta$ emission lines are
detected, the observed Br$\beta$/Br$\alpha$ luminosity ratios are
compared with the value predicted by case B theory ($\sim$0.64 for
10$^{4}$ K; Wynn-Williams 1984), and are found to agree to within a factor
of 2 (Tables 7, 8).
The lowest and highest observed values deviate from the theoretical
value by a factor of $\sim$2 (VV 114E, IC 694) and $\sim$1.5 (NGC 232, 
NGC 6285, Mrk 273), respectively. 
Given the possible presence of flux measurement uncertainties (due to
the faintness of Br$\beta$), we see no clear evidence that the
Br$\beta$/Br$\alpha$ luminosity ratio deviates from the case B
prediction.
Even if the factor of $\sim$2 for the lower value is due to dust extinction, 
assuming the dust extinction curve A$_{\lambda}$ 
$\propto$ $\lambda^{-1.75}$ \citep{car89}, the estimated dust extinction
is A$_{\rm V}$ $<$ 16 mag.
Thus, the strong Br emission lines are most likely to originate from
modestly obscured (A$_{\rm V}$ $<$ 30 mag) starbursts, as probed by 
the 3.3 $\mu$m PAH emission features ($\S$ 5.1).

For this reason, the Br$\alpha$ emission luminosity could also furnish an
independent probe for modestly obscured starburst magnitudes.
In starburst galaxies, the luminosity ratio of the optical H$\alpha$
and far-infrared (40--500 $\mu$m) emission is estimated to be 0.57
$\times$ 10$^{-2}$ \citep{ken98}. 
Given that the far-infrared luminosity is only slightly smaller than the
infrared (8--1000 $\mu$m) luminosity in high-luminosity starbursts
\citep{sam96}, and assuming a Br$\alpha$/H$\alpha$ luminosity ratio of
$\sim$0.02, as predicted by case B (10$^{4}$ K; Wynn-Williams 1984),
we obtain a Br$\alpha$ to infrared luminosity ratio of L$_{\rm
Br\alpha}$/L$_{\rm IR}$ $\sim$ 1 $\times$ 10$^{-4}$ for a starburst.

Figure 7 compares the L$_{\rm Br\alpha}$/L$_{\rm IR}$ and 
L$_{\rm 3.3PAH}$/L$_{\rm IR}$ ratios for Br$\alpha$-detected LIRGs and
ULIRGs. 
The source distribution lies along the dotted line, with some scatter,
suggesting that both the Br$\alpha$ and 3.3 $\mu$m PAH emission provide
similar starburst contributions to the infrared luminosities of these  
LIRGs and ULIRGs.
Hence, both the Br$\alpha$ and 3.3 $\mu$m PAH emission can be used as
independent measures of the magnitudes of modestly-obscured starbursts.
There may be a greater number of sources above the dashed line
than that below the line, indicating that for some fraction of sources, 
Br$\alpha$ emission tends to provide a higher starburst contribution than
3.3 $\mu$m PAH emission.
The fraction of such sources is higher in ULIRGs (Fig. 7).
Although AGN contribution to the Br$\alpha$ emission line is a
possibility, this could be due to selection bias, because only ULIRGs with 
clearly detectable, strong Br$\alpha$ emission lines are plotted. 
In fact, for the modestly obscured (A$_{\rm V}$ $<$ 30 mag) starbursts
probed with 3.3 $\mu$m PAH and Br$\alpha$ emission, dust extinction 
reduces the L$_{\rm 3.3PAH}$/L$_{\rm Br\alpha}$ flux ratios by only
$\sim$30\%, if the dust extinction curve A$_{\lambda}$ $\propto$
$\lambda^{-1.75}$ \citep{car89} is adopted. 

\subsection{Dust extinction and intrinsic AGN luminosities} 

For LIRGs and ULIRGs with low EW$_{\rm 3.3PAH}$ values, we can
reasonably assume that the bulk of the observed 2.5--5 $\mu$m continuum
fluxes originate in AGN-heated hot dust continuum emission. 
By applying the correction for flux attenuation by dust extinction, 
we can estimate the extinction-corrected intrinsic luminosity of
AGN-heated hot dust emission. 
Conversion from this type of luminosity to the intrinsic primary energetic
radiation luminosity of an AGN is straightforward if the AGN is surrounded by 
a large column of dust in virtually all directions, because
energy is conserved (without escape) from the inner hot dust regions to 
the outer cool dust regions. 
In a pure AGN with this type of geometry, and without strong starbursts, the
luminosity of the AGN-heated hot dust emission must be comparable to the 
primary energetic radiation of the AGN (Fig. 2 of Imanishi et al. 2007a).  
Buried AGNs in optically non-Seyfert LIRGs and ULIRGs could be modeled 
by this geometry, as a first approximation. 
 
However, for Seyfert-type AGNs, which are thought to be surrounded by
torus-shaped dusty media, the discussion is more complex. 
If the solid angle of the torus, viewed from the central compact SMBH,
does not vary with the distance from the center, the AGN-originated 2.5--5
$\mu$m continuum luminosity (hot dust of the inner part) should be
similar to the luminosity at longer infrared wavelengths (cool dust on
the outside). 
However, if the torus is strongly flared \citep{wad09} or warped
\citep{san89}, this assumption is no longer valid, and the luminosity at
the longer wavelengths could be greater than the AGN-originated 2.5--5
$\mu$m continuum luminosity.  
Furthermore, a significant fraction of the primary AGN radiation may
escape without being absorbed by dust, and such an emission component
cannot be probed by infrared observations.
To estimate the primary energetic radiation luminosity of an AGN from the 
AGN-originated 2.5--5 $\mu$m continuum luminosity, we adopt a correction
factor of 5 (L$_{\rm AGN}$/L$_{2.5-5 \mu m}$ = 5) for sources
optically classified as Seyferts \citep{ris10}.

We choose $\lambda_{\rm rest}$ = 3.5 $\mu$m as representative of
the 2.5--5 $\mu$m continuum emission, and estimate 
$\nu$F$_{\rm \nu}$(3.5$\mu$m), or equivalently 
$\lambda$F$_{\rm \lambda}$(3.5$\mu$m), for sources with EW$_{\rm 3.3PAH}$
$<$ 40 nm.
The 3.5 $\mu$m continuum flux attenuation is estimated from the 
$\tau_{3.4}$ and $\tau_{3.1}$ values, rather than the continuum slope
$\Gamma$, because the intrinsic $\Gamma$ value for unobscured AGNs could
have high uncertainty ($\S$5.4). 
The dust extinction is derived from the formula 
A$_{\rm V}$ = $\tau_{3.1}$/0.06 + $\tau_{3.4}$/0.006 (see $\S$5.4). 
The flux attenuation at 3.5 $\mu$m, relative to the optical $V$-band 
(0.55 $\mu$m), is observationally found to be 
A$_{\rm 3.5 \mu m}$/A$_{\rm V}$ = 0.03--0.06 \citep{rie85,nis08,nis09}. 
By adopting A$_{\rm 3.5 \mu m}$/A$_{\rm V}$ = 0.05, we obtain 
A$_{\rm 3.5 \mu m}$ = 0.83 $\times$ $\tau_{3.1}$ + 8.3 $\times$
$\tau_{3.4}$.

Figure 8 compares the extinction-corrected intrinsic AGN luminosities and
observed infrared luminosities of LIRGs and ULIRGs with 
EW$_{\rm 3.3PAH}$ $<$ 40 nm.
Only a small fraction of the sources exhibit intrinsically luminous
AGNs, which could quantitatively account for the observed infrared luminosities. 
However, we note that all but one of them are
3.4$\mu$m-absorption-detected sources.  
As the formula of A$_{\rm 3.5 \mu m}$ = 0.83 $\times$
$\tau_{3.1}$ + 8.3 $\times$ $\tau_{3.4}$ indicates, correction of the flux
attenuation at 3.5 $\mu$m is very sensitive to the $\tau_{3.4}$ values,
and yet the detection rate of this 3.4 $\mu$m bare carbonaceous dust
absorption feature is limited, partly because the 
feature is intrinsically weak (small oscillator strengths)
and/or can be veiled by the 3.4 $\mu$m PAH emission
sub-peak for PAH-detected sources.  
Thus, it is fair to assume that the estimated intrinsic AGN luminosities
for sources with non-detectable 3.4 $\mu$m absorption features are only 
lower limits, and the actual AGN luminosities could be much higher. 
To clarify this possible ambiguity, sources with detectable and
non-detectable 3.4 $\mu$m absorption features are distinguished by
different symbols in Figure 8.
For LIRGs and ULIRGs classified optically as Seyferts, the estimated AGN
luminosity is a significant fraction ($>$10\%) of the observed infrared 
luminosity in most cases. 
For buried AGN candidates in optically non-Seyfert LIRGs and ULIRGs
without detectable 3.4 $\mu$m absorption features, the buried AGNs could 
explain at least a few to 10\% of the observed infrared luminosities.

In this regard, the non-detection of the 3.4 $\mu$m bare carbonaceous
dust absorption feature in the ULIRG IRAS 00397$-$1312 is puzzling.
This galaxy displays a very red continuum, but no detectable 3.3 $\mu$m PAH
emission (Fig. 2), both of these factors indicating an energetically
important buried AGN.  
In this ULIRG, the abundance of carbonaceous dust (= the carrier 
of the 3.4 $\mu$m absorption feature) may be suppressed for some reason. 
At the same time, the intrinsic AGN luminosities for 
3.4$\mu$m-absorption-detected sources are often greater than the observed
infrared luminosities (Figure 8) where the carbonaceous dust abundance may be
enhanced and the $\tau_{3.4}$ value per unit dust extinction could be
higher than in the Galactic diffuse interstellar medium.
In this case, the intrinsic AGN luminosity could be overestimated, if we 
are based on the $\tau_{3.4}$ values and the Galactic extinction curve.
Possible variation of the abundance of carbonaceous dust as the carrier of 
the 3.4 $\mu$m absorption feature could introduce additional ambiguity 
into the estimates of the intrinsic AGN luminosities obtained from AKARI IRC
2.5--5 $\mu$m spectra.

\subsection{Comments on individual buried AGN candidates}

We selected buried AGN candidates based on low EW$_{\rm 3.3PAH}$
values, large $\tau_{3.1}$ and $\tau_{3.4}$ values, and large $\Gamma$
values in the AKARI IRC 2.5--5 $\mu$m spectra.
Although the presence of a buried AGN is a natural explanation for 
these observed properties, 
alternative scenarios are also possible, and cannot be ruled out completely.
For example, {\it extreme} starbursts which are exceptionally more
centrally concentrated than the surrounding molecular gas and dust
(Fig. 1e of Imanishi et al. 2007a), and consist of HII regions only,
with virtually no molecular gas or photo-dissociation regions, could also 
produce weak PAH emission, strong dust absorption features, and red
continuum slopes. 
It has been argued that extreme starbursts do not readily occur in ULIRGs
\citep{ima07a,ima09,ima10a}, because of the extremely high emission
surface brightness required, which is much higher than the value normally 
sustained by starburst phenomena \citep{tho05}.  
Additionally, a normal starburst nucleus with a mixed dust/source
geometry and a large amount of {\it foreground screen dust in an edge-on
host galaxy} (Fig. 1d of Imanishi et al. 2007a) could 
produce large $\tau_{3.1}$, $\tau_{3.4}$, and $\Gamma$ values, although 
a low EW$_{\rm 3.3PAH}$ value is not explained in this way. 
To investigate whether our buried AGN interpretation is 
generally persuasive, we compare our results with published data
for several selected well-studied sources.

Since LIRGs are generally closer than ULIRGs, high-quality data 
are available at other wavelengths, and can be used to detect the
presence of buried AGNs.
In Tables 5 and 9, 17 LIRG nuclei display AGN signatures. 
After excluding six optically identified AGNs (i.e., optical Seyferts;
MCG-03-34-064, NGC 5135, NGC 7469, IC 5298, NGC 7592W, NGC 7674), 
11 buried AGN candidates remain.
These are NGC 232, VV 114E, NGC 2623, IC 2810, IC 694, NGC 3690, IRAS
12224$-$0624, IC 860, NGC 5104, IRAS 15250$+$3609, and NGC 7771, of
which buried AGN signatures are found in VV 114E, NGC 2623, and NGC 3690 
by more than one of the methods based on AKARI IRC 2.5--5 $\mu$m
spectra.  
The nuclei of these three LIRGs are thus particularly strong buried AGN
candidates. 
In fact, for NGC 2623 and NGC 3690, X-ray observations strongly suggest
the presence of Compton thick (N$_{\rm H}$ $>$ 10$^{24}$ cm$^{-2}$) AGNs
\citep{del02,mai03,zez03,bal04}. 
The presence of an obscured AGN in NGC 2623 is also supported by the
detection of a high-excitation, mid-infrared [Ne V] 14.3 $\mu$m forbidden
emission line \citep{eva08}.  
For VV 114E, \citet{lef02} argued the presence of a luminous buried AGN
on the basis of the strong, featureless mid-infrared 15 $\mu$m continuum
emission. 
Thus, for all three of the particularly strong buried AGN candidates (VV 114E,
NGC 2623, NGC 3690), there are independent buried AGN arguments in other
works. 

Among other buried AGN candidates, IC 694, the pair galaxy of NGC 3690
in the Arp 299 merging system, may also contain an obscured 
X-ray-emitting AGN \citep{zez03,bal04}. 
IRAS 12224$-$0624 exhibits strong core
radio emission suggestive of an AGN \citep{hil01}. 
The infrared $>$5 $\mu$m spectrum of IRAS 15250$+$3609 is typical of
buried AGNs \citep{spo02,arm07}. 
These supporting results demonstrate the reliability of our
AKARI IRC 2.5--5 $\mu$m spectroscopic technique as a tool for finding
optically elusive buried AGNs.  

Unlike LIRGs, the literature contains relatively few references to
buried AGN signatures for members of our ULIRG sample.
Because most ULIRGs are more distant than LIRGs, the detection of
AGN signatures generally becomes difficult by X-ray observation (the
most powerful AGN search tool), simply due to the lack of
sensitivity of the currently available X-ray satellites. 
Even with this difficulty, \citet{ten08} detected AGN signatures 
in our buried AGN candidate IRAS 04103$-$2838 
by deep X-ray observations.
In another buried AGN candidate, IRAS 12127$-$1412, \citet{spo09} found
AGN signatures via high-velocity neon forbidden emission lines in
the mid-infrared range.

\subsection{Buried AGN fraction as a function of galaxy infrared luminosity}

The energetic importance of AGNs as a function of galaxy infrared
luminosity has previously been investigated by a number of researchers
\citep{tra01,ima09,vei09,nar09,val09,ima10a,nar10}. 
However, these studies focused on ULIRGs with 
L$_{\rm IR}$ $\geq$ 10$^{12}$L$_{\odot}$, simply because energy 
diagnostics of LIRGs with L$_{\rm IR}$ = 10$^{11-12}$L$_{\odot}$ 
are still relatively scarce \citep{bra06,ima09b,per10}.
The infrared emission of LIRGs is apparently more extended than that of
ULIRGs, so that slit spectroscopy using ground-based spectrographs with
large telescopes 
and the Spitzer IRS are not adequate to fully cover the emission from LIRGs. 
A second possible reason for the scarcity of LIRG studies is that while
there are numerous indications that luminous buried AGNs are present in at 
least some fraction of ULIRGs \citep{san88a,soi00}, and thus there is some
motivation for seeking AGNs through detailed infrared spectroscopy,
such AGN implications are generally rarer in optically non-Seyfert LIRGs
than in ULIRGs \citep{soi01}.

With the acquisition of AKARI IRC 2.5--5 $\mu$m slit-less spectra for
a large sample of LIRGs, systematic investigations of LIRG energy
sources are now possible, enabling us to understand the role of
both optically identified and optically elusive AGNs in
galaxies with a wide infrared luminosity range, from LIRGs to ULIRGs. 
It is known that the fraction of optically identified AGNs
(optical Seyferts) increases with increasing galaxy infrared luminosity 
\citep{vei99,got05}. 
Figure 9 shows a plot of the fraction of sources with detectable 
{\it buried} AGN signatures 
\footnote{
Strictly speaking, AGNs which have well-developed NLRs but their optical
emission is obscured by host galaxy dust, could be optically elusive,
but infrared detectable. Such AGNs are included in our buried AGN
category.
}
as a function of galaxy infrared luminosity,
including both LIRGs and ULIRGs. 
Only optically non-Seyfert LIRGs and ULIRGs are plotted, and optical
Seyferts are excluded. 
In the AKARI IRC results, the number of LIRGs with L$_{\rm IR}$ $<$ 
10$^{12}$L$_{\odot}$ is sufficient (50 sources), but the sample is not
statistically complete for ULIRGs with 
10$^{12}$L$_{\odot}$ $\leq$ L$_{\rm IR}$ $<$ 10$^{12.3}$L$_{\odot}$ 
(43 sources) and L$_{\rm IR}$ $\geq$ 10$^{12.3}$L$_{\odot}$ (30 sources).
In the Spitzer IRS results, although the sample is statistically complete 
for ULIRGs with 10$^{12}$L$_{\odot}$ $\leq$ L$_{\rm IR}$ $<$
10$^{12.3}$L$_{\odot}$ (54 sources) and L$_{\rm IR}$ $\geq$
10$^{12.3}$L$_{\odot}$ (31 sources), 
the number of LIRGs (L$_{\rm IR}$ $<$ 10$^{12}$L$_{\odot}$) is limited
(18 sources). 
Thus, these results complement each other.
Although the samples are not exactly the same, both the AKARI and Spitzer
results show that the energetic importance of buried AGNs gradually
increases with increasing galaxy infrared luminosity, ranging from LIRGs
to ULIRGs. 
Accordingly, the existing implications that the AGN-starburst connections are
luminosity-dependent in ULIRGs \citep{ima10a,nar10} can be extended to
LIRGs as well. 

The AGN luminosity increases when the mass accretion rate onto an
SMBH increases.
The luminosity dependence of AGN-starburst connections suggests that 
more (less) infrared luminous galaxies currently are actively
(sluggishly) accreting mass onto SMBHs.  
More luminous buried AGNs with higher SMBH mass accretion rates could
provide stronger feedback to the surrounding host galaxy, because the 
buried AGNs are still surrounded by a large amount of gas and dust. 
As \citet{ima10a} have argued, galaxies with greater infrared luminosities 
demonstrate the {\it relatively} higher energetic importance of buried AGNs, but 
also indicate higher {\it absolute} star formation rates, and thus 
are likely to be the progenitors of more massive galaxies with larger
stellar masses. 
The recently discovered galaxy downsizing phenomenon, in which massive
red galaxies have completed their major star formation more quickly in an
earlier cosmic age \citep{cow96,hea04,bun05}, is widely argued to have
resulted from stronger AGN feedback in the past \citep{gra04,sij07,boo09}.
Our new AKARI results further indicate a possible connection 
between buried AGN feedback and the origin of the galaxy downsizing
phenomenon.

\section{Summary}

We have reported the results of systematic AKARI IRC infrared 2.5--5 $\mu$m
slit-less spectroscopy of LIRGs (L$_{\rm IR}$ = 10$^{11-12}$L$_{\odot}$) 
and ULIRGs (L$_{\rm IR}$ $\geq$ 10$^{12}$L$_{\odot}$) in the local
universe at $z <$ 0.3. 
The slit-less spectroscopic capability of the AKARI IRC is particularly useful
for investigating the properties of LIRGs, which often display
spatially extended emission structures. 
When we combine these results with previously obtained AKARI IRC spectra
of ULIRGs, the total number of observed LIRG and ULIRG nuclei is $>$150.
This is by far the largest 2.5--5 $\mu$m spectral database of LIRGs and
ULIRGs, and thus is useful for investigating their general properties.
We draw the following major conclusions.

\begin{enumerate}

\item The 3.3 $\mu$m PAH emission features (the signatures of starburst
activity) were detected in the bulk of the observed LIRGs and ULIRGs. 
The AKARI IRC slit-less spectra clearly indicated that the upper
envelope of the 3.3 $\mu$m PAH to infrared luminosity ratios was 
L$_{\rm 3.3PAH}$/L$_{\rm IR}$ $\sim$ 10$^{-3}$.
Given that flux attenuation of the 3.3 $\mu$m PAH emission is
insignificant for dust extinction with A$_{\rm V}$ $<$ 30 mag, 
this ratio can be used as a canonical value for starburst activity with 
modest dust obscuration (A$_{\rm V}$ $<$ 30 mag), as has been 
widely assumed.

\item Starburst luminosities were estimated from the observed 
3.3 $\mu$m PAH emission features and Br$\alpha$ ($\lambda_{\rm rest}$ =
4.05 $\mu$m) emission lines. 
Both methods provided roughly consistent results, suggesting that 
3.3 $\mu$m PAH and Br$\alpha$ emission are both good indicators of
the absolute magnitudes of modestly obscured starbursts.

\item For sources with clearly detectable Br$\alpha$ 
($\lambda_{\rm rest}$ = 4.05 $\mu$m) and Br$\beta$ 
($\lambda_{\rm rest}$ = 2.63 $\mu$m) emission lines, it was confirmed
that the Br$\beta$/Br$\alpha$ flux ratios fell within the
expected range of case B (10$^{4}$ K) theory, within a factor of 
$\sim$2.  

\item The L$_{\rm 3.3PAH}$/L$_{\rm IR}$ ratios of LIRGs are 
$>$0.5 $\times$ 10$^{-3}$ in many cases for which the bulk of the
observed large infrared luminosities could be accounted for by 
detected modestly obscured starburst activity.
However, the L$_{\rm 3.3PAH}$/L$_{\rm IR}$ ratios of ULIRGs are mostly 
less than half the canonical ratio for modestly obscured starbursts,
and hence ULIRGs are generally PAH-deficient.
Additional energy sources, such as AGNs or very heavily obscured
starbursts (A$_{\rm V}$ $>$ 30 mag), would be required in the majority
of ULIRGs.

\item For the PAH-deficient ULIRGs, the equivalent widths of the 3.3
$\mu$m PAH emission, the optical depths of the dust absorption features
observed at 2.5--5 $\mu$m, and the continuum slopes were used to find 
potential hosts for AGNs (the energy sources of which should be more
centrally concentrated than dust), by differentiating them from heavily
obscured starbursts (with mixed dust/source geometry).
More than half of the observed ULIRGs that  are classified optically as
non-Seyferts displayed luminous buried AGN signatures. 
The detectable buried AGN fraction in optically non-Seyfert
galaxies was found to systematically increase with increasing galaxy
infrared luminosity, from LIRGs to ULIRGs.

\item The fraction of sources with strong dust absorption features
and/or very red continuum emission was substantially higher in ULIRGs
than in LIRGs, suggesting that the contributions from obscured emission
to the observed 2.5--5 $\mu$m fluxes are higher in ULIRGs.

\item For sources with clearly detectable AGN signatures, the dust
extinction toward the AGN-heated, PAH-free continuum emission was
derived from the optical depths of the absorption features at 
$\lambda_{\rm rest}$ = 3.1 $\mu$m by ice-covered dust grains, and 
at $\lambda_{\rm rest}$ = 3.4 $\mu$m by bare carbonaceous dust grains. 
The intrinsic, extinction-corrected AGN luminosities were then
estimated, and were quantitatively found to be comparable to or even
larger than the observed infrared luminosities when 3.4 $\mu$m bare
carbonaceous dust absorption features were detected. 
For these sources, the AGNs could be energetically dominant. 
However, for sources with AGN signatures, but without detectable 3.4
$\mu$m dust absorption features, the estimated intrinsic AGN
luminosities were far below the observed infrared luminosities. 
The intrinsically weak 3.4 $\mu$m dust absorption features could be
diluted by PAH sub-peaks at the same wavelength, and the dust
extinction of the AGN-heated dust emission might be underestimated for
these sources.

\item The AKARI IRC spectra are unique in that CO$_{2}$ absorption features
at $\lambda_{\rm rest}$ = 4.26 $\mu$m, inaccessible from the ground,
were detected in many of the observed LIRGs and ULIRGs.  
The CO$_{2}$-absorption-detected sources often accompany clear 3.1
$\mu$m H$_{2}$O ice absorption features, as is expected from the 
formation scenario for CO$_{2}$ molecules through the photolysis of
ice-covered dust grains.

\end{enumerate}


This work is based on observations made with AKARI, a JAXA project, with
the participation of ESA. We thank the AKARI IRC instrument team for making 
this study possible. 
We thank the anonymous referee for his/her useful comments. 
M.I. was supported by a Grant-in-Aid for Scientific Research
(19740109). Part of the data analysis was performed using a computer
system operated by the Astronomical Data Analysis Center (ADAC) and the
Subaru Telescope of the National Astronomical Observatory, Japan.
This research made use of the SIMBAD database, operated at CDS,
Strasbourg, France, and the NASA/IPAC Extragalactic Database (NED)
operated by the Jet Propulsion Laboratory, California Institute of
Technology, under contract with the National Aeronautics and Space
Administration.

\clearpage


\clearpage
\begin{deluxetable}{lcrrrrcll}
\tabletypesize{\scriptsize}
\tablecaption{LIRGs observed with AKARI IRC, and their {\it IRAS}-based
infrared emission properties and optical spectral classifications
\label{tbl-1}}
\tablewidth{0pt}
\tablehead{
\colhead{Object} & \colhead{z} &  
\colhead{f$_{\rm 12}$}  & \colhead{f$_{\rm 25}$}  & 
\colhead{f$_{\rm 60}$}  & \colhead{f$_{\rm 100}$}  & 
\colhead{log L$_{\rm IR}$} & \colhead{f$_{25}$/f$_{60}$} & \colhead{Optical} \\
 &    & \colhead{(Jy)} & \colhead{(Jy)}  & \colhead{(Jy)} 
& \colhead{(Jy)}  & \colhead{(L$_{\odot}$)} &  & \colhead{Class} \\
\colhead{(1)} & \colhead{(2)} & \colhead{(3)} & \colhead{(4)} & 
\colhead{(5)} & \colhead{(6)} & \colhead{(7)} & \colhead{(8)} &
\colhead{(9)} 
}
\startdata
IRAS 00085$-$1223 (NGC 34) & 0.020 & 0.35 & 2.39 & 17.05 & 16.86 & 11.5 & 0.14 (C) & Sy2 $^{a}$\\  
IRAS 00163$-$1039 (MCG-02-01-051) & 0.027 & 0.28 & 1.20 & 7.48 & 9.66 & 11.4 & 0.16 (C) & HII $^{a}$\\  
IRAS 00402$-$2349 (NGC 232) & 0.022 & 0.36 & 1.28 & 10.05 & 17.14 & 11.3 & 0.13 (C) & HII $^{a}$\\  
IRAS 01053$-$1746 (IC 1623A/B, VV 114E/W) & 0.020 & 1.03 & 3.65 & 22.93 & 31.55 & 11.7 & 0.16 (C) & HII + HII  $^{a}$\\  
IRAS 01076$-$1707 (MCG-03-04-014) & 0.035 & 0.34 & 0.90 & 7.25 & 10.33 & 11.7 & 0.12 (C) & HII $^{a}$\\  
IRAS 01159$-$4443 (ESO 244-G012) & 0.023 & 0.39 & 1.95 & 9.27 & 11.76 & 11.4 & 0.21 (W) & HII $^{b}$ \\
IRAS 01173$+$1405 (CGCG 436-030, MCG+02-04-025) & 0.031 & 0.21 & 1.54 & 10.71 & 9.67 & 11.6 & 0.14 (C) & HII $^{a}$\\  
IRAS 01484$+$2220 (NGC 695) & 0.033 & 0.50 & 0.83 & 7.59 & 13.56 & 11.7 & 0.11 (C) & HII $^{a}$\\
\enddata
\end{deluxetable}

\clearpage

\begin{deluxetable}{lcrrrrcrl}
\tabletypesize{\scriptsize}
\tablecaption{
ULIRGs observed with AKARI IRC, and their {\it IRAS}-based
infrared emission properties and optical spectral classifications
\label{tbl-2}} 
\tablewidth{0pt}
\tablehead{
\colhead{Object} & \colhead{z} &  
\colhead{f$_{\rm 12}$}  & \colhead{f$_{\rm 25}$}  & 
\colhead{f$_{\rm 60}$}  & \colhead{f$_{\rm 100}$}  & 
\colhead{log L$_{\rm IR}$} & \colhead{f$_{25}$/f$_{60}$} & \colhead{Optical} \\
 &    & \colhead{(Jy)} & \colhead{(Jy)}  & \colhead{(Jy)} 
& \colhead{(Jy)}  & \colhead{(L$_{\odot}$)} &  & \colhead{Class} \\
\colhead{(1)} & \colhead{(2)} & \colhead{(3)} & \colhead{(4)} & 
\colhead{(5)} & \colhead{(6)} & \colhead{(7)} & \colhead{(8)} &
\colhead{(9)}
}
\startdata
 IRAS 00188$-$0856 & 0.128 & $<$0.12 & 0.37 & 2.59 & 3.40 & 12.3 & 0.14 (C) & LI  $^{a}$ (Sy2 $^{b}$) \\  
 IRAS 04103$-$2838 & 0.118 & 0.08 & 0.54 & 1.82 & 1.71 & 12.2 & 0.30 (W) & LI  $^{a}$ (Sy2 $^{b}$)\\
 IRAS 10378+1108 & 0.136 & $<$0.11 & 0.24 & 2.28 & 1.82 & 12.3 & 0.11 (C) &LI  $^{a}$ (Sy2 $^{b}$)\\  
 IRAS 10485$-$1447 & 0.133 & $<$0.11 & 0.25 & 1.73 & 1.66 & 12.2 & 0.14 (C)& LI  $^{a}$ (Sy2 $^{b}$)\\  
 IRAS 11095$-$0238 & 0.106 & 0.06 & 0.42 & 3.25 & 2.53 & 12.2 & 0.13 (C)& LI  $^{a}$ \\
\enddata

\tablecomments{
Col.(1): Object name.
Col.(2): Redshift. 
Cols.(3)--(6): f$_{12}$, f$_{25}$, f$_{60}$, and f$_{100}$ are 
{\it IRAS} fluxes at 12 $\mu$m, 25 $\mu$m, 60 $\mu$m, and 100 $\mu$m,
respectively, taken from \citet{kim98} or the IRAS Faint Source Catalog.
Col.(7): Decimal logarithm of infrared (8$-$1000 $\mu$m) luminosity in
units of solar luminosity (L$_{\odot}$), calculated from $L_{\rm IR} =
2.1 \times 10^{39} \times$ D(Mpc)$^{2}$ $\times$ (13.48 $\times$
$f_{12}$ + 5.16 $\times$ $f_{25}$ + $2.58 \times f_{60} + f_{100}$) ergs
s$^{-1}$ \citep{sam96}. 
For sources that have upper limits in some {\it IRAS} bands, we can derive
upper and lower limits on the infrared luminosity by assuming that the
actual flux is equal to the {\it IRAS}-upper limit and the zero value,
respectively. The difference between the upper and lower values is
usually very small, less than 0.2 dex. We assume that the infrared
luminosity is the average of these values.  
Col.(8): {\it IRAS} 25 $\mu$m to 60 $\mu$m flux ratio.
ULIRGs with f$_{25}$/f$_{60}$ $<$ 0.2 and $>$ 0.2 are
classified as cool and warm sources (denoted as "C" and "W"),
respectively \citep{san88b}.
Col.(9): Optical spectral classification.
$^{a}$: \citet{vei99}.
$^{b}$: \citet{yua10}.
$^{c}$: \citet{arm89}.
$^{d}$: \citet{duc97}.
}

\end{deluxetable}

\begin{deluxetable}{lll}
\tabletypesize{\scriptsize}
\tablecaption{AKARI IRC observation log for LIRGs \label{tbl-3}}
\tablewidth{0pt}
\tablehead{
\colhead{Object} & \colhead{Observation ID} & \colhead{Observation date} \\
\colhead{(1)} & \colhead{(2)} & \colhead{(3)} 
}
\startdata
NGC 34 & 1120100-001, 002 & 2008 June 17, 18 \\
MCG-02-01-051 & 1120101-001, 002 & 2008 June 20, 2009 June 21 \\
NGC 232 & 1120102-002 & 2008 December 20 \\
IC 1623A/B (VV 114) & 1120103-002 & 2008 June 29 \\
MCG-03-04-014 & 1120104-001, 002 & 2008 June 29 \\
ESO 244-G012 & 1122223-001, 002, 003, 004 & 2009 December 16 \\
CGCG 436-030 & 1120105-002 & 2008 July 15 \\
NGC 695 & 1120108-001 & 2009 July 25 \\
\enddata

\tablecomments{
Col.(1): Object name.
Col.(2): Observation ID.
Col.(3): Observation date in UT.
}

\end{deluxetable}

\clearpage

\begin{deluxetable}{lll}
\tabletypesize{\scriptsize}
\tablecaption{AKARI IRC observation log for ULIRGs \label{tbl-4}}
\tablewidth{0pt}
\tablehead{
\colhead{Object} & \colhead{Observation ID} & \colhead{Observation date} \\
\colhead{(1)} & \colhead{(2)} & \colhead{(3)} 
}
\startdata
 IRAS 00188$-$0856 & 1120008-001, 002, 003 & 2008 June 22 \\
 IRAS 04103$-$2838 & 1120092-001, 002, 003 & 2008 August 14 \\
 IRAS 10378+1108 & 1120155-001, 002 & 2009 May 28 \\
 IRAS 10485$-$1447 & 1120156-001, 002, 003 & 2009 June 10, 11 \\
 IRAS 11095$-$0238 & 1122075-001, 002 & 2009 December 12 \\
 IRAS 11582+3020 & 1120088-001, 002, 003 & 2008 June 8 \\
 IRAS 12032+1707 & 1120089-001, 002 & 2009 June 14 \\
 IRAS 12112+0305 & 1120031-001, 002, 003 & 2009 June 22, 23 \\
\enddata

\tablecomments{
Col.(1): Object name.
Col.(2): Observation ID.
Col.(3): Observation date in UT.
}

\end{deluxetable}

\clearpage

\begin{deluxetable}{lrrcrcrc}
\tabletypesize{\scriptsize}
\tablecaption{The 3.3 $\mu$m PAH emission, continuum slopes, and
AGN signatures of LIRGs derived from AKARI IRC spectra \label{tbl-5}}
\tablewidth{0pt}
\tablehead{
\colhead{Object} &  \colhead{f$_{3.3 \rm PAH}$} & \colhead{L$_{3.3 \rm PAH}$} & 
\colhead{L$_{3.3 \rm PAH}$/L$_{\rm IR}$} & 
\colhead{rest EW$_{3.3 \rm PAH}$} & \colhead{AGN} & \colhead{Continuum}  & \colhead{Obscured} \\
 & \colhead{($\times$10$^{-14}$ ergs s$^{-1}$ cm$^{-2}$)} & \colhead{($\times$10$^{41}$ergs s$^{-1}$)} 
& \colhead{($\times$10$^{-3}$)} & \colhead{(nm)} & \colhead{sign} & \colhead{slope ($\Gamma$)} & \colhead{AGN sign} \\  
\colhead{(1)} & \colhead{(2)} & \colhead{(3)} & \colhead{(4)} &
\colhead{(5)} & \colhead{(6)} & \colhead{(7)} & \colhead{(8)} 
}
\startdata
NGC 34 & 45.6 & 3.6 & 0.32 & 41 & X & $-$0.1 & X \\
MCG-02-01-051 & 63.6 & 9.2 & 0.91 & 180 & X & $-$0.1 & X \\
NGC 232 & 32.6 & 2.6 & 0.33 & 30 & $\bigcirc$ & $-$0.2 & X \\
VV 114E (IC 1623A) & 53.6 & 4.2 & 0.24 & 36 & $\bigcirc$ & 1.7 & $\bigcirc$ \\
VV 114W (IC 1623B) & 74.5 & 5.9 & 0.34 & 156 & X & 0.04 & X \\
MCG-03-04-014 & 72.7 & 17.9 & 1.1 & 116 & X & $-$0.1 & X \\
ESO 244-G012 (N) & 63.1 & 6.6 & 0.69 & 104 & X & 0.07 & X \\
CGCG 436-030 & 24.2 & 4.7 & 0.28 & 44 & X & 0.9 & X \\
\enddata

\tablecomments{
Col. (1): Object name. 
Col. (2): Observed flux of 3.3 $\mu$m PAH emission in 10$^{-14}$ [ergs
          s$^{-1}$ cm$^{-2}$].  
Col. (3): Observed luminosity of 3.3 $\mu$m PAH emission in 10$^{41}$
          [ergs s$^{-1}$]. 
Col. (4): Observed 3.3 $\mu$m PAH-to-infrared luminosity ratio in units
          of 10$^{-3}$. 
          Typical ratios for modestly obscured starbursts are 
          $\sim$10$^{-3}$ (Mouri et al. 1990; Imanishi 2002; Section 5.1
          of this paper).
Col. (5): Rest-frame equivalent width of 3.3 $\mu$m PAH emission 
          (EW$_{\rm 3.3PAH}$).  
          The values for starbursts are typically $\sim$100 nm
          \citep{moo86,imd00}.   
Col. (6): AGN signatures from EW$_{\rm 3.3PAH}$.
$\bigcirc$: present (EW$_{\rm 3.3PAH}$ $<$ 40 nm). X: none 
(EW$_{\rm 3.3PAH}$ $\geq$ 40 nm). 
Col. (7): Continuum slope $\Gamma$ (f$_{\nu}$ $\propto$ $\lambda^{\Gamma}$).
Larger $\Gamma$ values mean redder continua in the AKARI IRC 2.5--5 $\mu$m
spectra of Figures 1 and 2.  
Col. (8): Obscured AGN signatures from the continuum slope $\Gamma$.
$\bigcirc$: present ($\Gamma$ $>$ 1.0). X: none ($\Gamma$ $\leq$ 1.0). 
}

\end{deluxetable}

\clearpage

\begin{deluxetable}{lrrrrcrc}
\tabletypesize{\scriptsize}
\tablecaption{
The 3.3 $\mu$m PAH emission, continuum slopes, and
AGN signatures of ULIRGs derived from AKARI IRC spectra \label{tbl-6}}
\tablewidth{0pt}
\tablehead{
\colhead{Object} &  \colhead{f$_{3.3 \rm PAH}$} & \colhead{L$_{3.3 \rm PAH}$} & 
\colhead{L$_{3.3 \rm PAH}$/L$_{\rm IR}$} & \colhead{rest EW$_{3.3 \rm PAH}$} & \colhead{AGN} 
& \colhead{Continuum} & \colhead{Obscured} \\
& \colhead{($\times$10$^{-14}$ ergs s$^{-1}$ cm$^{-2}$)} & 
\colhead{($\times$10$^{41}$ ergs s$^{-1}$)} & \colhead{($\times$10$^{-3}$)} & \colhead{(nm)} & \colhead{sign} 
& \colhead{slope ($\Gamma$)} & \colhead{AGN sign} \\  
\colhead{(1)} & \colhead{(2)} & \colhead{(3)} & \colhead{(4)} & \colhead{(5)}
& \colhead{(6)} & \colhead{(7)} & \colhead{(8)} 
}
\startdata
 IRAS 00188$-$0856 & 2.1 & 8.0 & 0.09 & 32 & $\bigcirc$ & 2.4 &
$\bigcirc$  \\
 IRAS 04103$-$2838 & 4.8 & 15.0 & 0.26 & 46 & X & 1.5 & $\bigcirc$ \\
 IRAS 10378+1108 & 2.8 & 11.8 & 0.16 & 125 & X & 0.6 & X \\
 IRAS 10485$-$1447 & 1.0 & 4.1 & 0.07 & 44 & X & 0.9 & X \\
 IRAS 11095$-$0238 & 3.3 & 8.3 & 0.13 & 141 & X & 1.5 & $\bigcirc$ \\
 IRAS 11582+3020 & 2.3 & 28.7 & 0.24 & 161 & X & 0.2 & X \\
 IRAS 12032+1707 & 1.1 & 13.5 & 0.09 & 61  & X & 0.1 & X \\
 IRAS 12112+0305 & 10.2 & 11.6 & 0.16 & 105 & X & 0.6 & X \\
\enddata

\tablecomments{
Cols. (1)-(8): Same as in Table 5.
}

\end{deluxetable}

\clearpage

\begin{deluxetable}{lrrcccc}
\tabletypesize{\scriptsize}
\tablecaption{Hydrogen recombination emission lines in LIRGs \label{tbl-7}}
\tablewidth{0pt}
\tablehead{
\colhead{Object} & \colhead{f(Br$\alpha$)} & \colhead{f(Br$\beta$)} &
\colhead{L(Br$\alpha$)} & \colhead{L(Br$\beta$)} &
\colhead{Br$\beta$/Br$\alpha$} & \colhead{L(Br$\alpha$)/L$_{\rm IR}$} \\
  & \colhead{($\times$10$^{-14}$ ergs s$^{-1}$ cm$^{-2}$)} &
\colhead{($\times$10$^{-14}$ ergs s$^{-1}$ cm$^{-2}$)} & \colhead{($\times$10$^{41}$ ergs s$^{-1}$)} &
\colhead{($\times$10$^{41}$ ergs s$^{-1}$)} & & \colhead{($\times$10$^{-4}$)} \\
\colhead{(1)} & \colhead{(2)} & \colhead{(3)} & \colhead{(4)} &
\colhead{(5)} & \colhead{(6)} & \colhead{(7)} 
}
\startdata
NGC 34 & 3.0 & --- & 0.24 & --- & --- & 0.22 \\
MCG-02-01-051 & 7.2 & --- & 1.05 & --- & --- & 1.03 \\ 
NGC 232 & 3.8 & 3.6 & 0.36 & 0.35 & 0.95 & 0.47 \\
VV 114E & 12.9 & 4.1 & 1.02 & 0.32 & 0.32 & 0.58 \\
VV 114W & 15.0 & 7.3 & 1.18 & 0.57 & 0.49 & 0.68 \\
MCG-03-04-014 & 6.0 & 3.0 & 1.48 & 0.74 & 0.50 & 0.87\\
ESO 244-G012 (N) & 9.8 & 4.8 & 1.02 & 0.50 & 0.49 & 1.06 \\
CGCG 436-030 & 6.1 & 2.8 & 1.17 & 0.54 & 0.46 & 0.70 \\
\enddata

\tablecomments{
Col. (1): Object name. 
Col. (2): Flux of Br$\alpha$ emission line ($\lambda_{\rm rest}$ = 4.05
$\mu$m) in 10$^{-14}$ [ergs s$^{-1}$ cm$^{-2}$].
Col. (3): Flux of Br$\beta$ emission line ($\lambda_{\rm rest}$ = 2.63
$\mu$m) in 10$^{-14}$ [ergs s$^{-1}$ cm$^{-2}$].
Col. (4): Br$\alpha$ luminosity in 10$^{41}$ [ergs s$^{-1}$].
Col. (5): Br$\beta$ luminosity in 10$^{41}$ [ergs s$^{-1}$].
Col. (6): Br$\beta$ to Br$\alpha$ flux (luminosity) ratio.
Col. (7): Br$\alpha$ to infrared luminosity ratio in 10$^{-4}$.
}

\end{deluxetable}

\clearpage

\begin{deluxetable}{lcccccc}
\tabletypesize{\scriptsize}
\tablecaption{Hydrogen recombination emission lines in ULIRGs
\label{tbl-8}}
\tablewidth{0pt}
\tablehead{
\colhead{Object} & \colhead{f(Br$\alpha$)} & \colhead{f(Br$\beta$)} &
\colhead{L(Br$\alpha$)} & \colhead{L(Br$\beta$)} & 
\colhead{Br$\beta$/Br$\alpha$} & \colhead{L(Br$\alpha$)/L$_{\rm IR}$} \\
  & \colhead{($\times$10$^{-14}$ ergs s$^{-1}$ cm$^{-2}$)} &
\colhead{($\times$10$^{-14}$ ergs s$^{-1}$ cm$^{-2}$)} & \colhead{($\times$10$^{41}$ ergs s$^{-1}$)} &
\colhead{($\times$10$^{41}$ ergs s$^{-1}$)} & & \colhead{($\times$10$^{-4}$)} \\
\colhead{(1)} & \colhead{(2)} & \colhead{(3)} & \colhead{(4)} &
\colhead{(5)} & \colhead{(6)} & \colhead{(7)} 
}
\startdata
IRAS 12112+0305 & 1.8 & 1.5 & 2.09 & 1.72 & 0.82 & 0.28 \\
IRAS 12359$-$0725 & 1.4 & 0.6 & 6.06 & 2.66 & 0.44 & 1.17 \\
IRAS 14348$-$1447 & 2.3 & 2.1 & 3.35 & 3.09 & 0.92 & 0.44 \\
Arp 220 & 4.3 & 2.5 & 0.28 & 0.16 & 0.58 & 0.057 \\
IRAS 13509+0442 & 0.90 & --- & 3.85 & --- & --- & 0.59 \\
IRAS 14060+2919 & 1.5 & 1.0 & 4.52 & 3.16 & 0.70 & 1.03 \\
IRAS 15206+3342 & 2.8 & 1.9 & 9.80 & 6.58 & 0.67 & 1.64 \\
IRAS 16474+3430 (S) & 1.6 & 1.1 & 4.29 & 2.97 & 0.69 & 0.82 \\
IRAS 02021$-$2103 & 0.78 & --- & 2.36 & --- & --- & 0.58 \\
Mrk 273 & 4.4 & 4.2 & 1.24 & 1.17 & 0.96 & 0.27 \\
IRAS 15001+1433 & 2.2 & --- & 13.8 & --- & --- & 1.40 \\
IRAS 20100$-$4156 & 2.0 & --- & 7.90 & --- & --- & 0.53 \\
IRAS 20551$-$4250 & 1.4 & 1.2 & 0.53 & 0.47 & 0.88 & 0.14 \\
IRAS 23128$-$5919 & 8.3 & 7.8 & 3.44 & 3.20 & 0.93 & 0.92 \\ 
\enddata

\tablecomments{
Cols. (1)-(7): Same as in Table 7.
}

\end{deluxetable}

\clearpage

\begin{deluxetable}{lccccc}
\tabletypesize{\small}
\tablecaption{Absorption features and AGN signatures \label{tbl-9}}
\tablewidth{0pt}
\tablehead{
\colhead{Object} & \colhead{$\tau_{3.1}$} & \colhead{$\tau_{3.4}$} &
\colhead{$\tau_{CO2}$} & \colhead{$\tau_{CO}$} & \colhead{AGN sign} \\
\colhead{(1)} & \colhead{(2)} & \colhead{(3)} & \colhead{(4)} &
\colhead{(5)} & \colhead{(6)} 
}
\startdata
NGC 34 & 0.2 & --- & 0.1 & 0.25 & X \\
NGC 232 & 0.15 & --- & 0.15 & --- & X \\
VV 114E & 0.4 & --- & 0.3 & 0.1 & $\bigcirc$ \\
CGCG 436-030 & 0.25 & --- & 0.25 & 0.2 & X \\
NGC 695 & --- & --- & --- & 0.3 & X \\
UGC 2238 & --- & --- & 0.15 & --- & X \\
NGC 2623 & 0.4 & --- & 0.3 & 0.3  & $\bigcirc$ \\
NGC 3110 & --- & --- & 0.1 & 0.15 & X \\
\enddata

\tablecomments{
Col. (1): Object name. 
LIRGs and ULIRGs are separated by the central horizontal solid line.
The additional interesting sources are included in each category. 
Col. (2): Observed optical depth of the 3.1 $\mu$m H$_{2}$O ice absorption
feature.
Col. (3): Observed optical depth of the 3.4 $\mu$m bare carbonaceous
dust absorption feature.
Col. (4): Observed optical depth of the 4.26 $\mu$m CO$_{2}$ absorption
feature.
The 4.26 $\mu$m CO$_{2}$ absorption feature is intrinsically
so narrow \citep{van96,whi98} that it may not be fully resolvable in AKARI
IRC spectra with R $\sim$ 120.
In this case, the $\tau_{CO2}$ value could be slightly underestimated
for sources with large $\tau_{CO2}$ values (e.g., IRAS 15250+3609).     
Col. (5): Observed optical depth of the 4.67 $\mu$m CO absorption feature.
Col. (6): Obscured AGN signature, based on $\tau_{3.1}$ $>$ 0.3 
and/or $\tau_{3.4}$ $>$ 0.2 (see $\S$5.4).
}

\end{deluxetable}

\clearpage

\begin{deluxetable}{lr|lr}
\tabletypesize{\small}
\tablecaption{Continuum slopes ($\Gamma$) of ULIRGs, previously observed with
AKARI IRC by Imanishi et al. (2008). \label{tbl-3}}
\tablewidth{0pt}
\tablehead{
\colhead{Object} &  \colhead{$\Gamma$} & \colhead{Object} & \colhead{$\Gamma$} \\
\colhead{(1)} & \colhead{(2)} & \colhead{(3)} & \colhead{(4)} 
}
\startdata
IRAS 03521+0028 & 0.07 & IRAS 09539+0857 & $-$0.01 \\
IRAS 04074$-$2801 & 0.2 & IRAS 10494+4424 & $-$0.01 \\
IRAS 05020$-$2941 & 0.2 & IRAS 16468+5200 & 0.4 \\
IRAS 09463+8141 & $-$0.4 & IRAS 16487+5447 & $-$0.2 \\
IRAS 10091+4704 & $-$1.6 & IRAS 17028+5817W & $-$0.2 \\
IRAS 11028+3130 & $-$1.4 & IRAS 17044+6720 & 2.3 \\
IRAS 11180+1623 &$-$0.07 & IRAS 00456$-$2904 & $-$0.2 \\
IRAS 14121$-$0126 & 0.8  & IRAS 01298$-$0744 & 1.9 \\
IRAS 16333+4630 & 0.1    & IRAS 01569$-$2939 & $-$0.8 \\
IRAS 21477+0502 & $-$0.6 & IRAS 11387+4116 & $-$0.6 \\
IRAS 23129+2548 & $-$0.7 & IRAS 13539+2920 & $-$0.2 \\
IRAS 01199$-$2307 & 1.6  & IRAS 17028+5817E & $-$3.5 \\
IRAS 01355$-$1814 & $-$0.5 & IRAS 01494$-$1845 & $-$0.08 \\
IRAS 03209$-$0806 & 0.2  & IRAS 02480$-$3745 & $-$1.7 \\
IRAS 10594+3818 & 0.03   & IRAS 04313$-$1649 & $-$1.6 \\
IRAS 12447+3721 & 0.8    & IRAS 08591+5248 & $-$0.5 \\
IRAS 13469+5833 & $-$0.6 & IRAS 10035+2740 & $-$0.3 \\
IRAS 14202+2615 & 0.9    & IRAS 05189$-$2524 & 1.3 \\
IRAS 15043+5754 & $-$0.2 & IRAS 14394+5332 & 2.0 \\
IRAS 17068+4027 & 0.7    & IRAS 17179+5444 & 0.8 \\
IRAS 22088$-$1831 & $-$0.4 & IRAS 23498+2423 & 1.3 \\
IRAS 00482$-$2721 & 0.3  & UGC 5101 &  1.0 \\
IRAS 08572+3915 & 4.7    & IRAS 19254$-$7245 & 2.0 \\
\enddata

\tablecomments{
Col. (1): Object name.
Col. (2): Continuum slope ($\Gamma$).
Col. (3): Object name.
Col. (4): Continuum slope ($\Gamma$).
}

\end{deluxetable}

\clearpage

\begin{figure}
\includegraphics[angle=0,scale=.27]{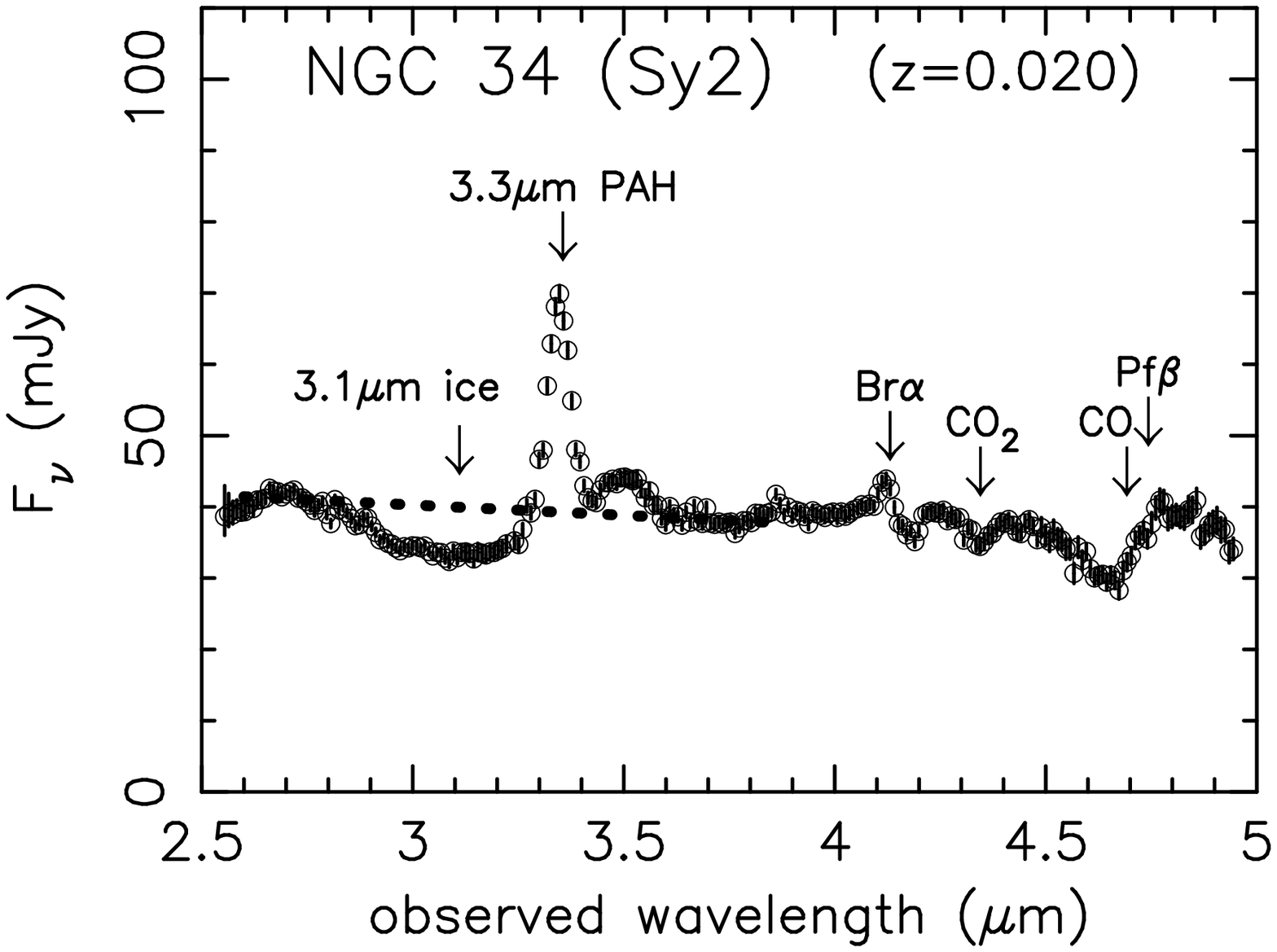}
\includegraphics[angle=0,scale=.27]{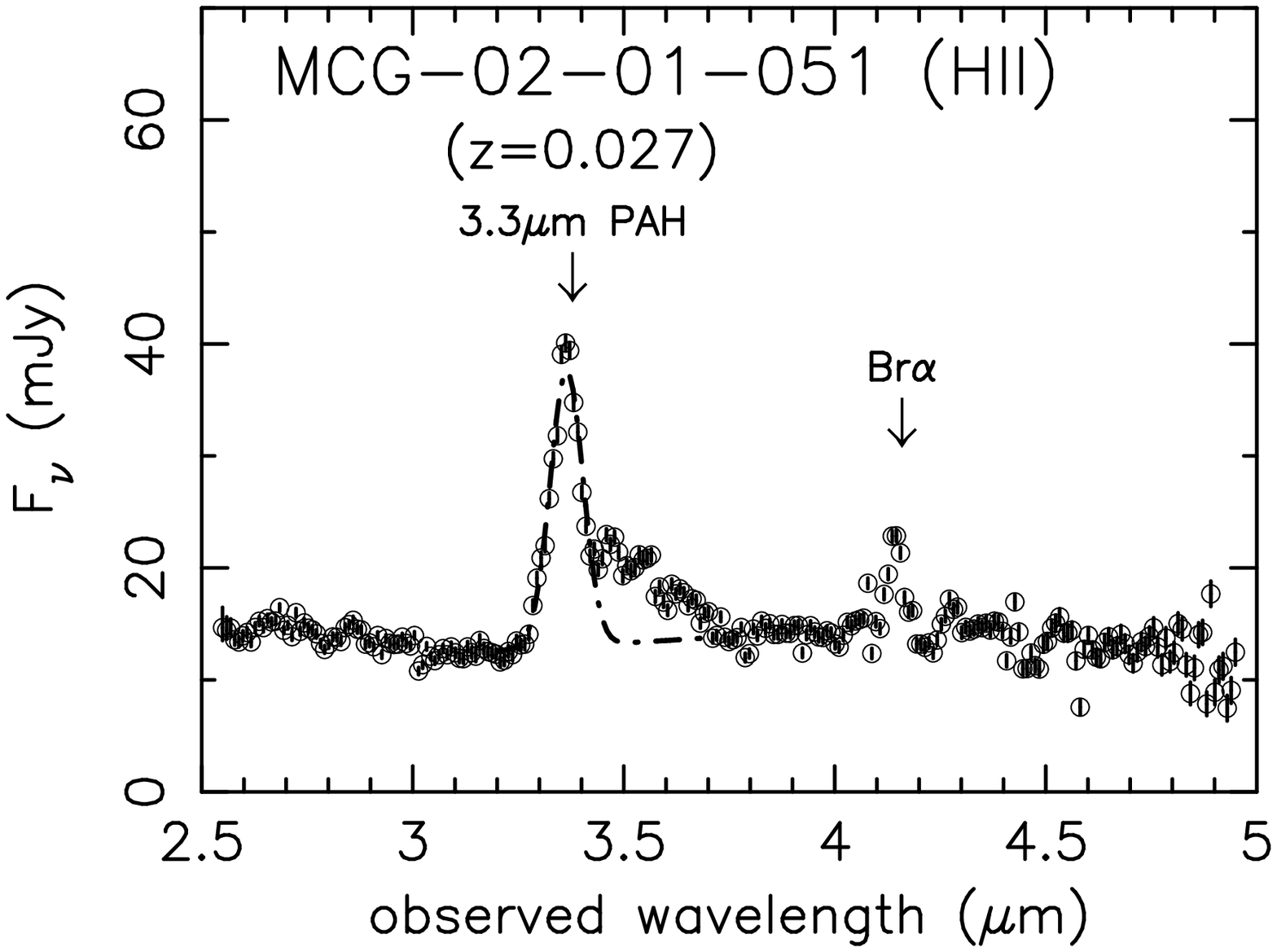}
\includegraphics[angle=0,scale=.27]{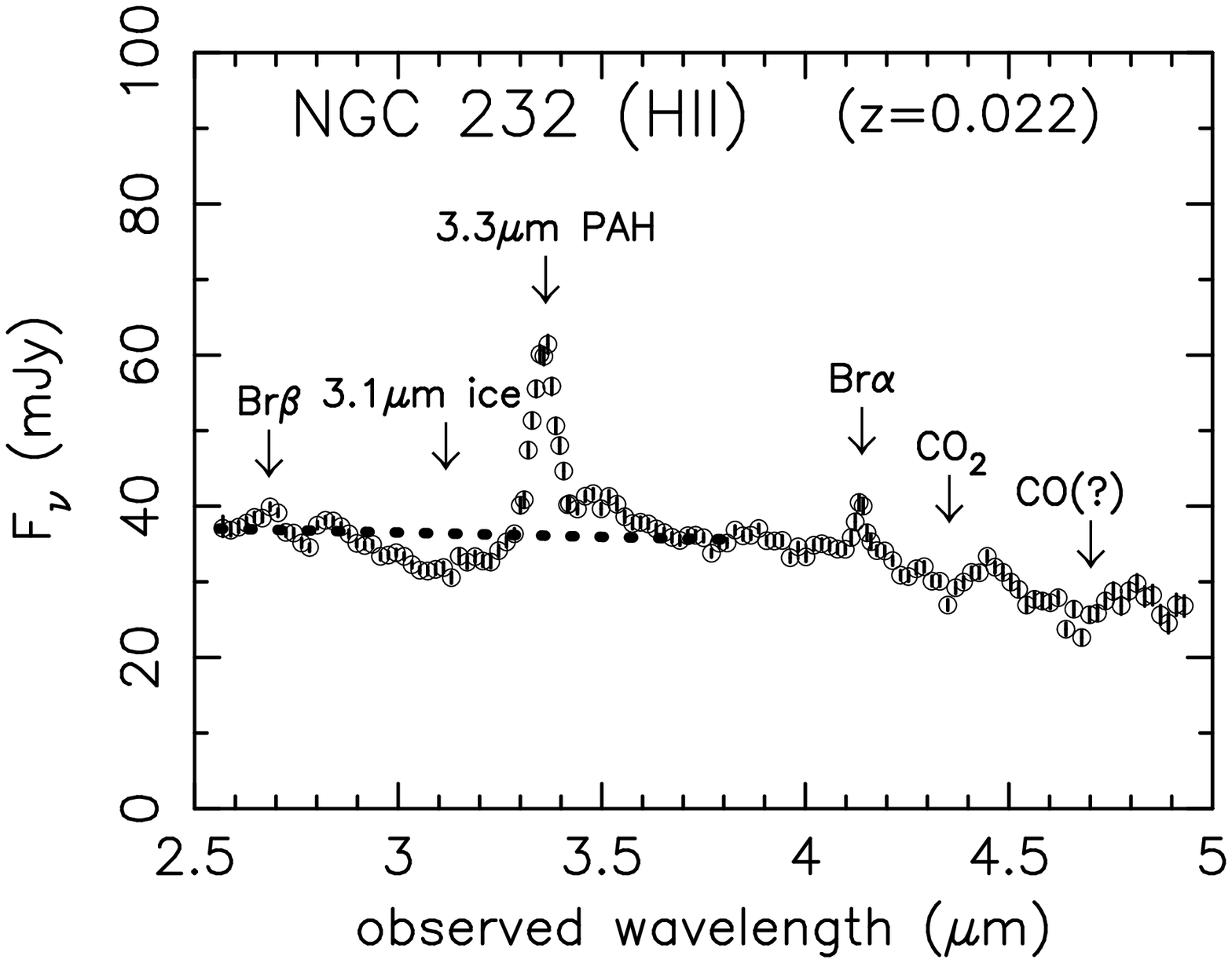} \\
\includegraphics[angle=0,scale=.27]{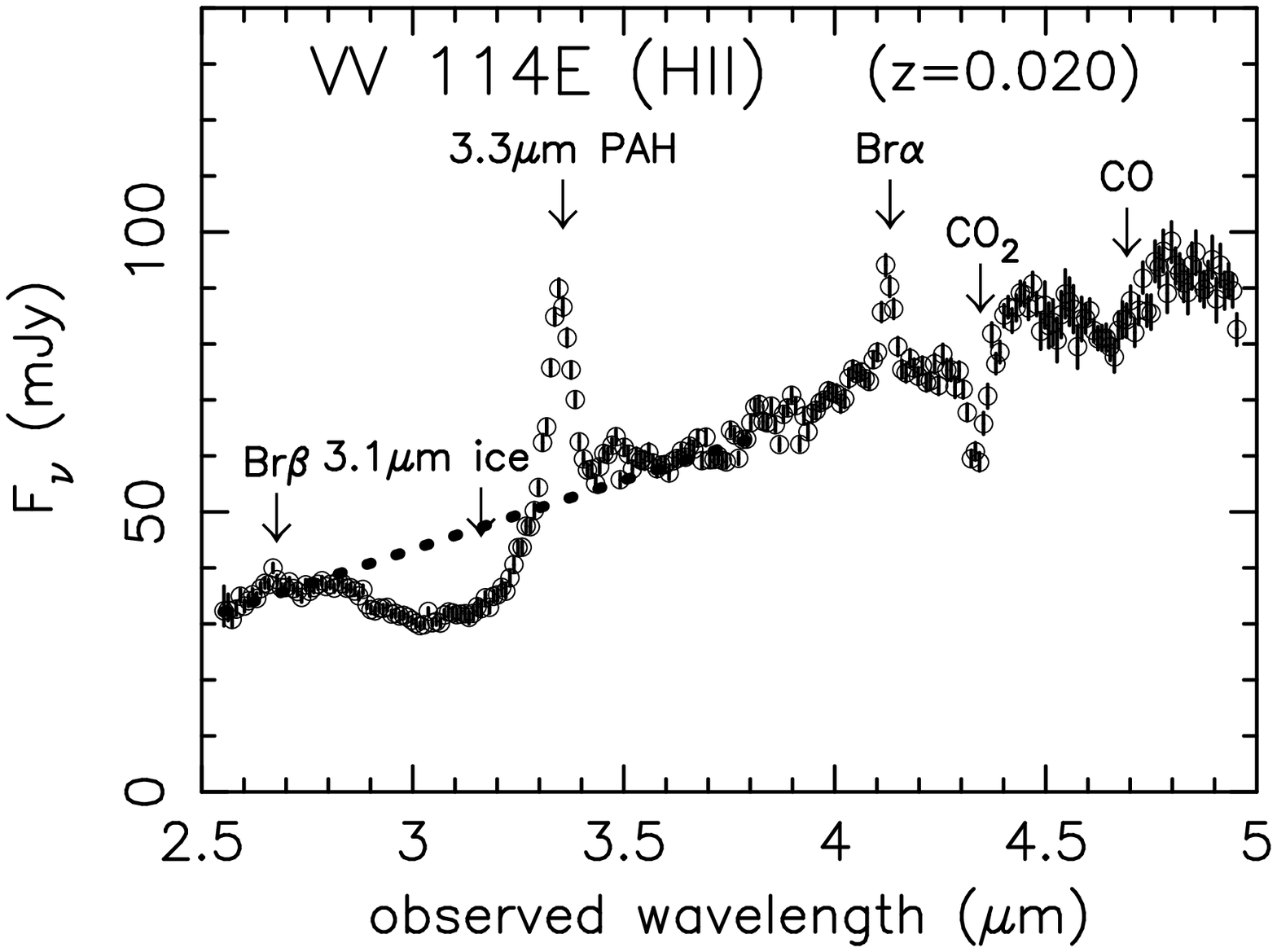}
\includegraphics[angle=0,scale=.27]{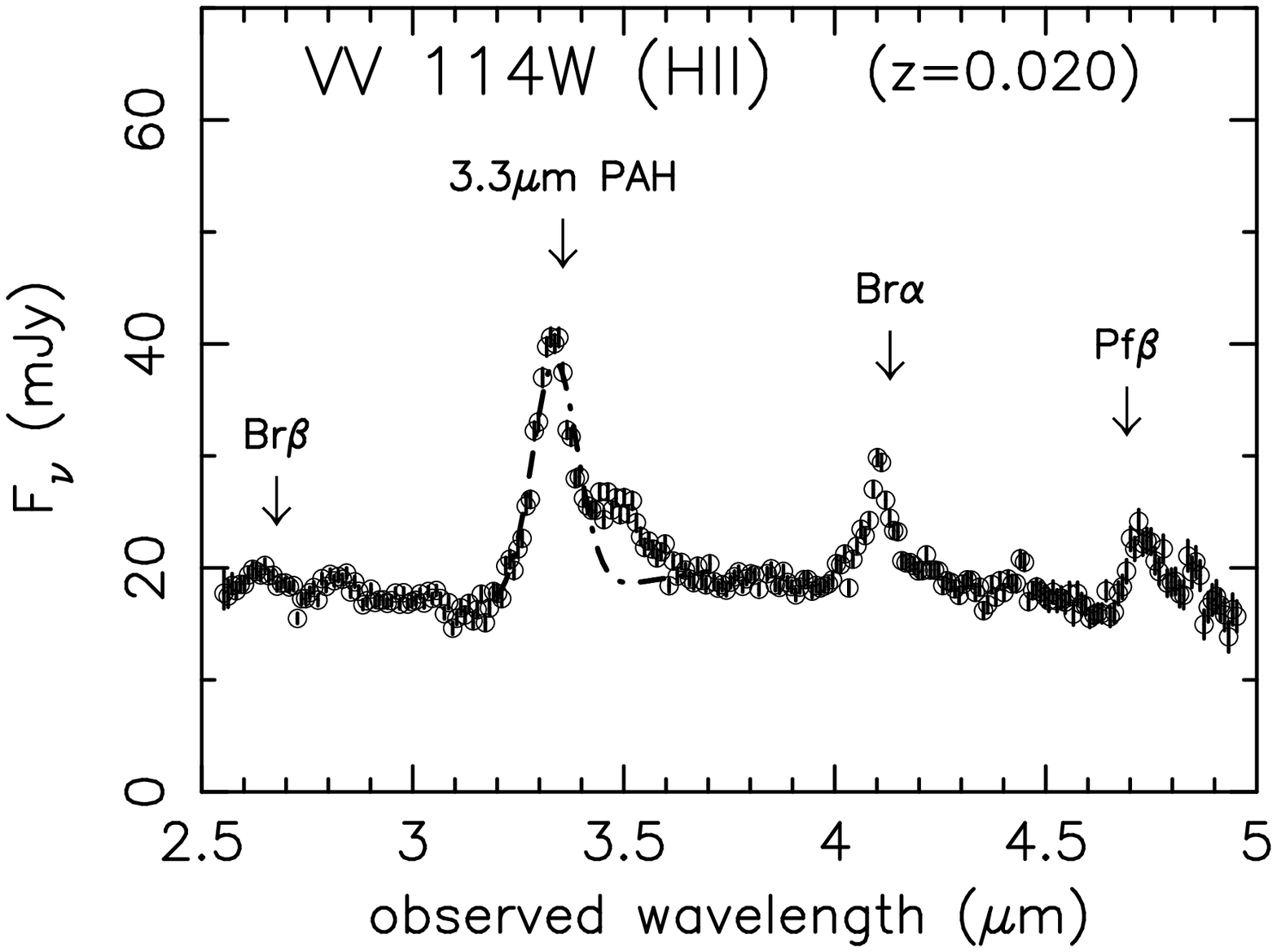}
\includegraphics[angle=0,scale=.27]{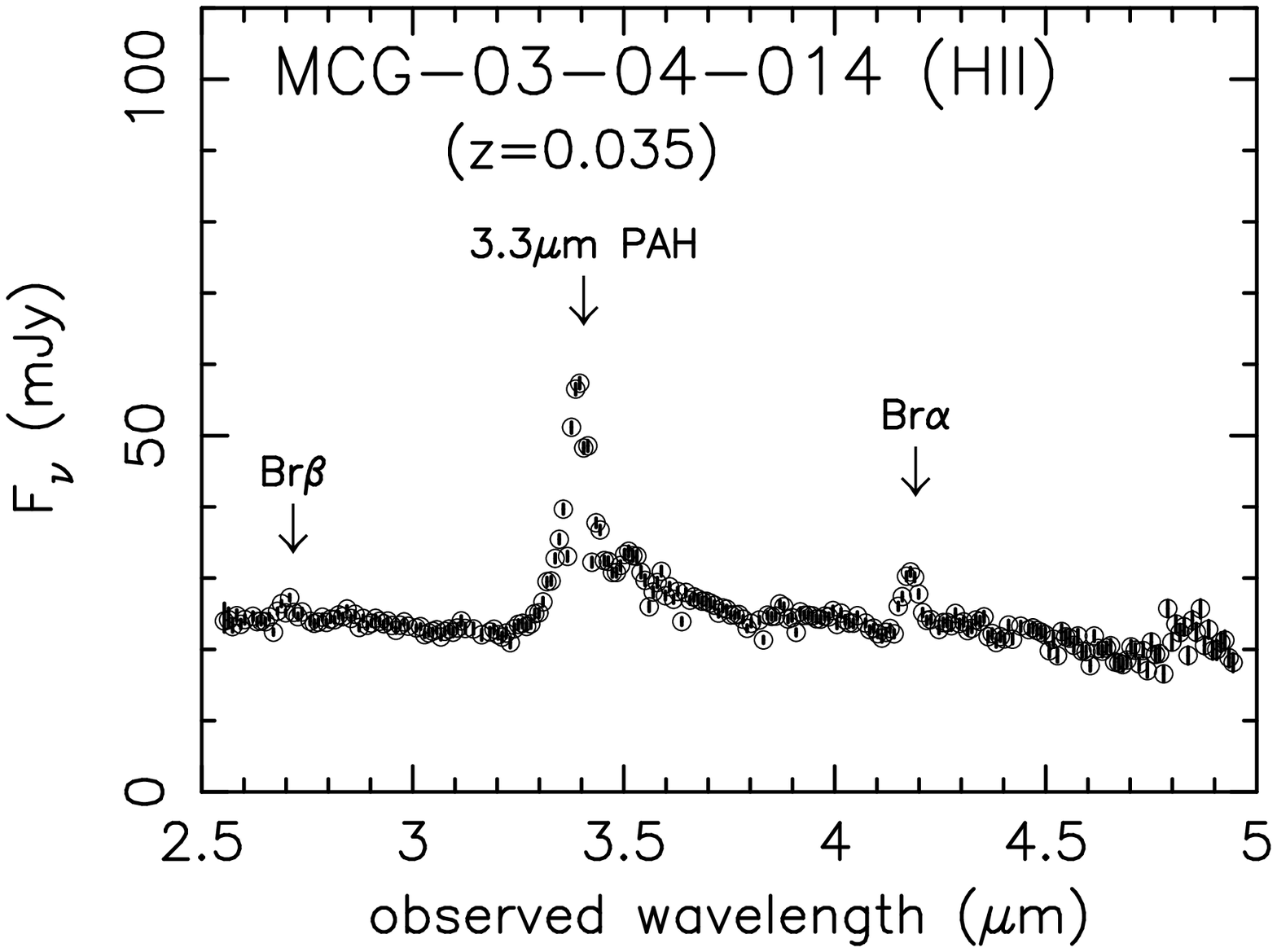} \\
\includegraphics[angle=0,scale=.27]{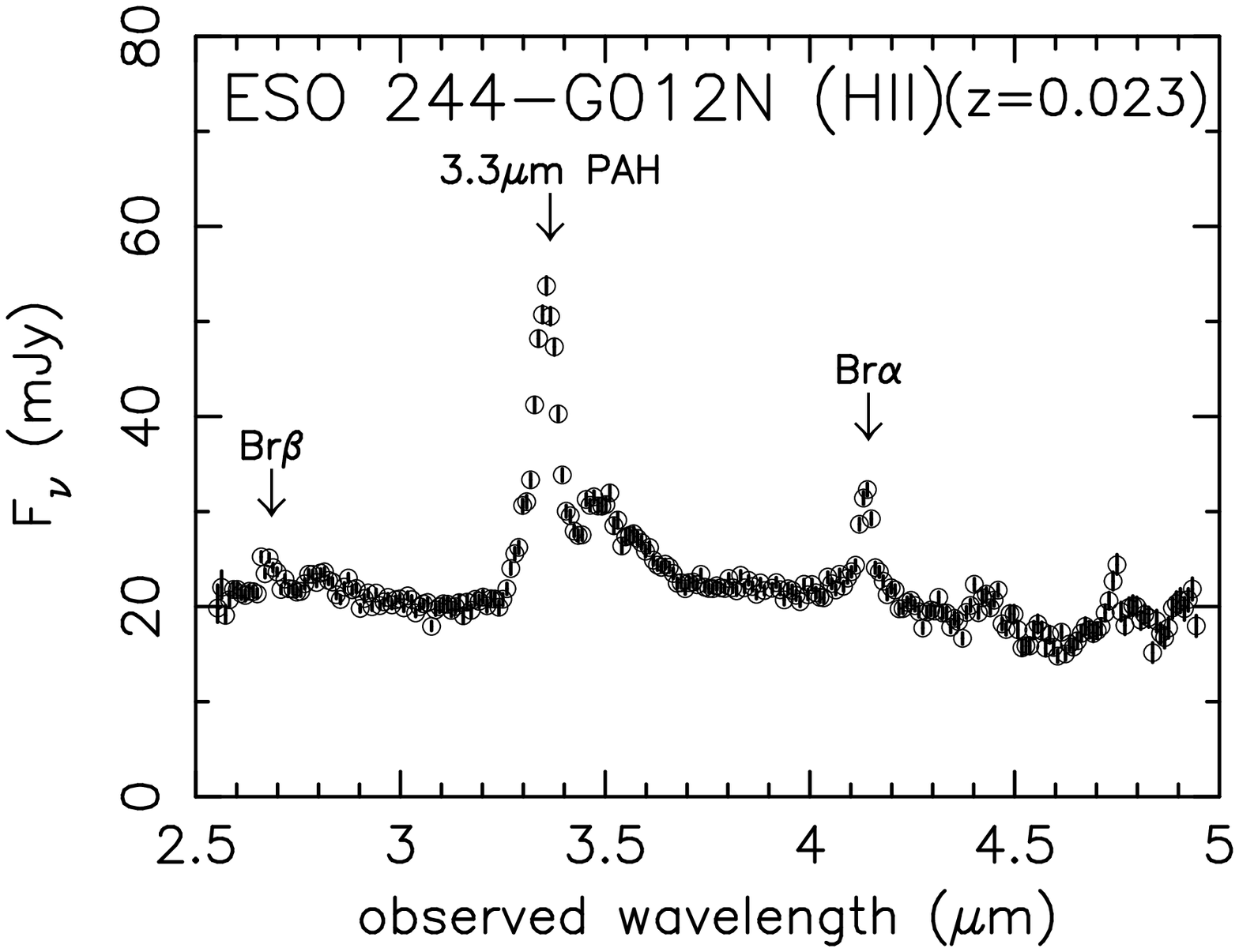}
\includegraphics[angle=0,scale=.27]{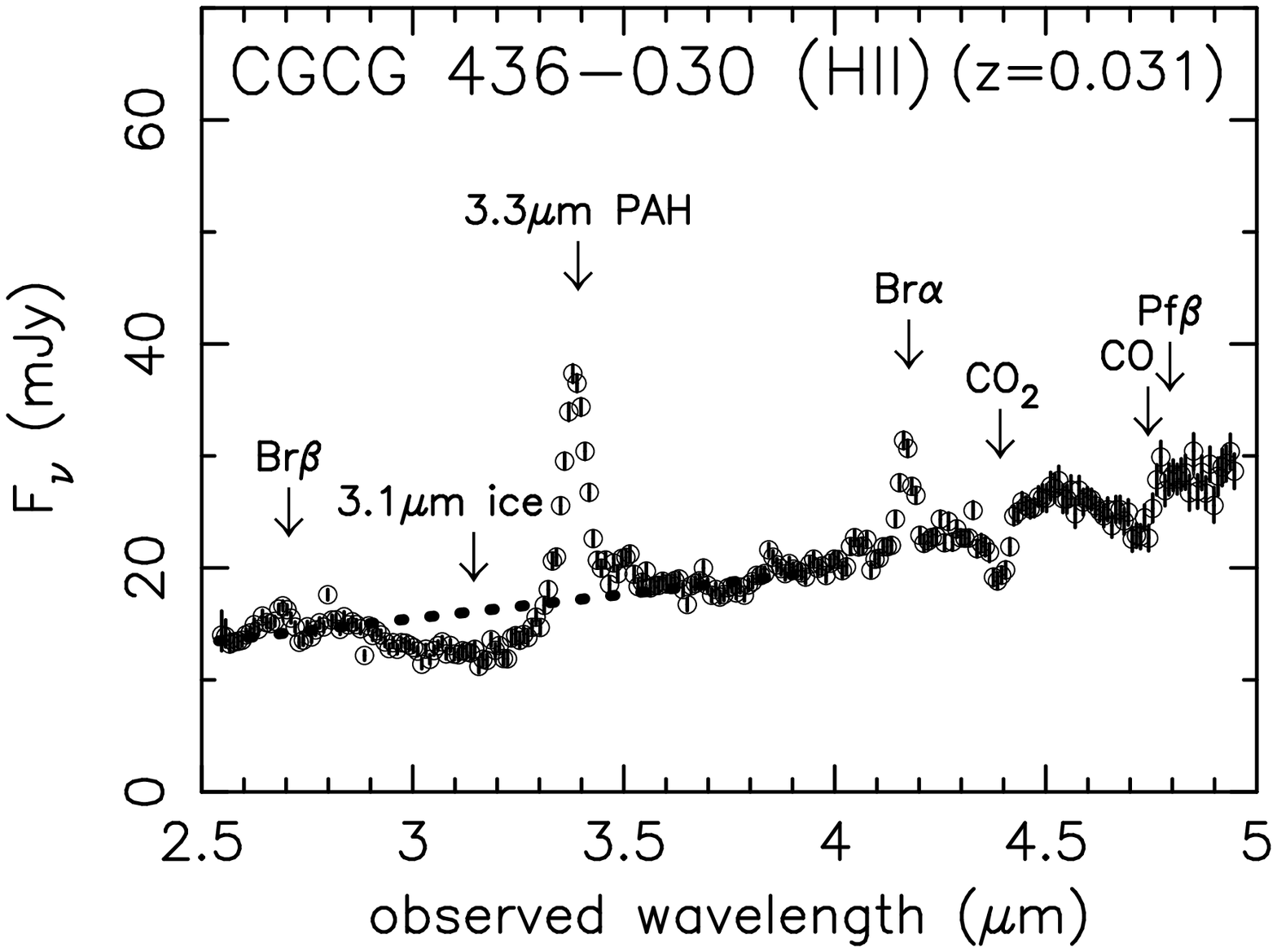}
\includegraphics[angle=0,scale=.27]{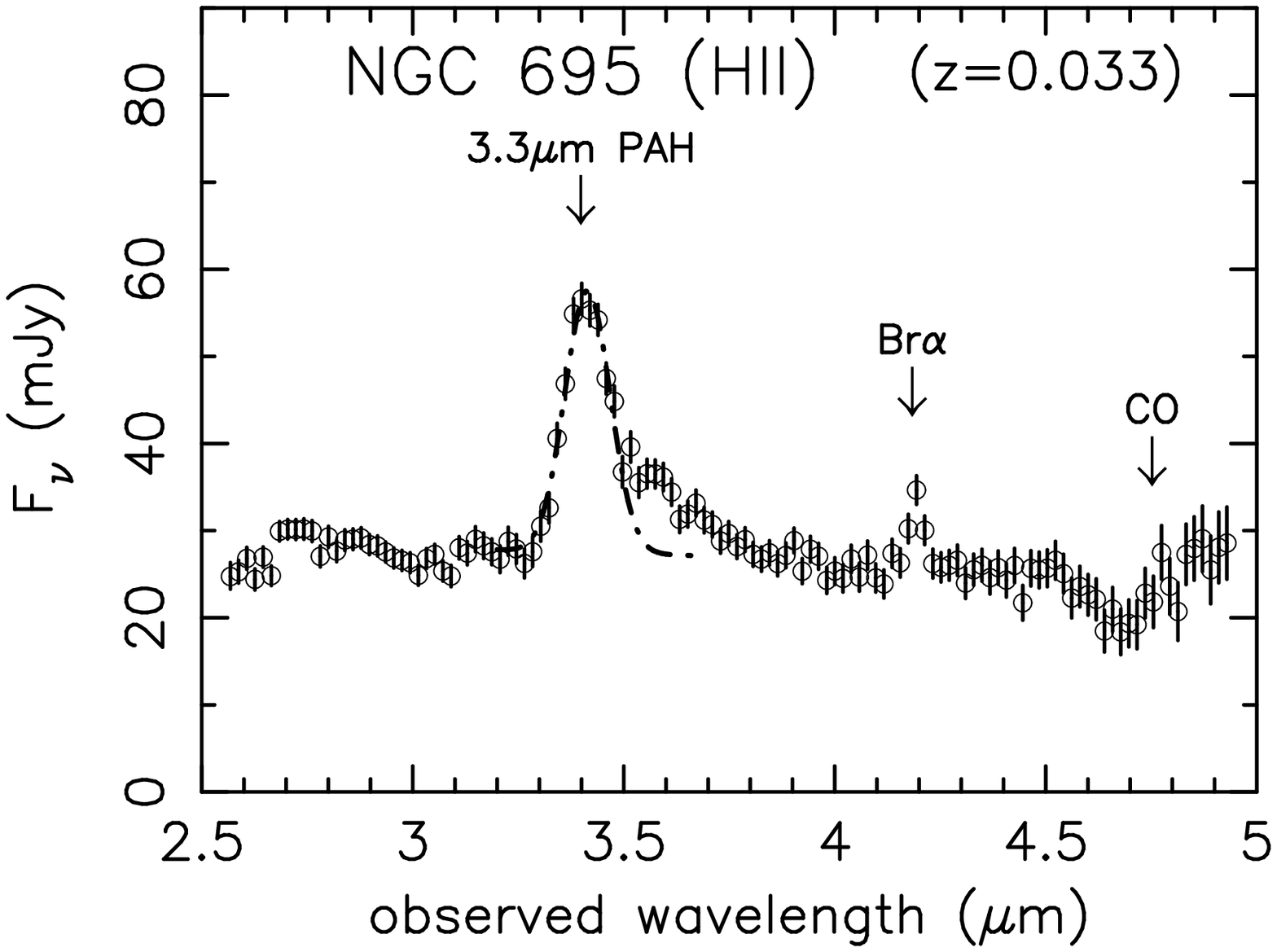} \\
\includegraphics[angle=0,scale=.27]{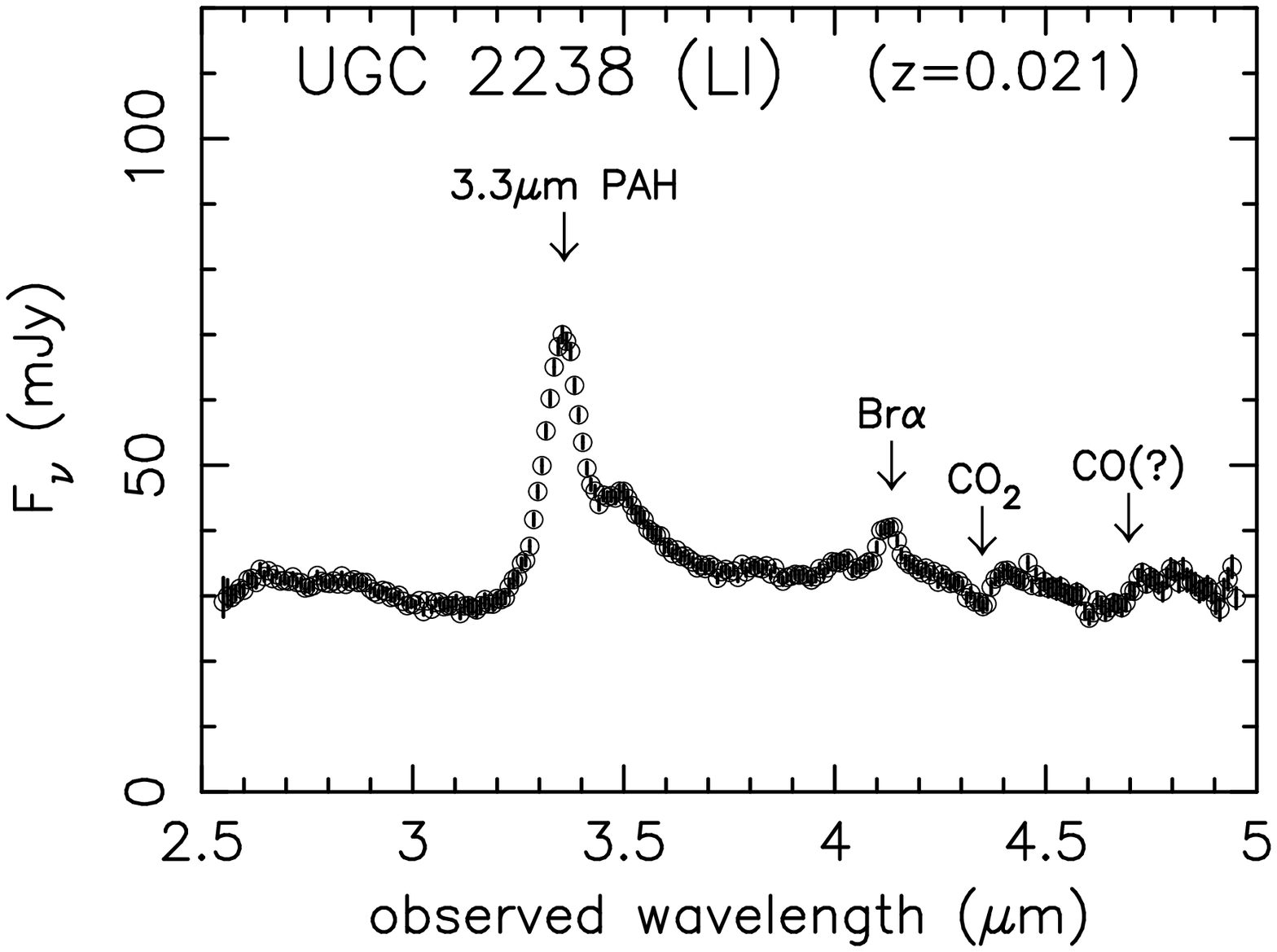} 
\includegraphics[angle=0,scale=.27]{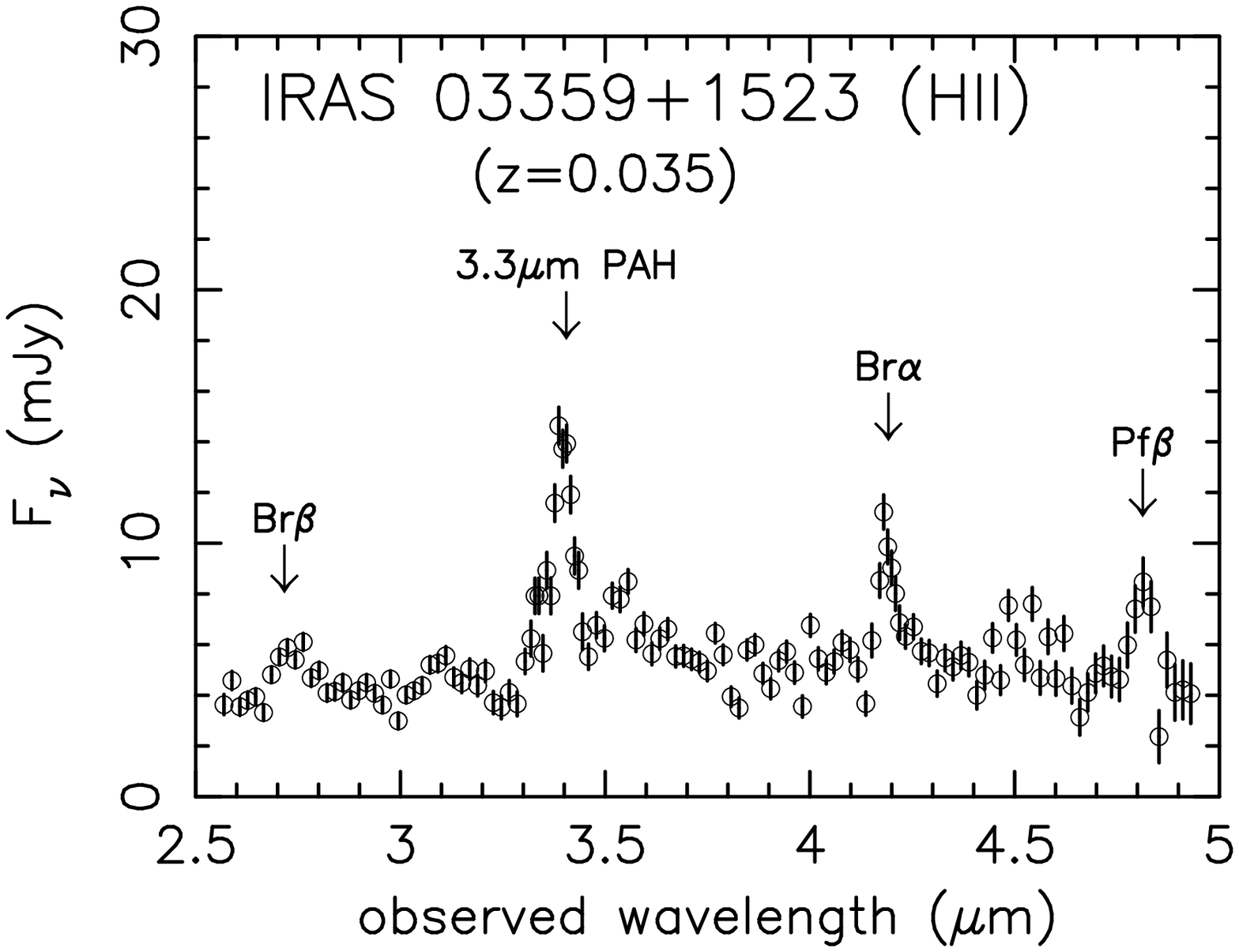} 
\includegraphics[angle=0,scale=.27]{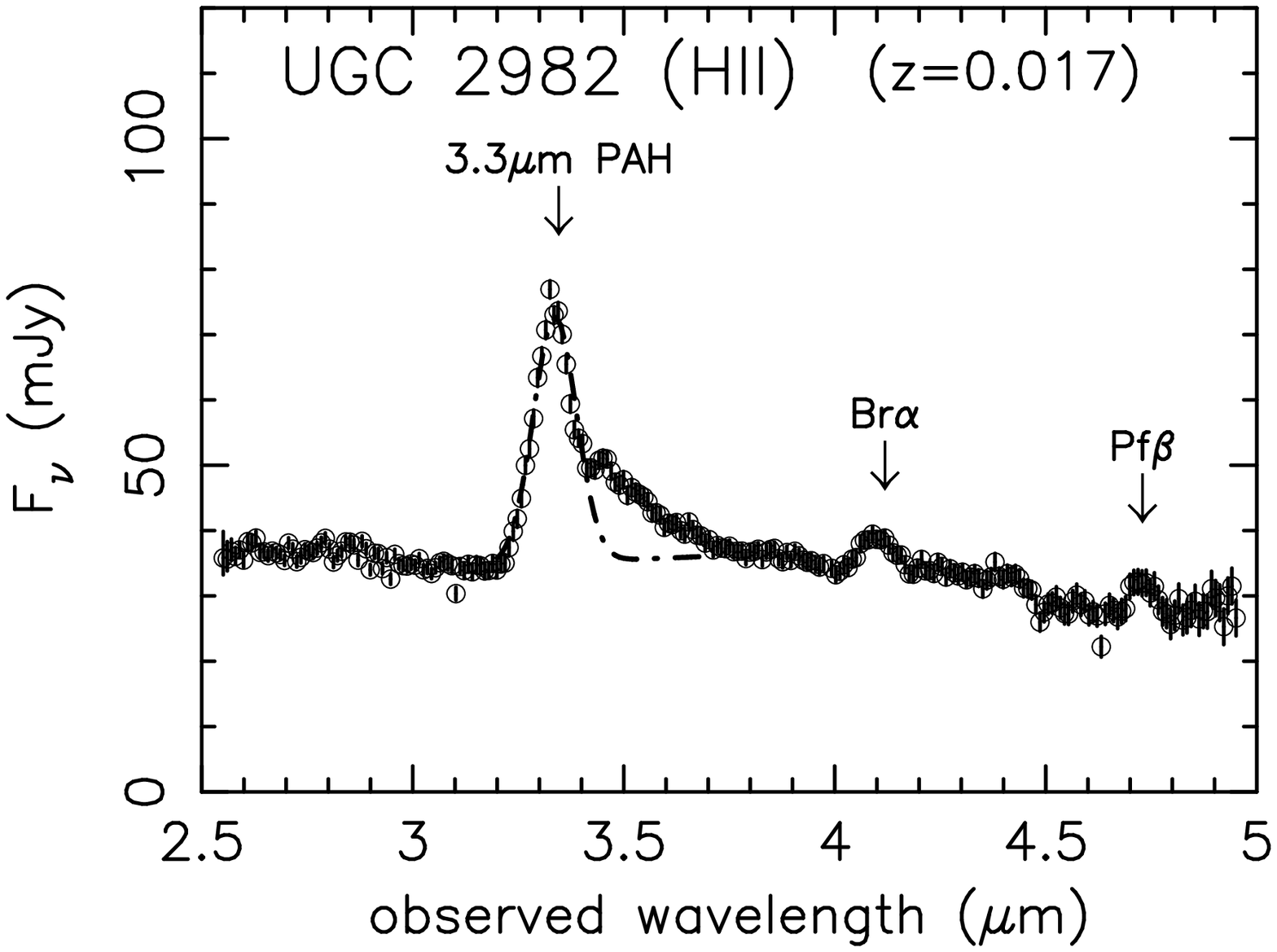} \\
\includegraphics[angle=0,scale=.27]{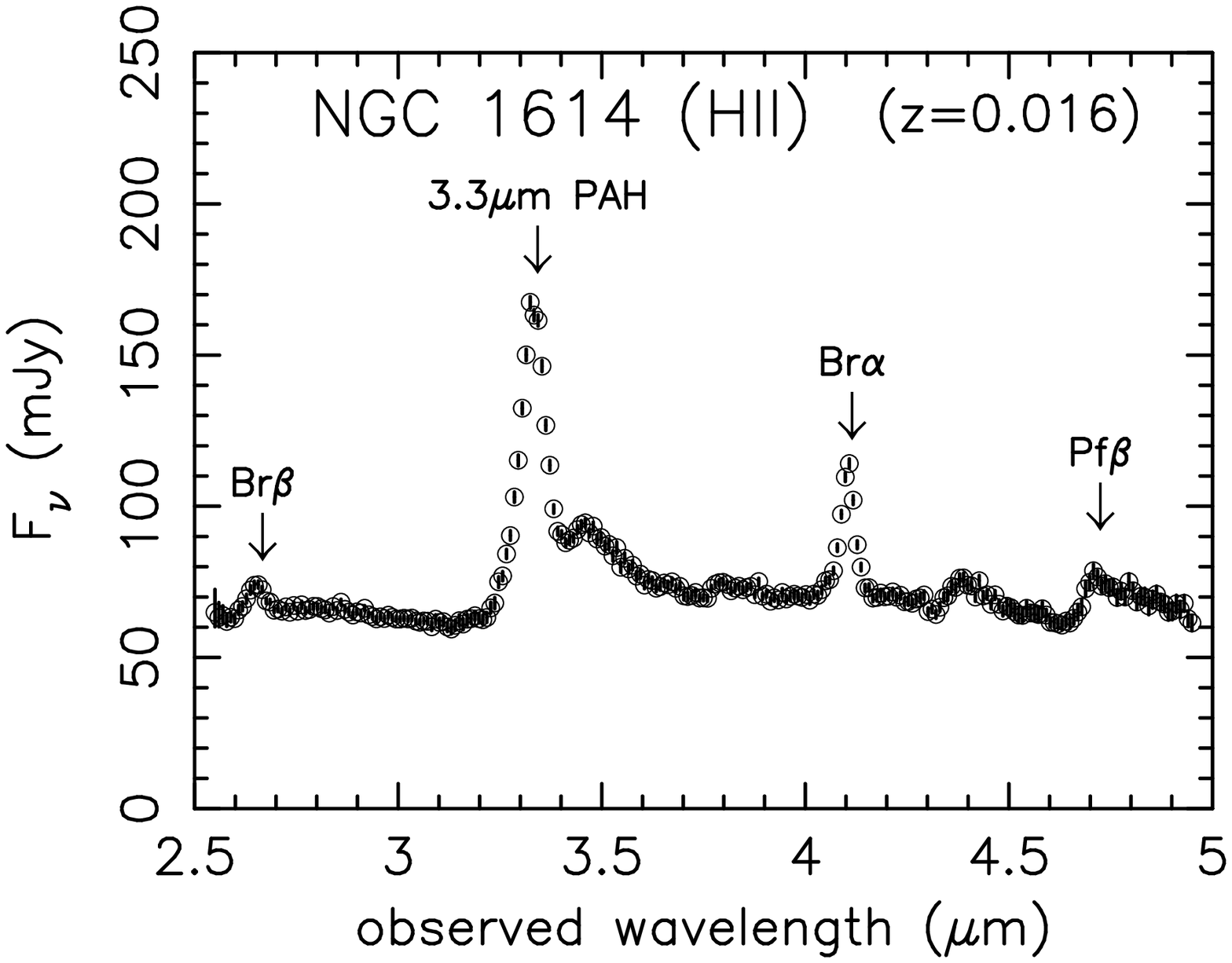}
\includegraphics[angle=0,scale=.27]{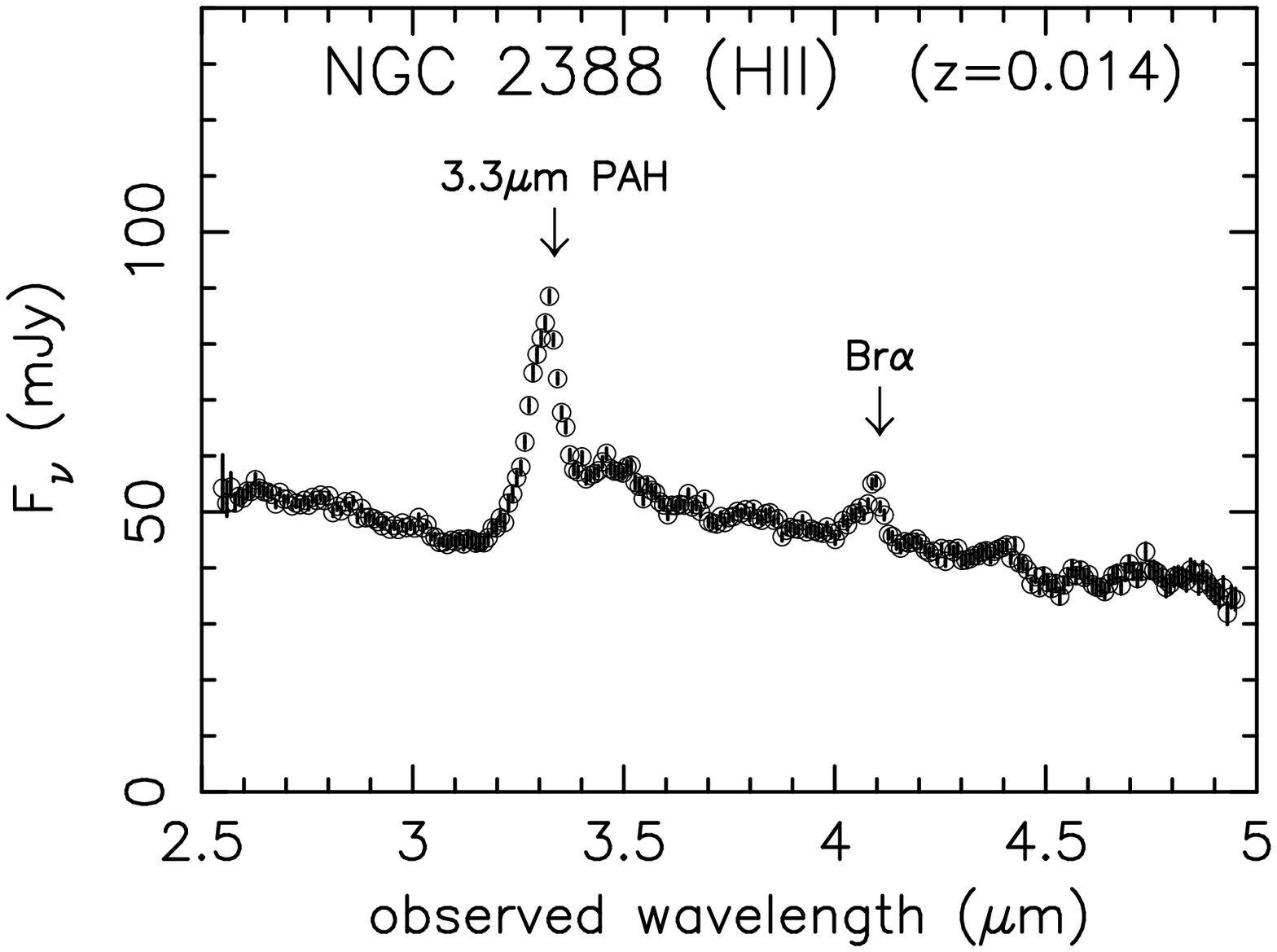}
\includegraphics[angle=0,scale=.27]{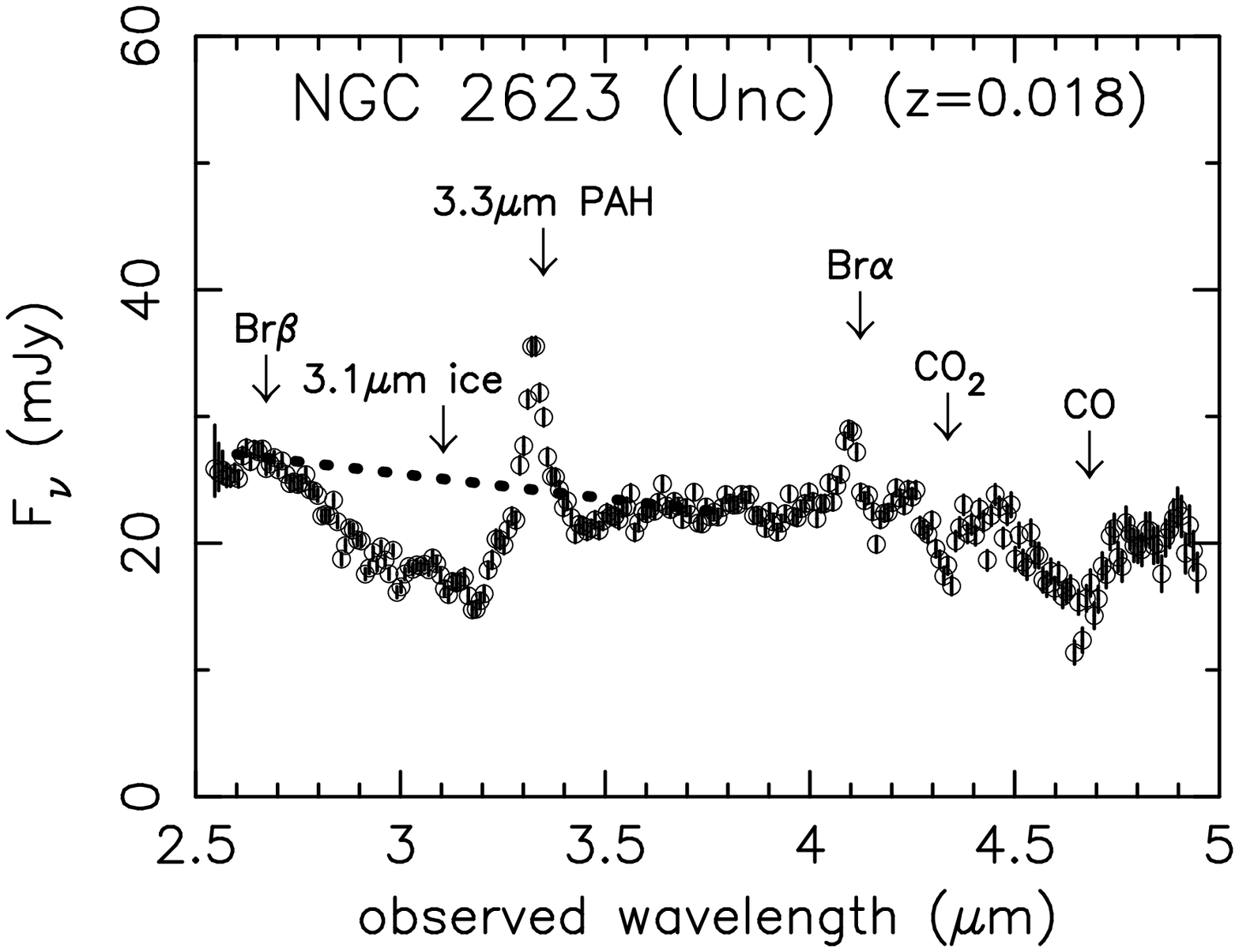} \\
\end{figure}

\clearpage

\begin{figure}
\includegraphics[angle=0,scale=.27]{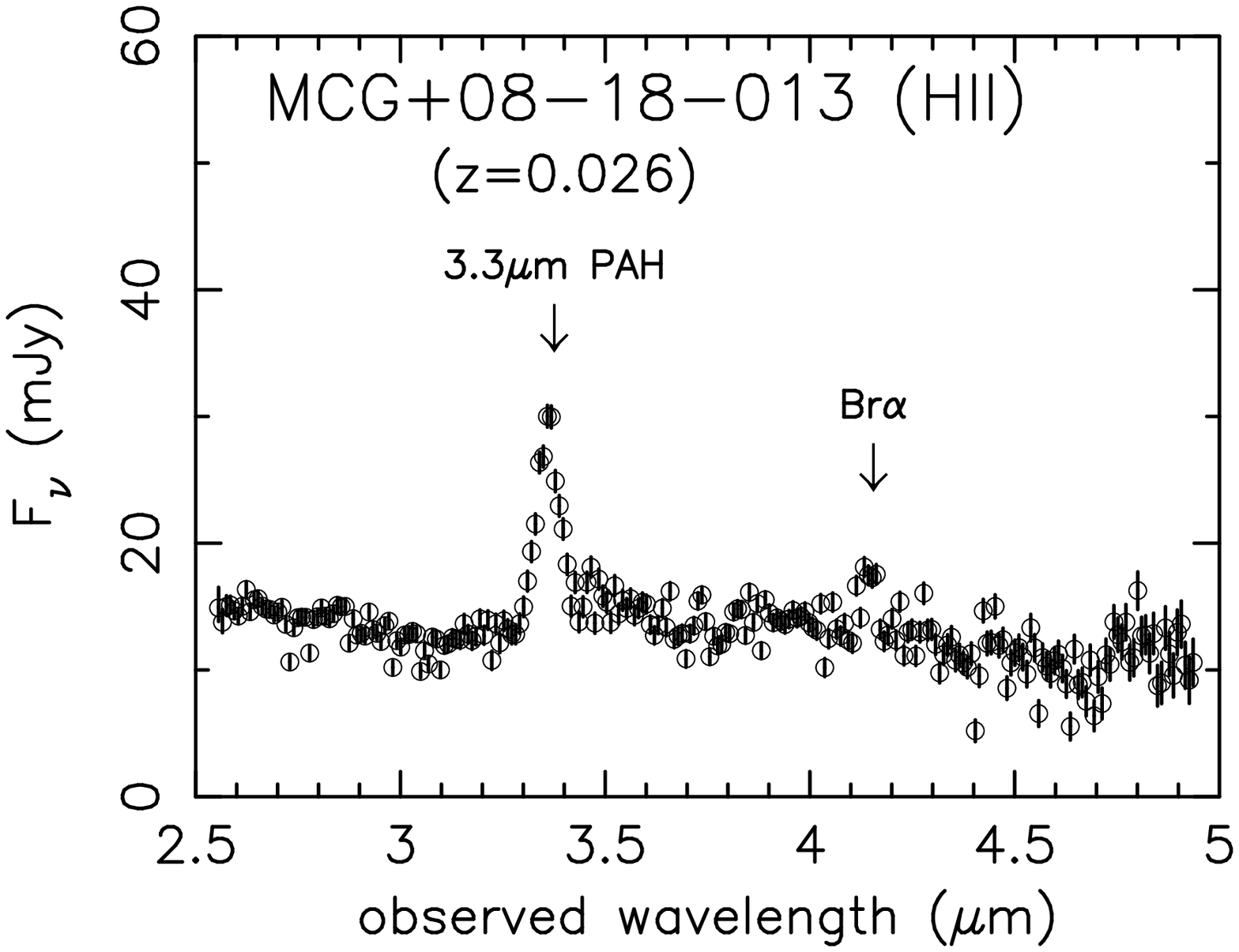}
\includegraphics[angle=0,scale=.27]{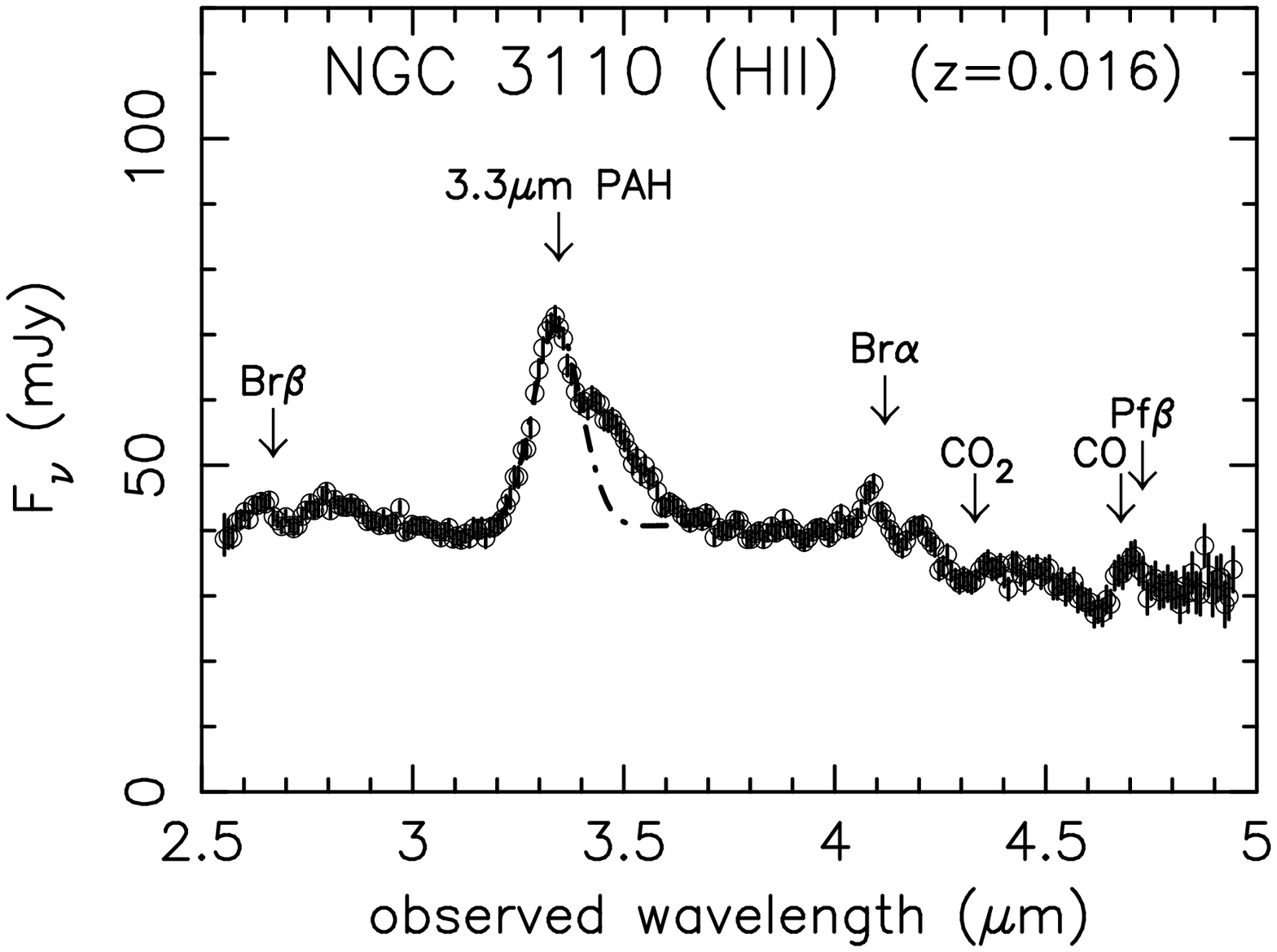}
\includegraphics[angle=0,scale=.27]{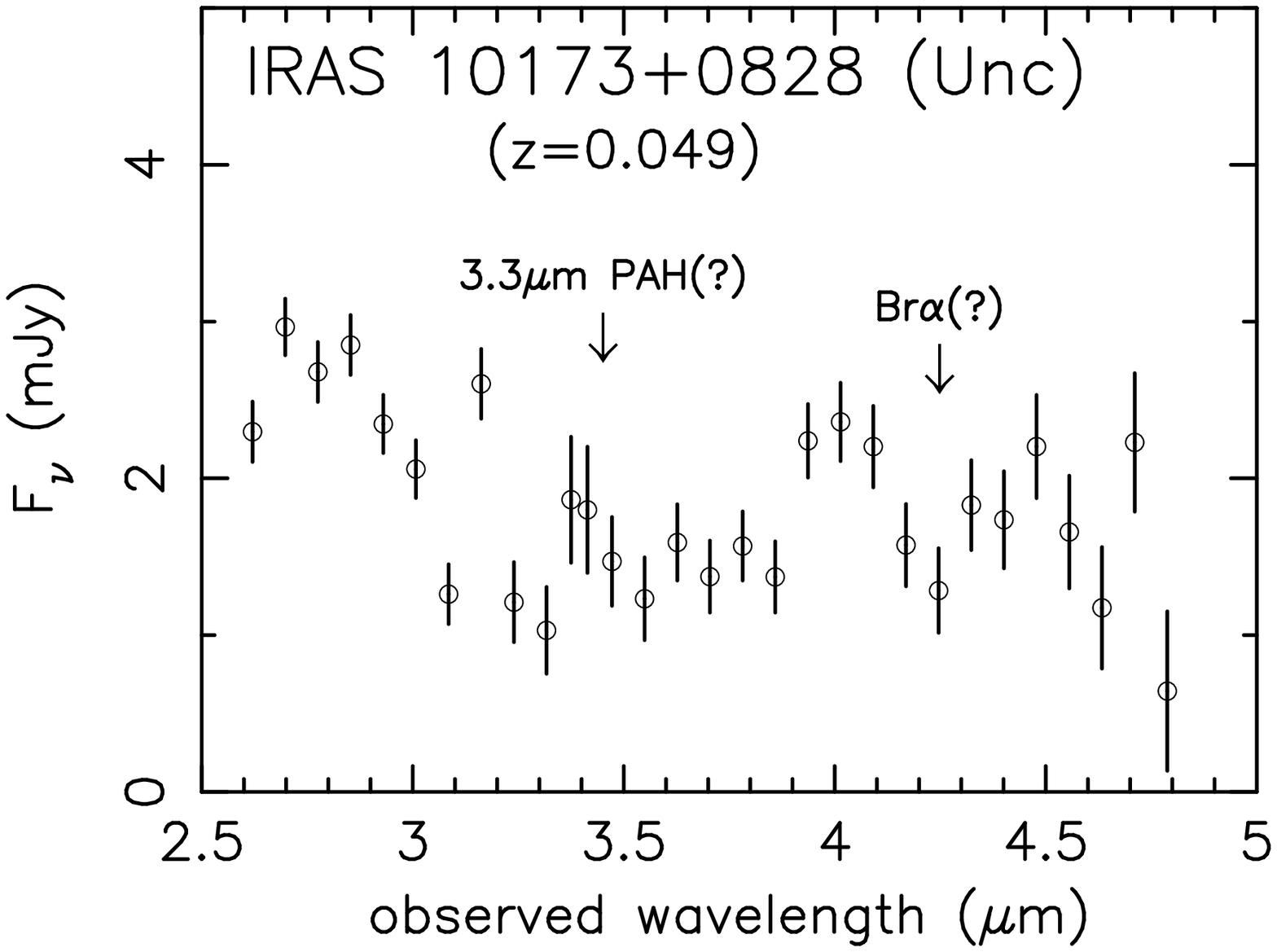} \\
\includegraphics[angle=0,scale=.27]{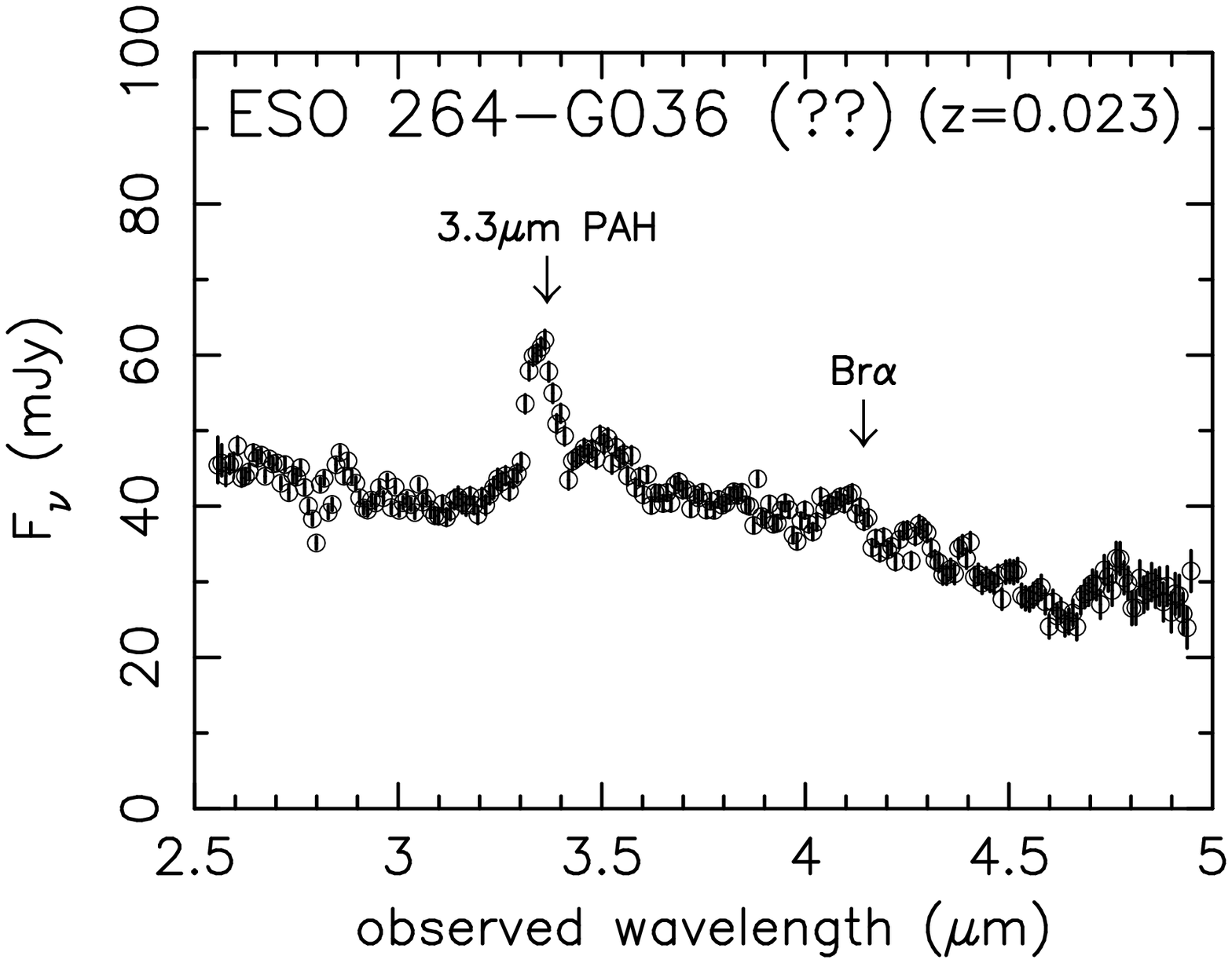}
\includegraphics[angle=0,scale=.27]{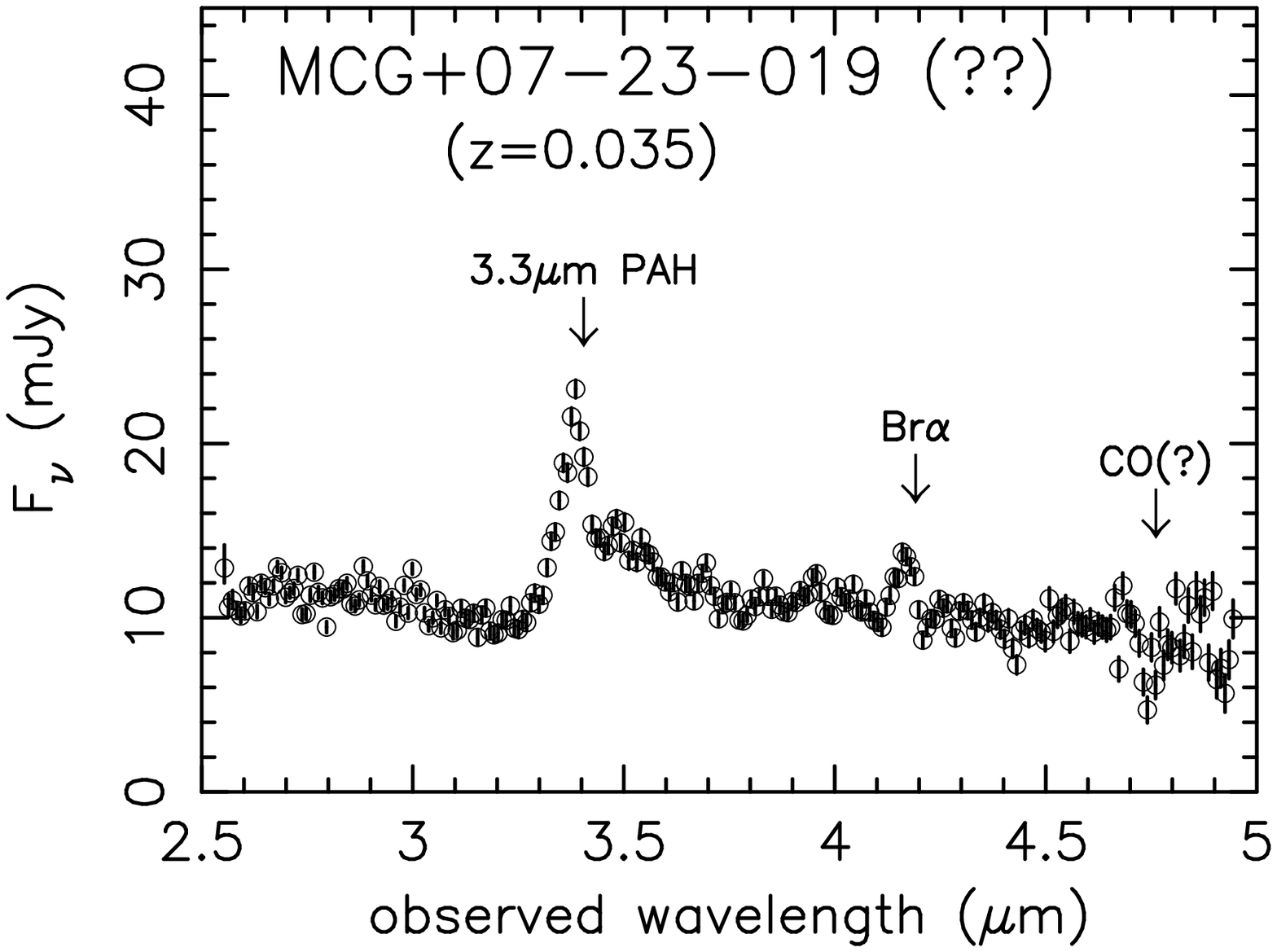}
\includegraphics[angle=0,scale=.27]{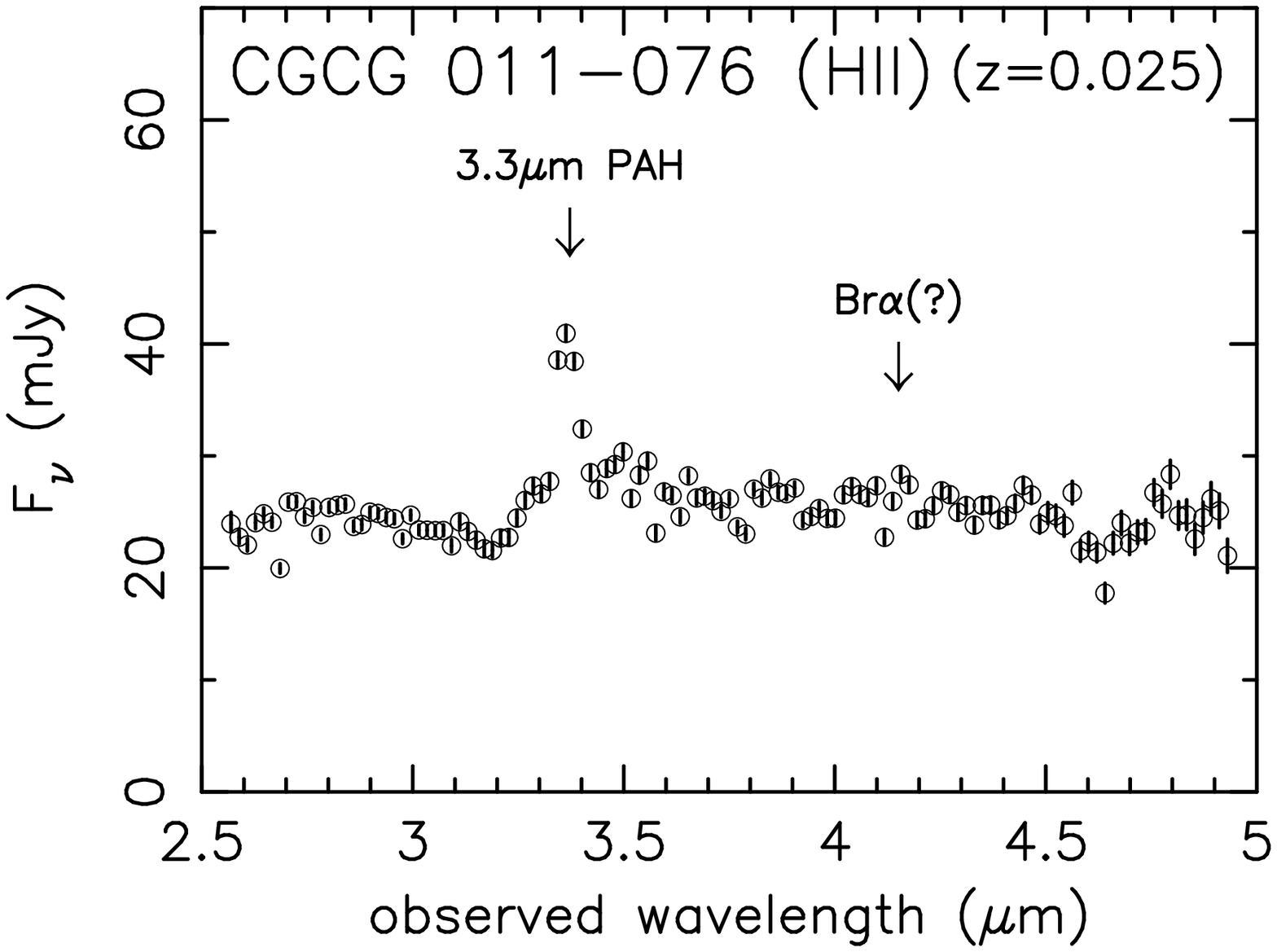} \\
\includegraphics[angle=0,scale=.27]{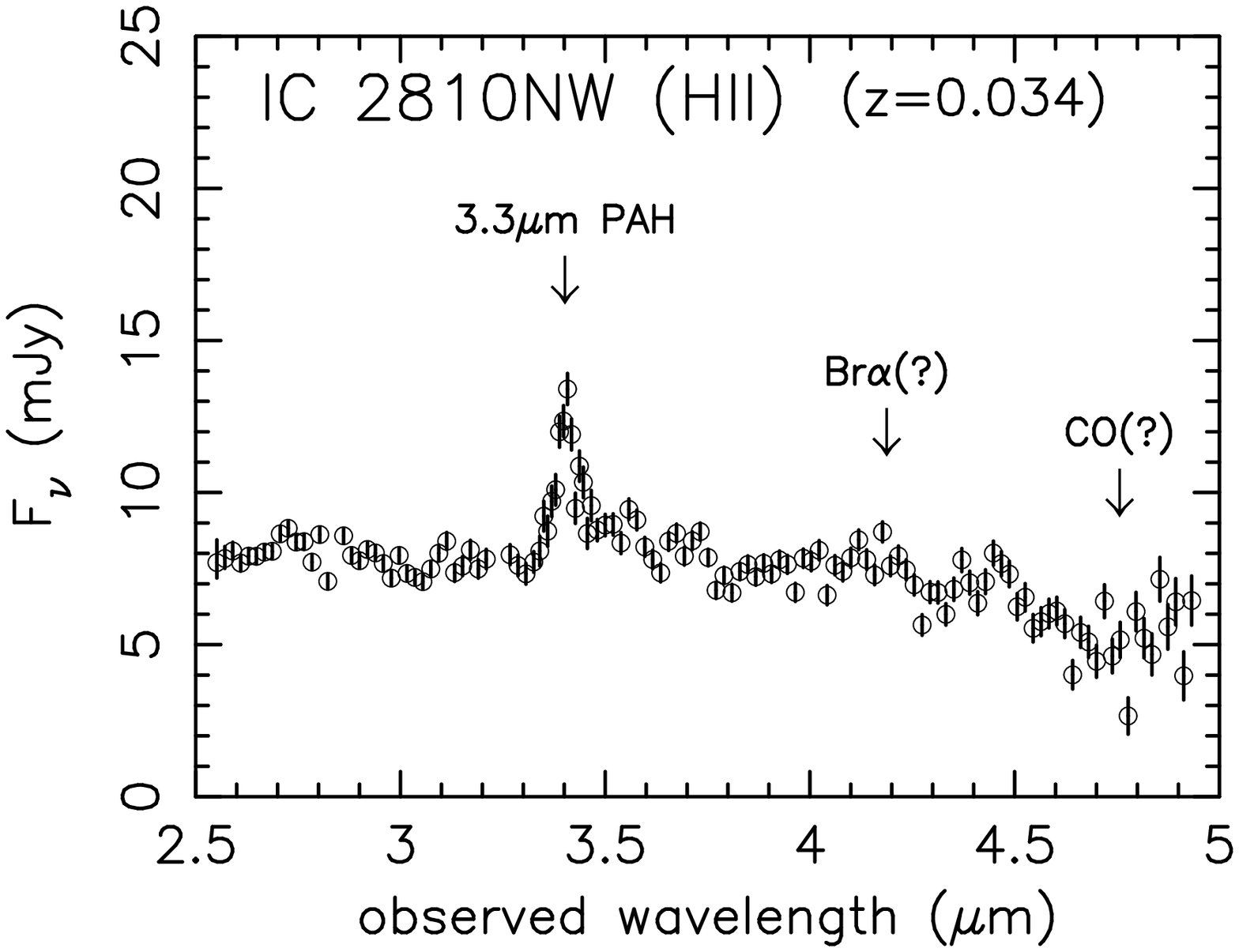}
\includegraphics[angle=0,scale=.27]{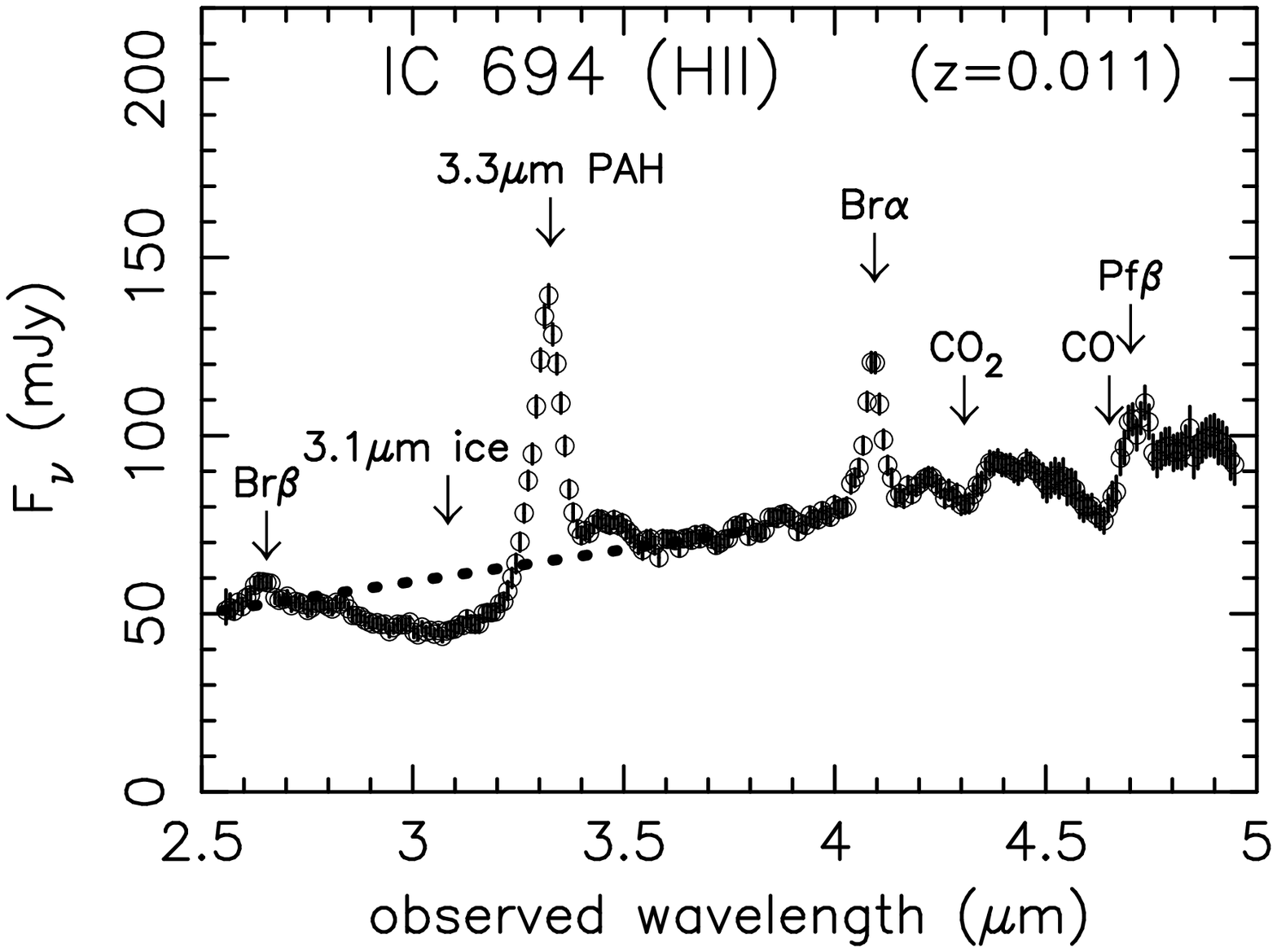}
\includegraphics[angle=0,scale=.27]{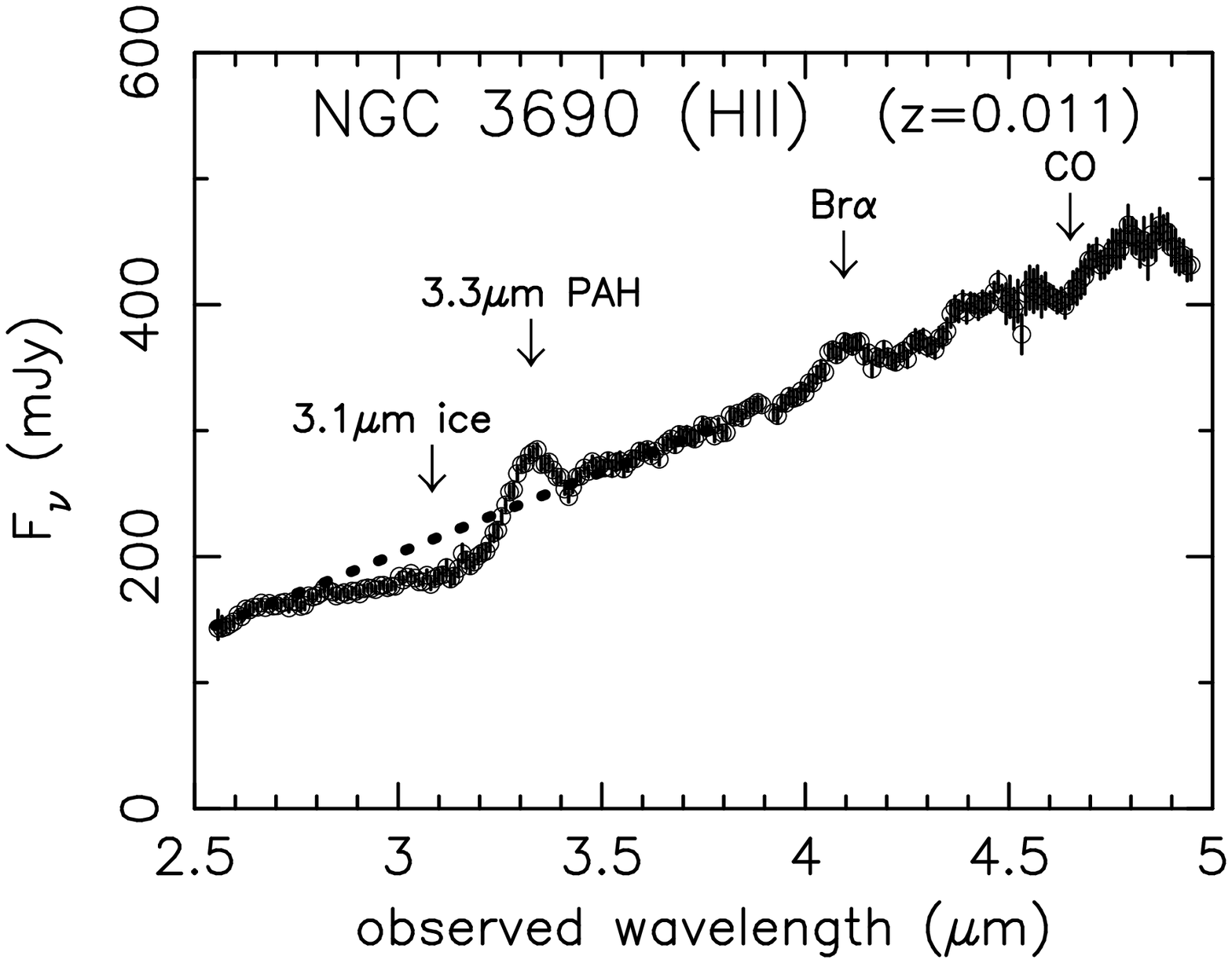} \\
\includegraphics[angle=0,scale=.27]{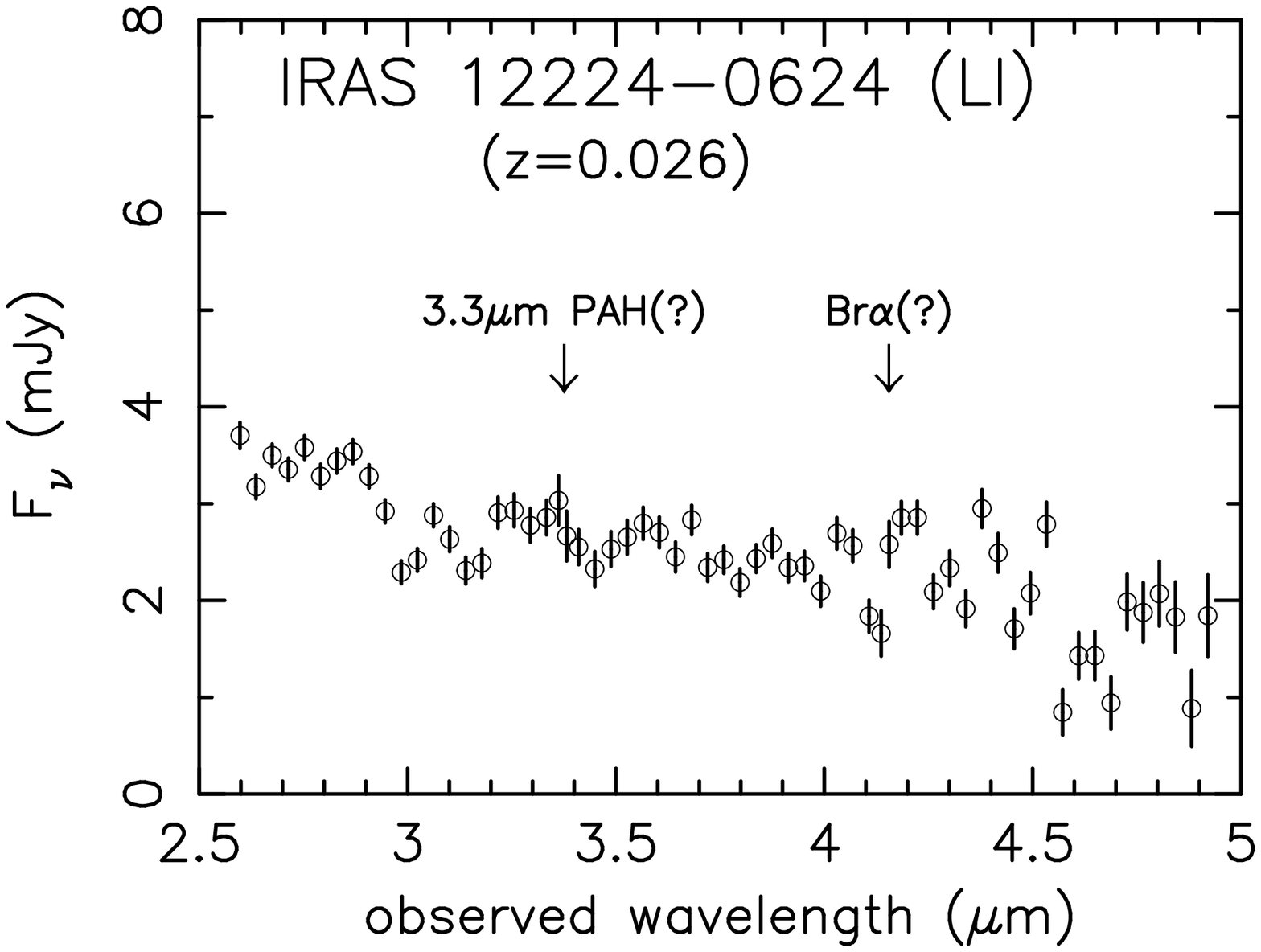}
\includegraphics[angle=0,scale=.27]{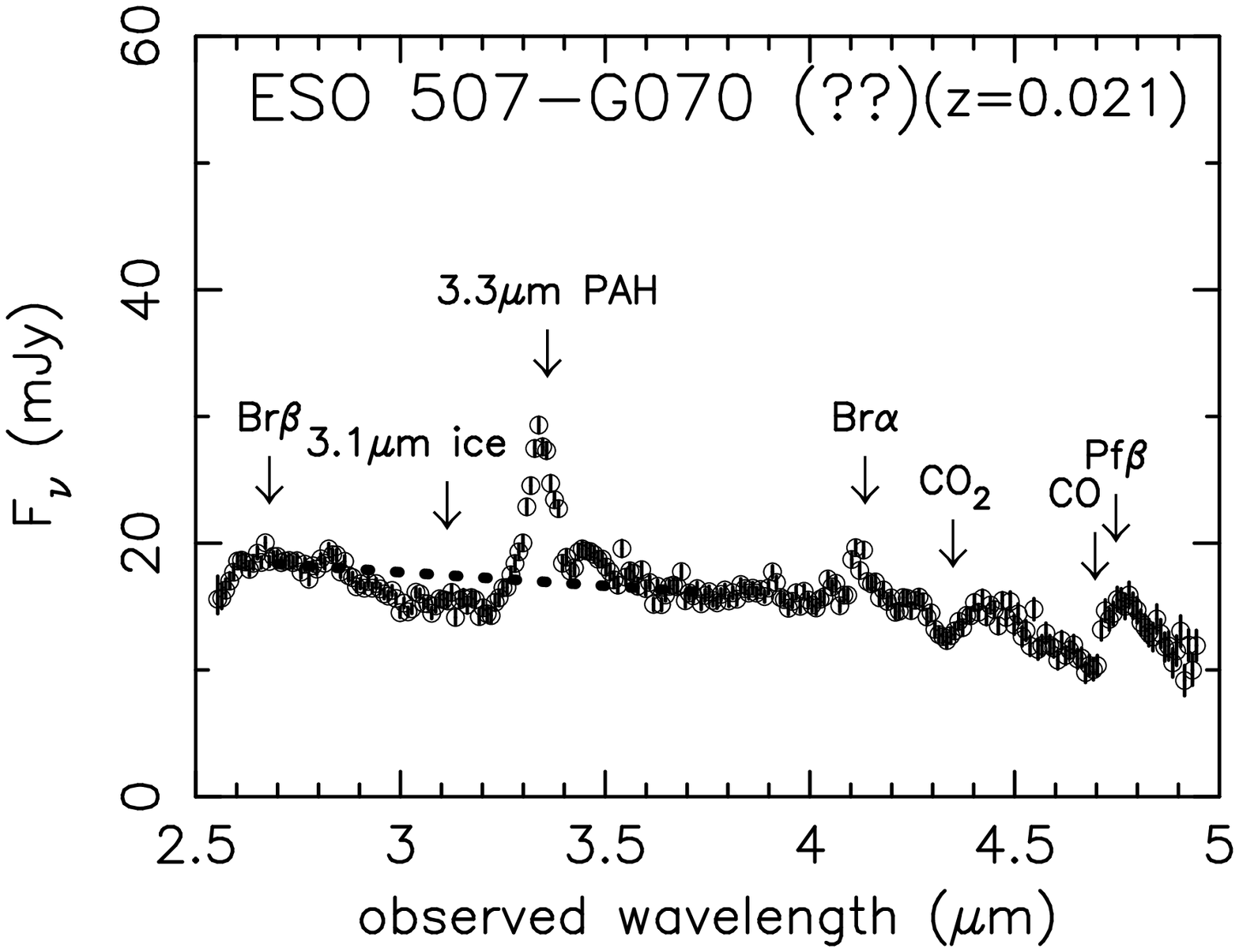}
\includegraphics[angle=0,scale=.27]{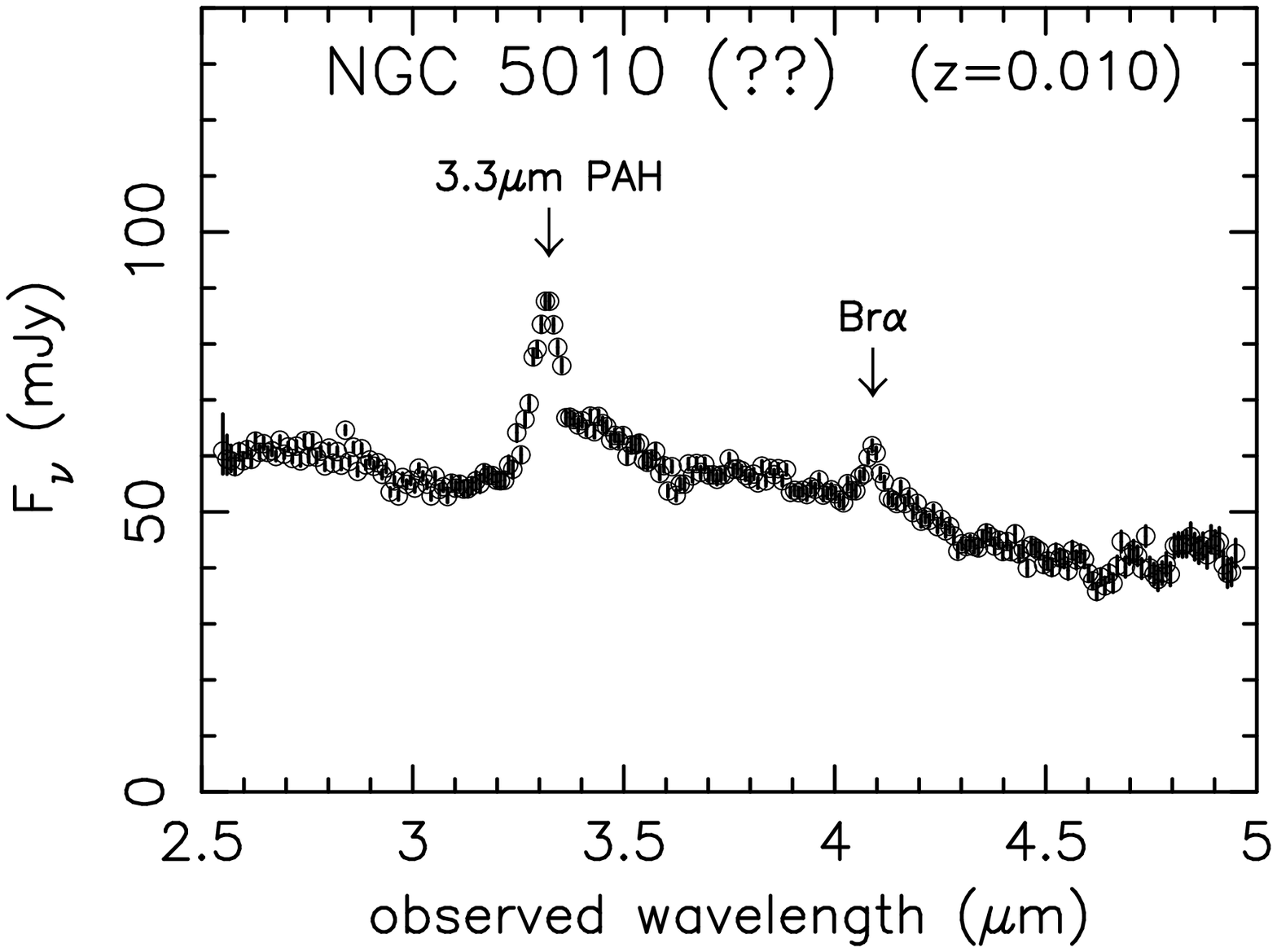} \\
\includegraphics[angle=0,scale=.27]{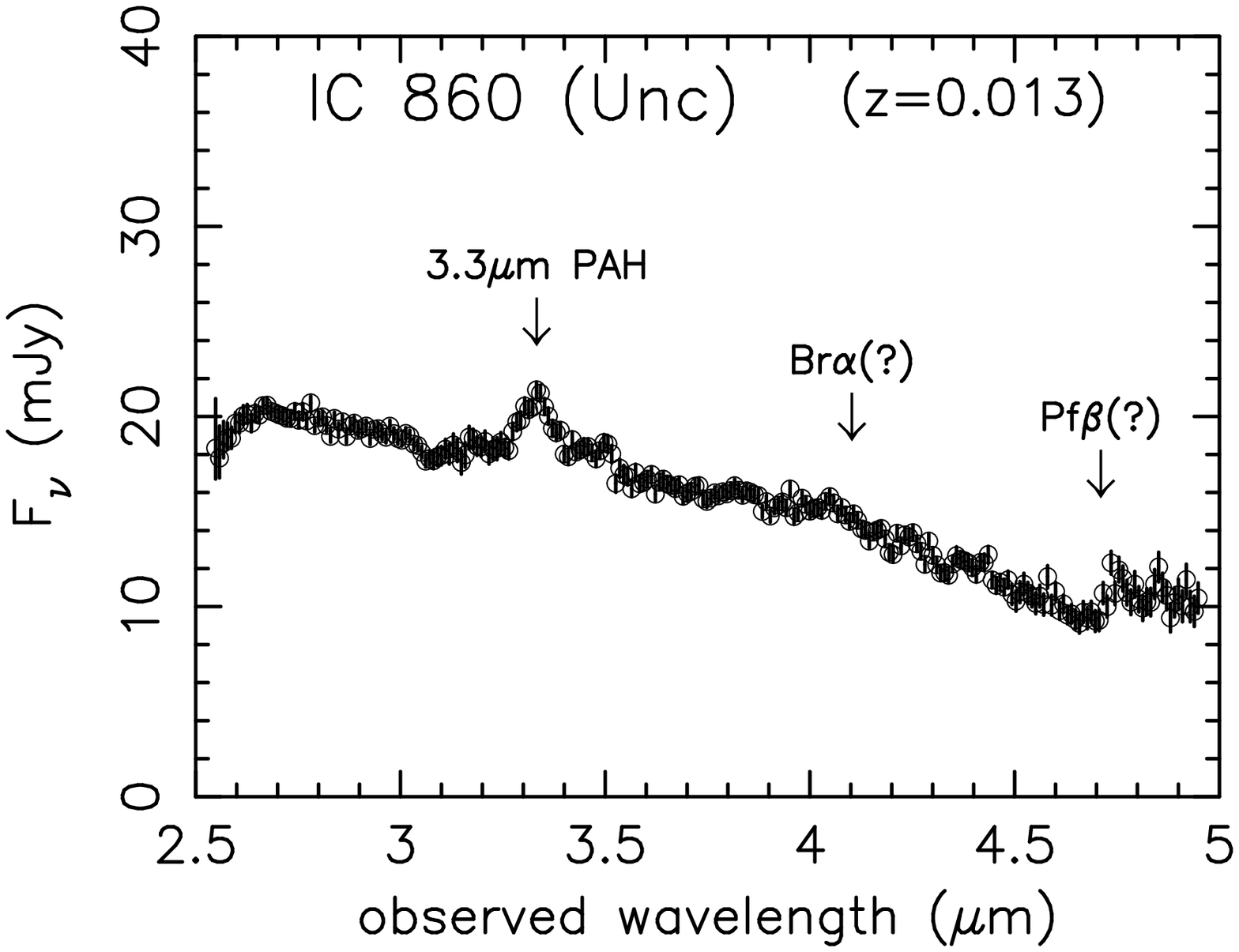}
\includegraphics[angle=0,scale=.27]{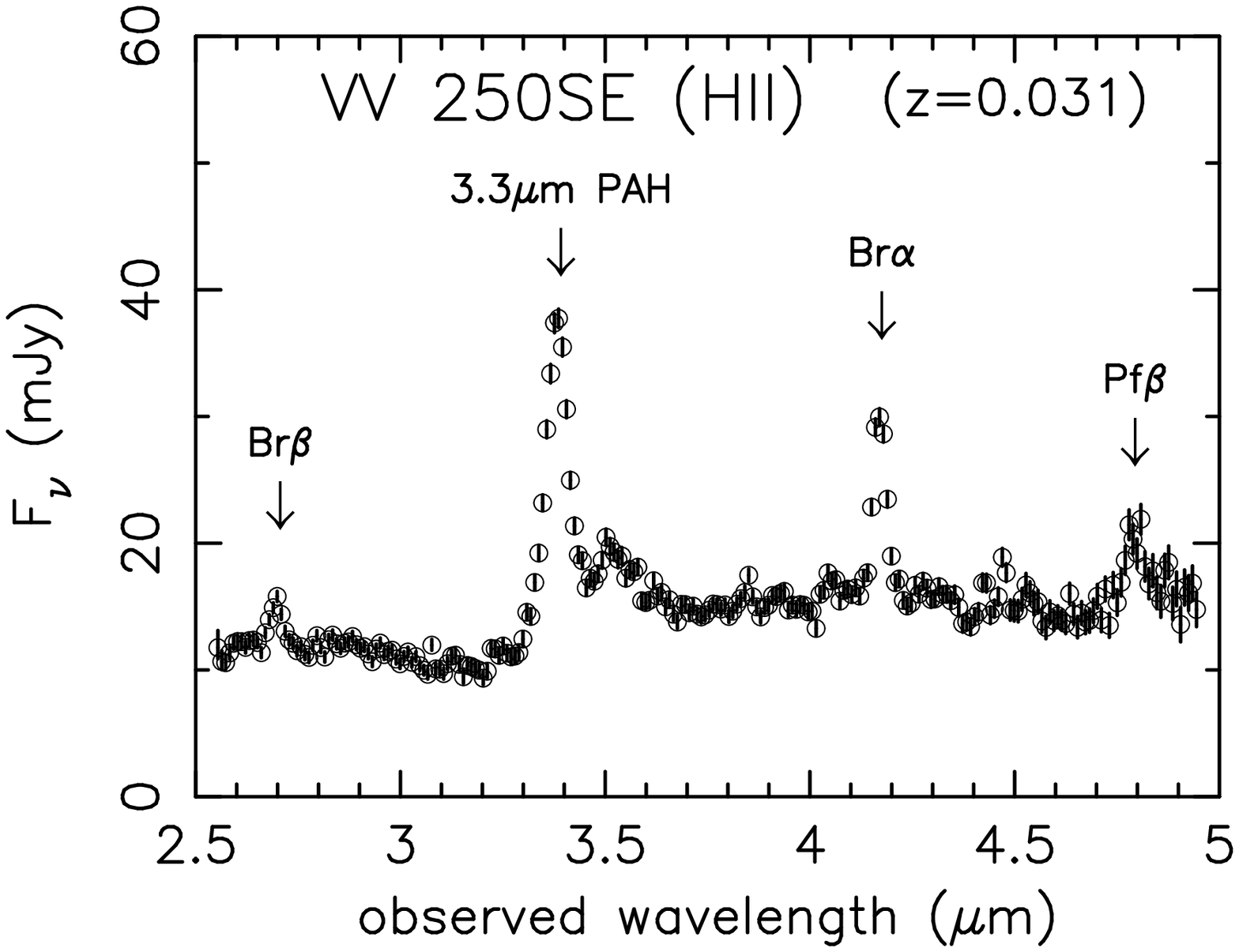}
\includegraphics[angle=0,scale=.27]{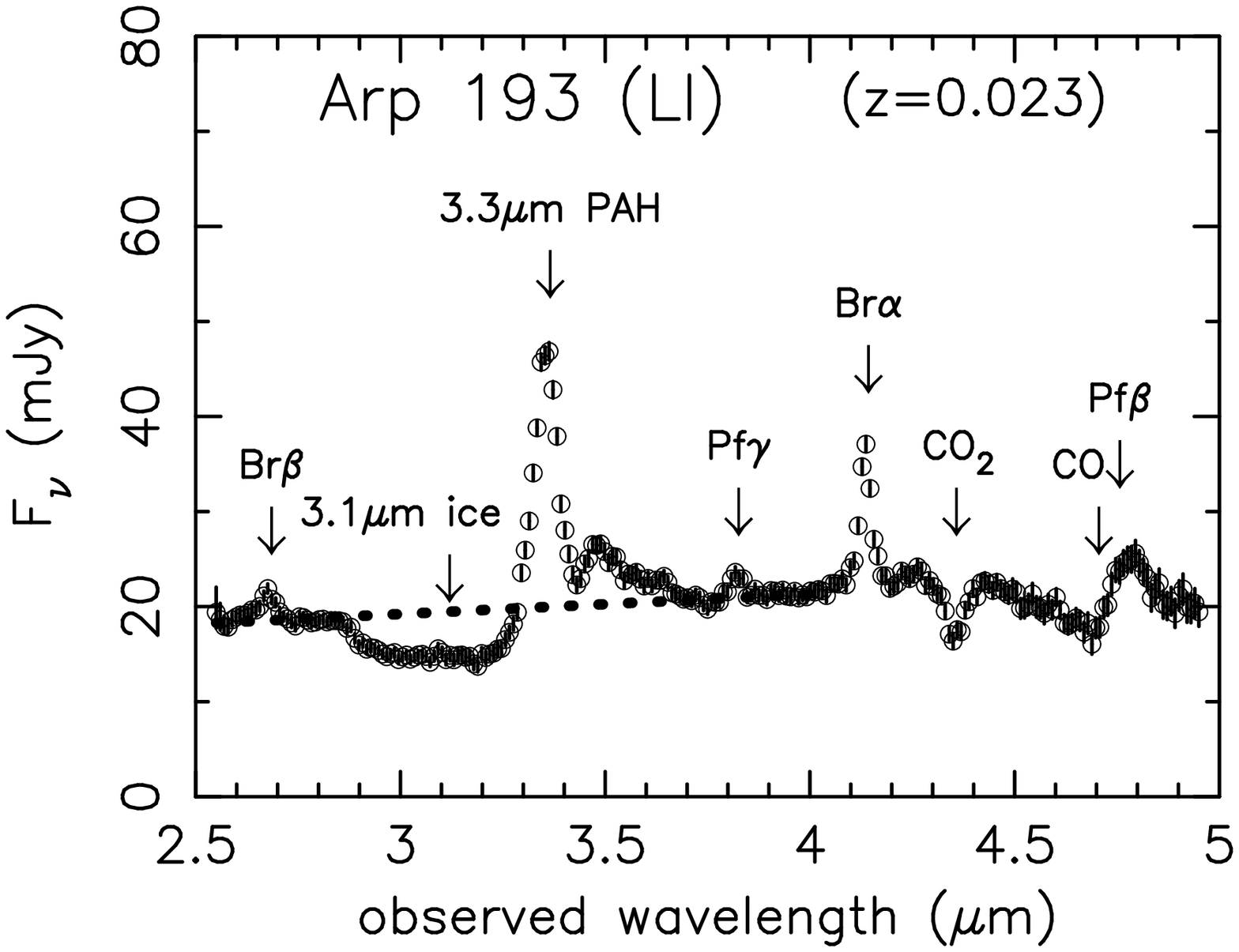} \\
\end{figure}

\clearpage

\begin{figure}
\includegraphics[angle=0,scale=.27]{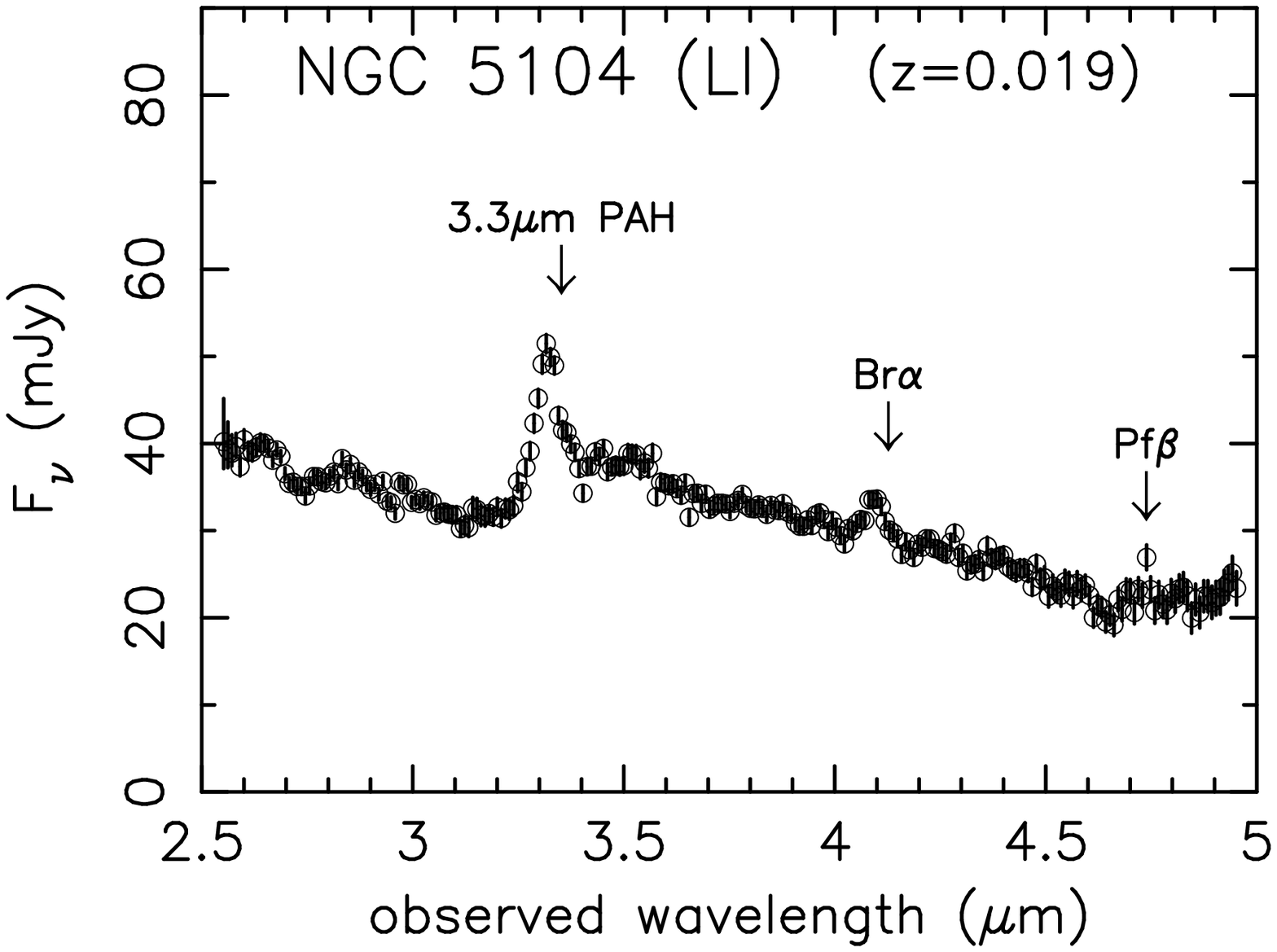} 
\includegraphics[angle=0,scale=.27]{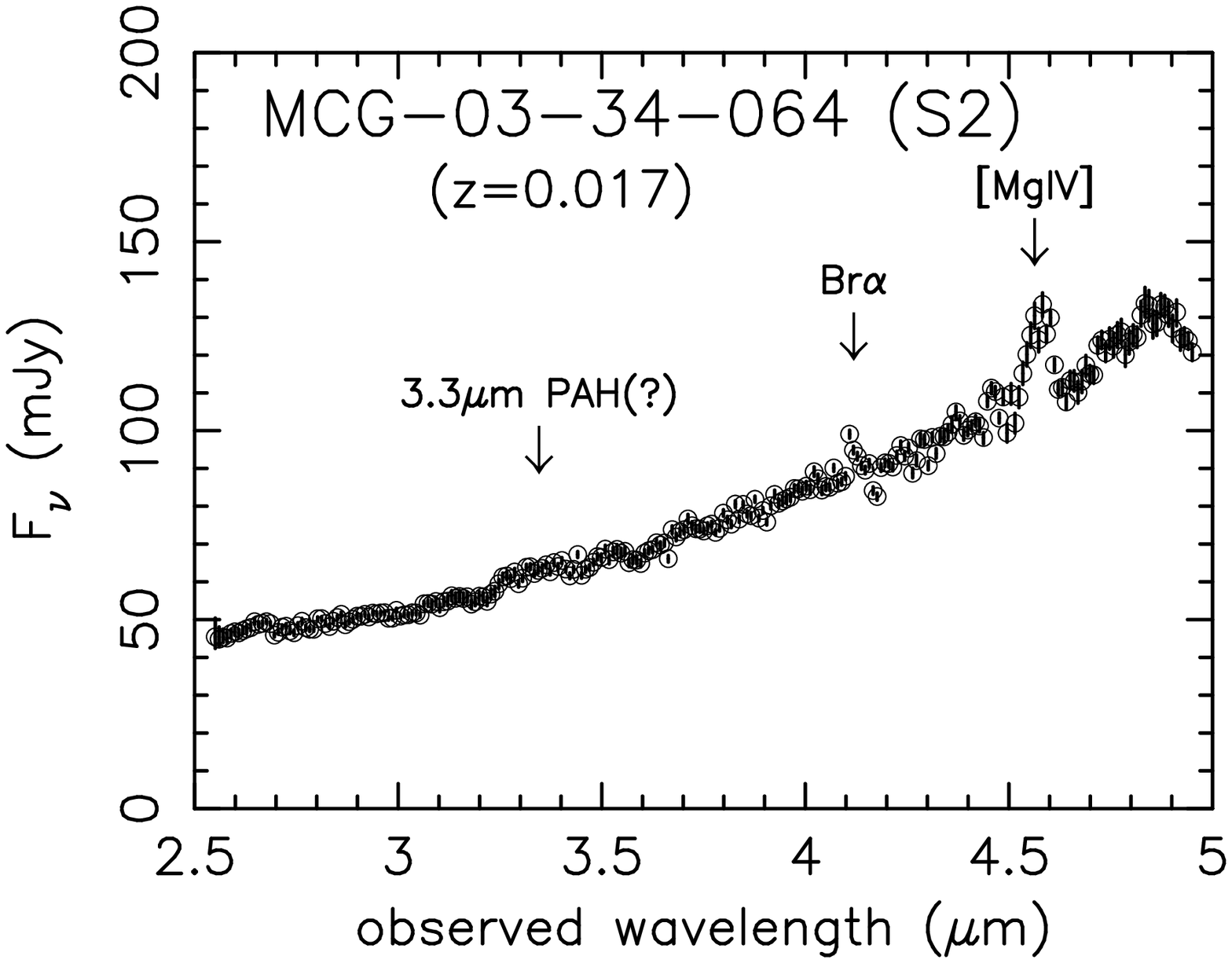} 
\includegraphics[angle=0,scale=.27]{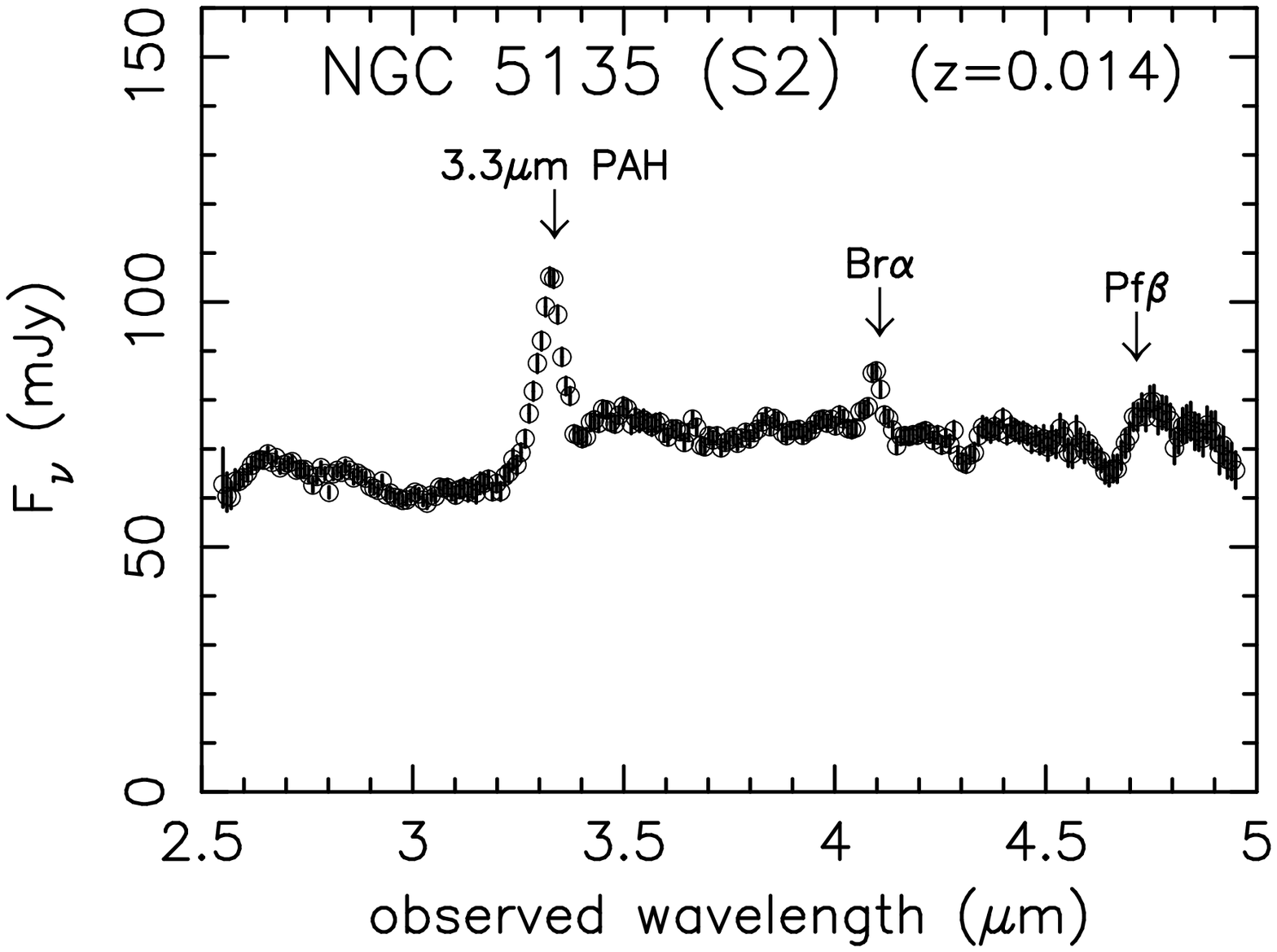} \\ 
\includegraphics[angle=0,scale=.27]{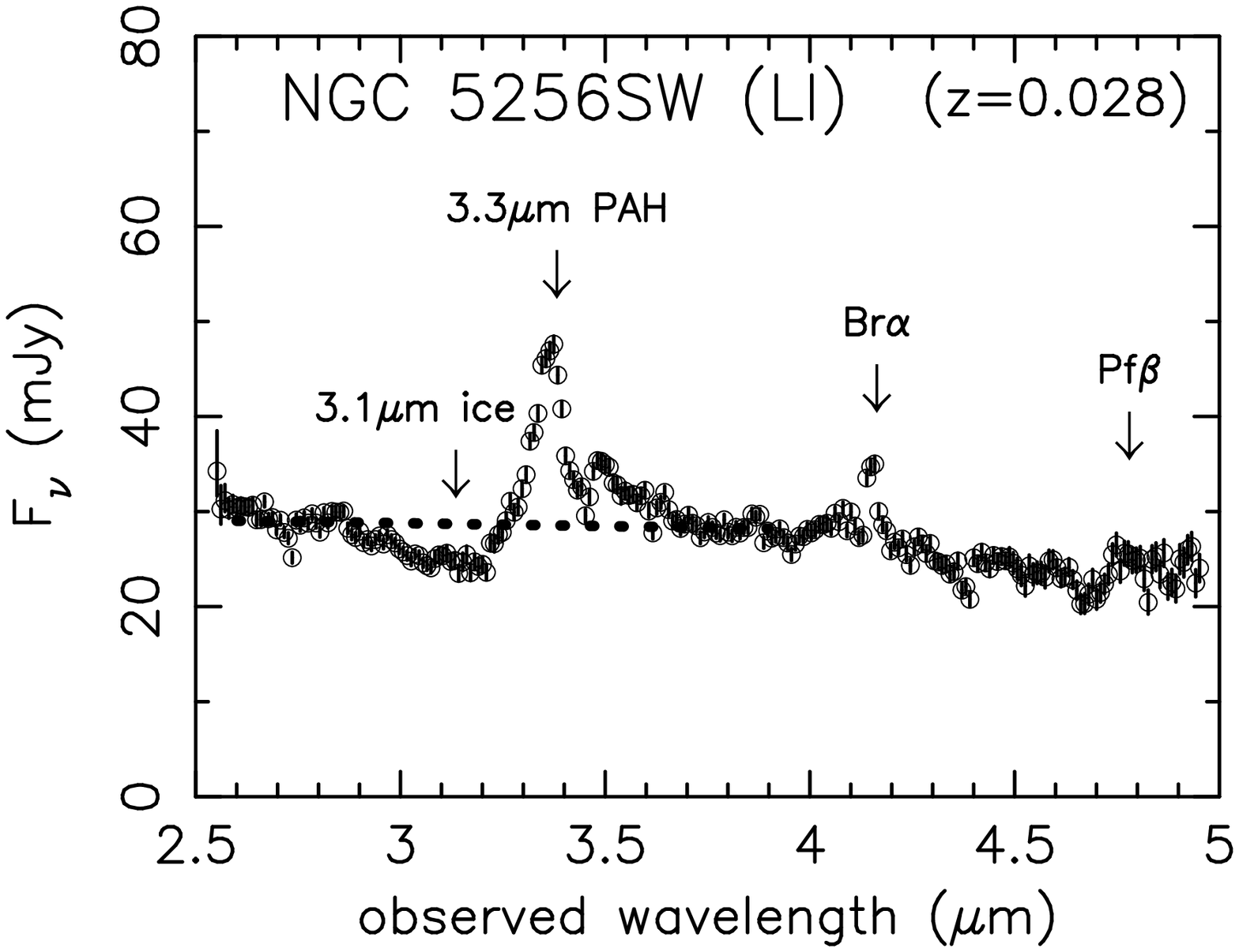}
\includegraphics[angle=0,scale=.27]{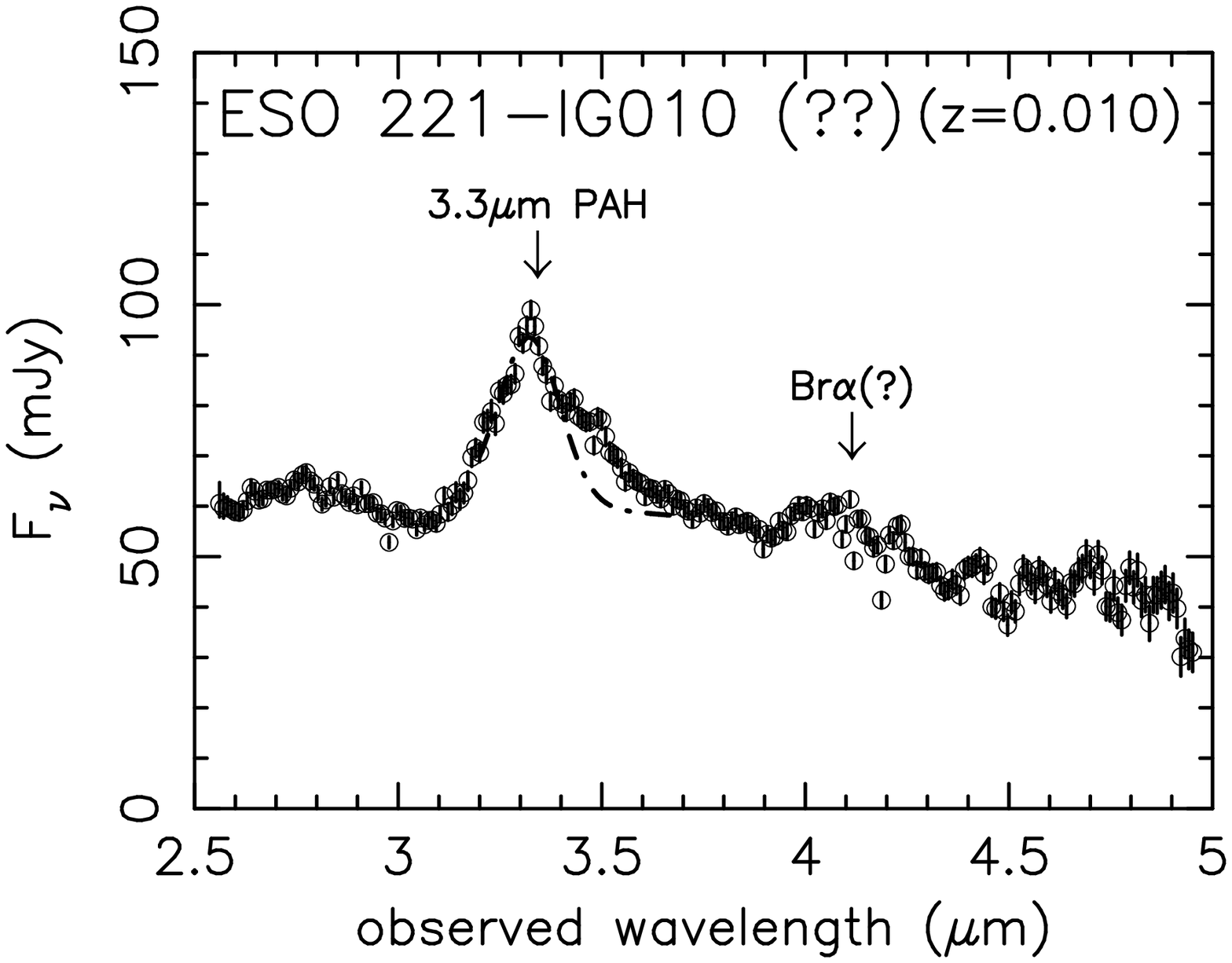}
\includegraphics[angle=0,scale=.27]{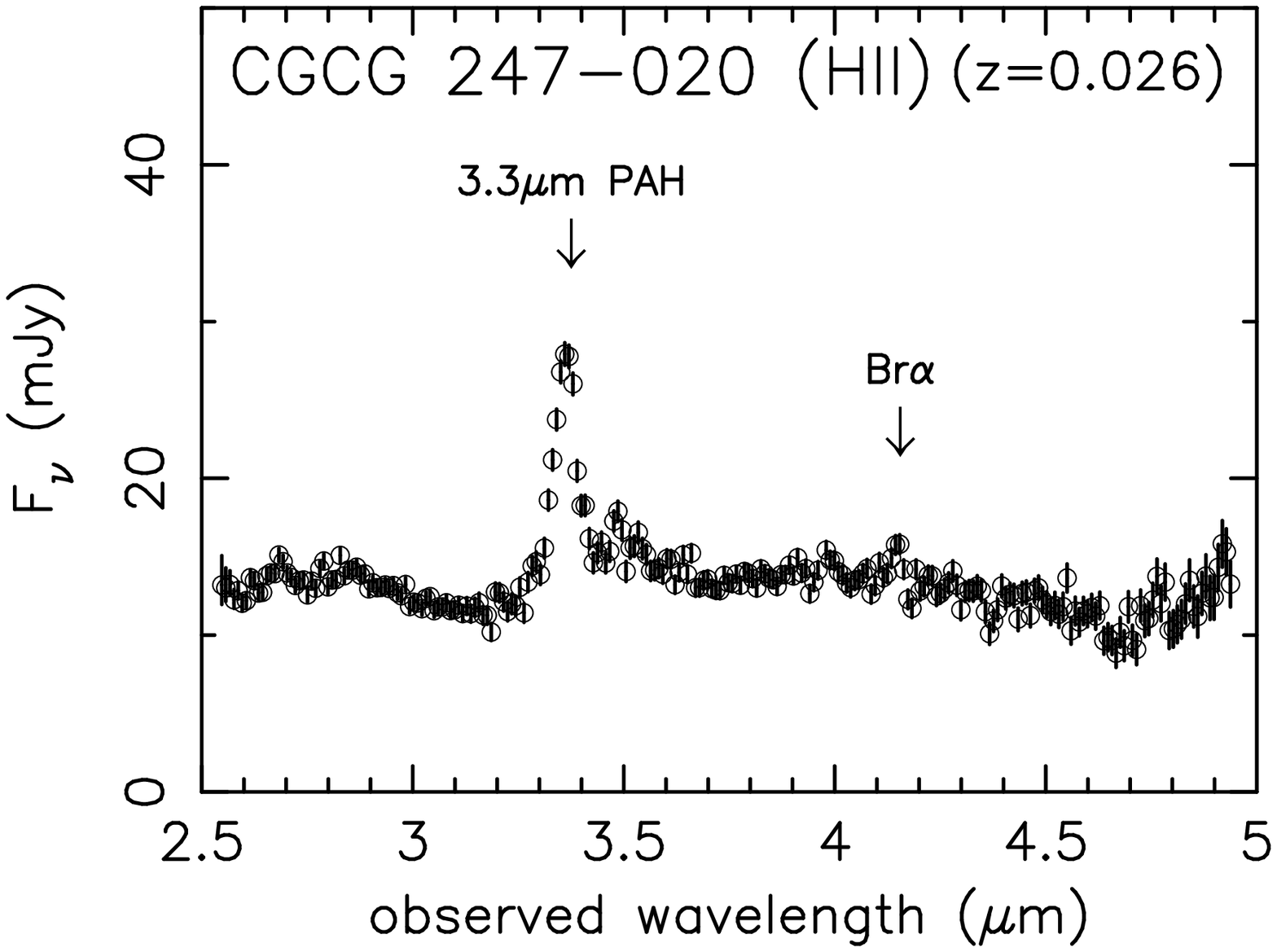} \\
\includegraphics[angle=0,scale=.27]{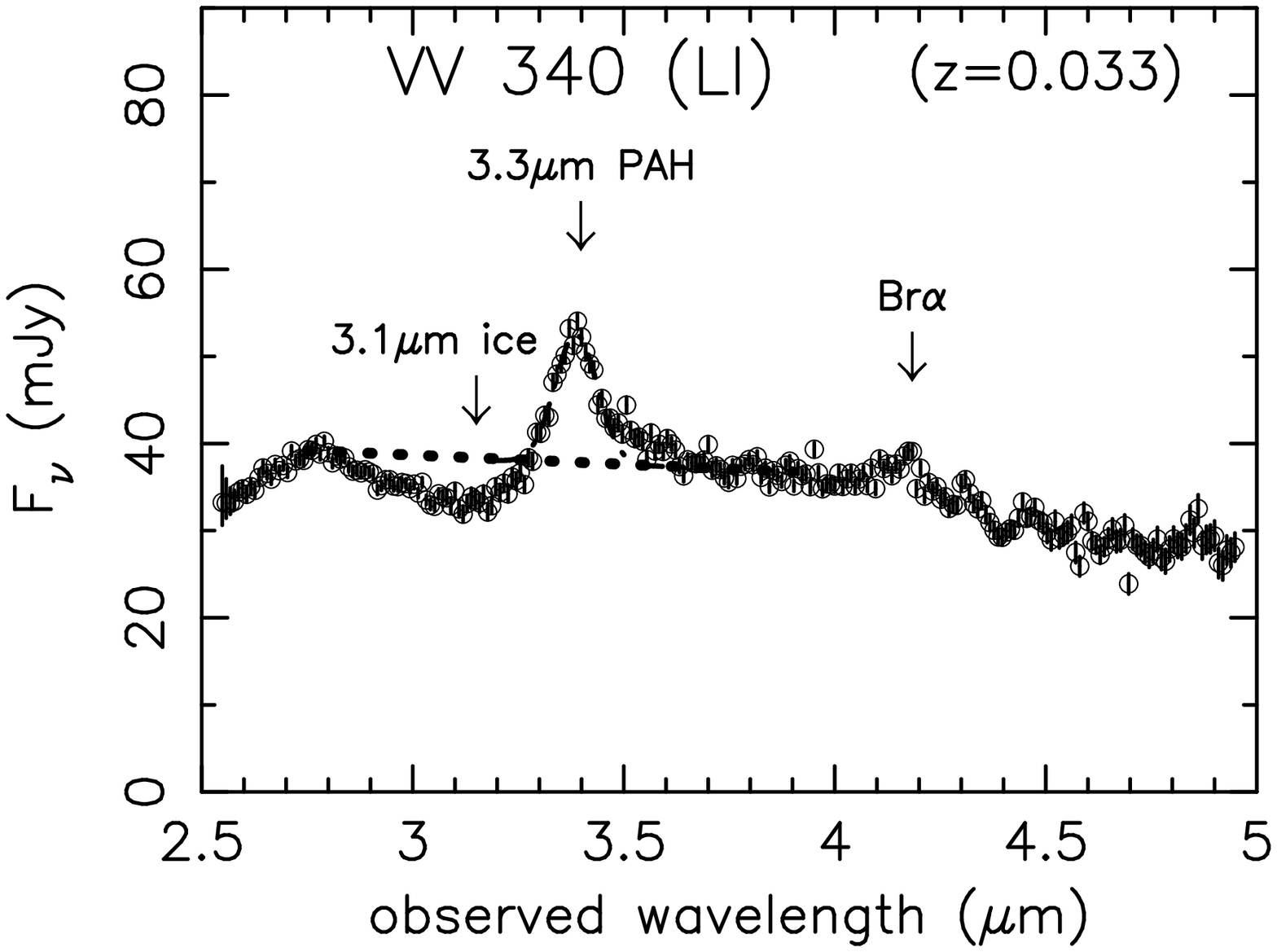}
\includegraphics[angle=0,scale=.27]{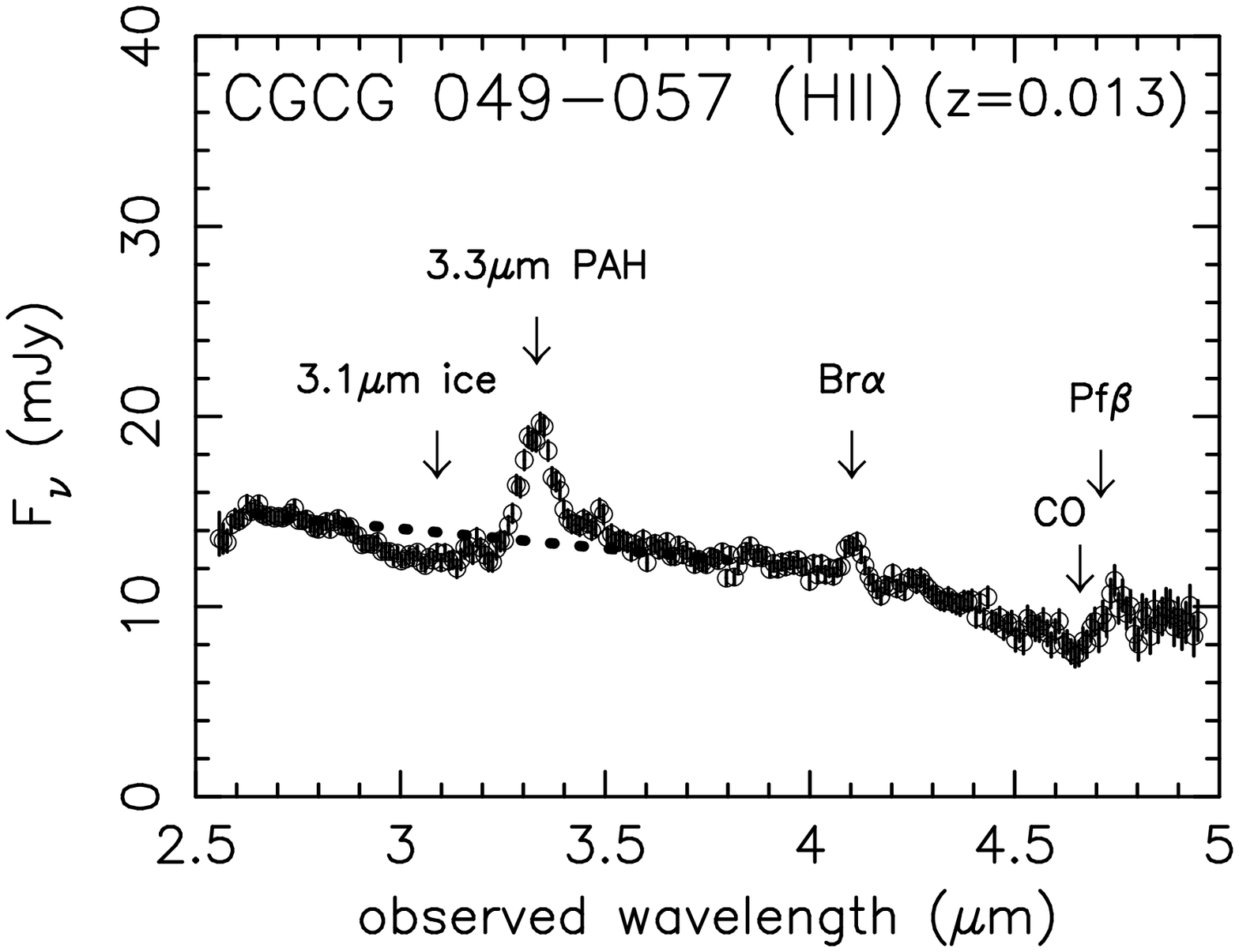}
\includegraphics[angle=0,scale=.27]{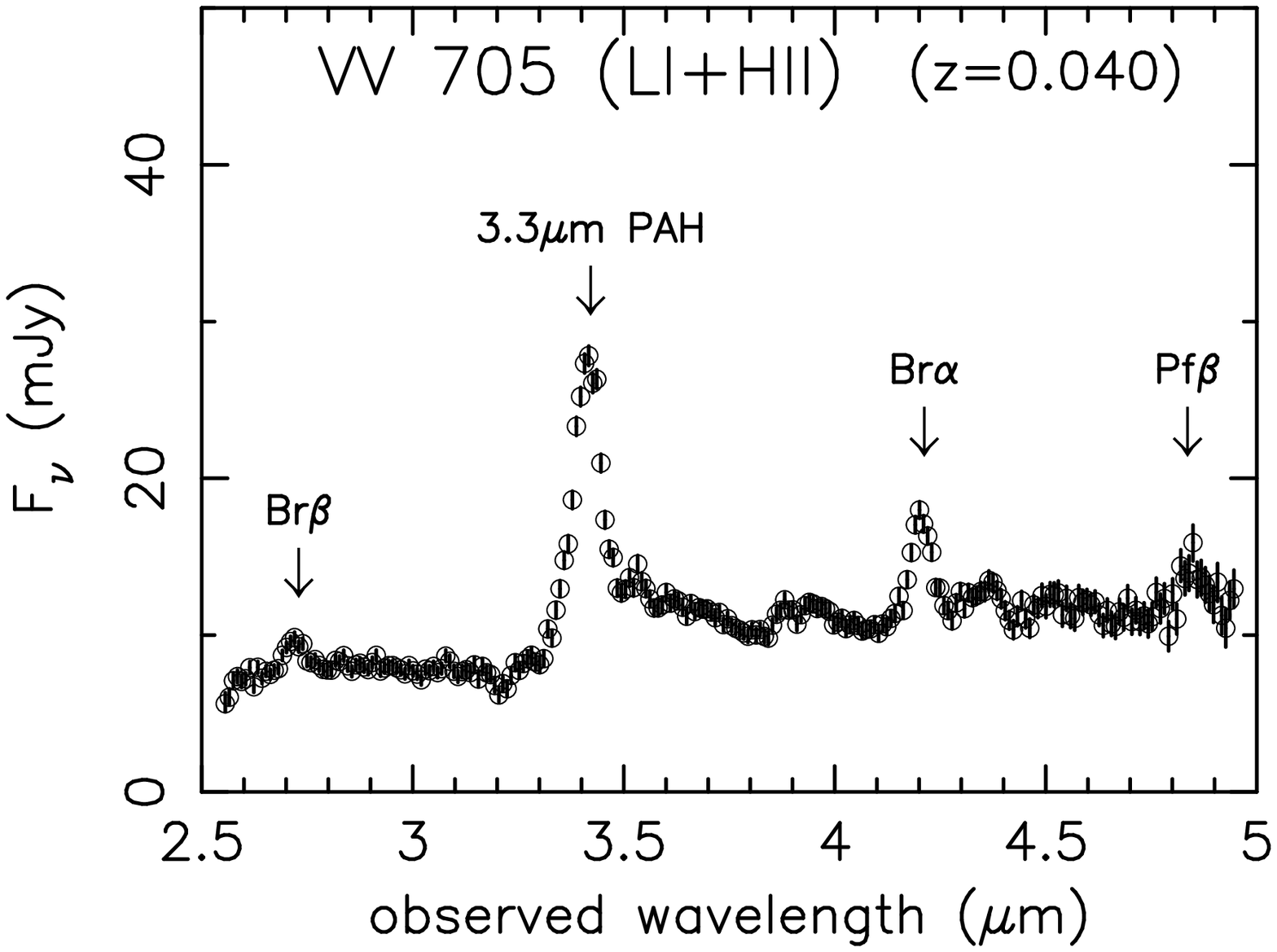} \\
\includegraphics[angle=0,scale=.27]{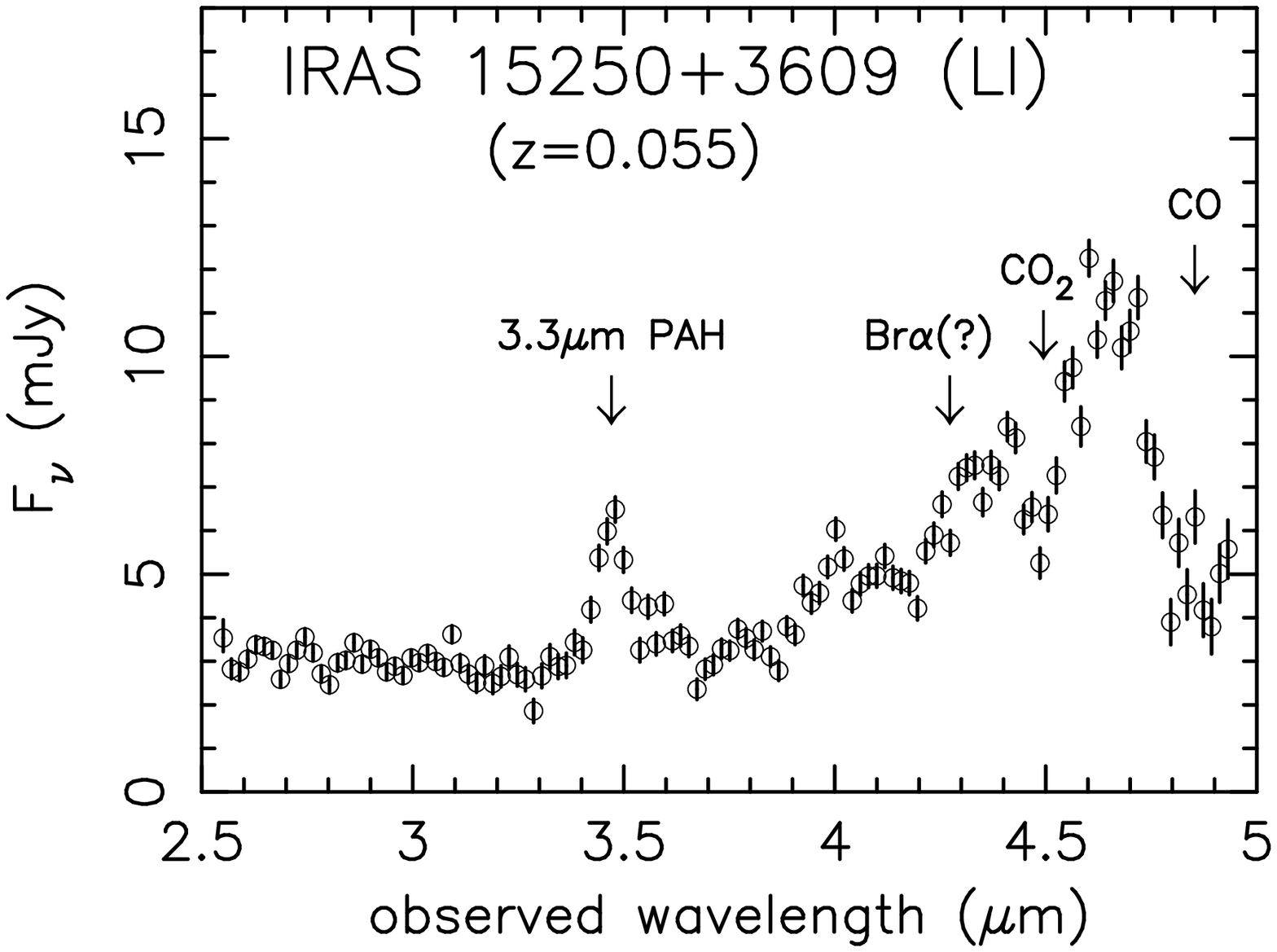}
\includegraphics[angle=0,scale=.27]{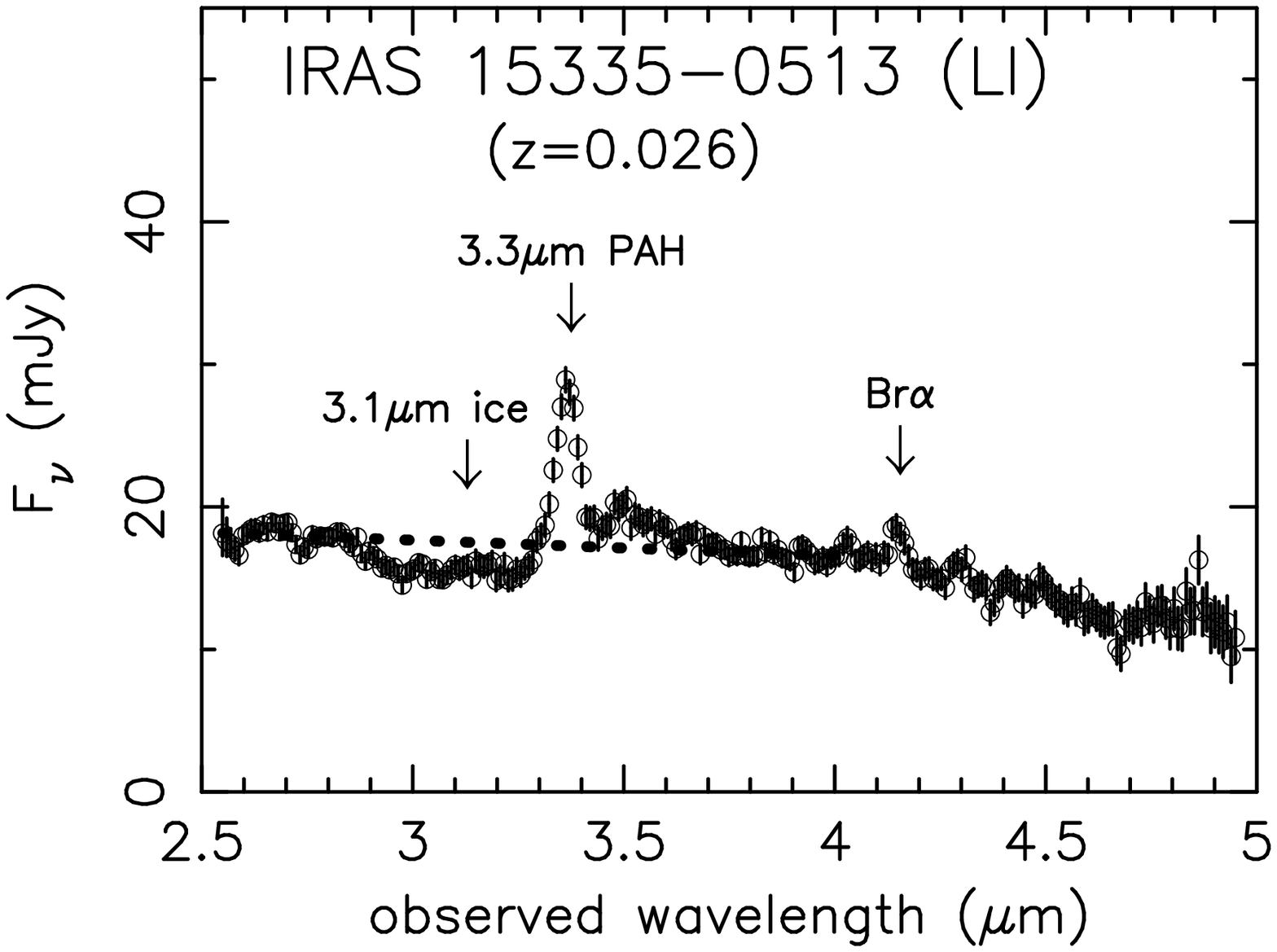}
\includegraphics[angle=0,scale=.27]{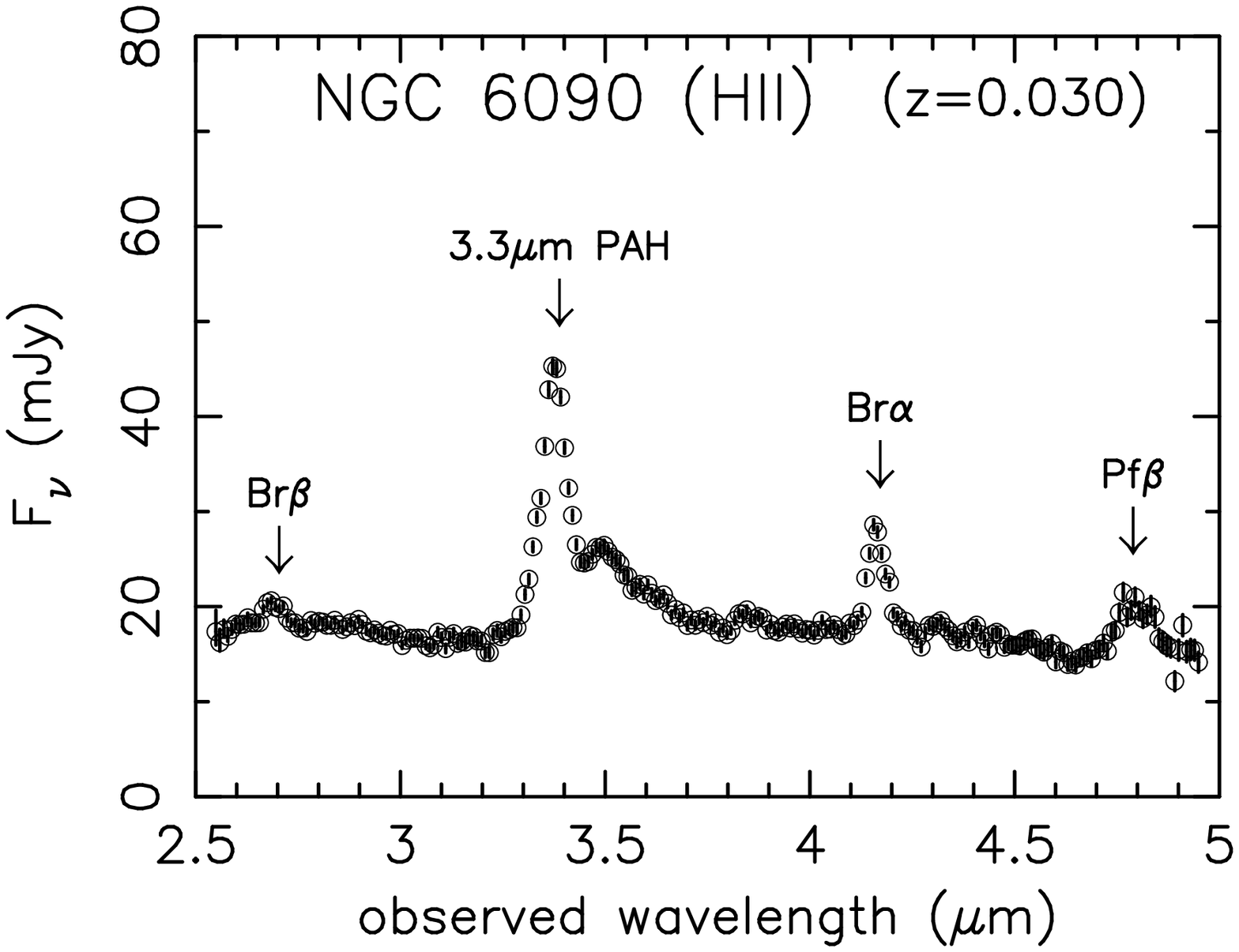} \\ 
\includegraphics[angle=0,scale=.27]{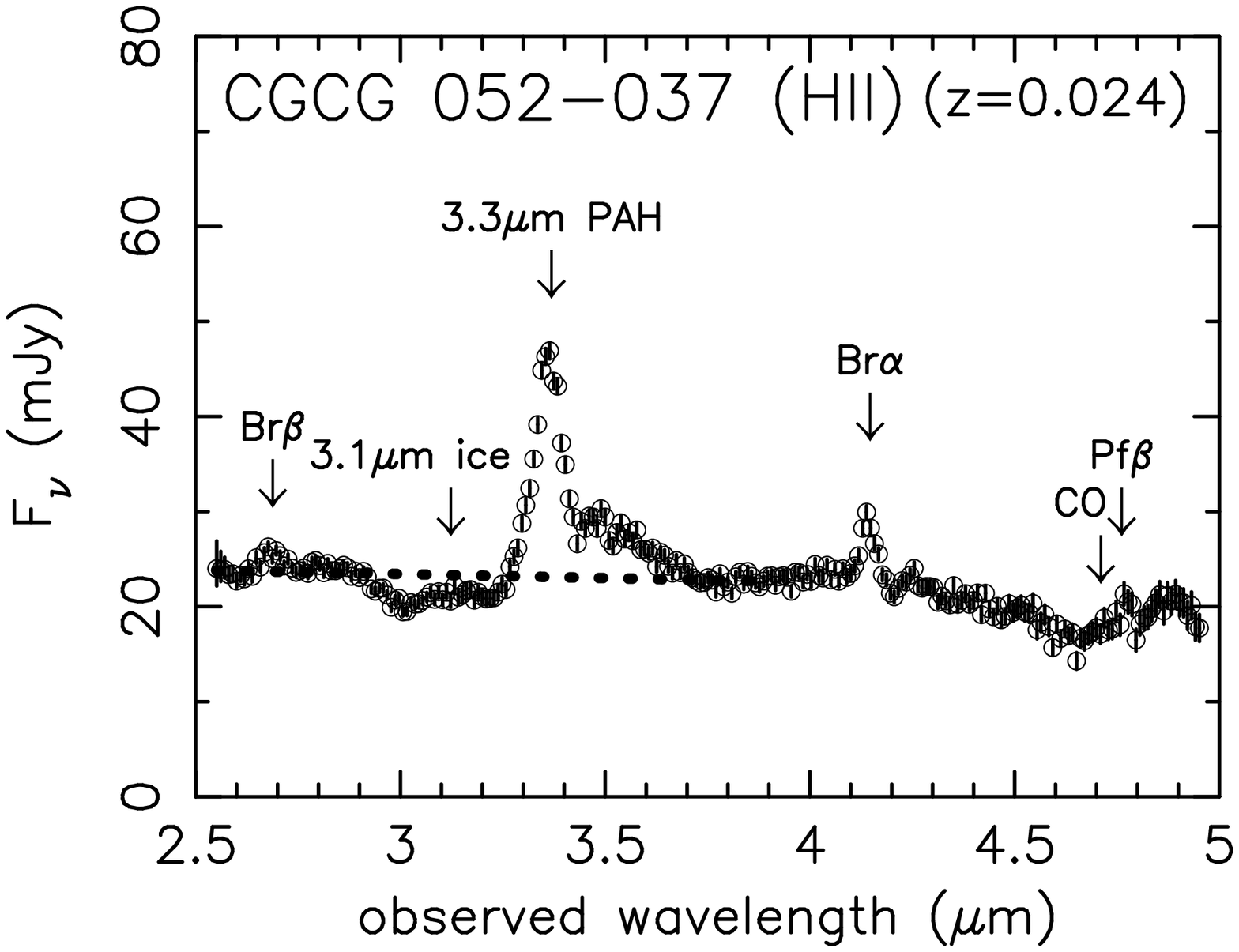}
\includegraphics[angle=0,scale=.27]{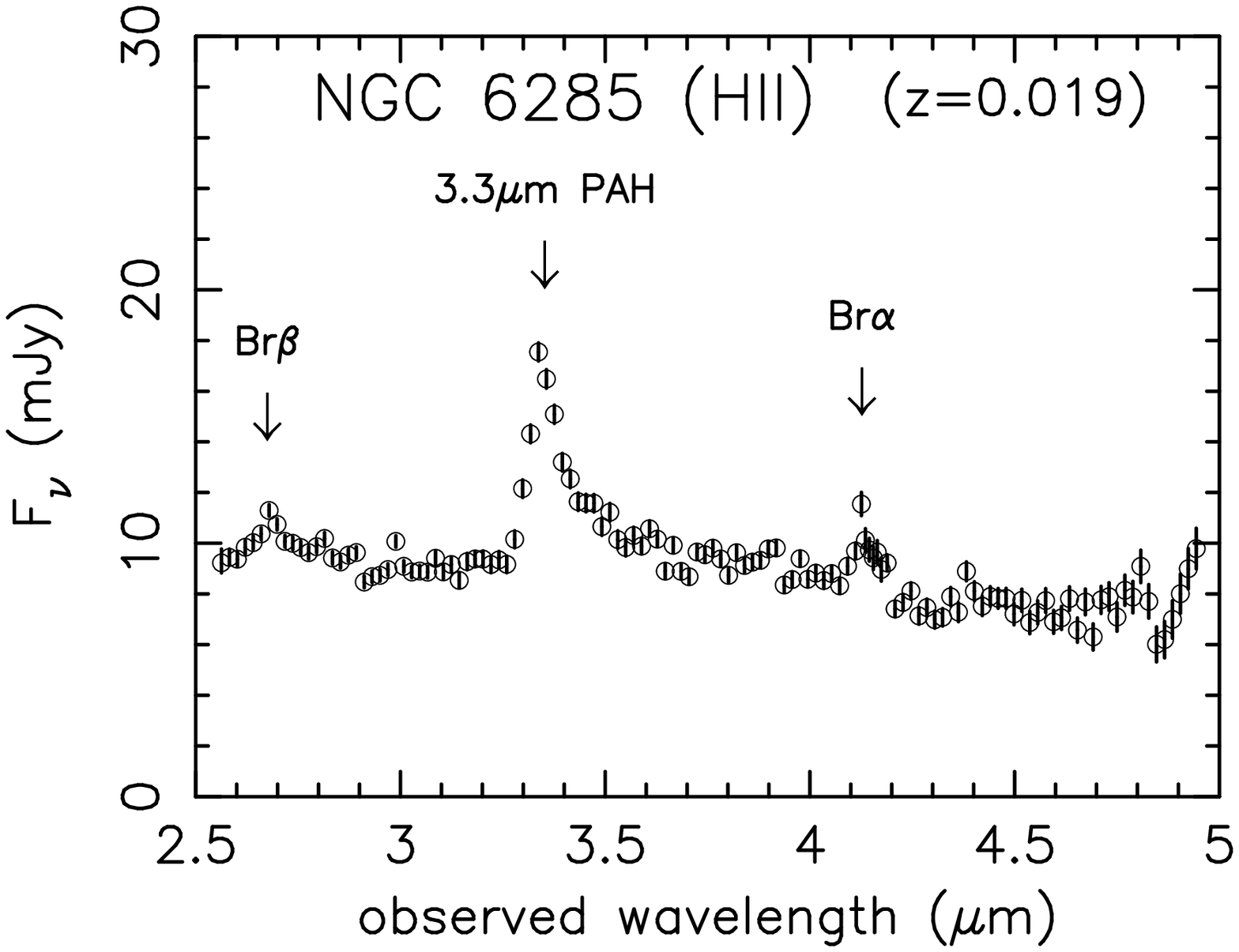}
\includegraphics[angle=0,scale=.27]{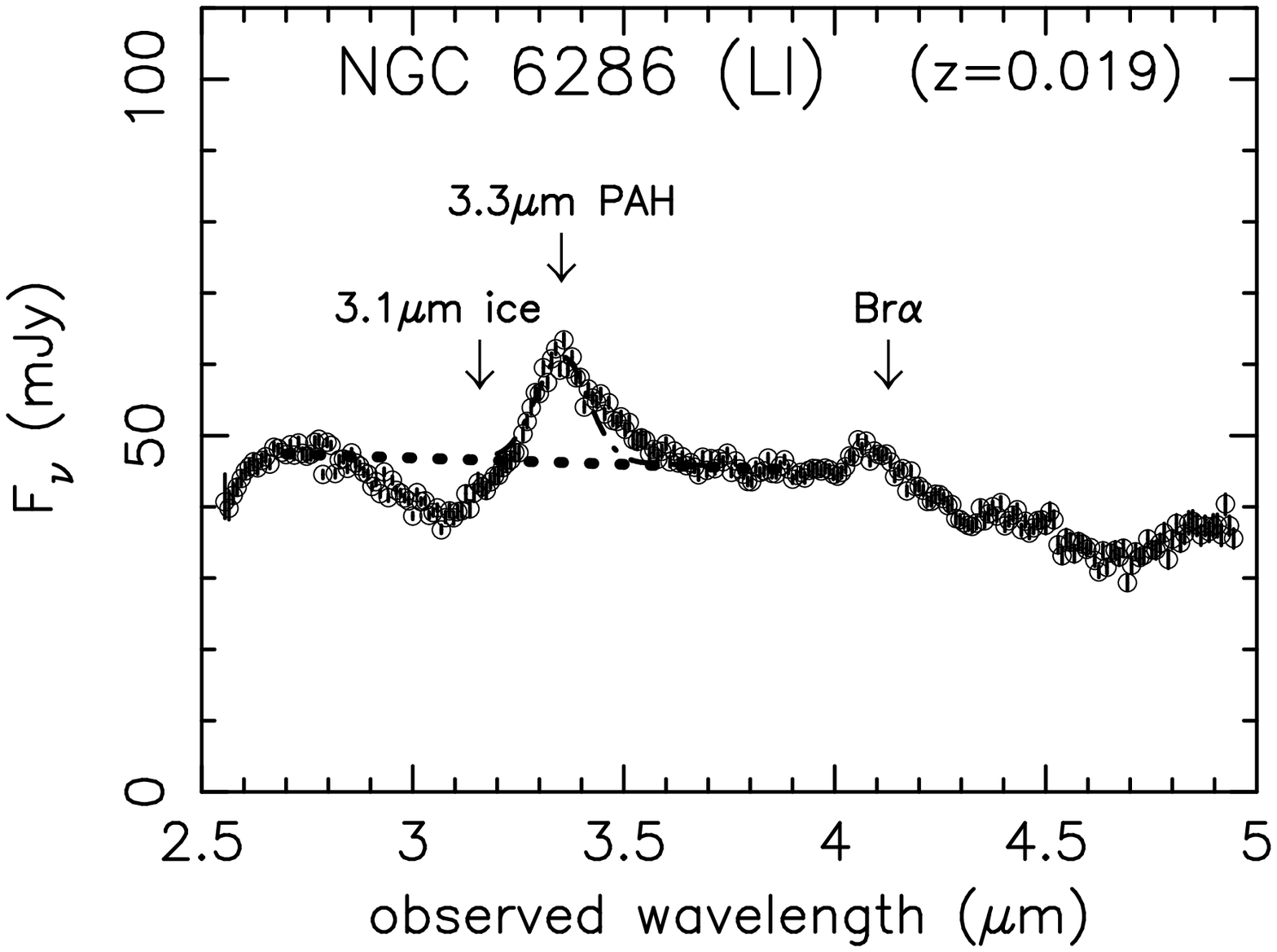} \\
\end{figure}

\clearpage

\begin{figure}
\includegraphics[angle=0,scale=.27]{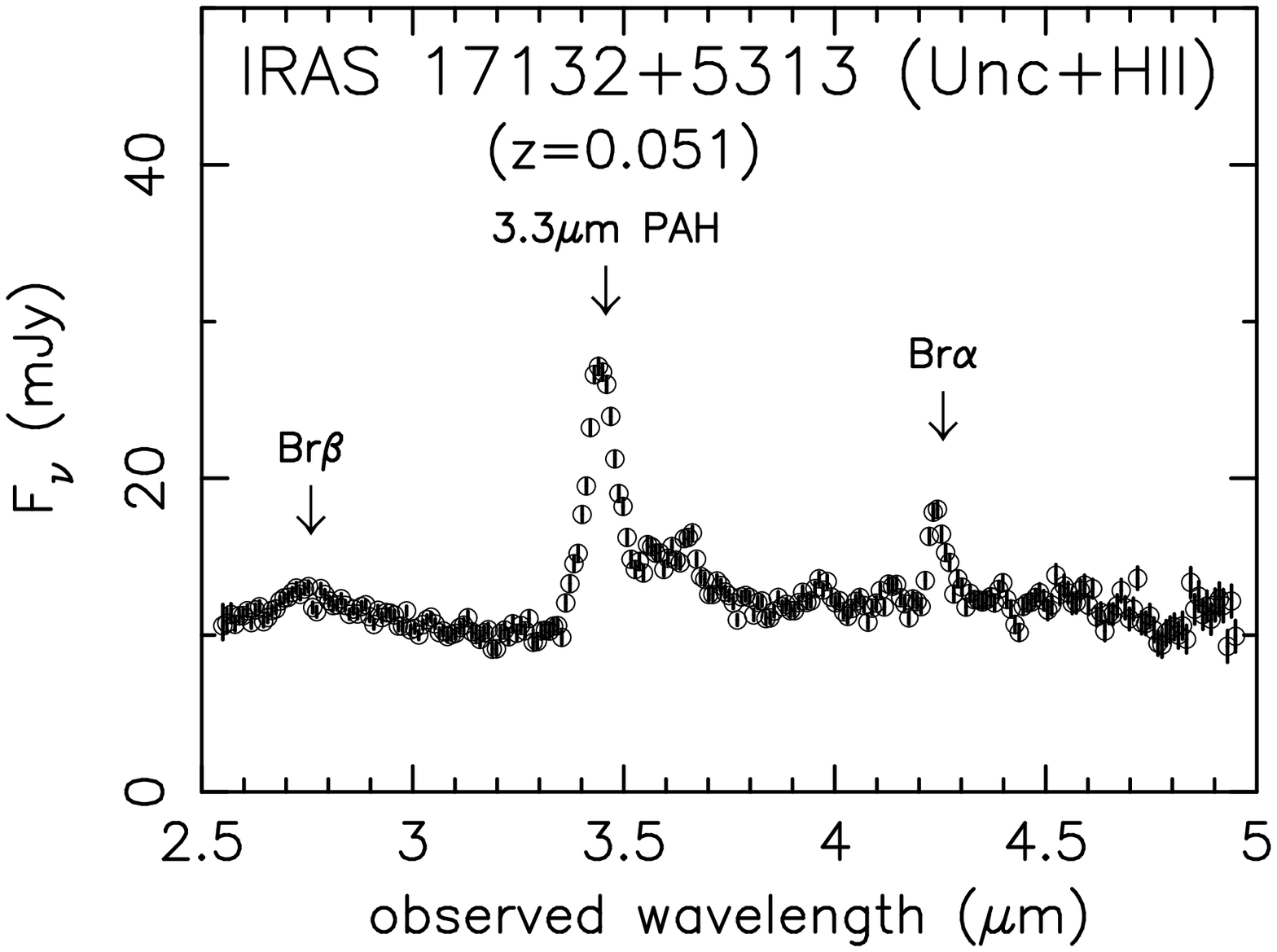}
\includegraphics[angle=0,scale=.27]{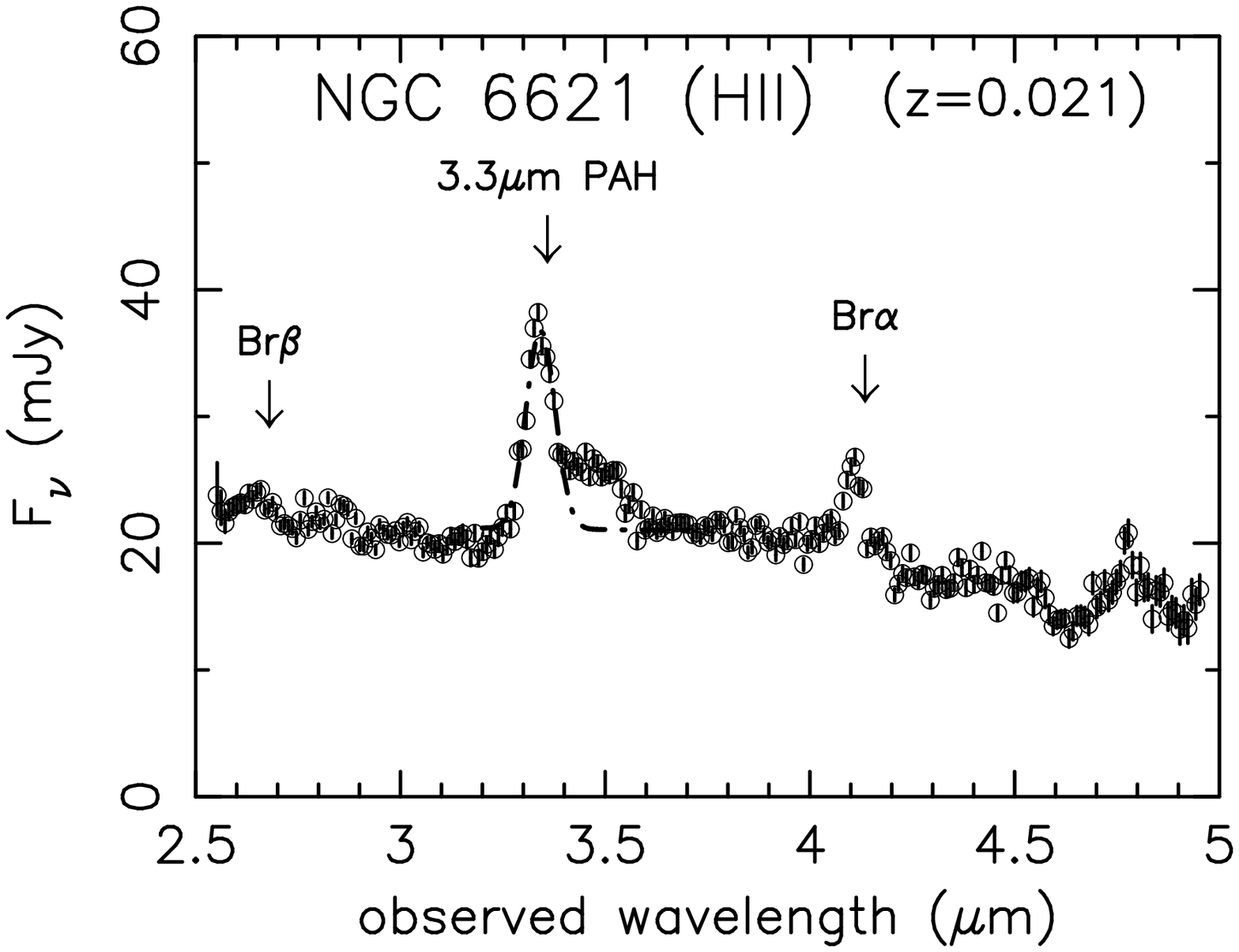}
\includegraphics[angle=0,scale=.27]{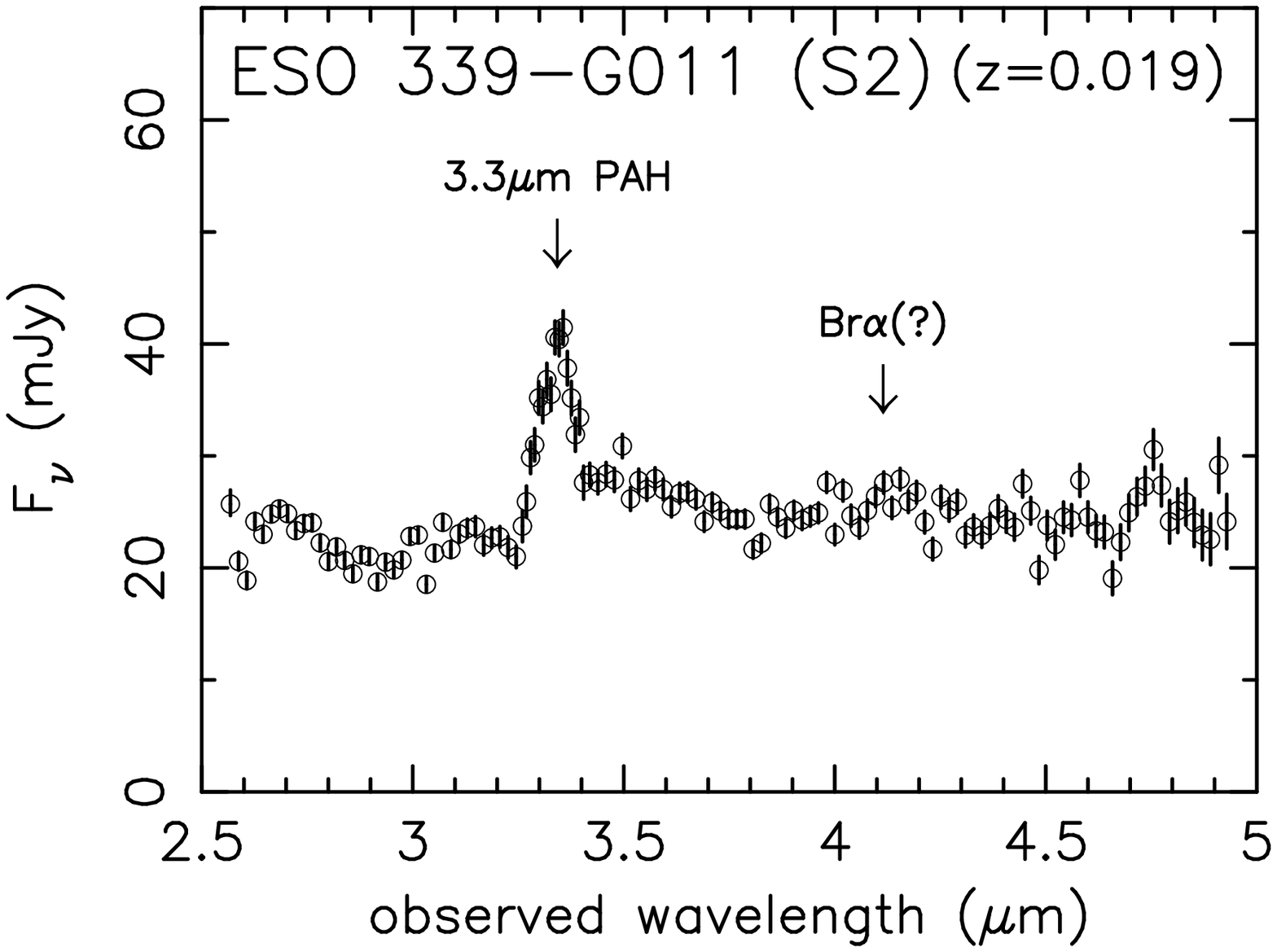} \\
\includegraphics[angle=0,scale=.27]{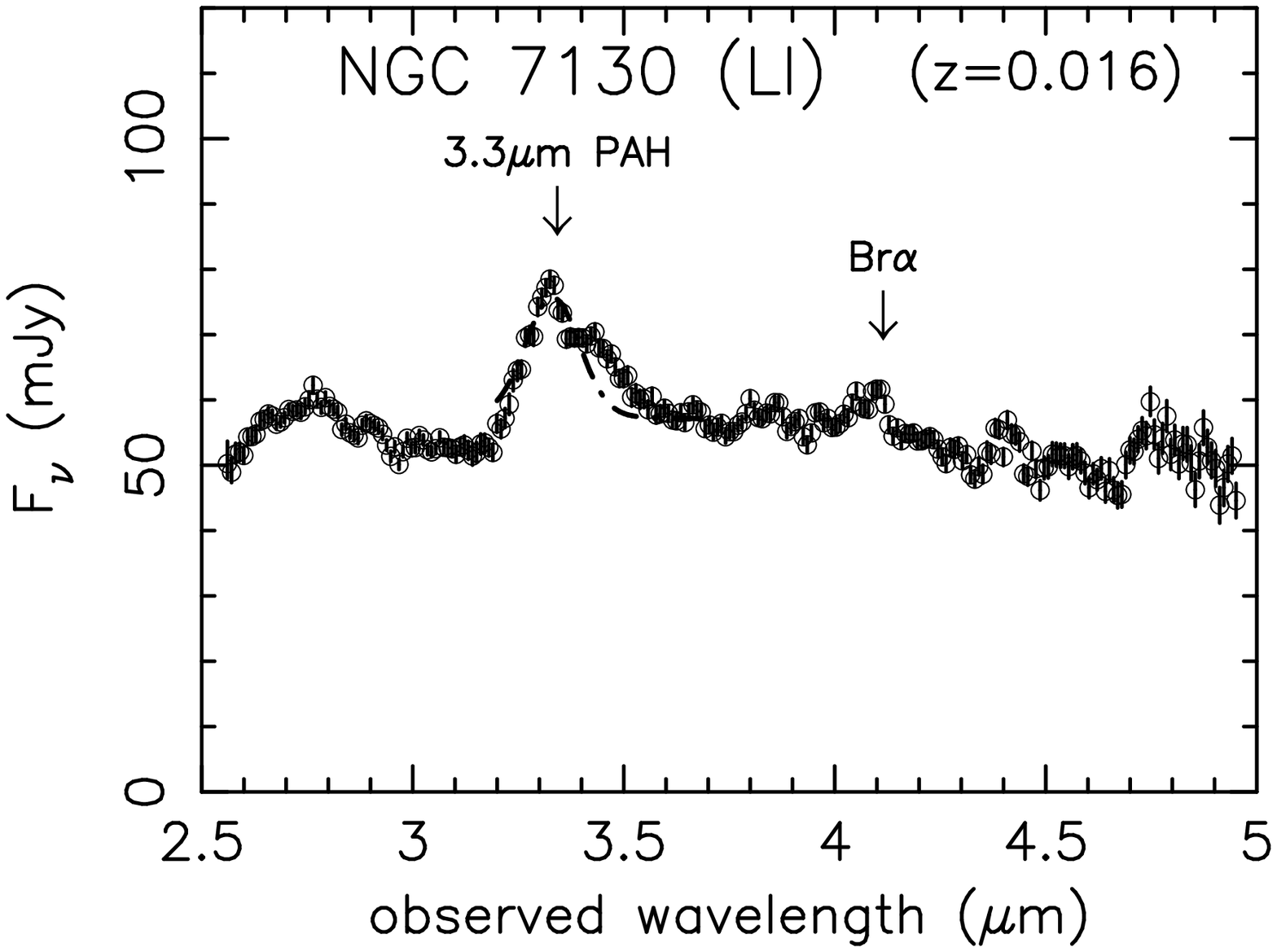}
\includegraphics[angle=0,scale=.27]{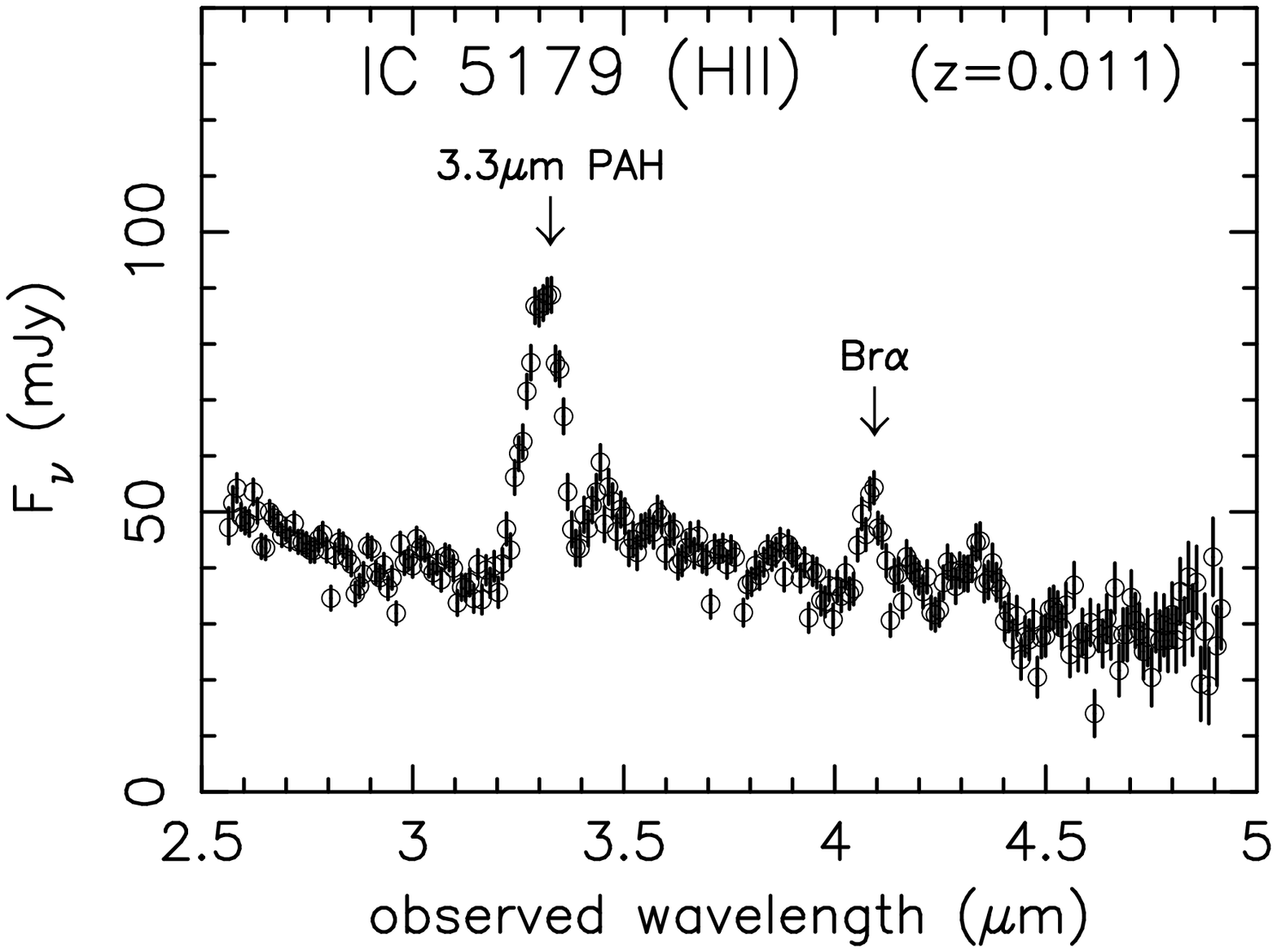}
\includegraphics[angle=0,scale=.27]{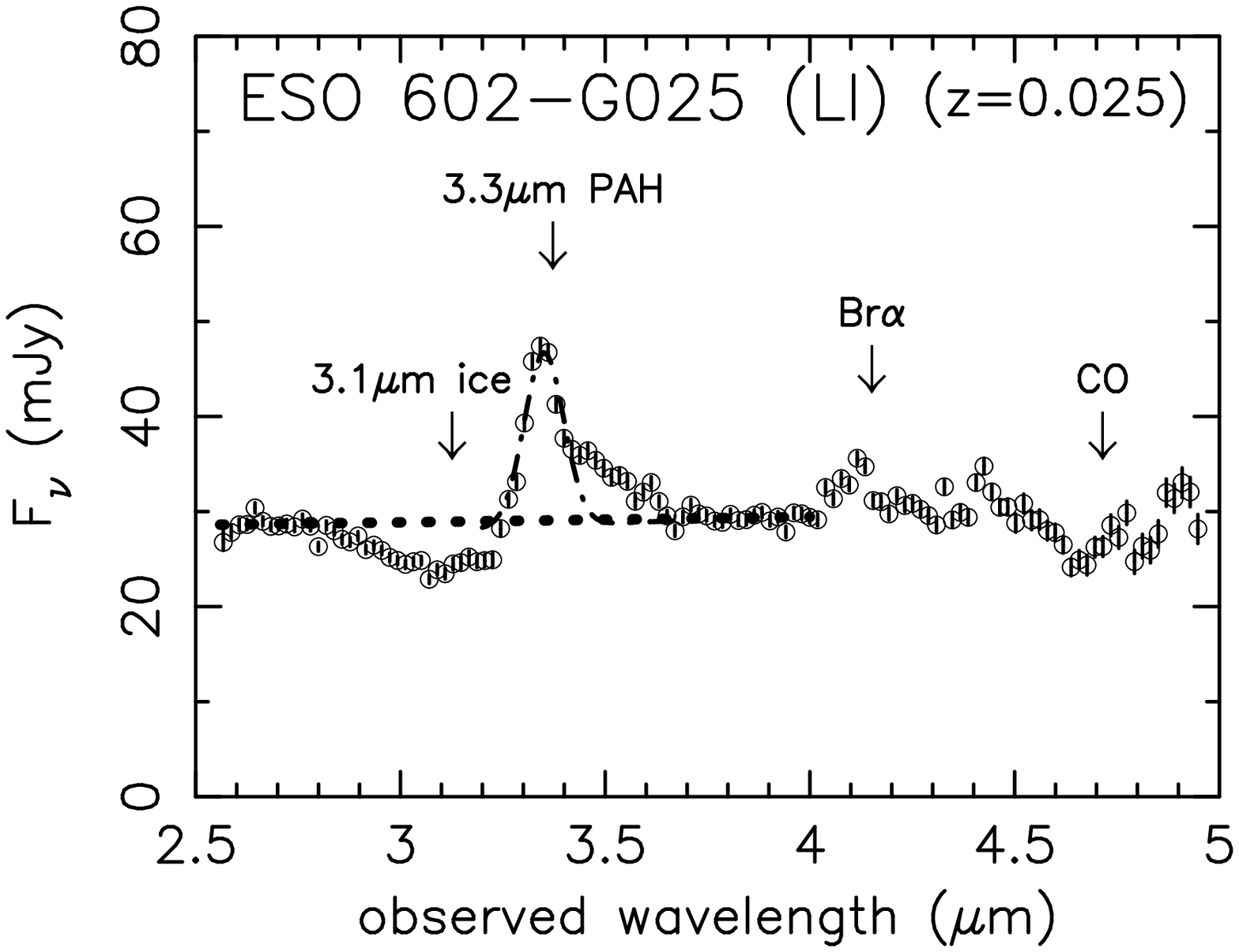} \\
\includegraphics[angle=0,scale=.27]{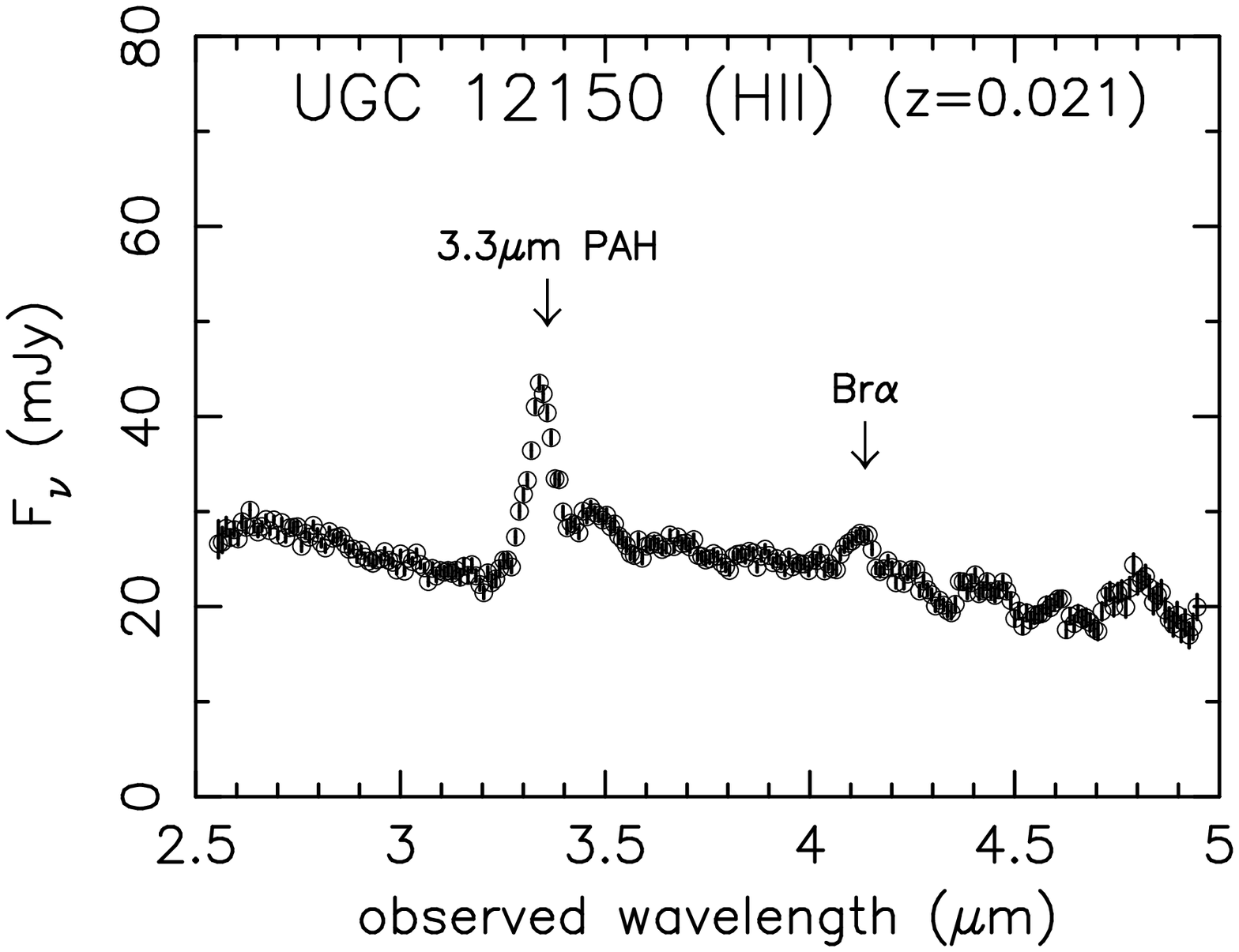}
\includegraphics[angle=0,scale=.27]{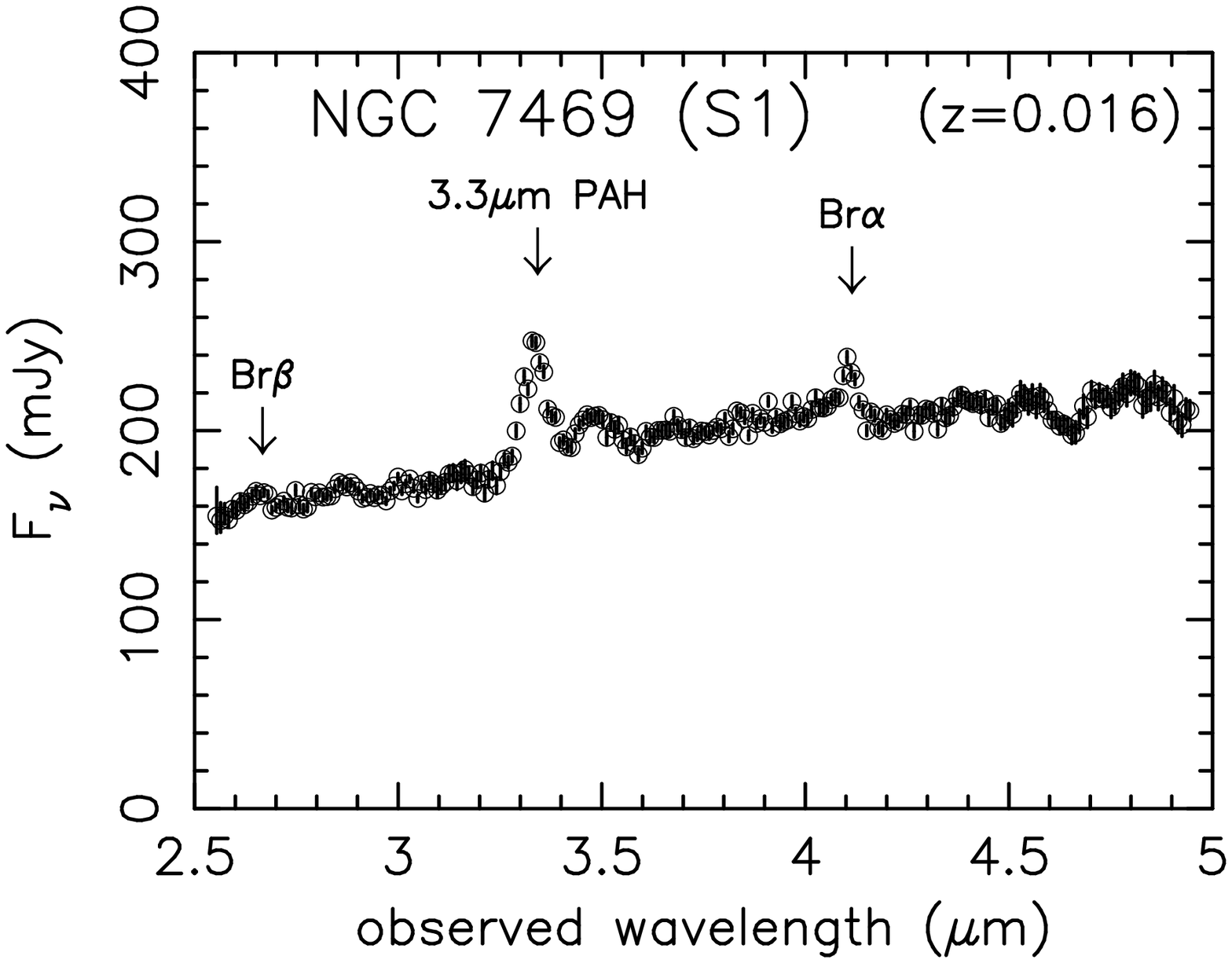}
\includegraphics[angle=0,scale=.27]{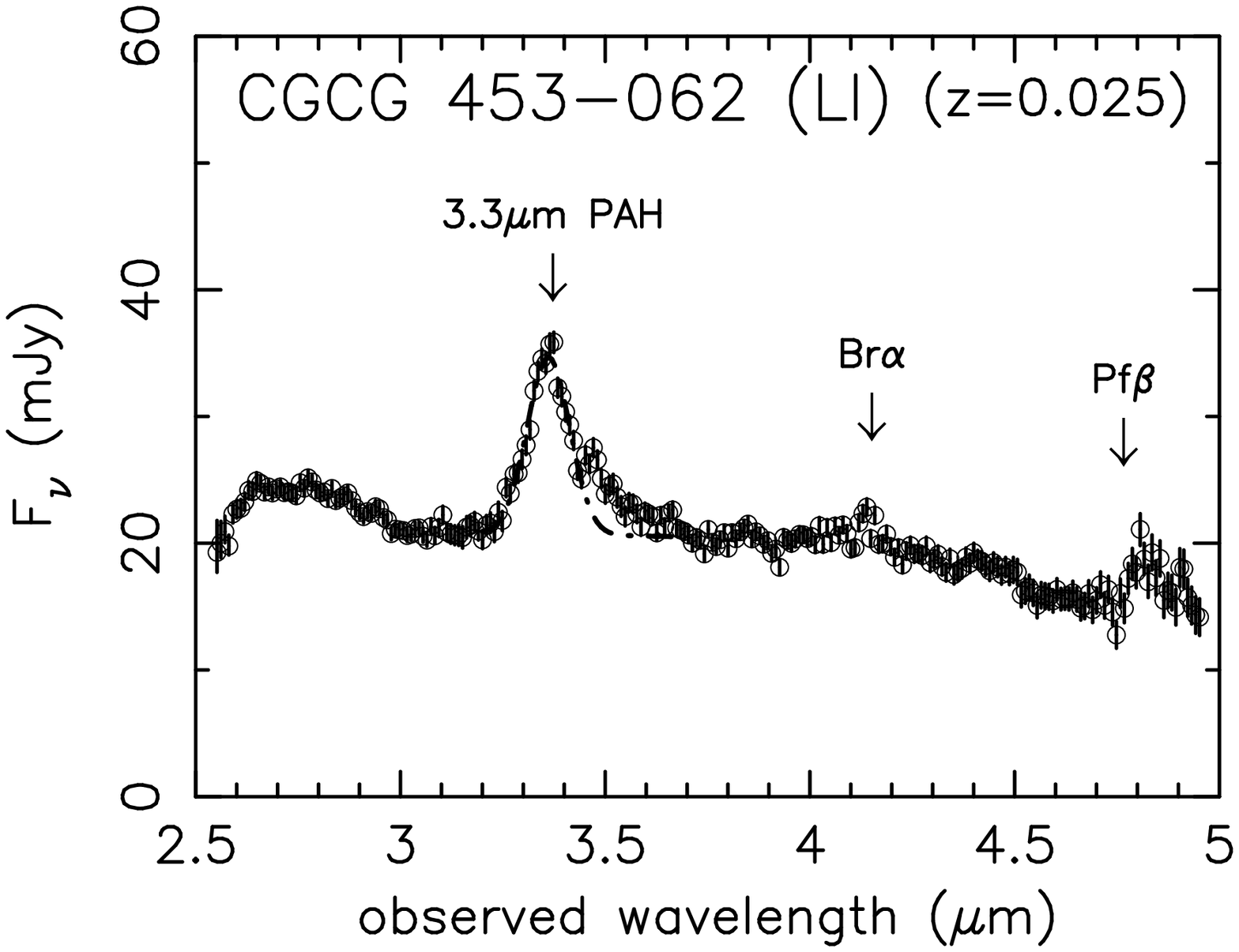} \\
\includegraphics[angle=0,scale=.27]{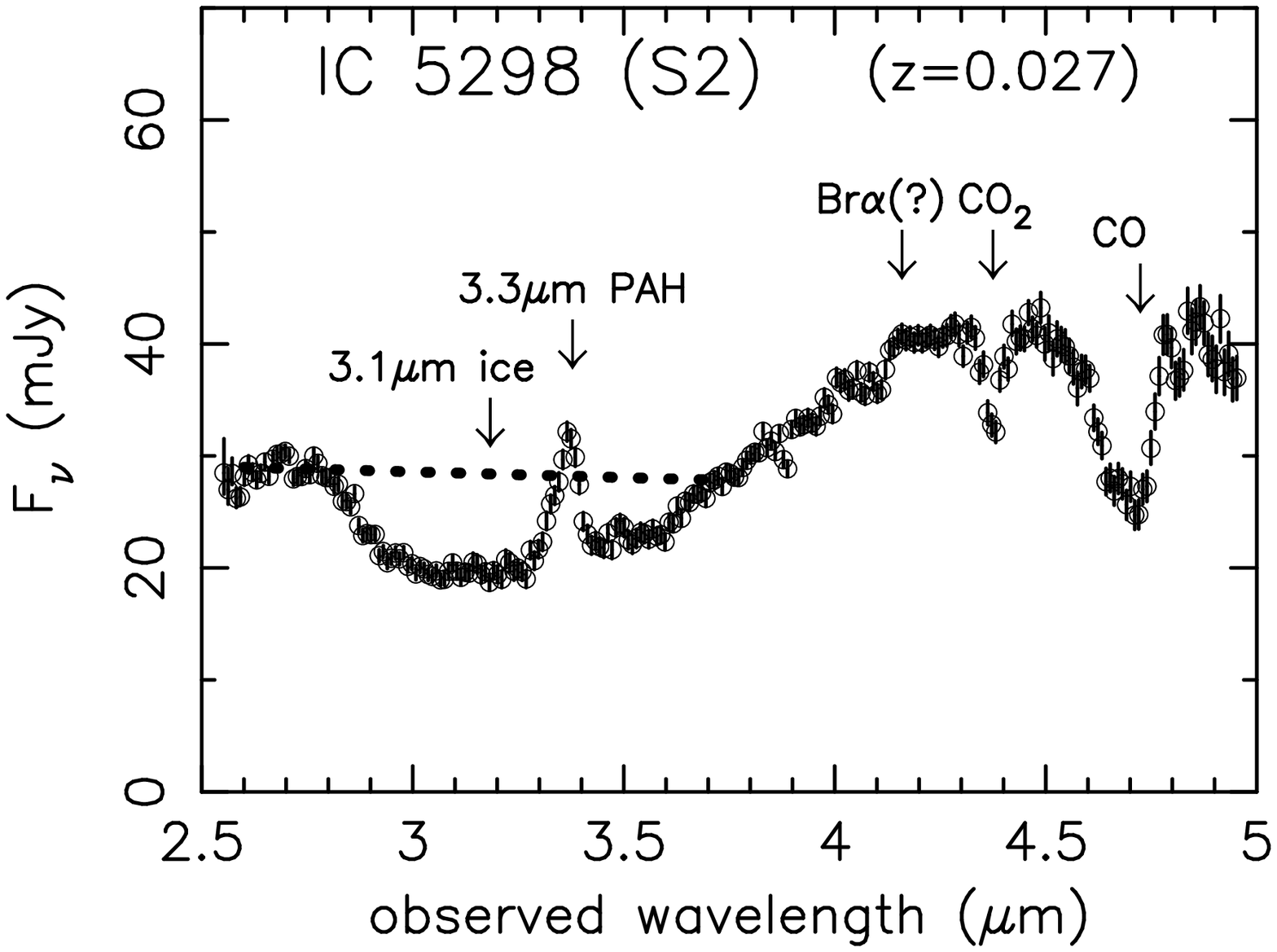}
\includegraphics[angle=0,scale=.27]{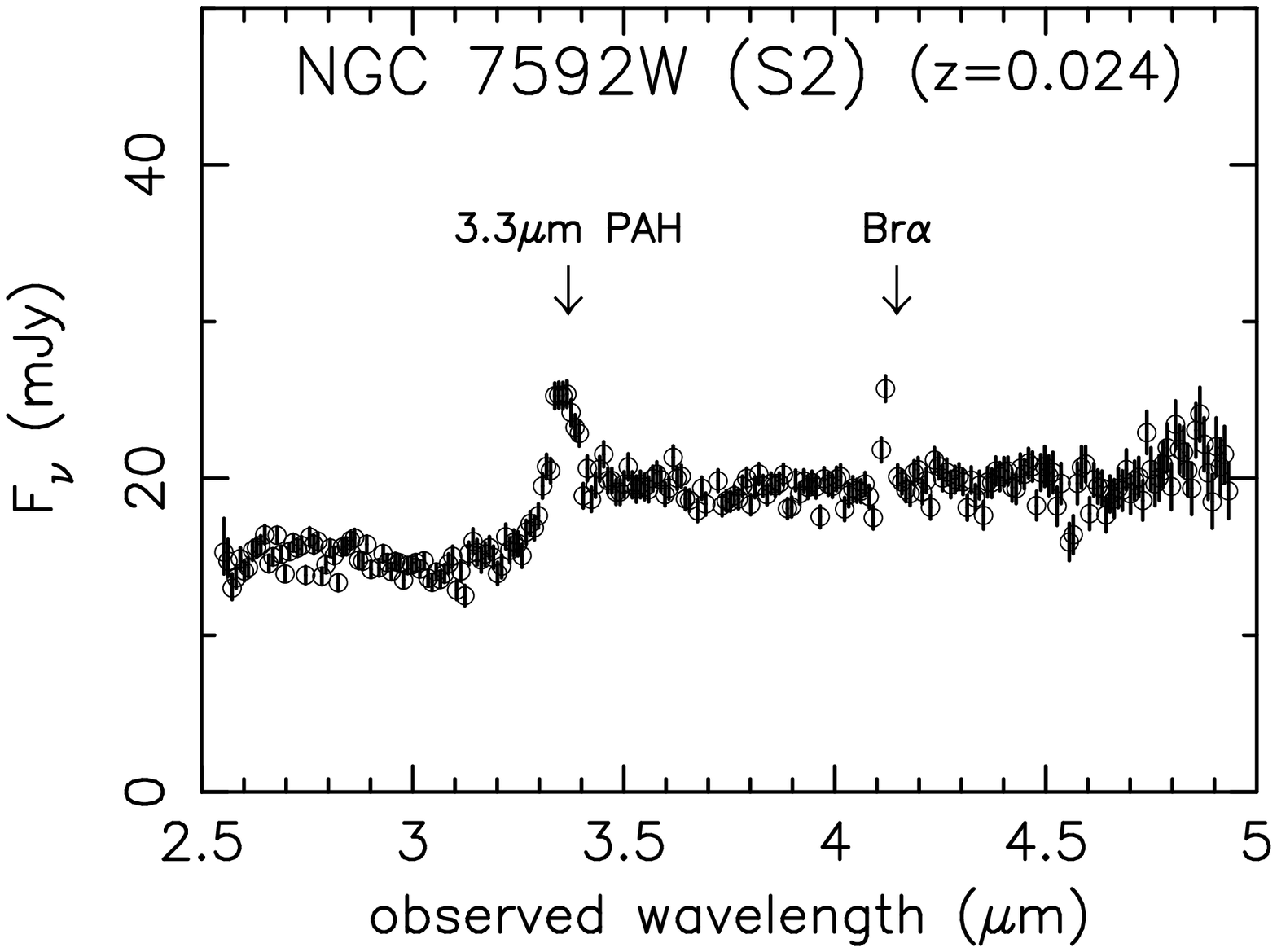}
\includegraphics[angle=0,scale=.27]{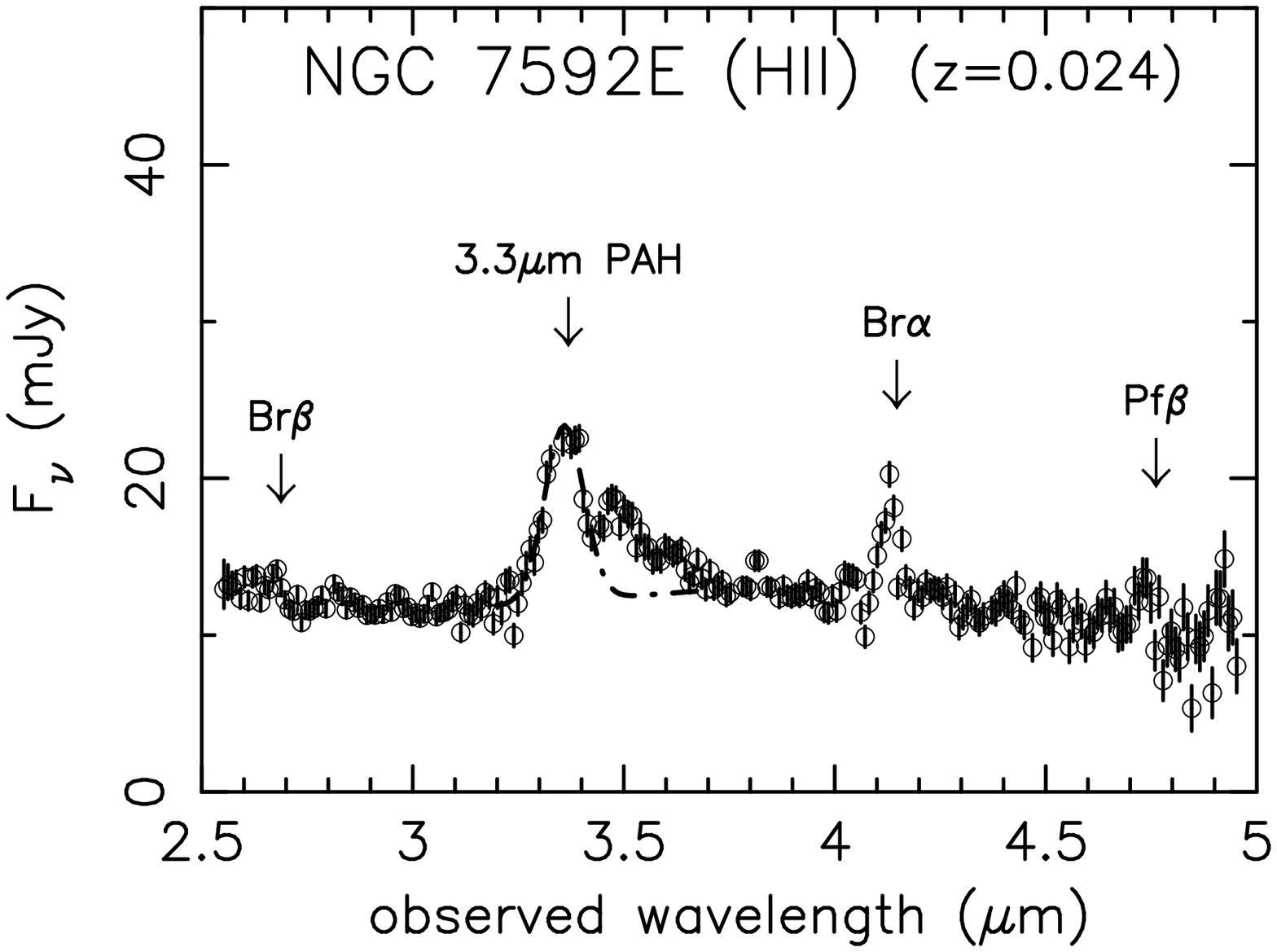} \\
\includegraphics[angle=0,scale=.27]{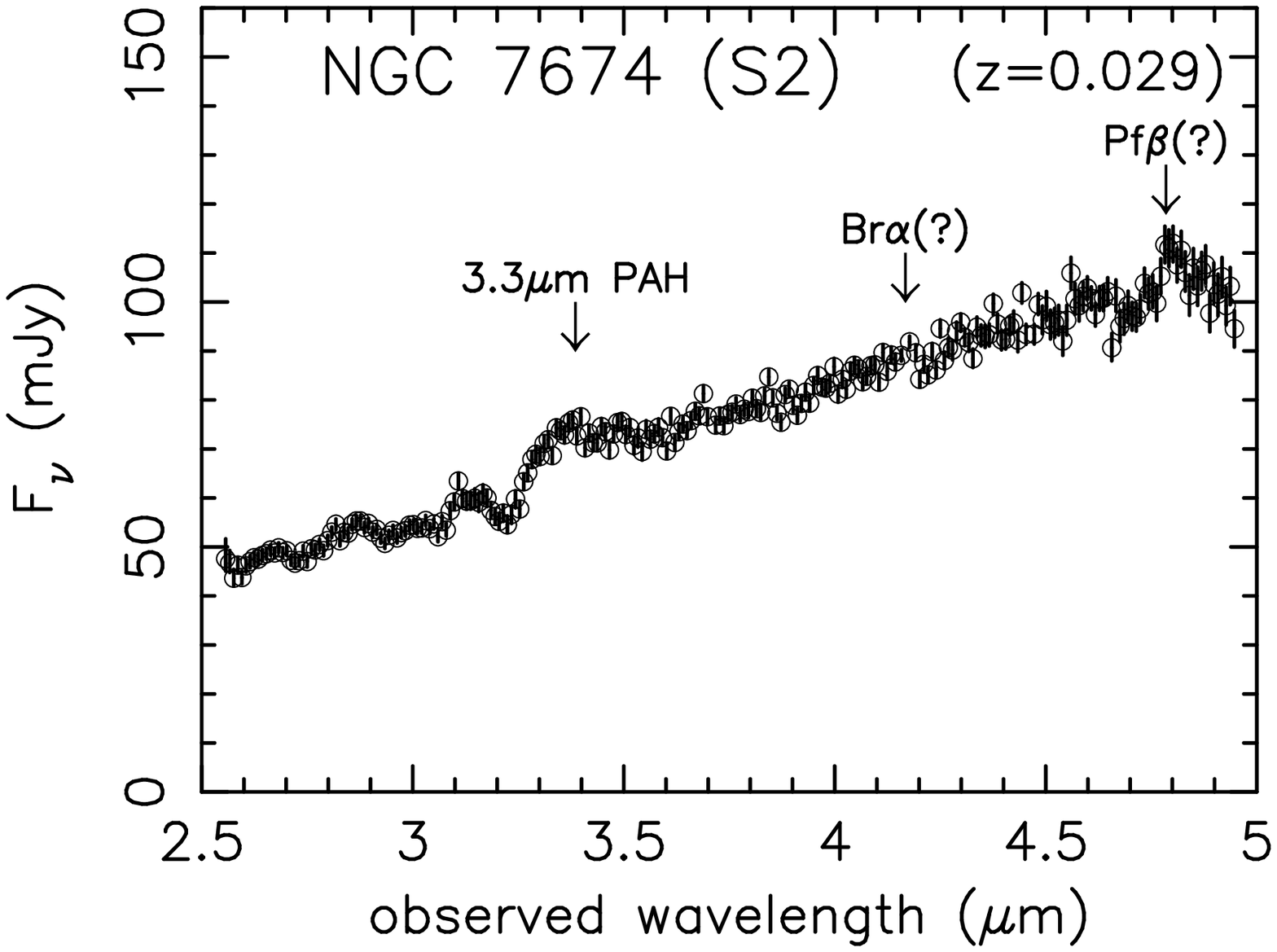}
\includegraphics[angle=0,scale=.27]{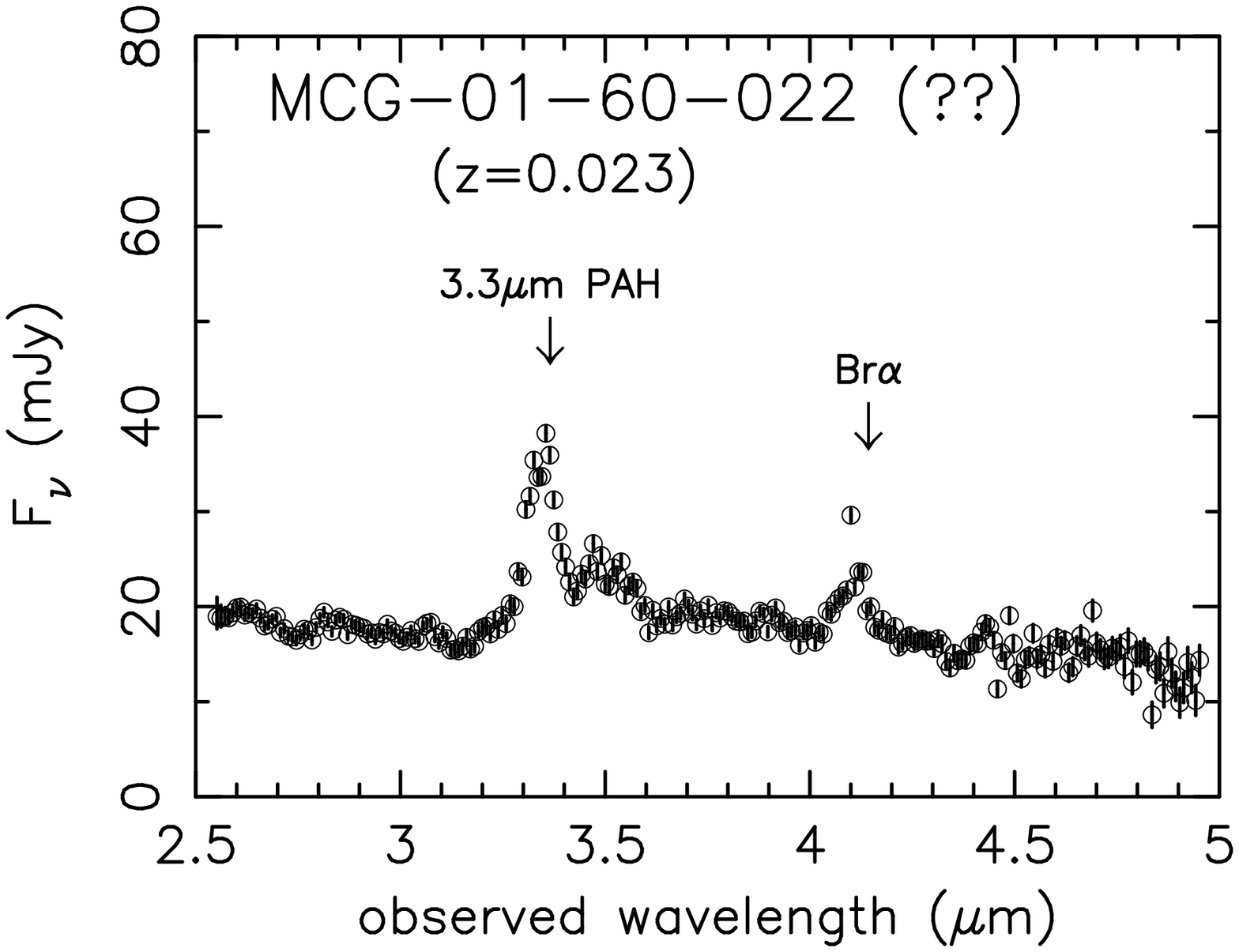}
\includegraphics[angle=0,scale=.27]{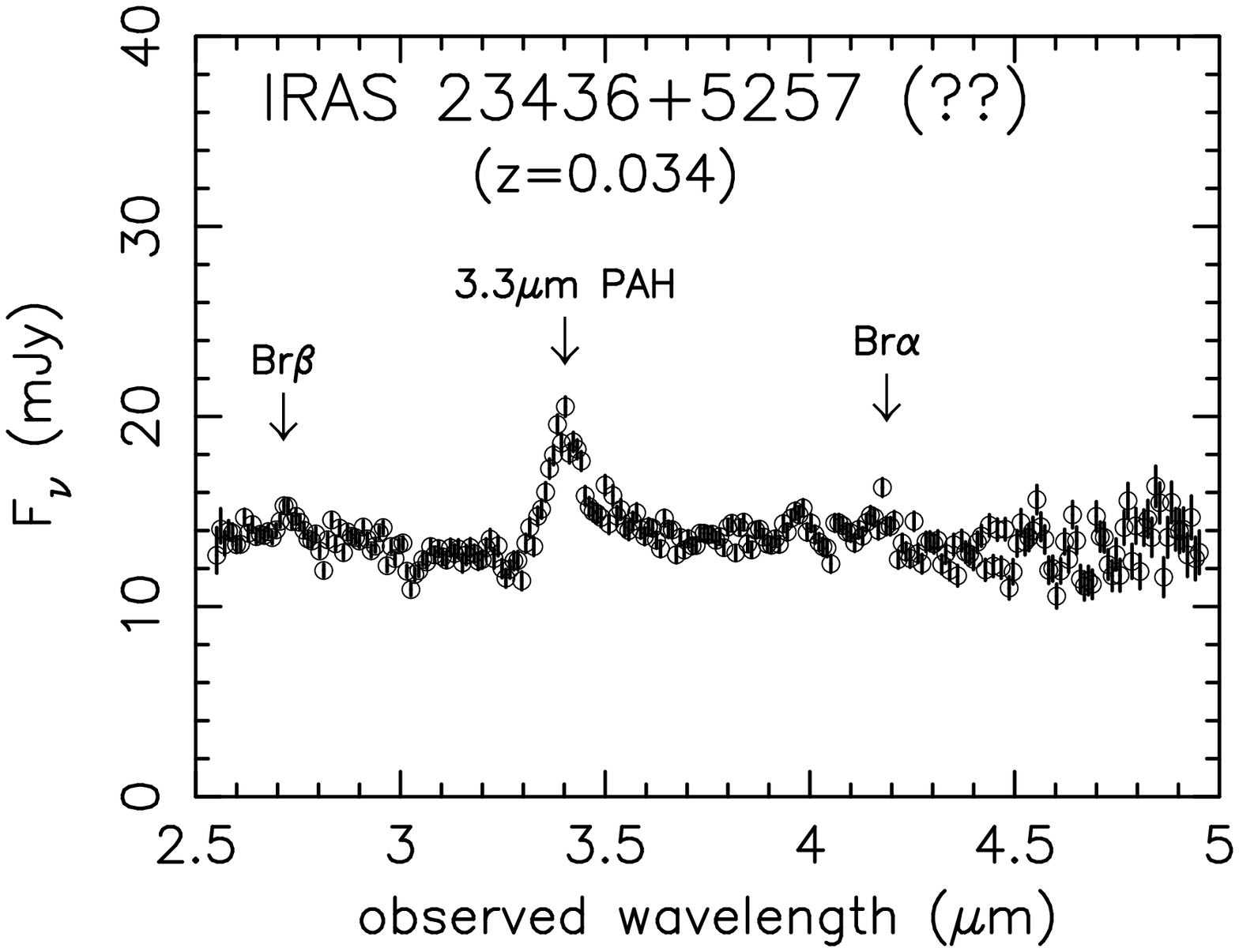} \\
\end{figure}

\clearpage

\begin{figure}
\includegraphics[angle=0,scale=.27]{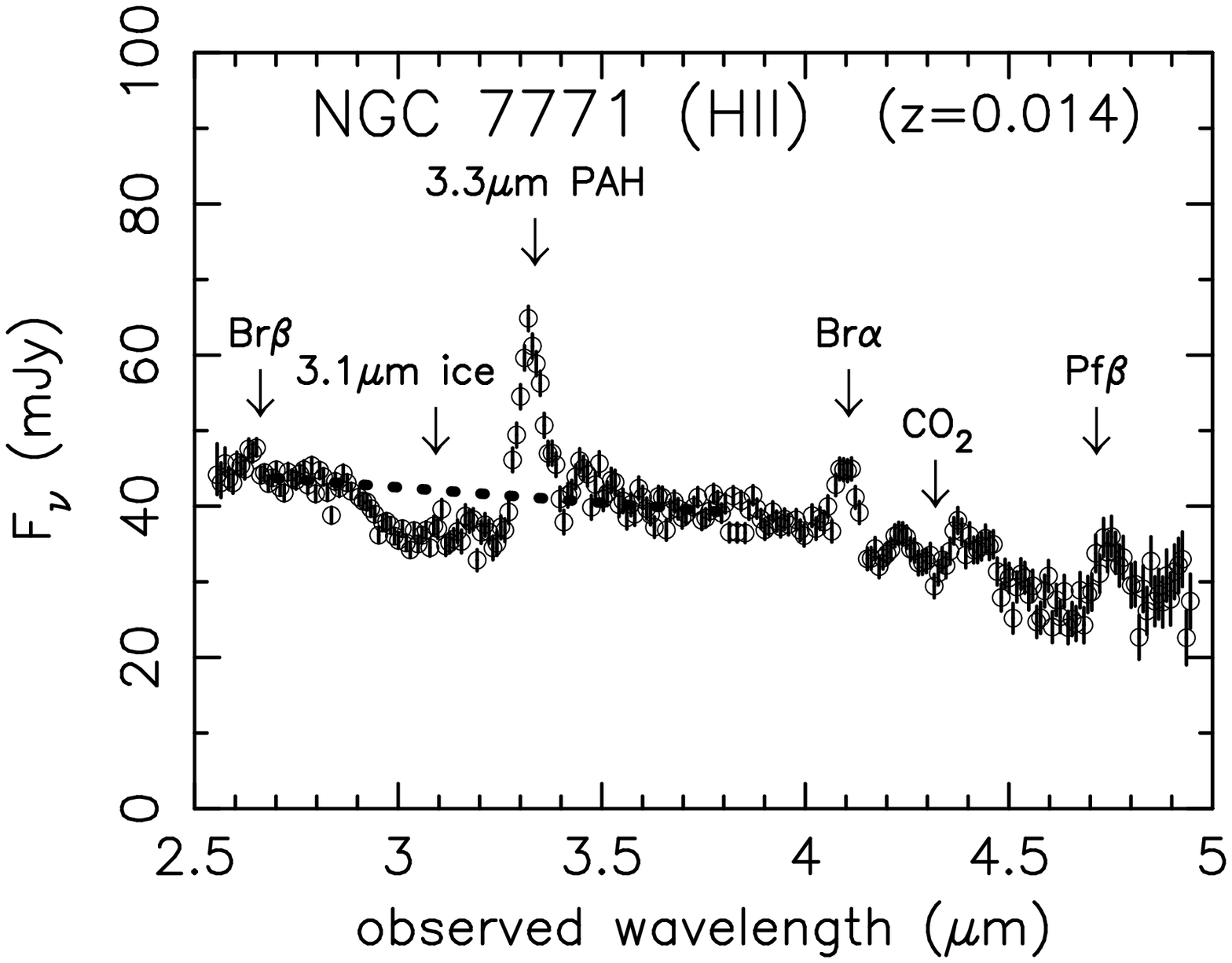}
\includegraphics[angle=0,scale=.27]{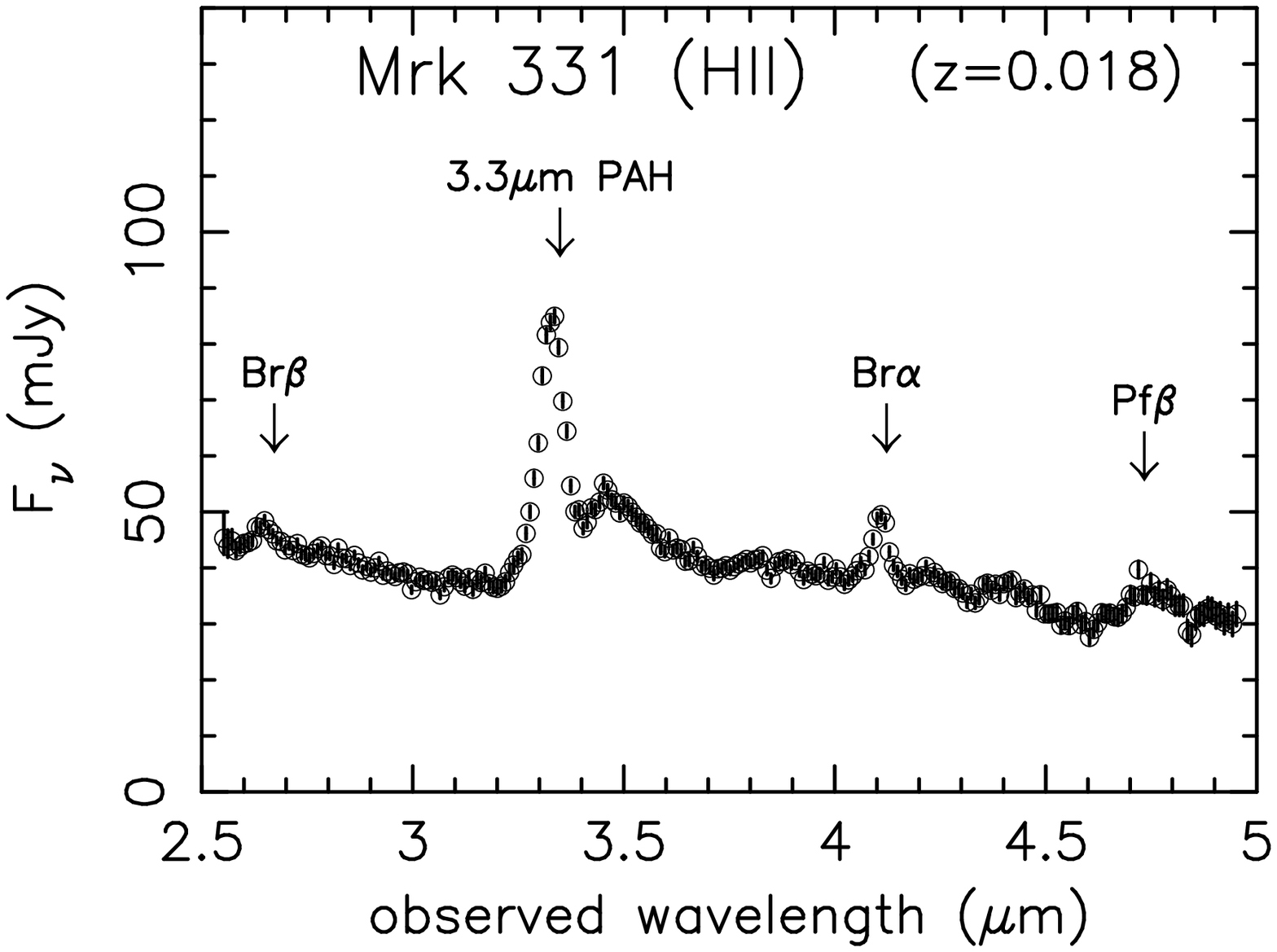}
\includegraphics[angle=0,scale=.27]{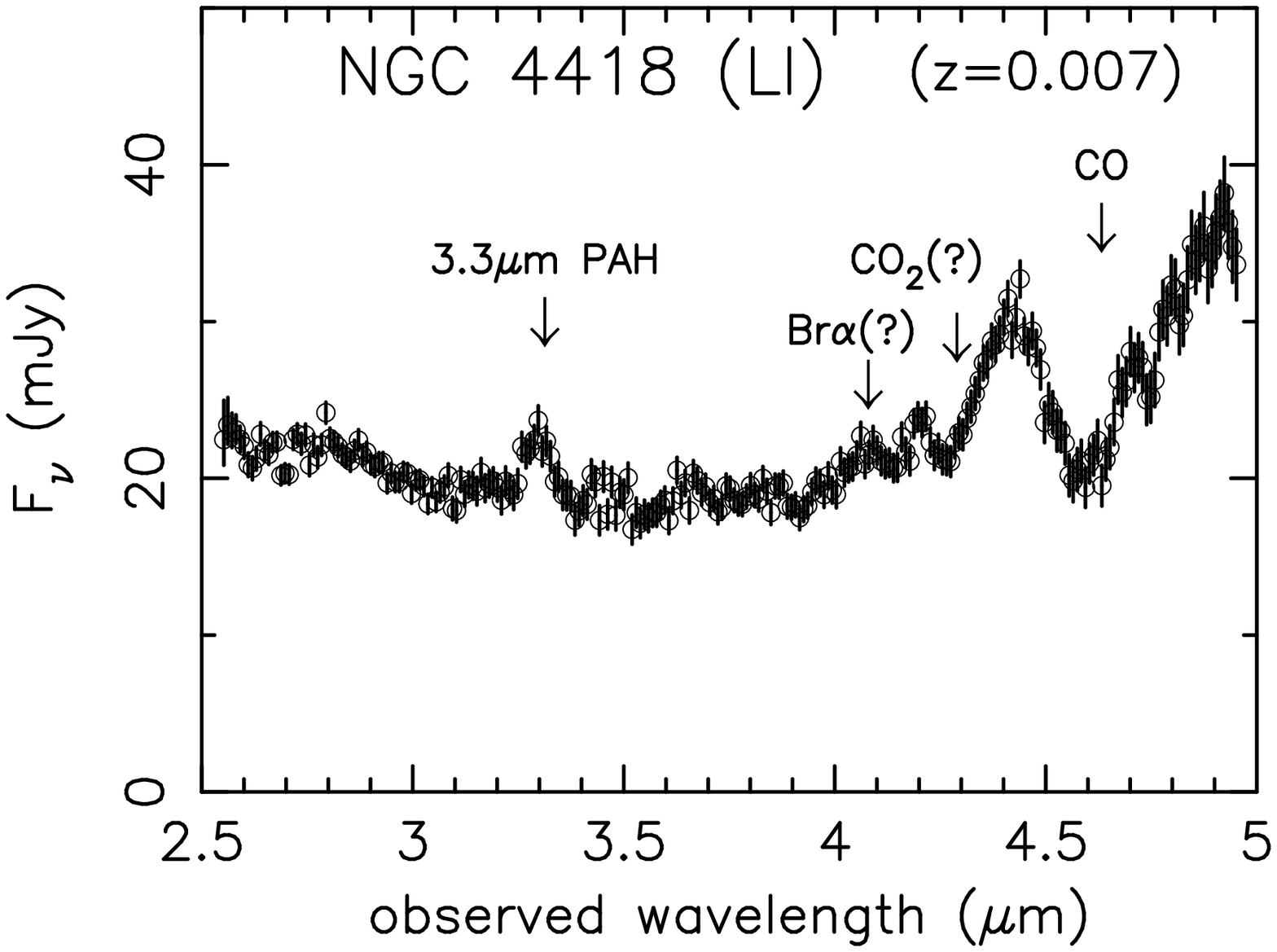} \\
\includegraphics[angle=0,scale=.27]{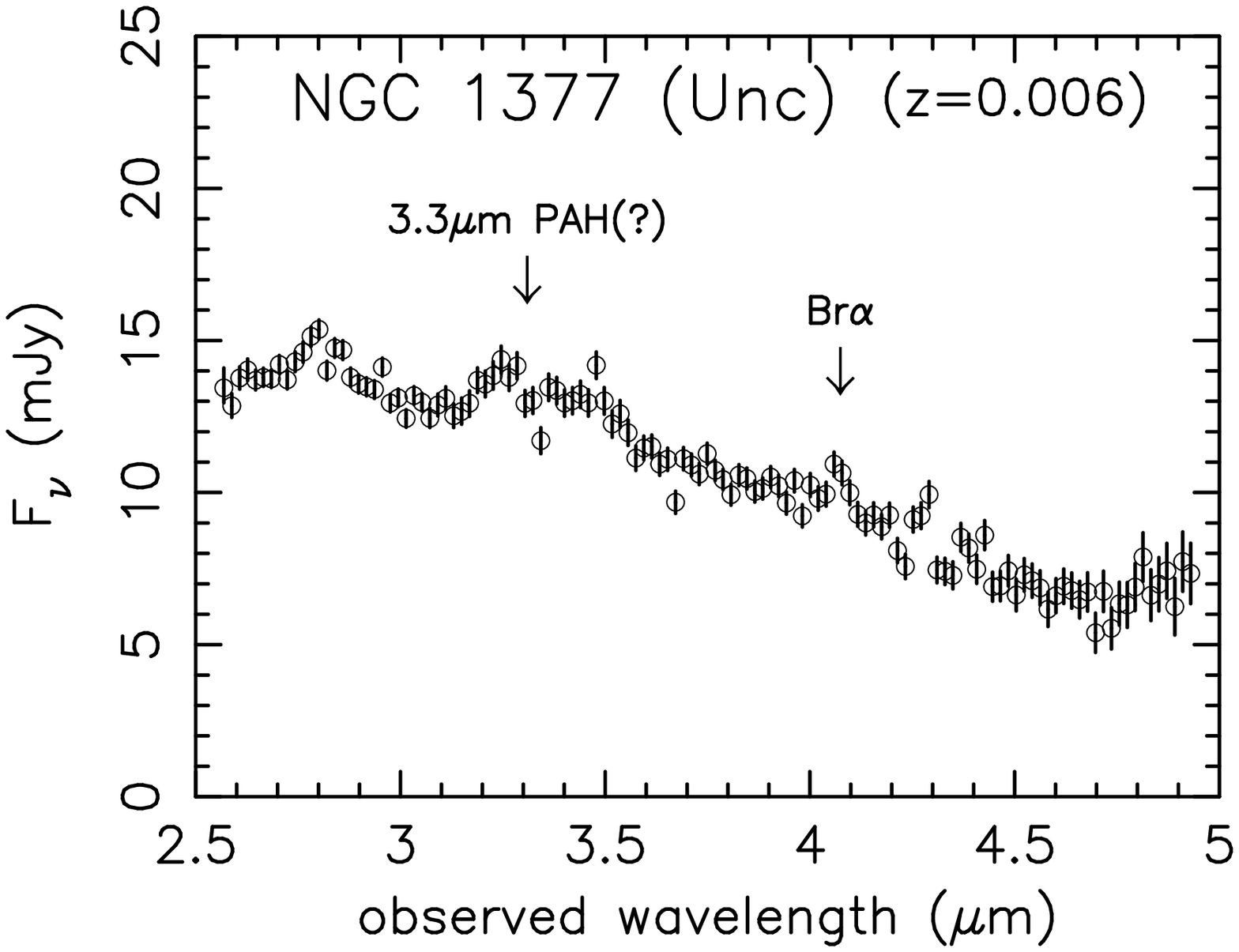}
\caption{
AKARI IRC infrared 2.5--5 $\mu$m spectra of LIRGs. 
The abscissa is the observed wavelength in [$\mu$m] and the
ordinate is the flux F$_{\nu}$ in [mJy].
For each object, the optical classification and redshift are shown. 
"LI", "HII", "S2", "S1", "Unc", and "?" denote LINER, HII region, 
Seyfert 2, Seyfert 1, unclassified, and no optical classification,
respectively.
The dotted lines indicate the continuum levels for measuring the optical
depths of the broad 3.1 $\mu$m H$_{2}$O ice absorption features.
The expected wavelengths of the 3.3 $\mu$m PAH emission features and 
Br$\alpha$ emission lines ($\lambda_{\rm rest}$ = 4.05 $\mu$m) are
indicated for all sources, provided they are covered by $\lambda_{\rm obs}$
$<$ 4.8 $\mu$m in the observed frame. 
The expected wavelengths of Br$\beta$ emission ($\lambda_{\rm rest}$ =
2.63 $\mu$m), Pf$\beta$ emission ($\lambda_{\rm rest}$ = 4.65 $\mu$m), and 
Pf$\gamma$ emission ($\lambda_{\rm rest}$ = 3.74 $\mu$m), as well as 
the broad 3.1 $\mu$m H$_{2}$O ice absorption, 3.4 $\mu$m bare
carbonaceous dust absorption,  4.26 $\mu$m CO$_{2}$ absorption, and 
4.67 $\mu$m CO absorption features, are also added for clearly detected
sources.
For sources whose 3.3 $\mu$m PAH emission and 3.4 $\mu$m PAH sub-peak
are highly blended, our adopted choices of the 3.3 $\mu$m PAH peak
components are shown as dashed-dotted lines.
In MCG-03-34-064 (an optical Seyfert 2), the [MgIV] line at 
$\lambda_{\rm rest}$ = 4.487 $\mu$m is detected and indicated. 
The "?" indicates that detection is unclear.
}
\end{figure}

\clearpage

\begin{figure}
\includegraphics[angle=0,scale=.27]{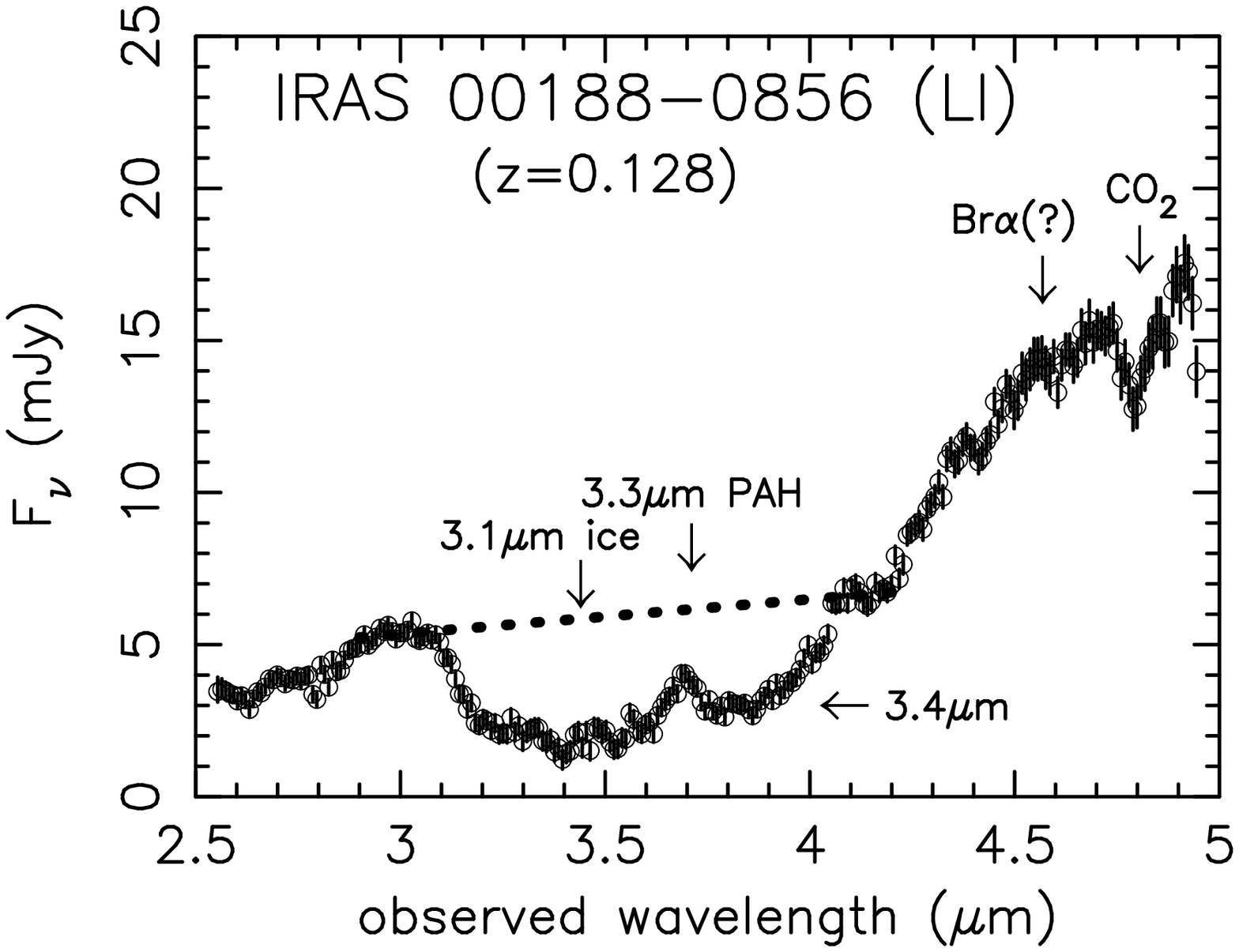}
\includegraphics[angle=0,scale=.27]{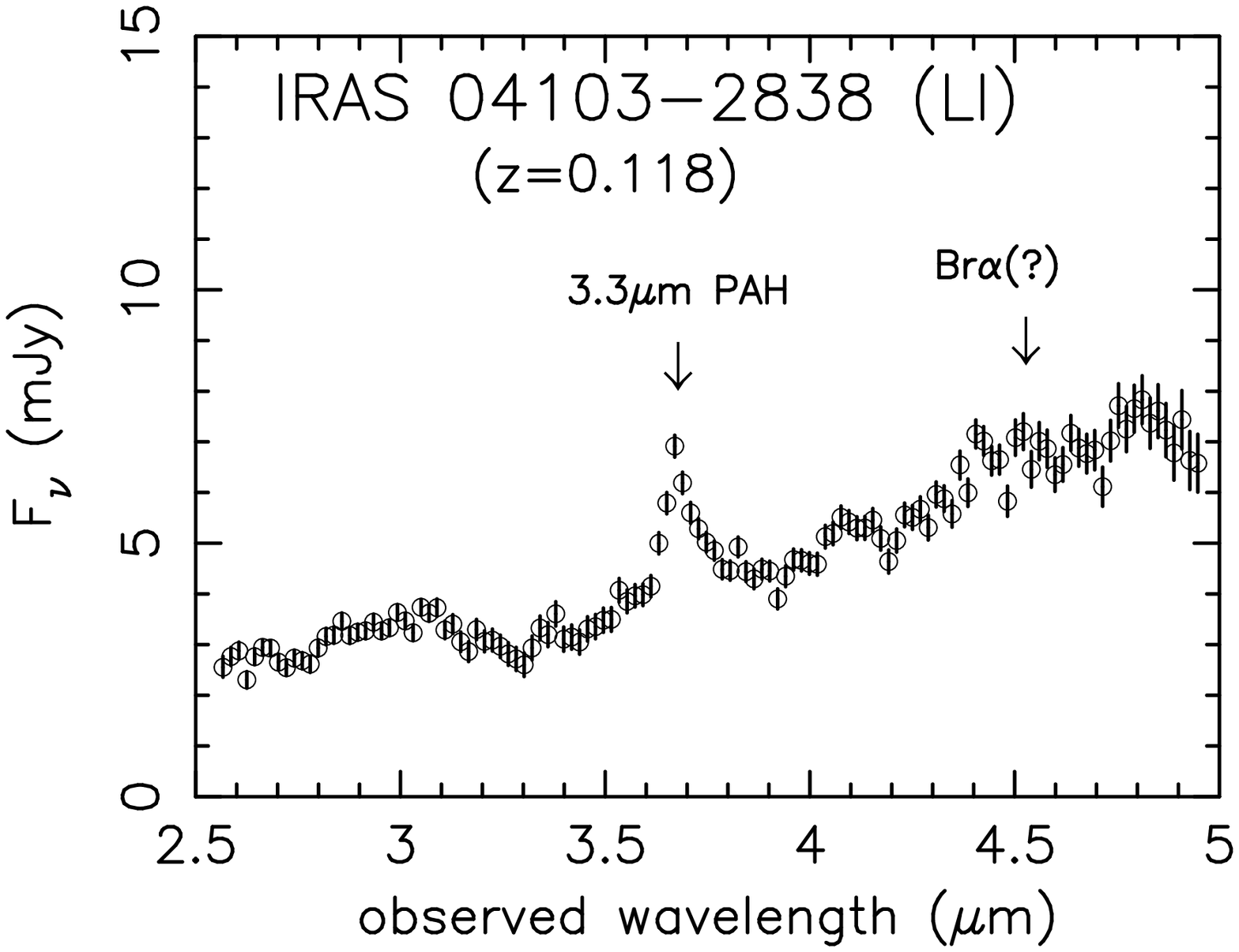}
\includegraphics[angle=0,scale=.27]{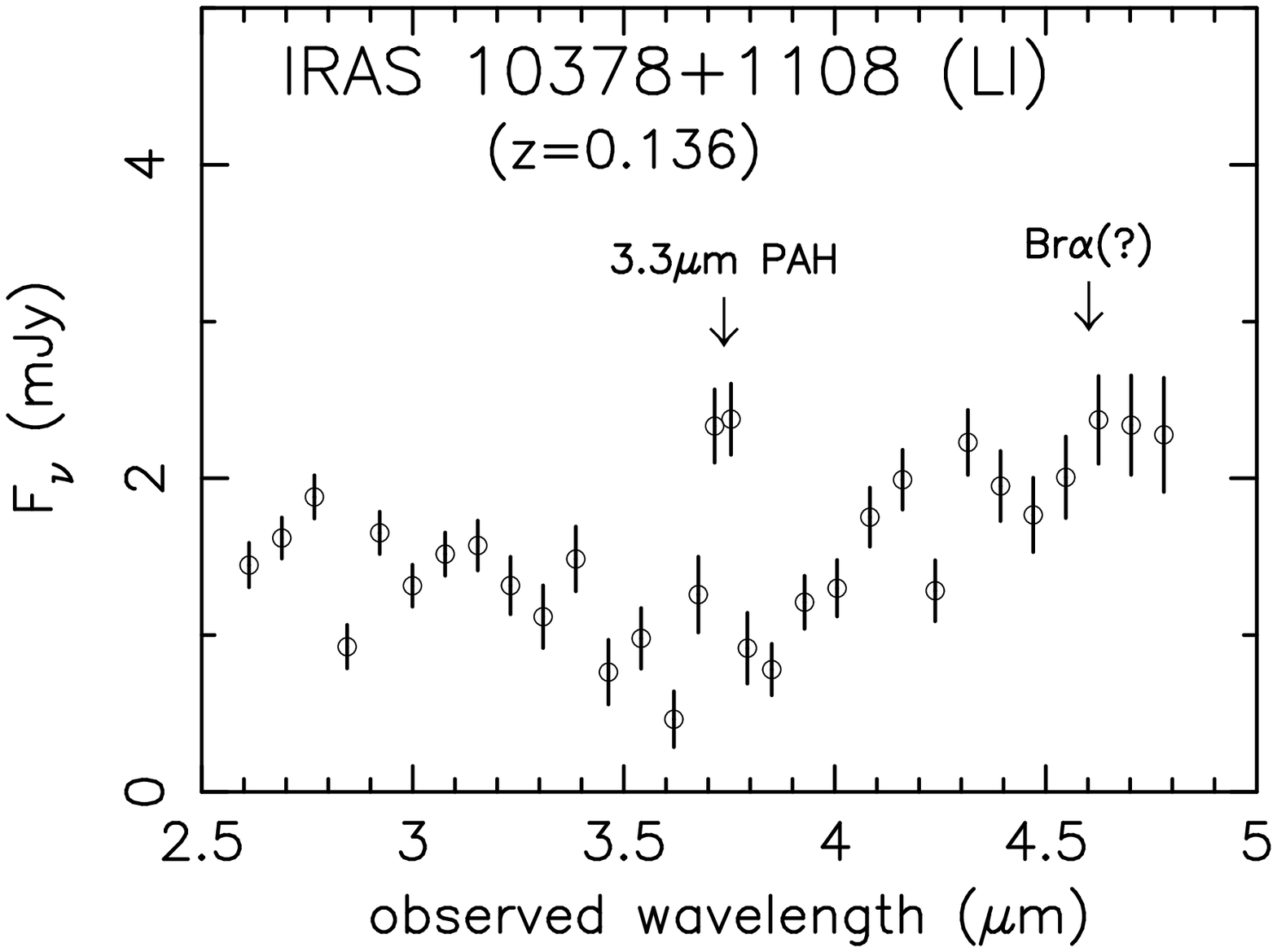} \\
\includegraphics[angle=0,scale=.27]{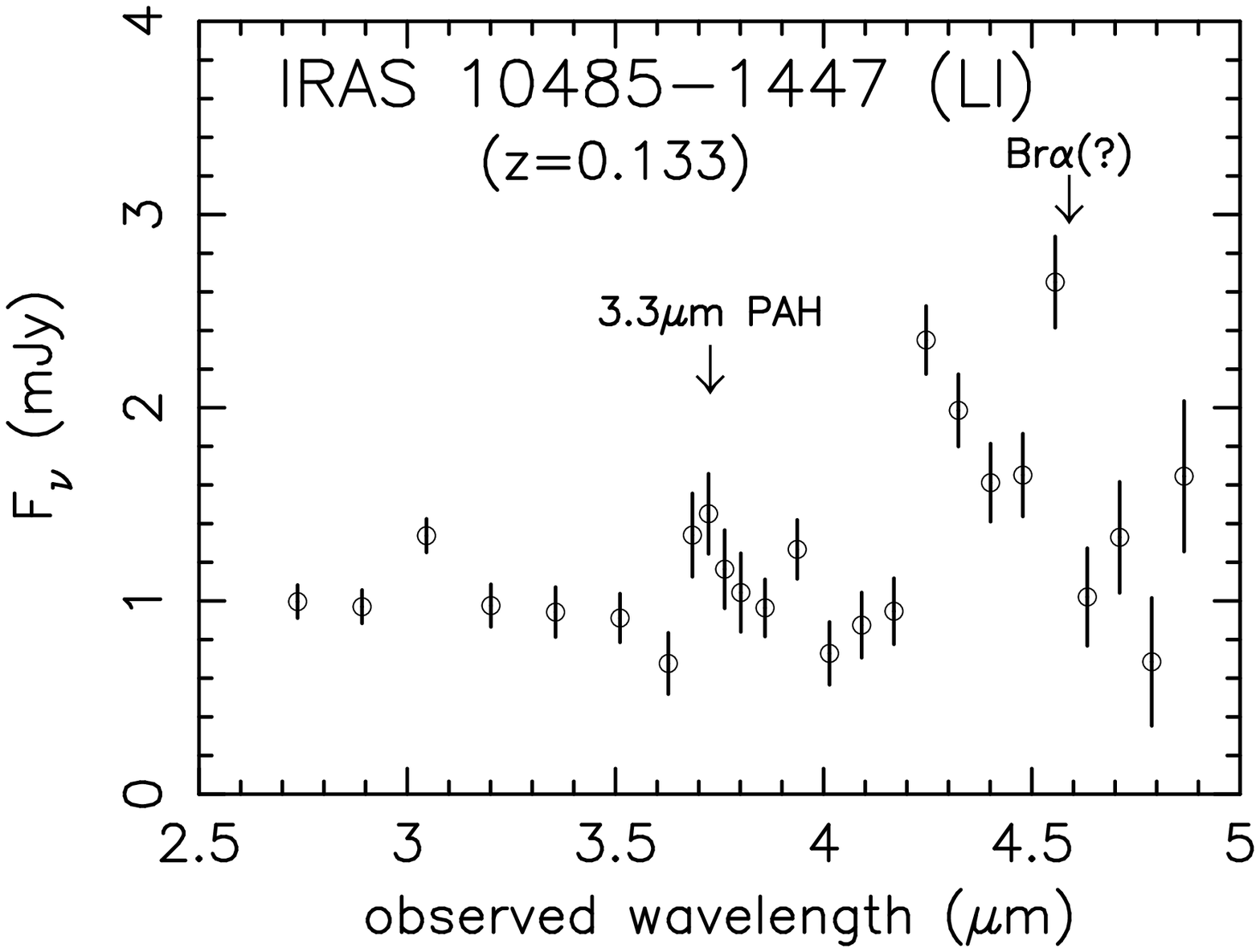}
\includegraphics[angle=0,scale=.27]{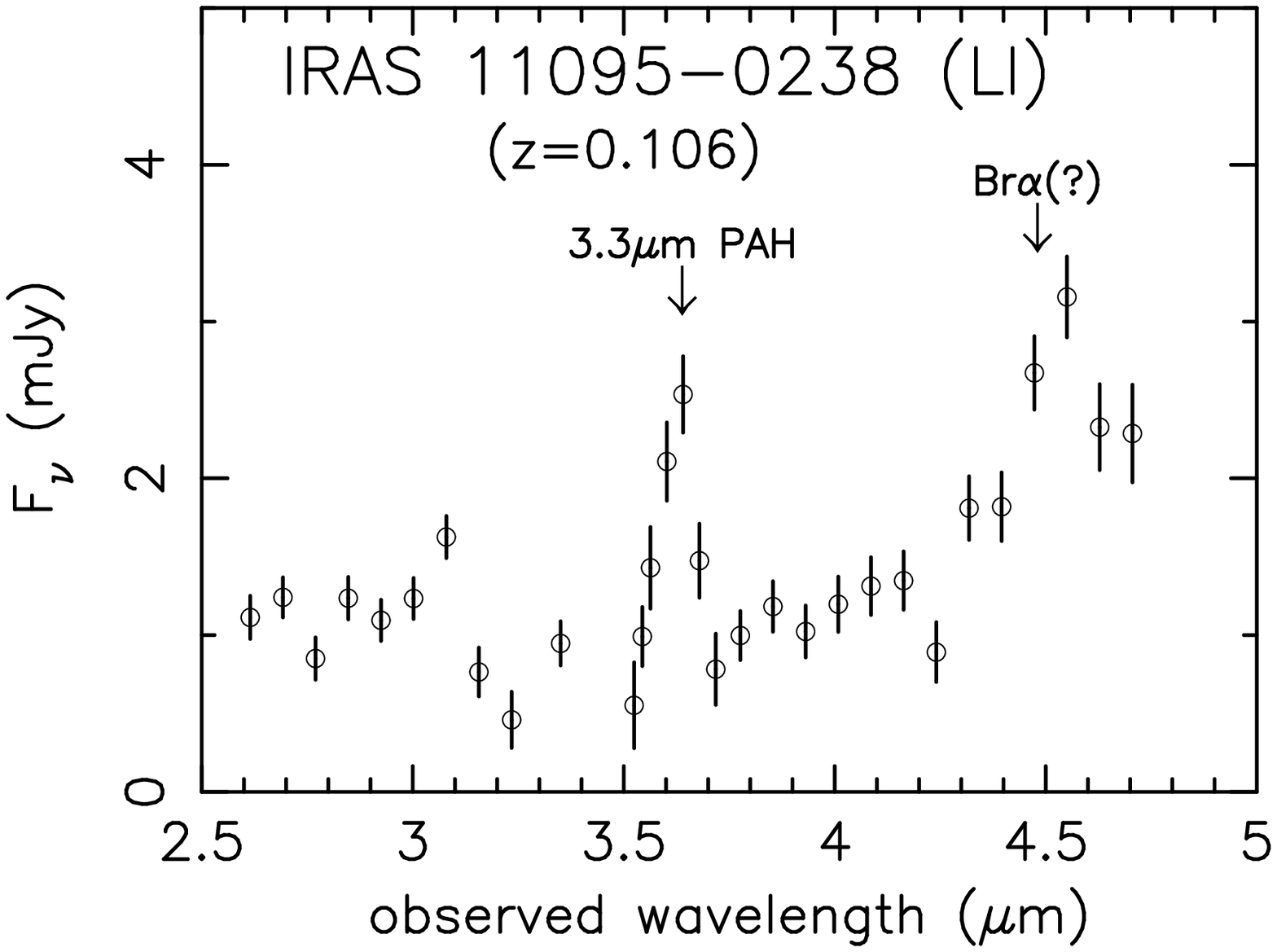}
\includegraphics[angle=0,scale=.27]{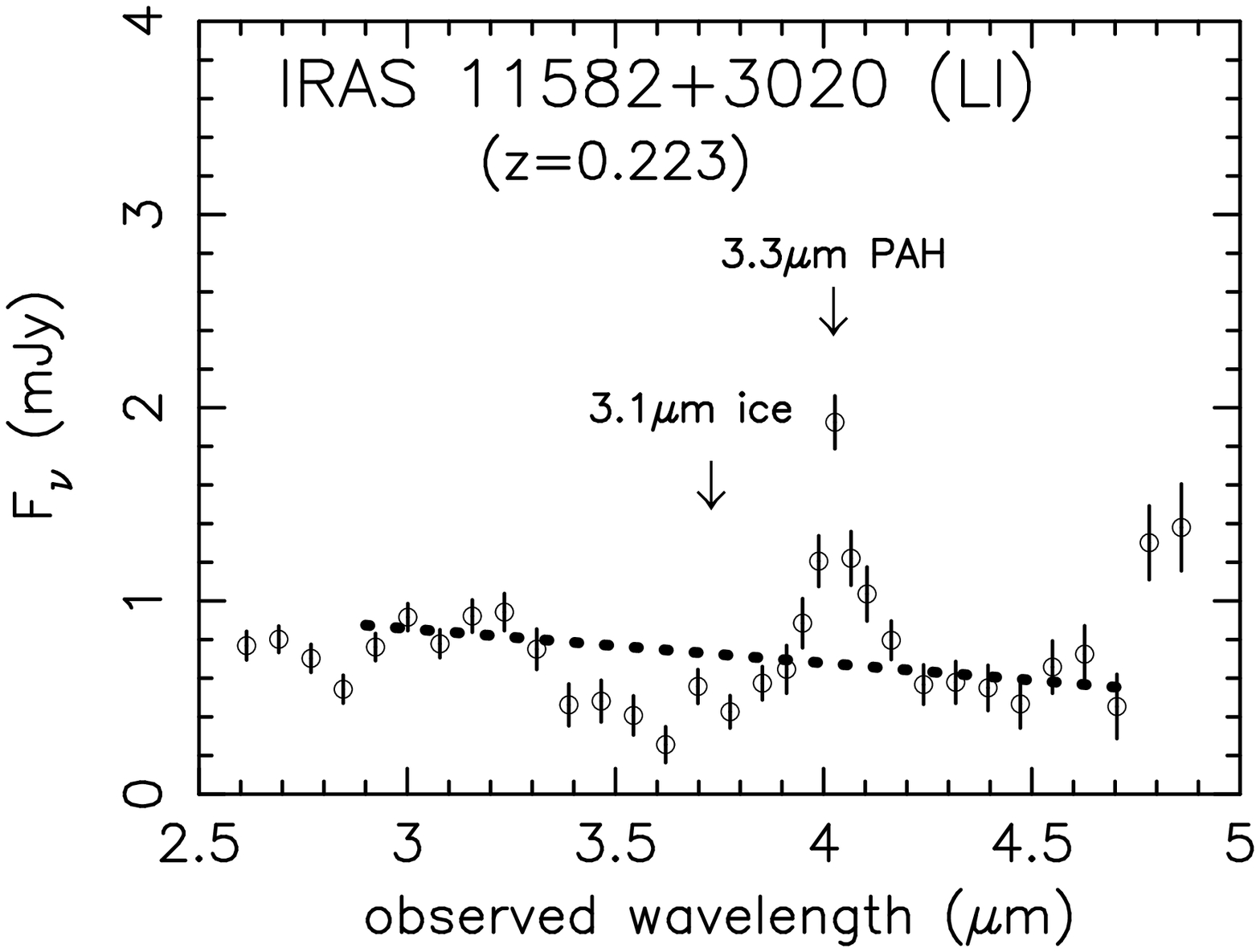} \\
\includegraphics[angle=0,scale=.27]{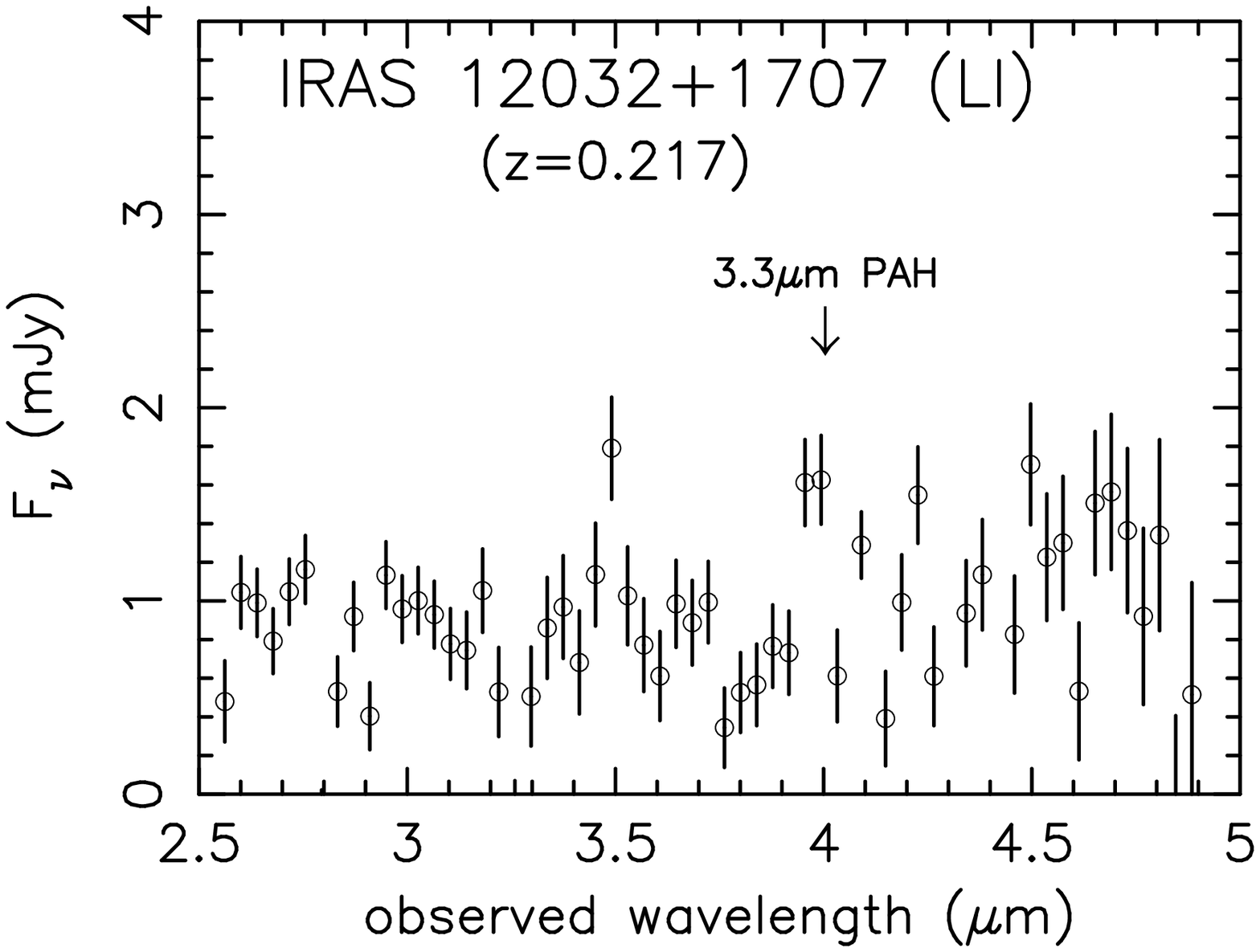}
\includegraphics[angle=0,scale=.27]{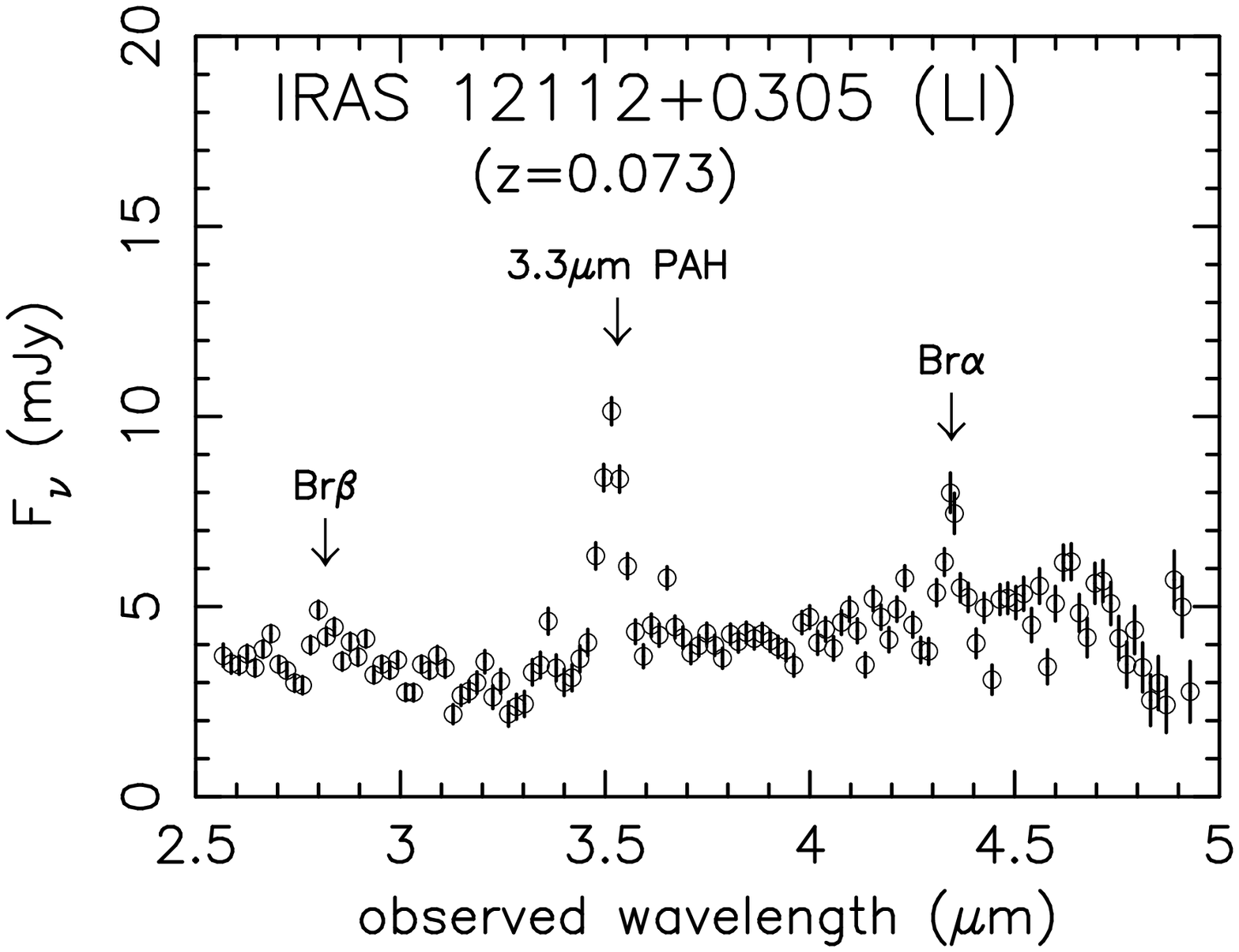}
\includegraphics[angle=0,scale=.27]{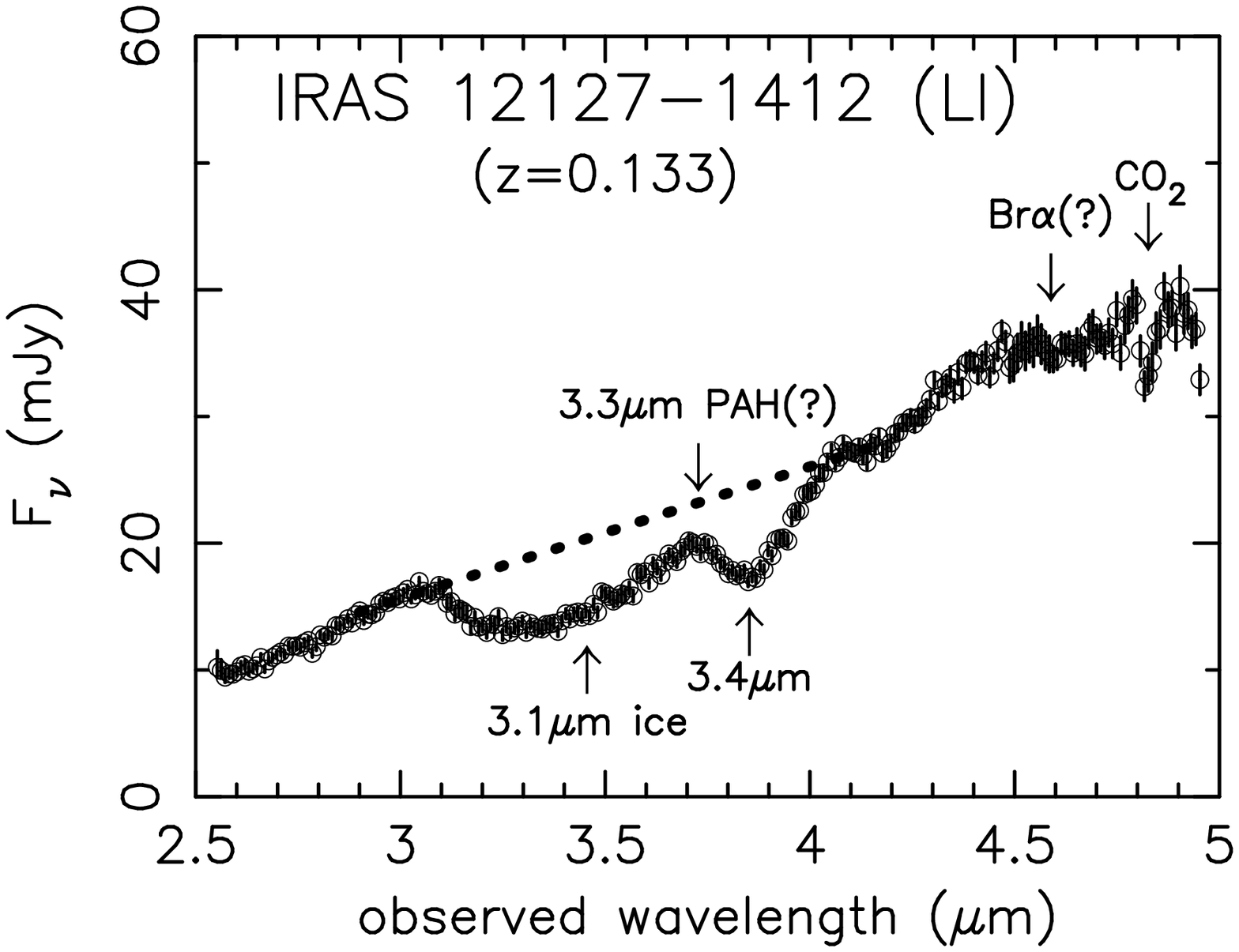} \\
\includegraphics[angle=0,scale=.27]{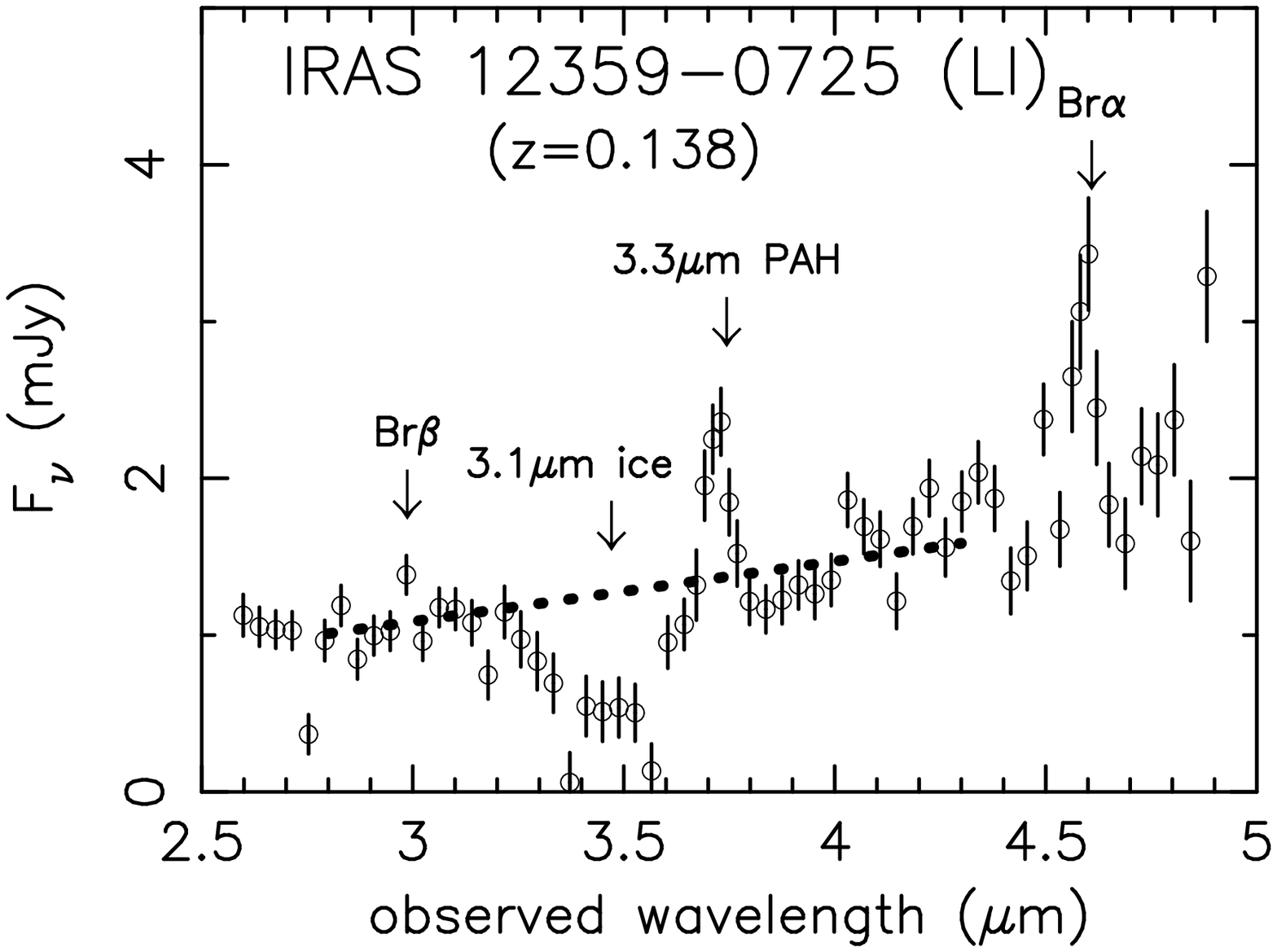}
\includegraphics[angle=0,scale=.27]{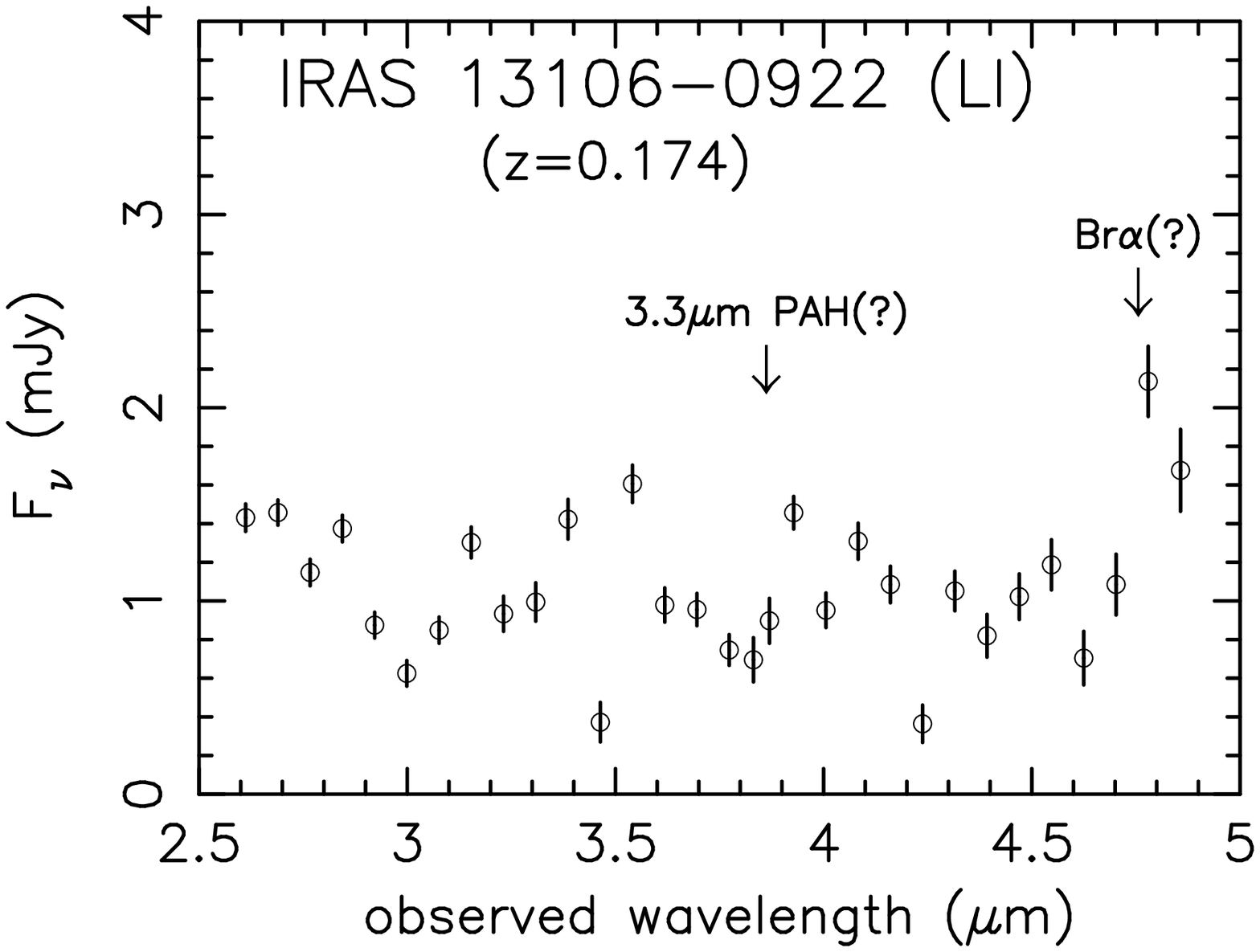}
\includegraphics[angle=0,scale=.27]{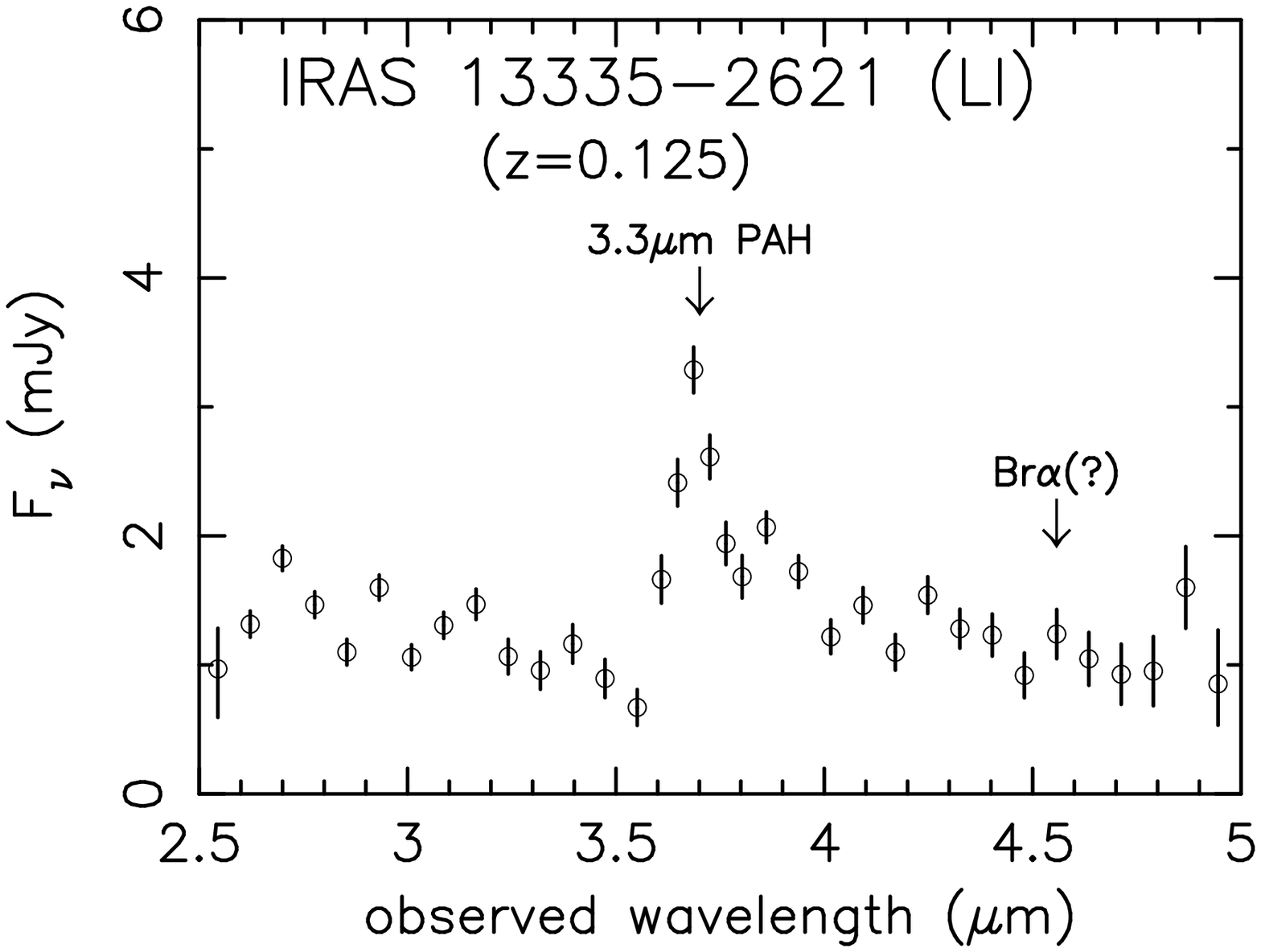} \\
\includegraphics[angle=0,scale=.27]{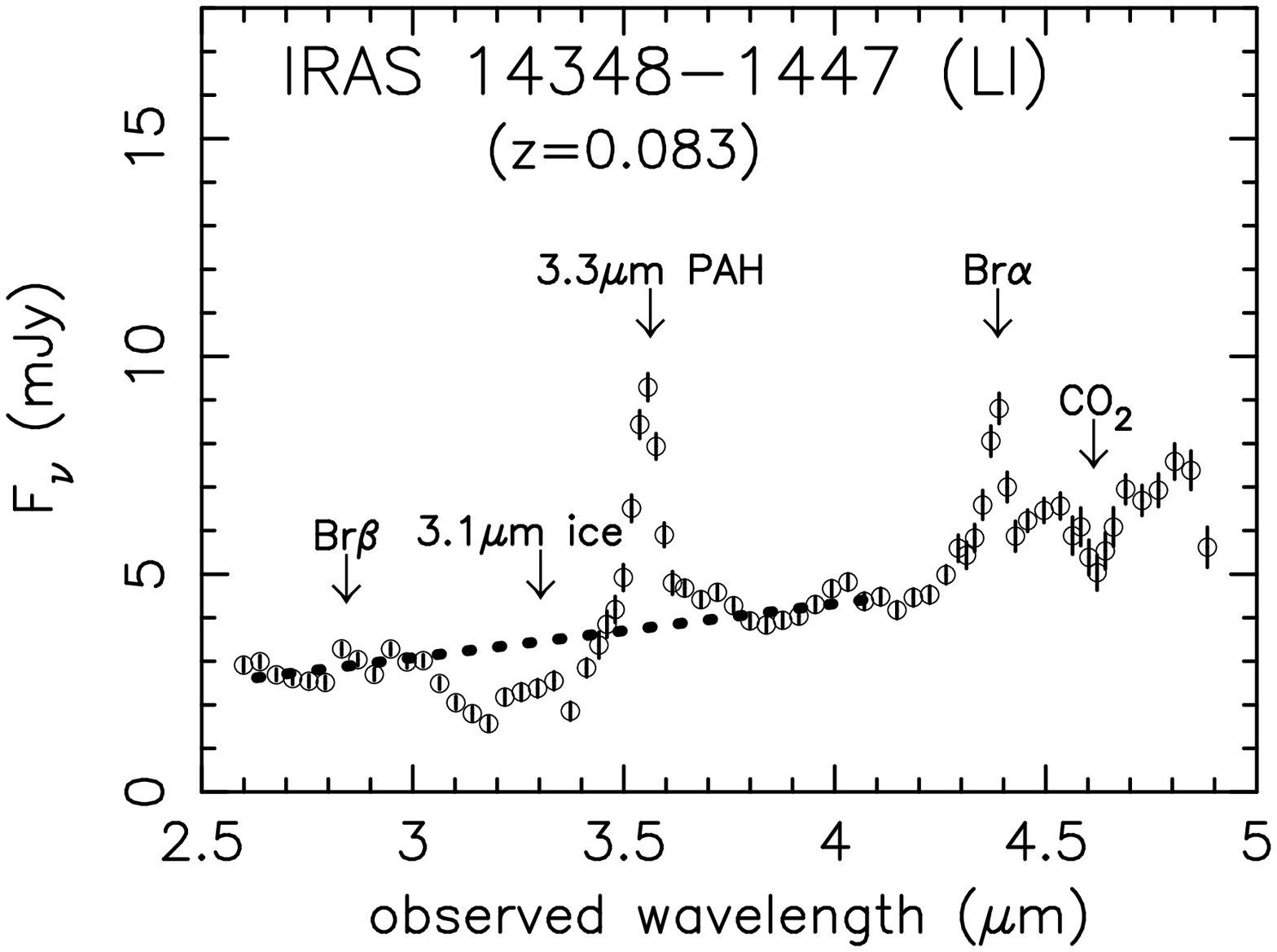}
\includegraphics[angle=0,scale=.27]{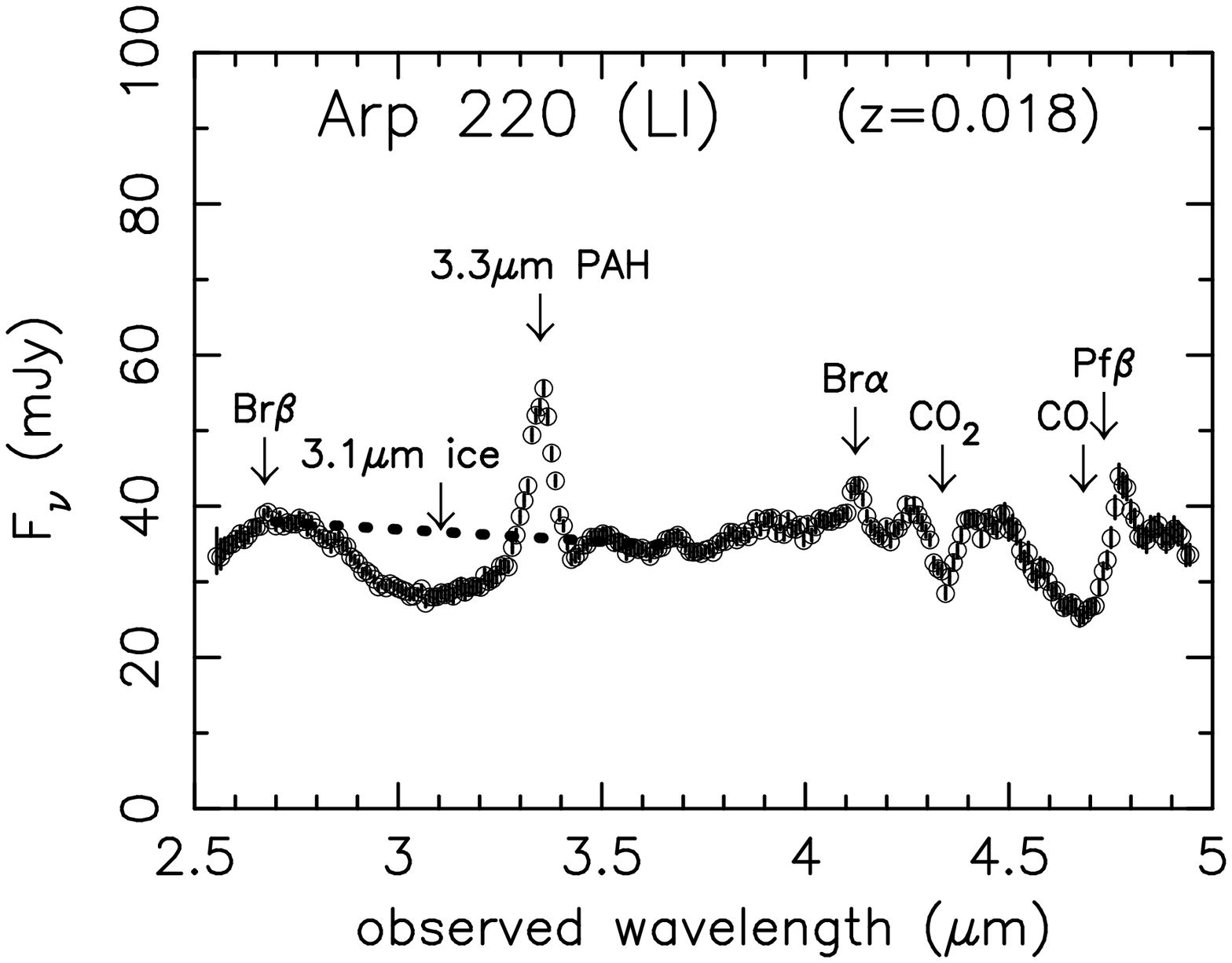} 
\includegraphics[angle=0,scale=.27]{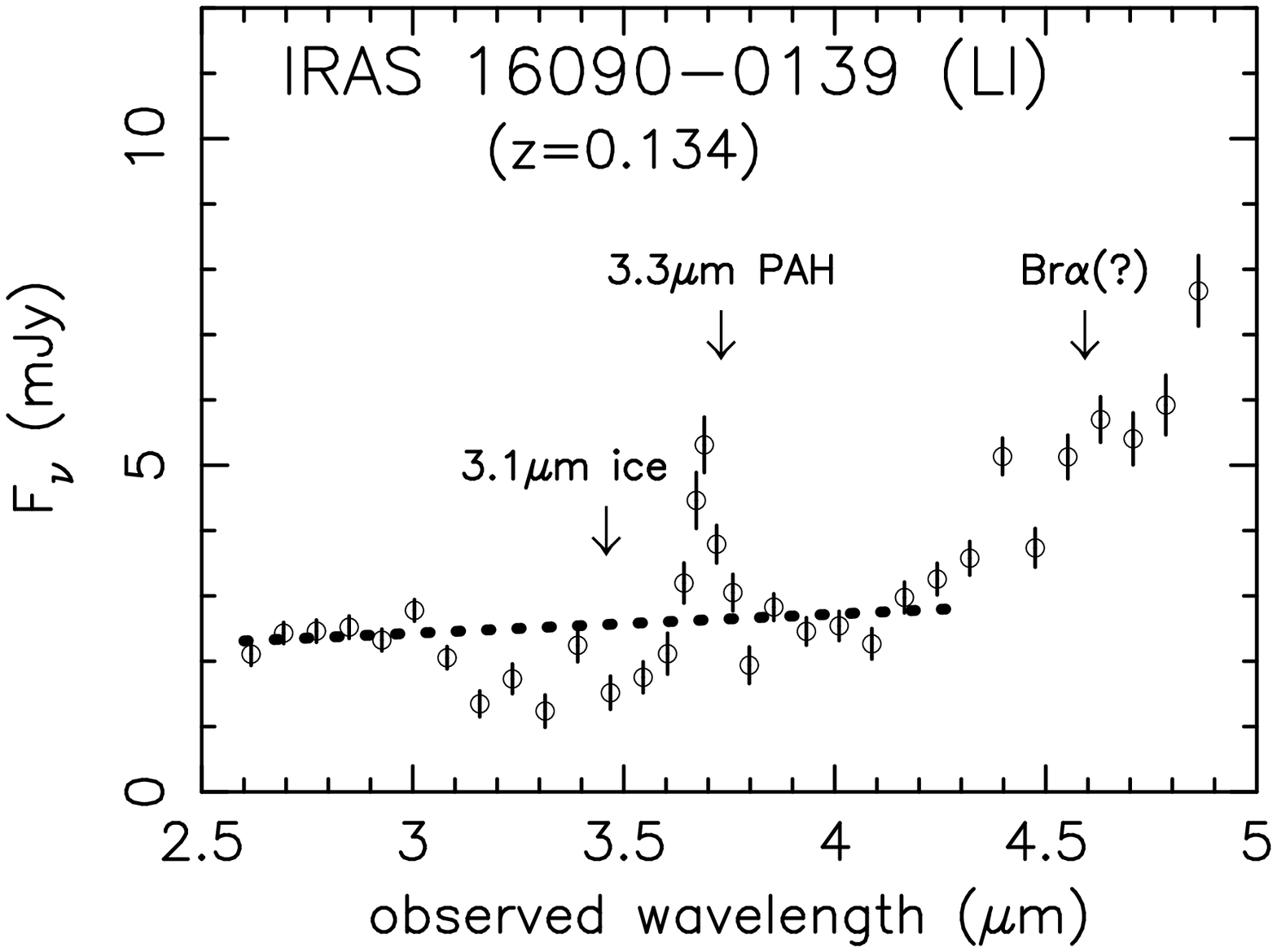} \\
\end{figure}

\clearpage

\begin{figure}
\includegraphics[angle=0,scale=.27]{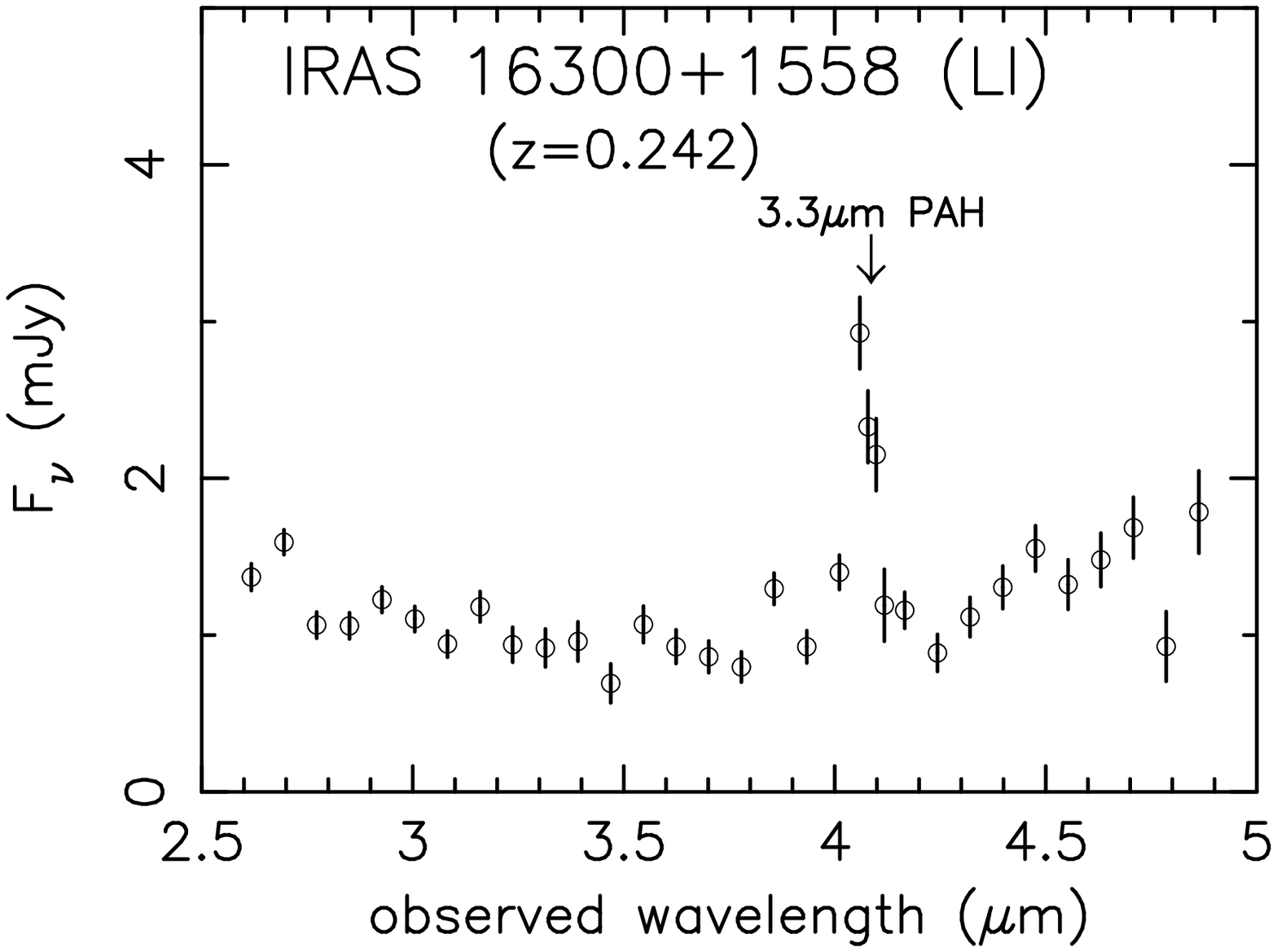}
\includegraphics[angle=0,scale=.27]{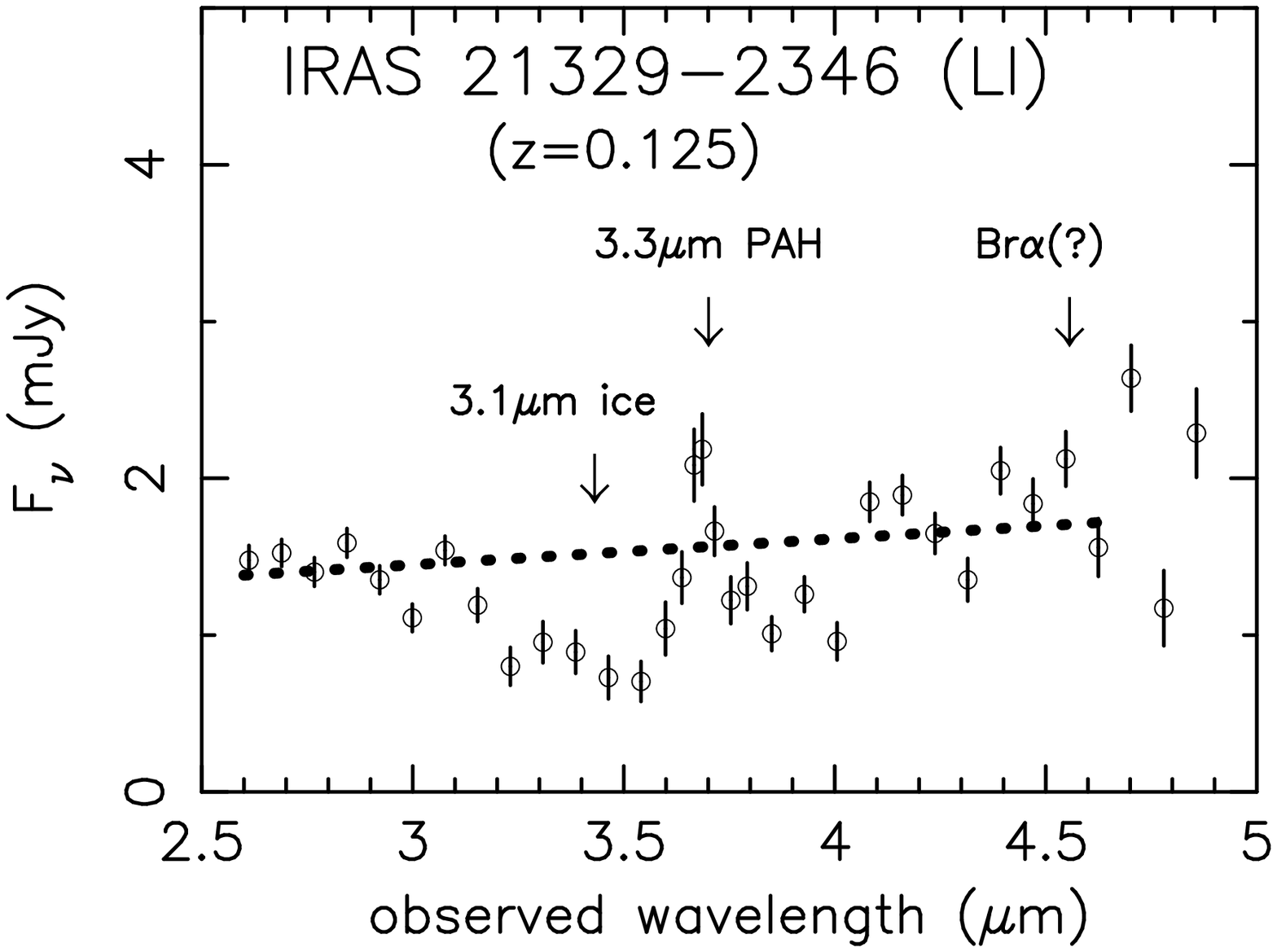}
\includegraphics[angle=0,scale=.27]{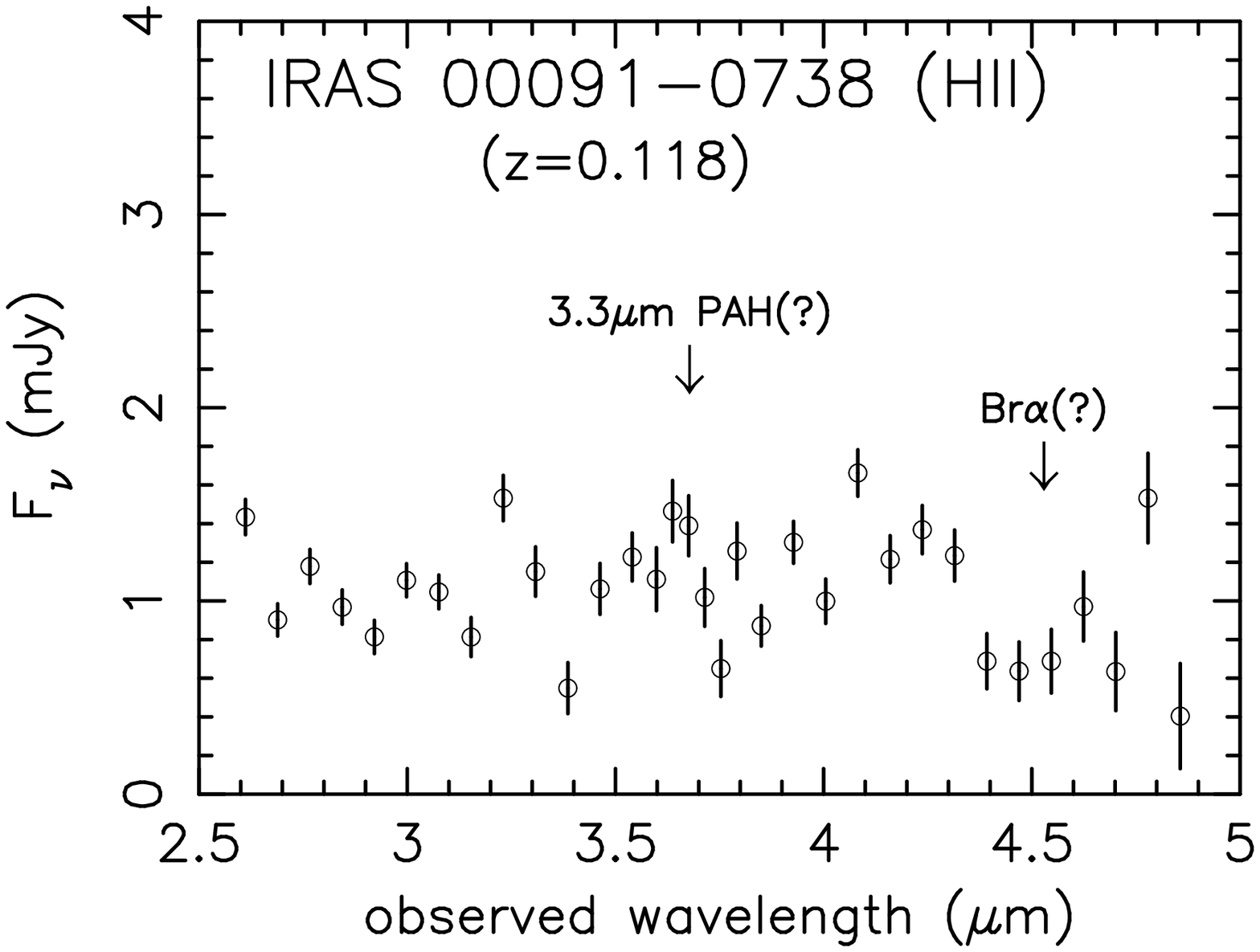} \\
\includegraphics[angle=0,scale=.27]{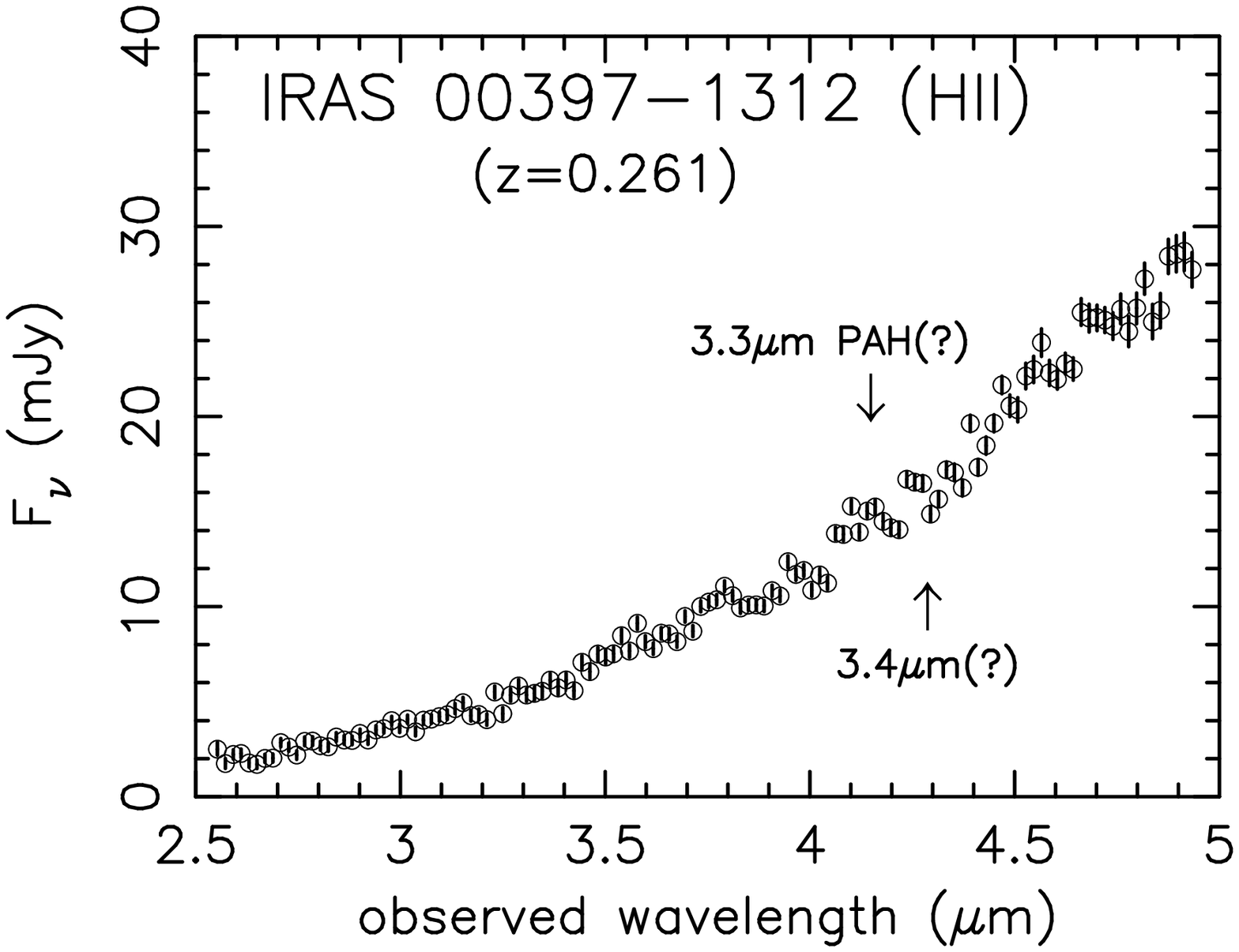}
\includegraphics[angle=0,scale=.27]{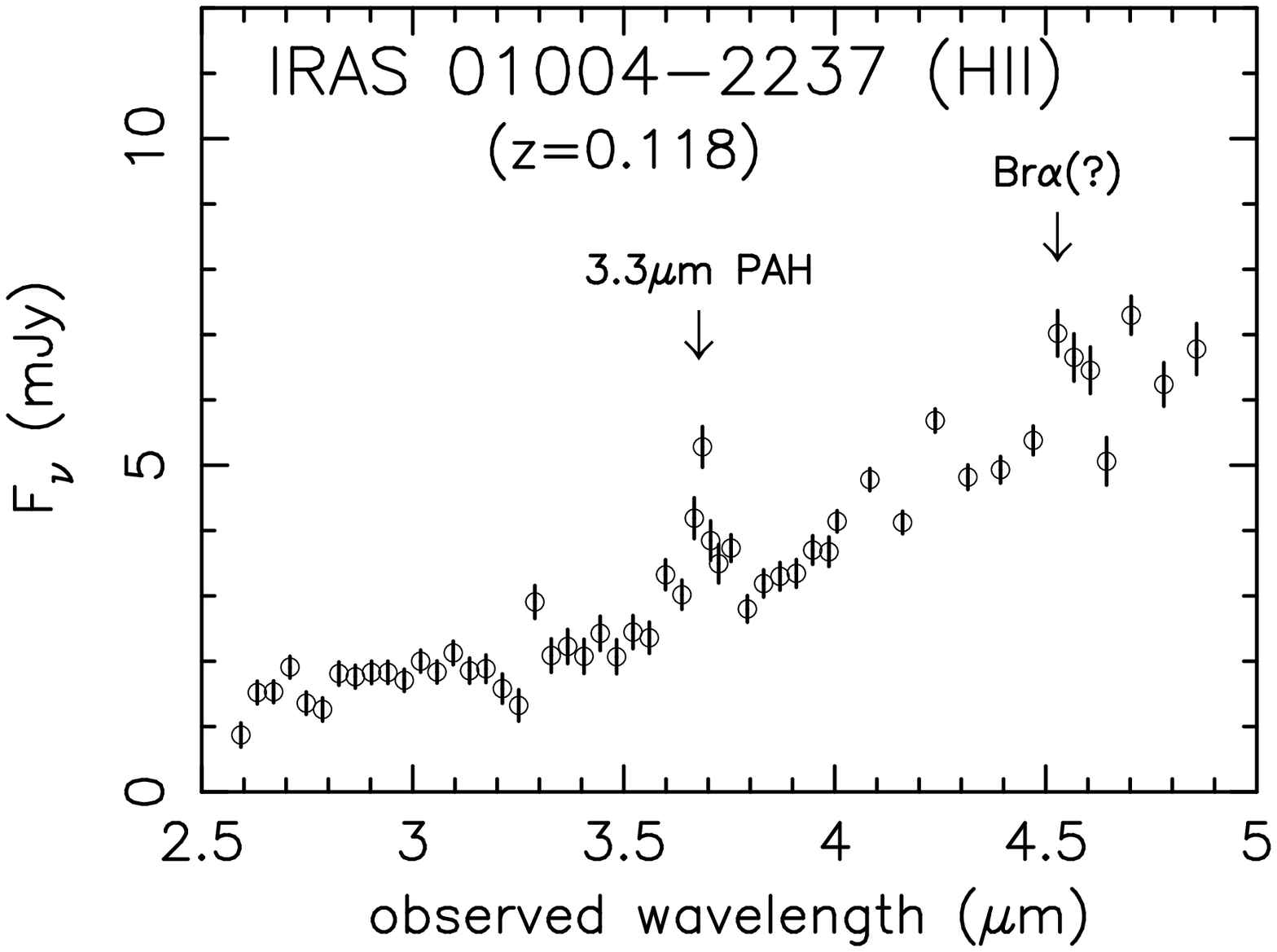}
\includegraphics[angle=0,scale=.27]{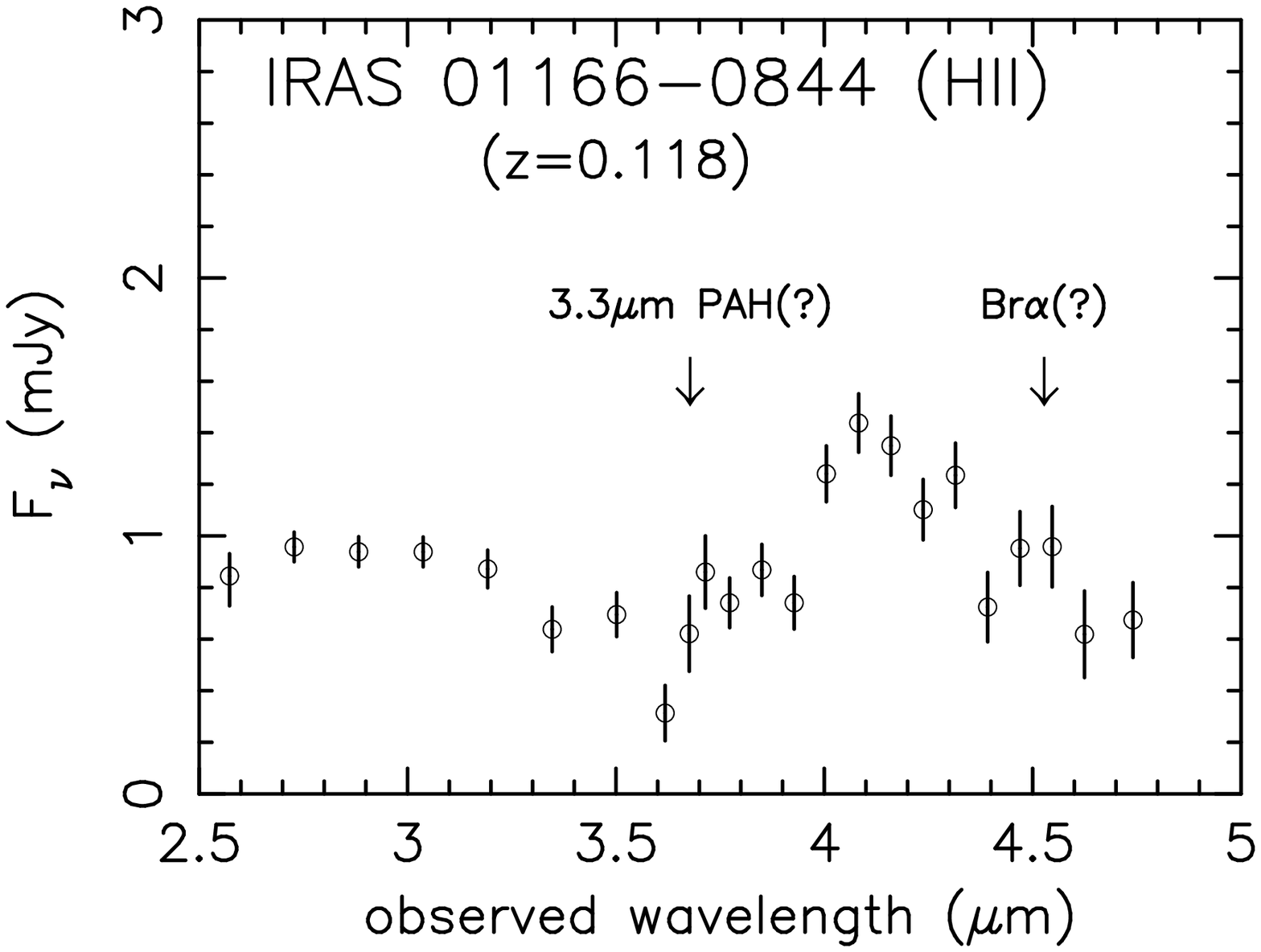} \\
\includegraphics[angle=0,scale=.27]{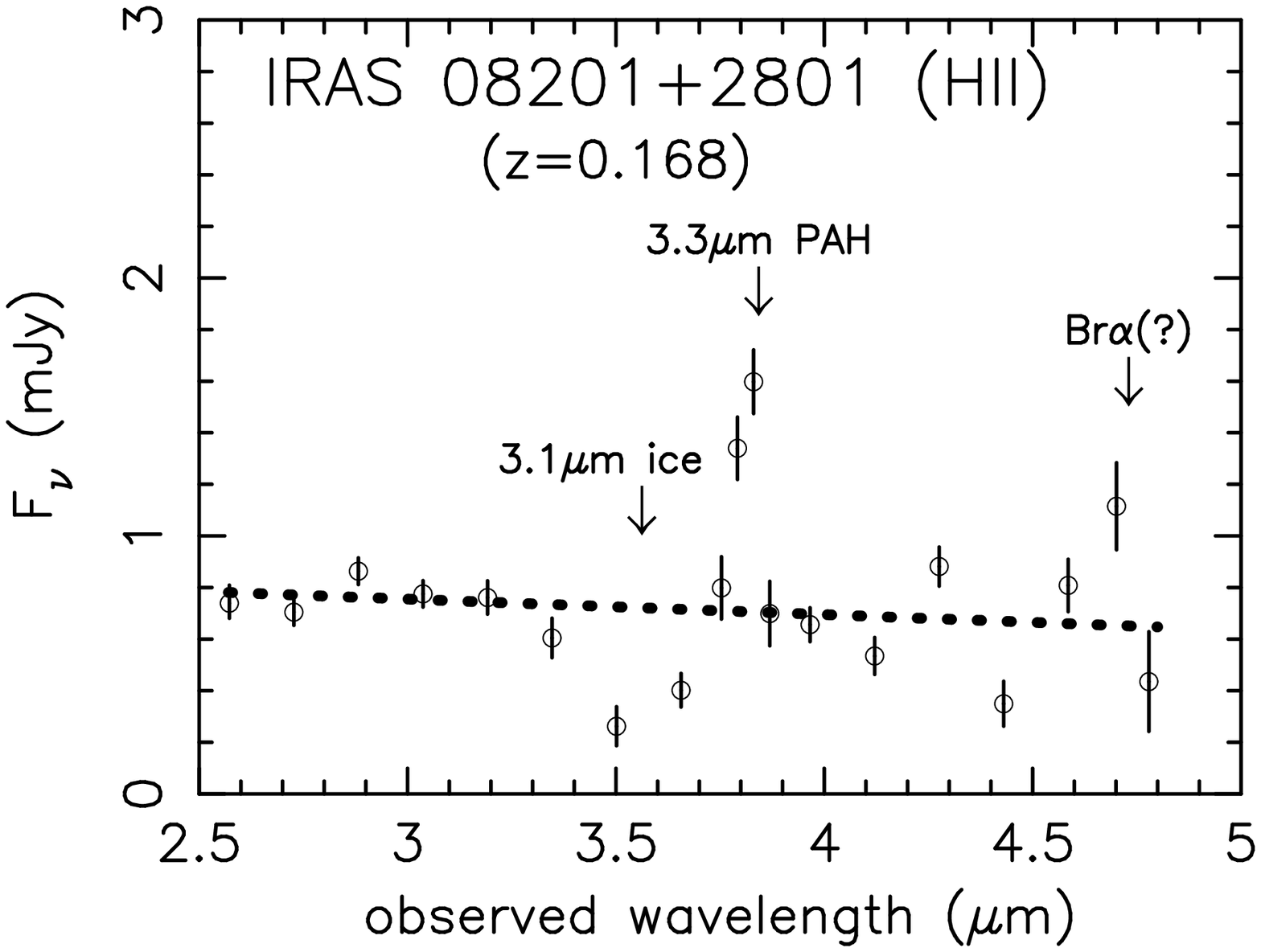}
\includegraphics[angle=0,scale=.27]{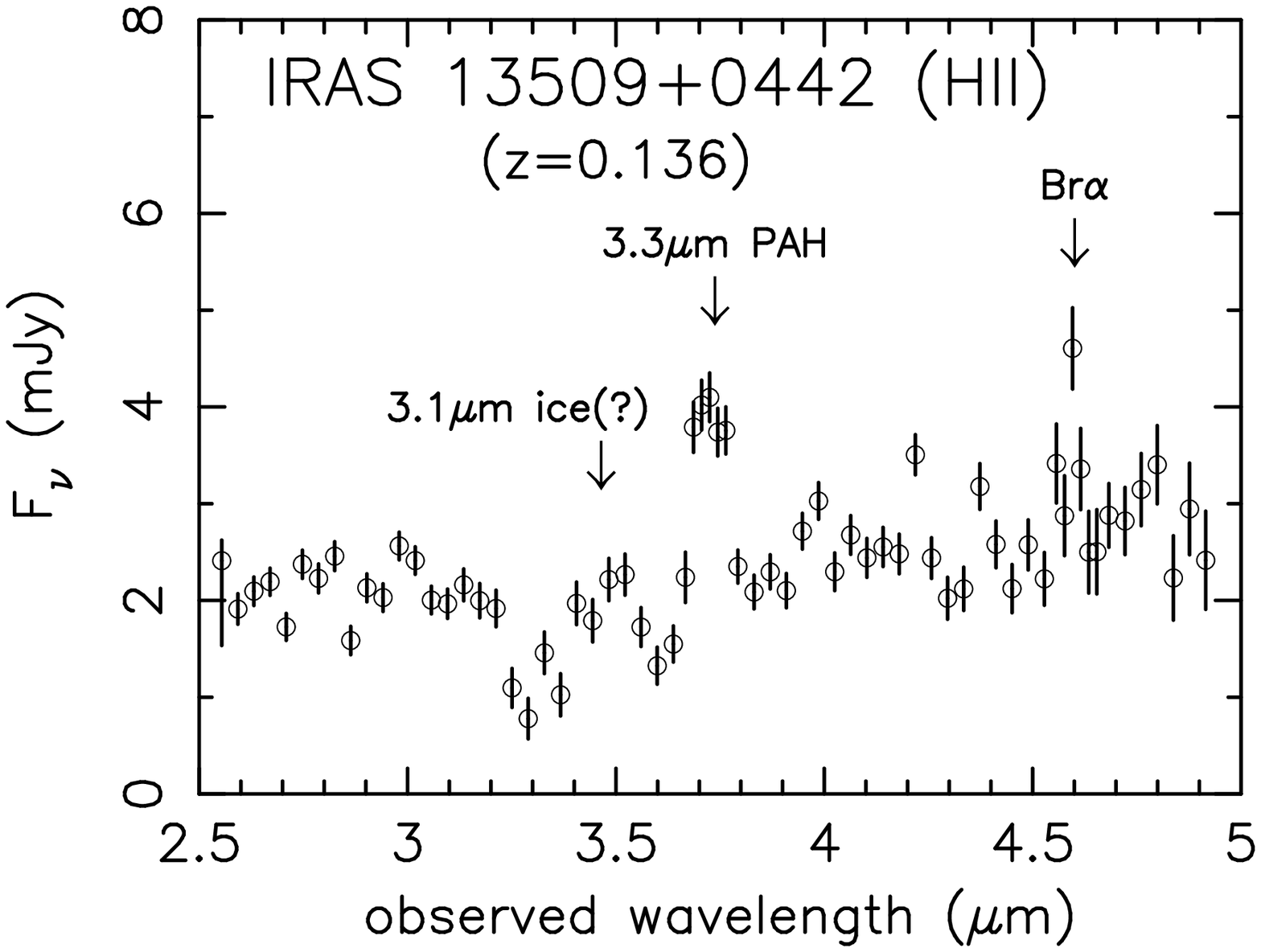}
\includegraphics[angle=0,scale=.27]{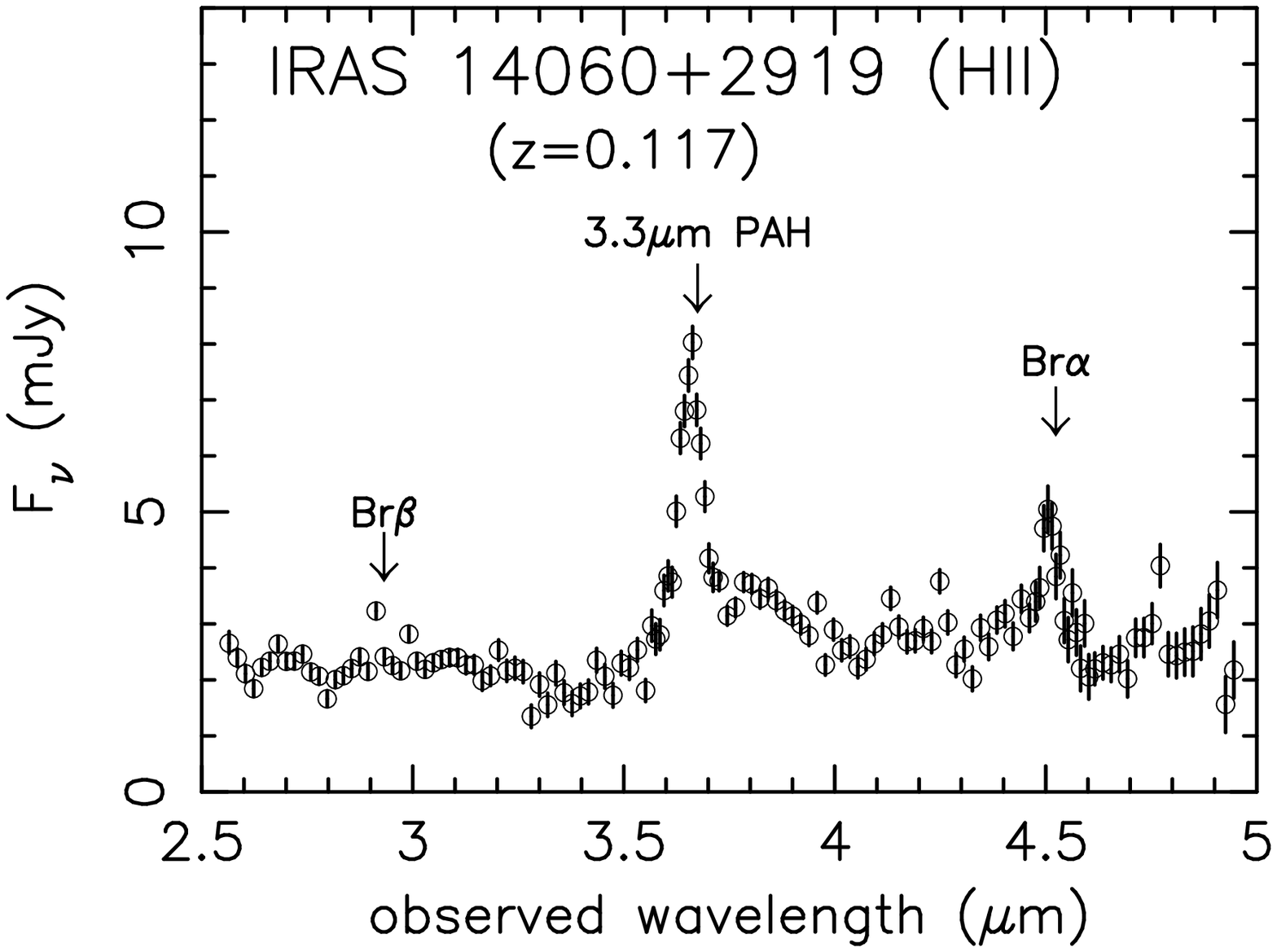} \\
\includegraphics[angle=0,scale=.27]{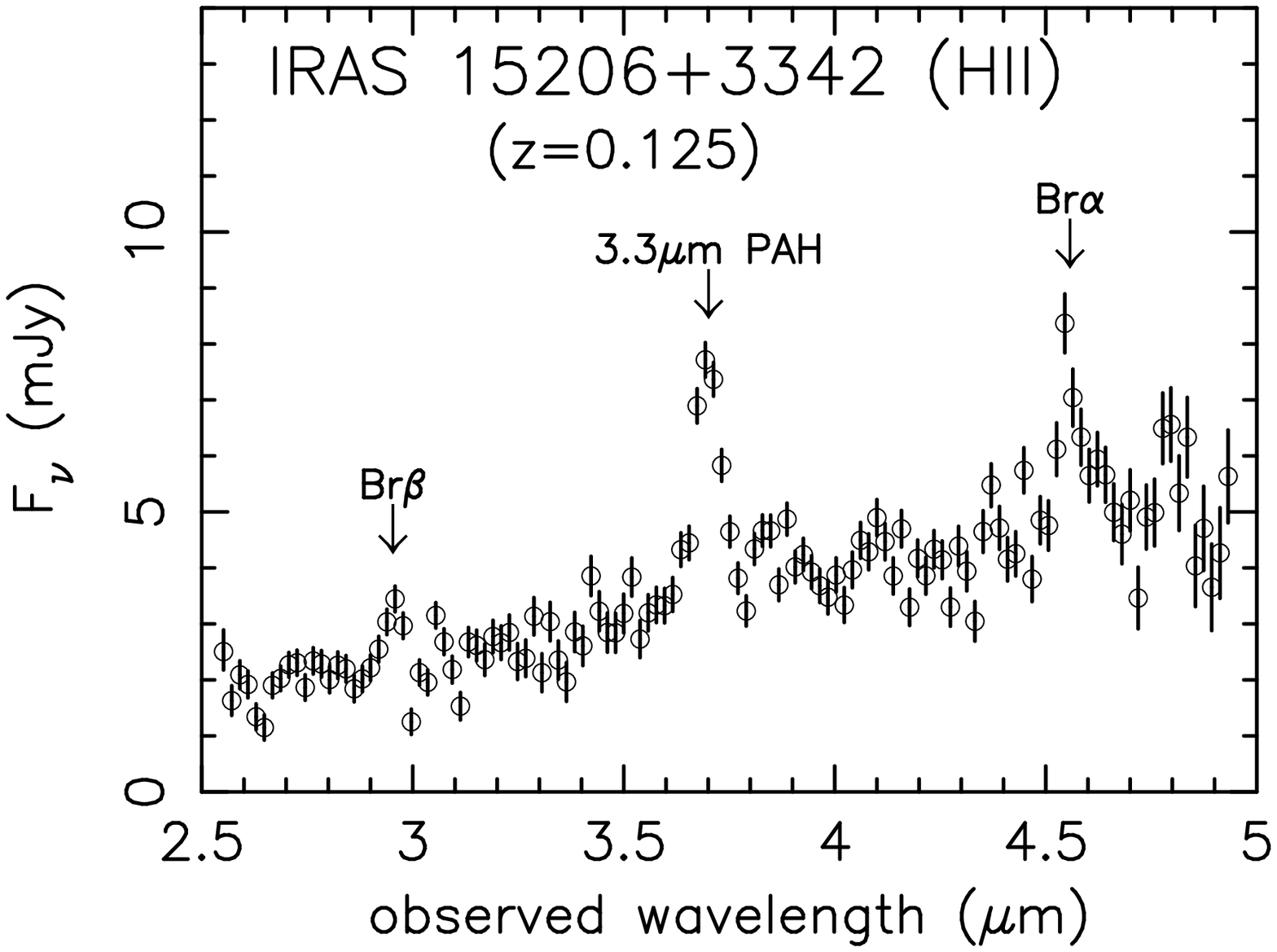}
\includegraphics[angle=0,scale=.27]{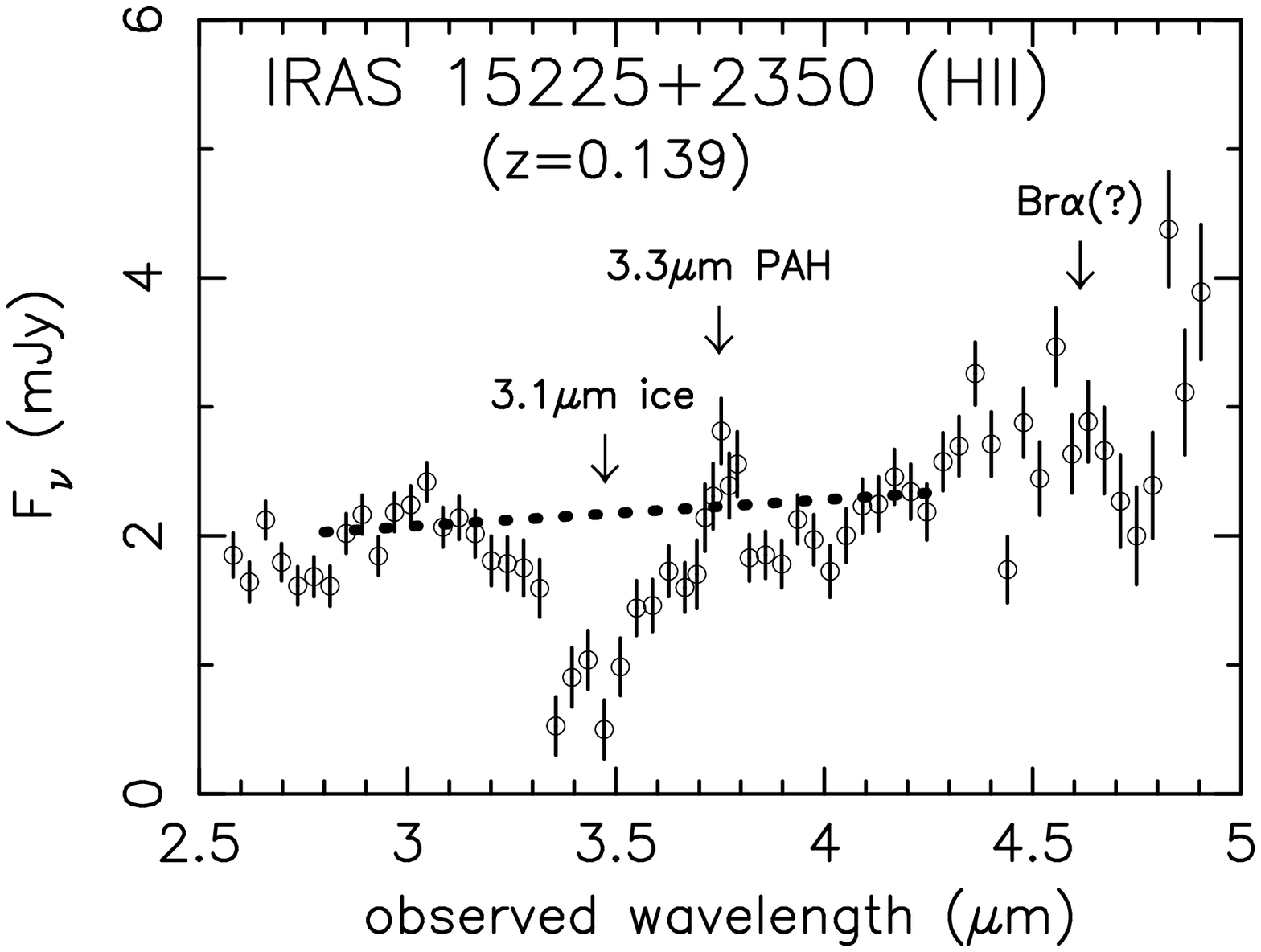}
\includegraphics[angle=0,scale=.27]{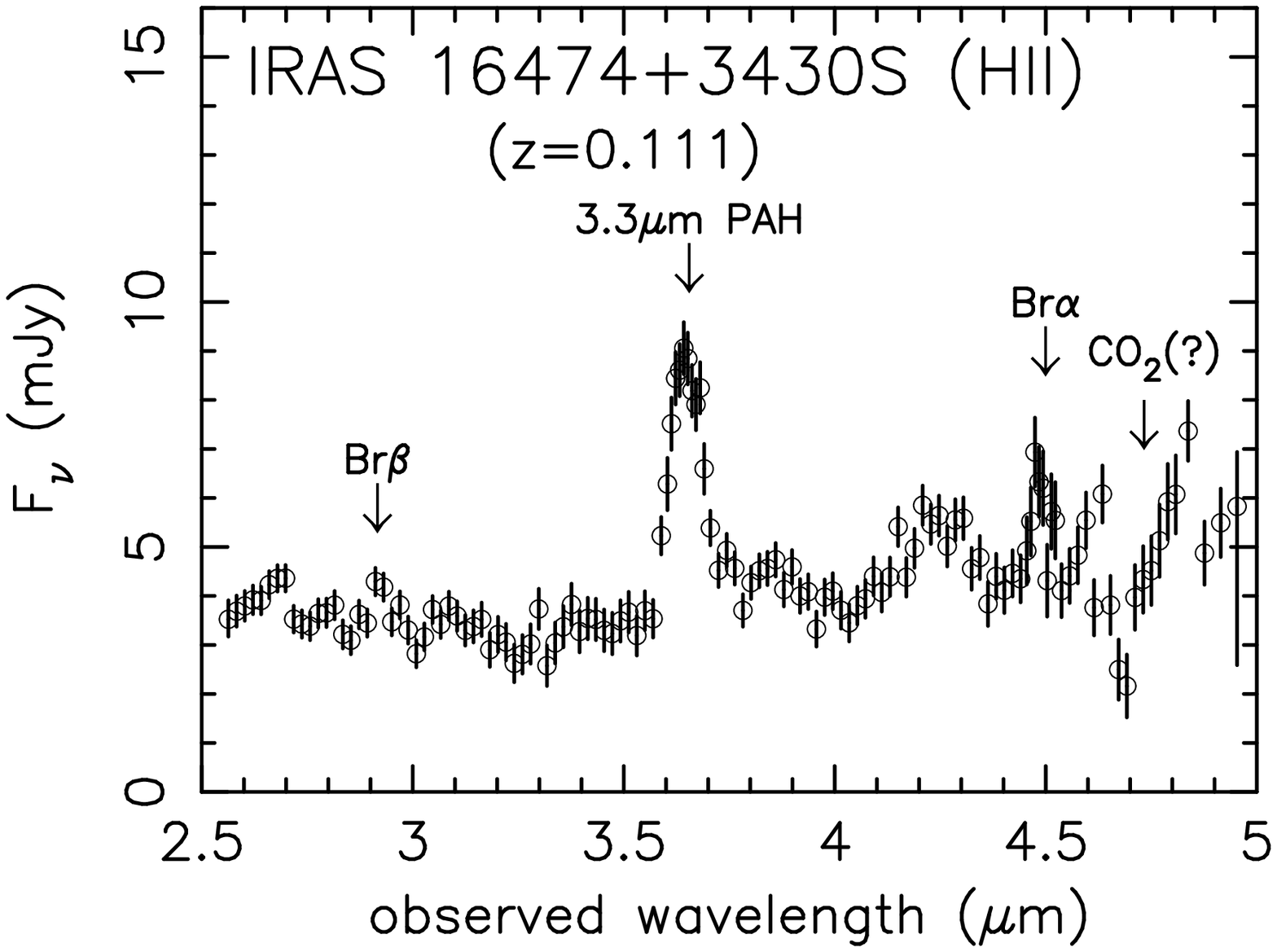} \\
\includegraphics[angle=0,scale=.27]{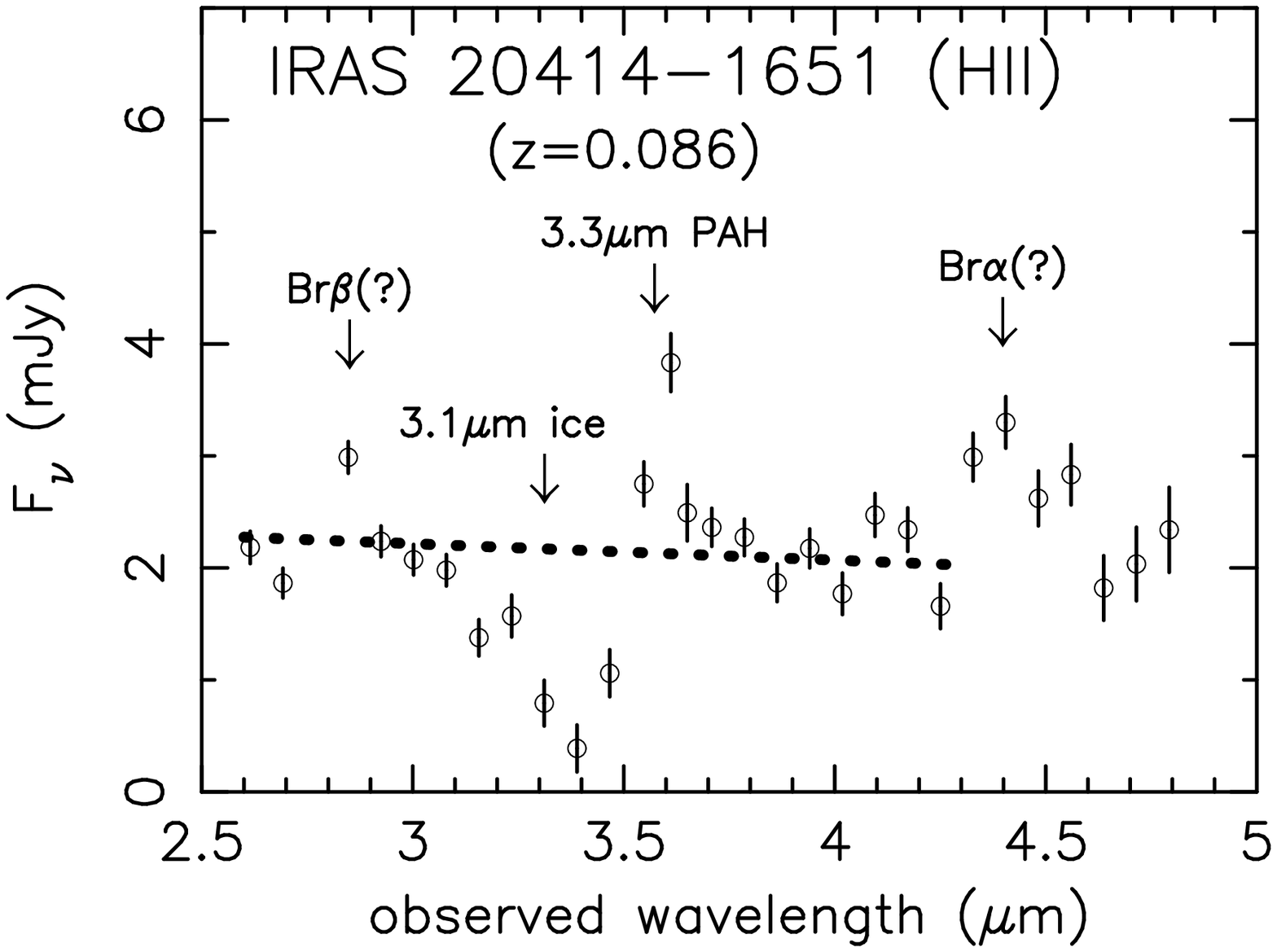}
\includegraphics[angle=0,scale=.27]{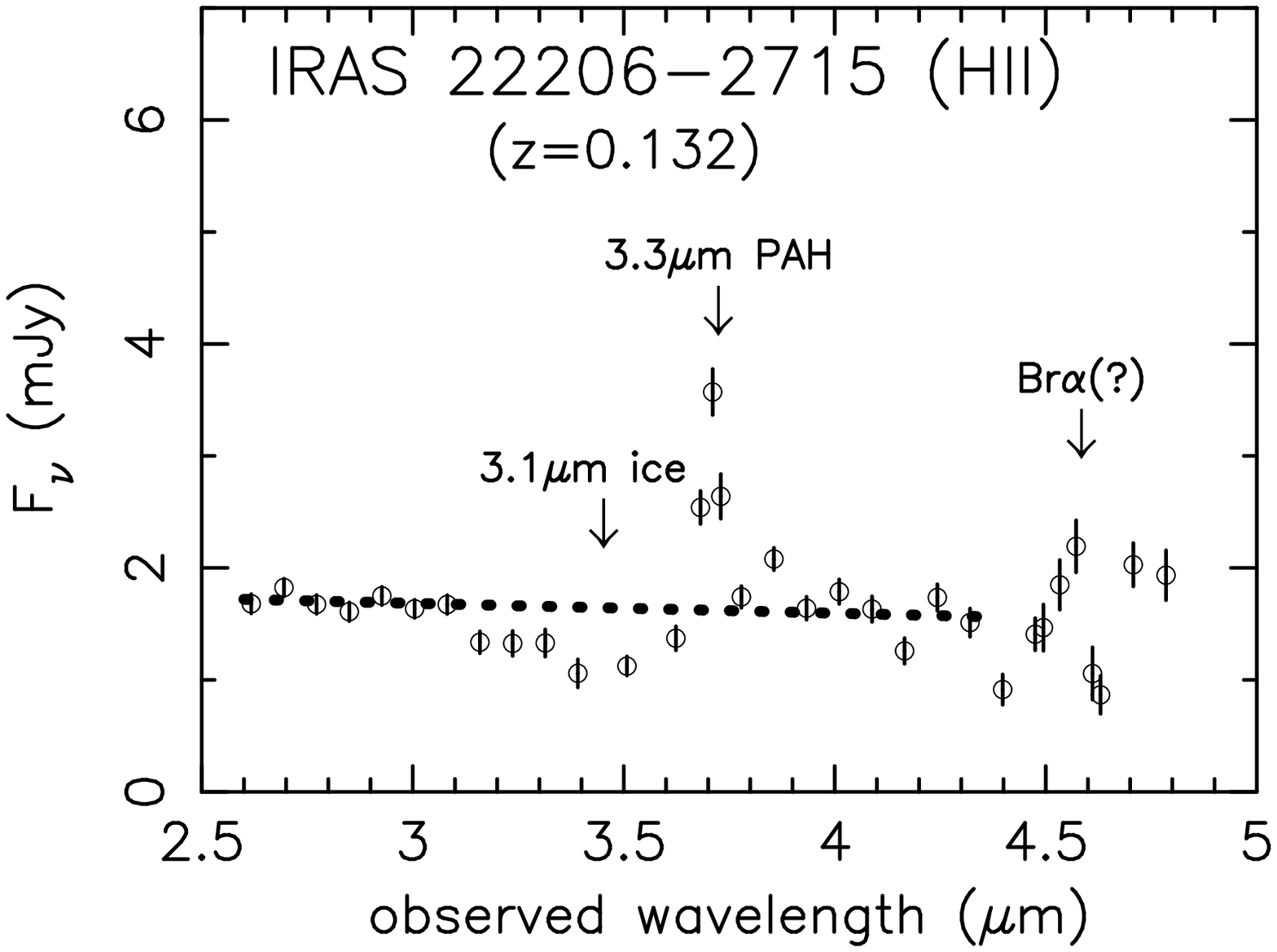}
\includegraphics[angle=0,scale=.27]{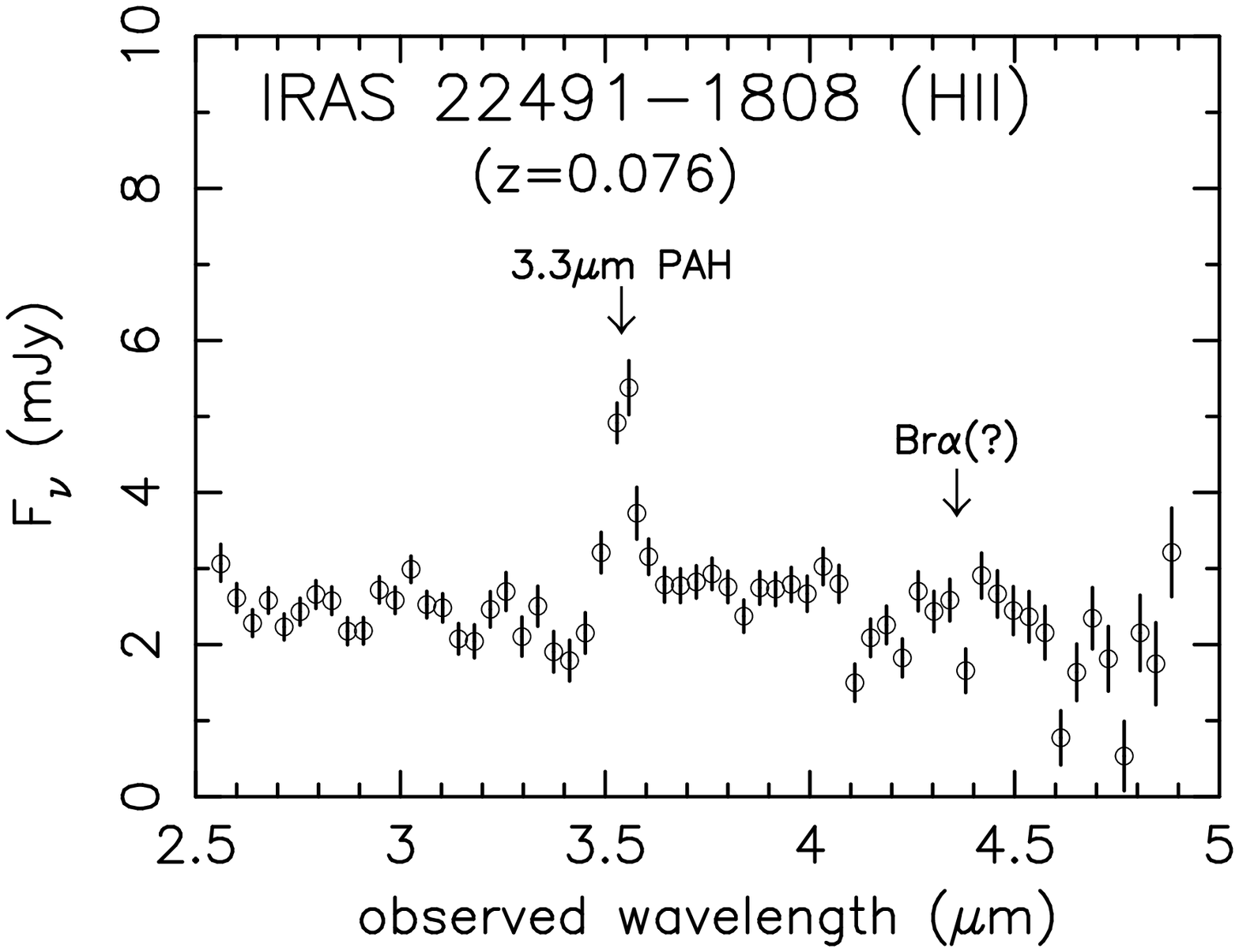} \\
\end{figure}

\clearpage

\begin{figure}
\includegraphics[angle=0,scale=.27]{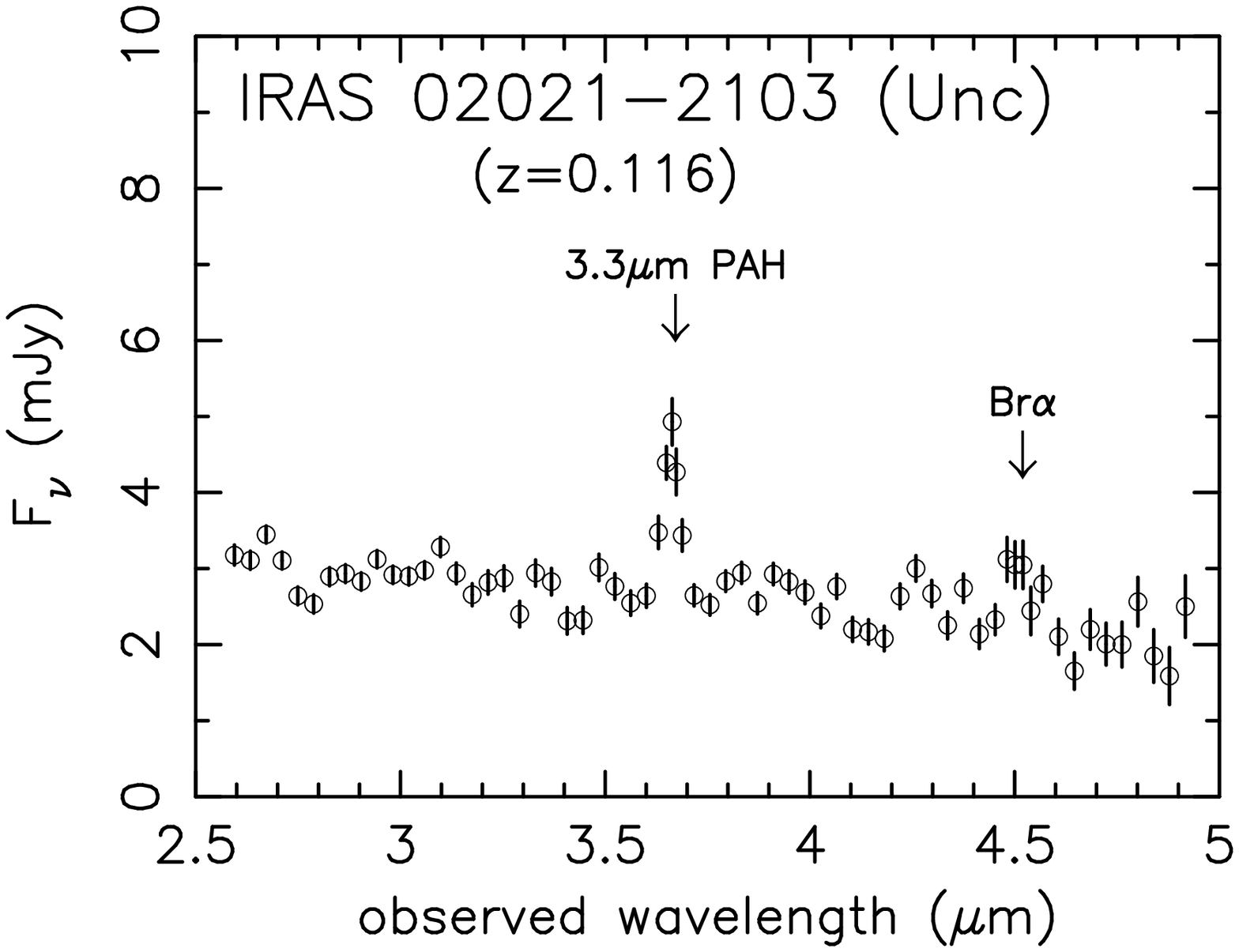}
\includegraphics[angle=0,scale=.27]{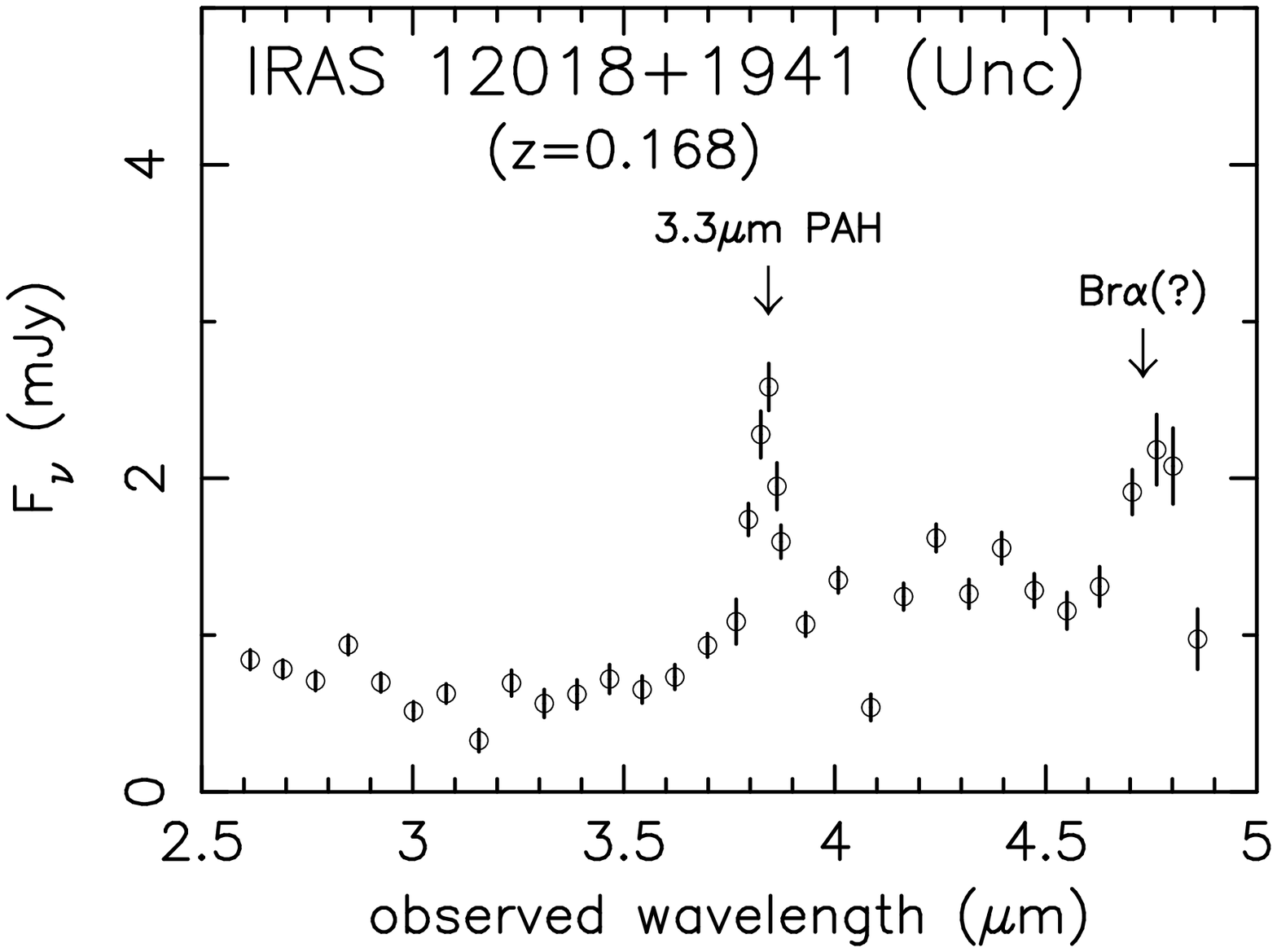}
\includegraphics[angle=0,scale=.27]{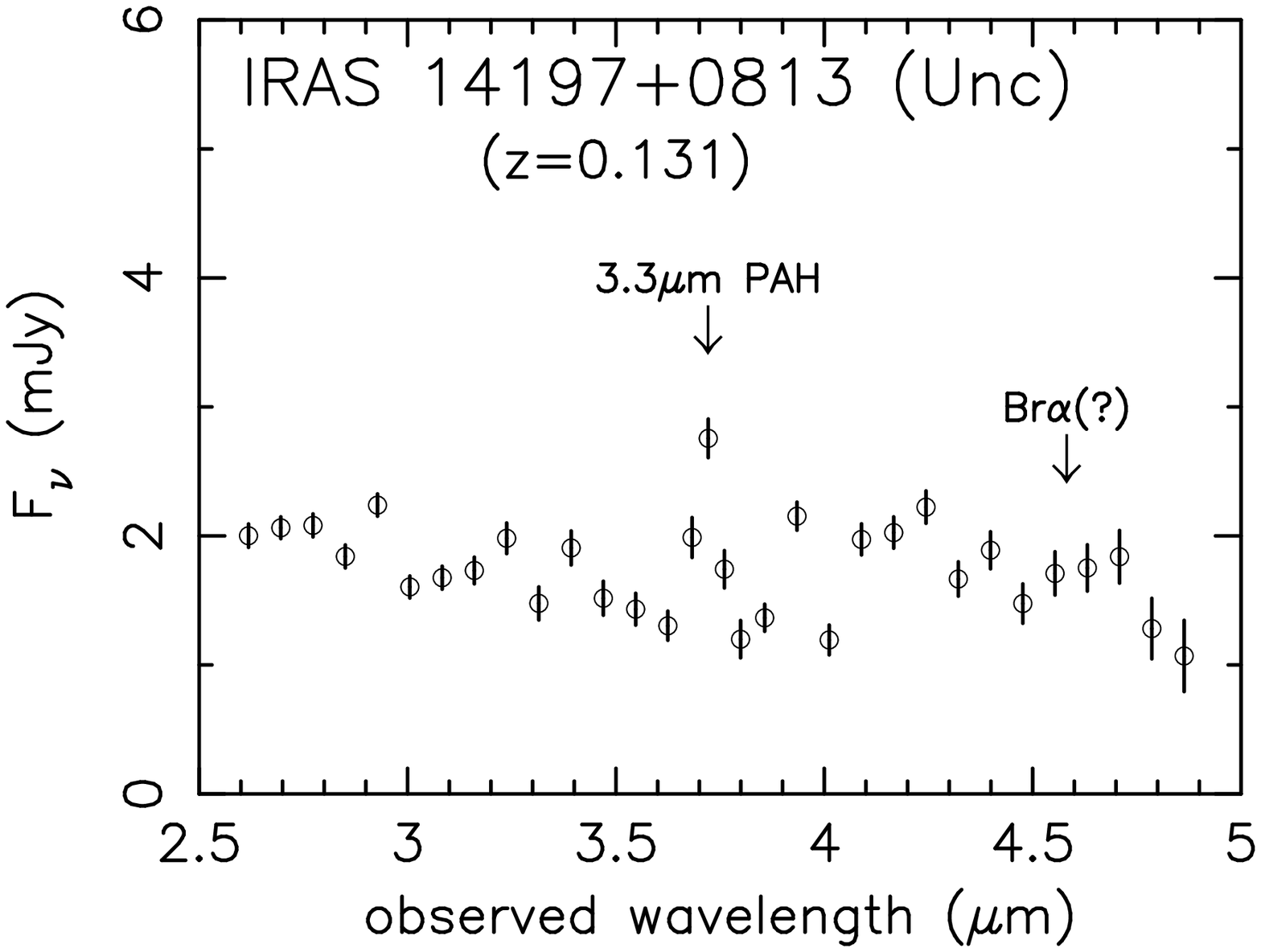} \\
\includegraphics[angle=0,scale=.27]{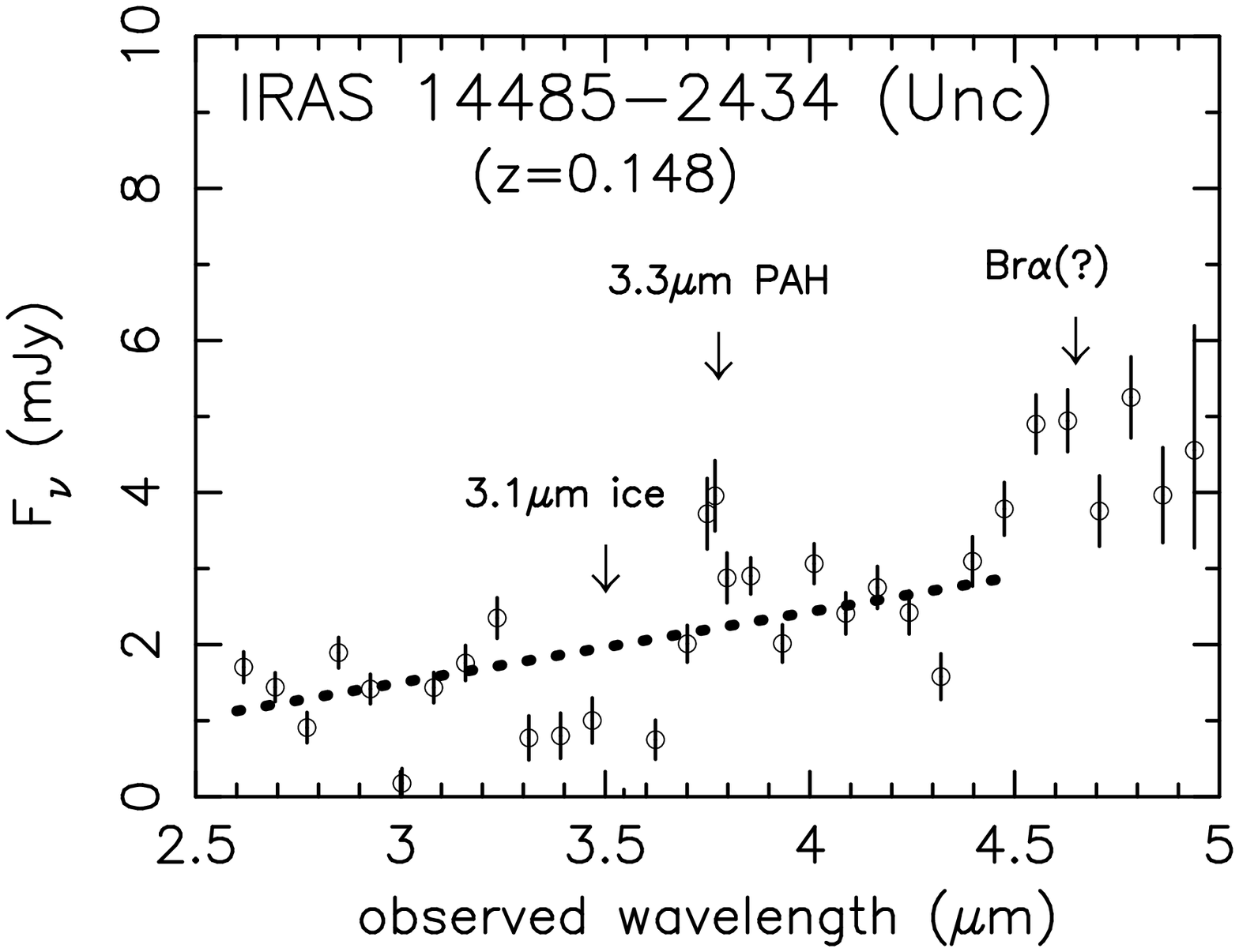}
\includegraphics[angle=0,scale=.27]{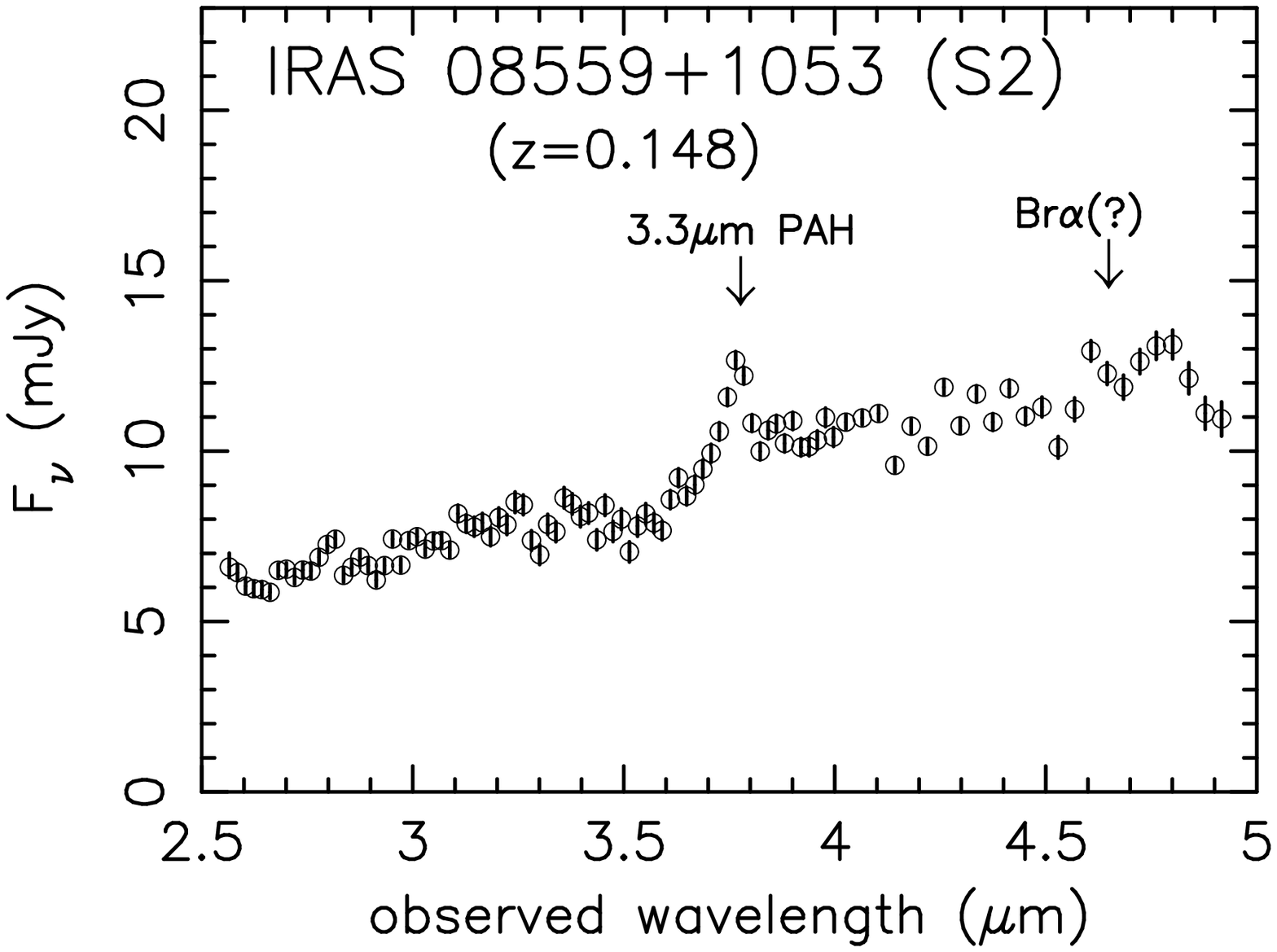}
\includegraphics[angle=0,scale=.27]{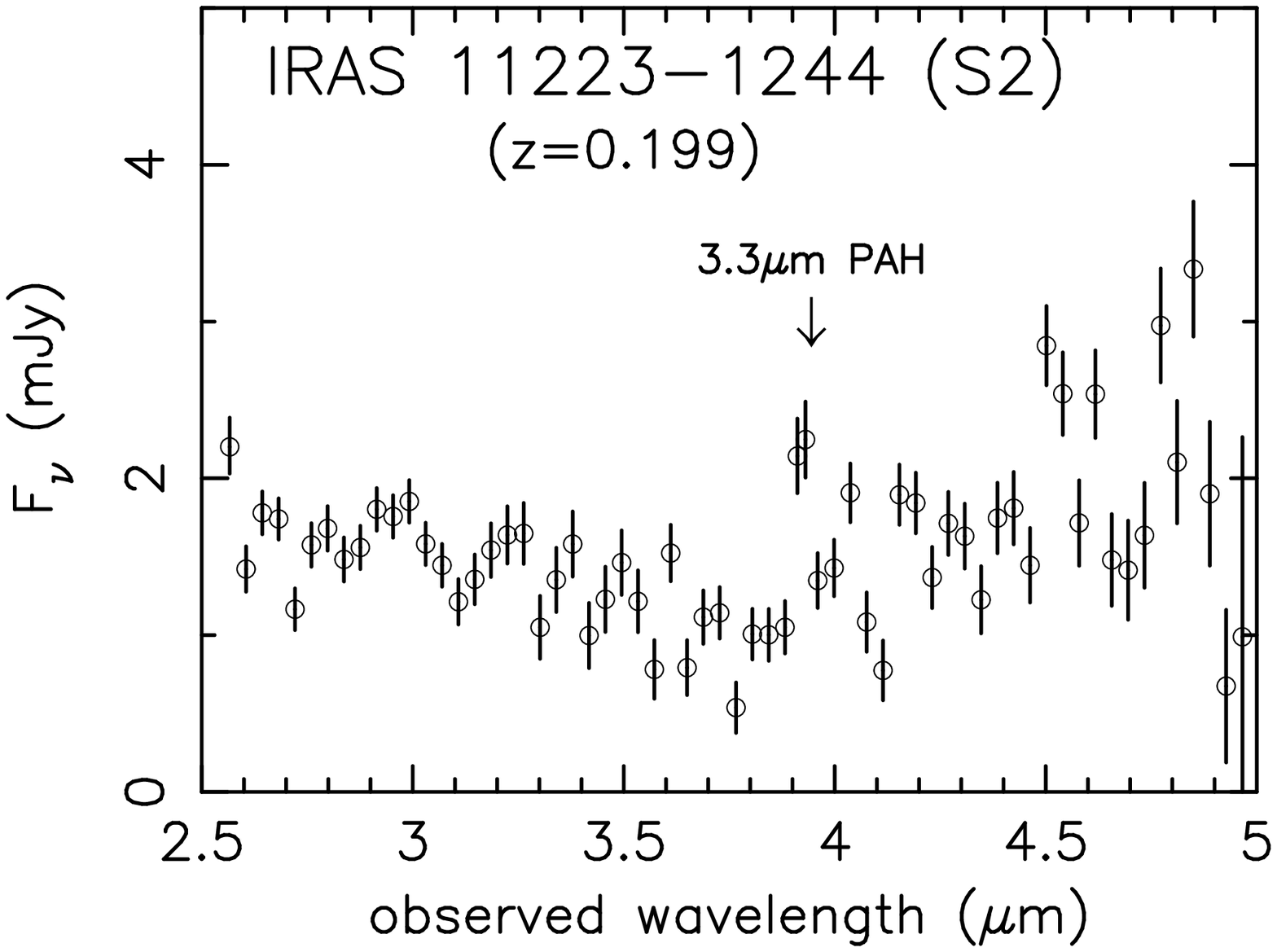} \\
\includegraphics[angle=0,scale=.27]{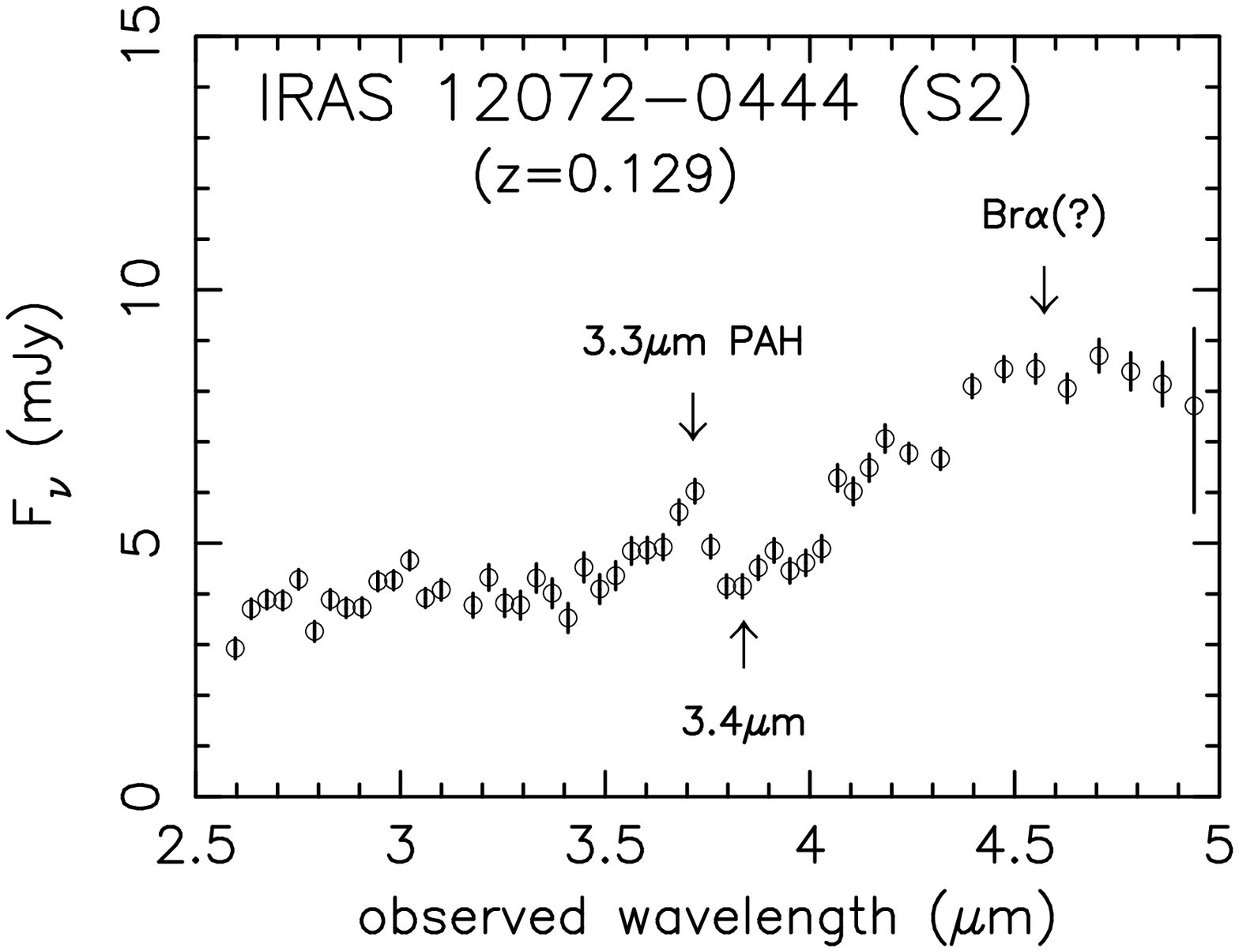}
\includegraphics[angle=0,scale=.27]{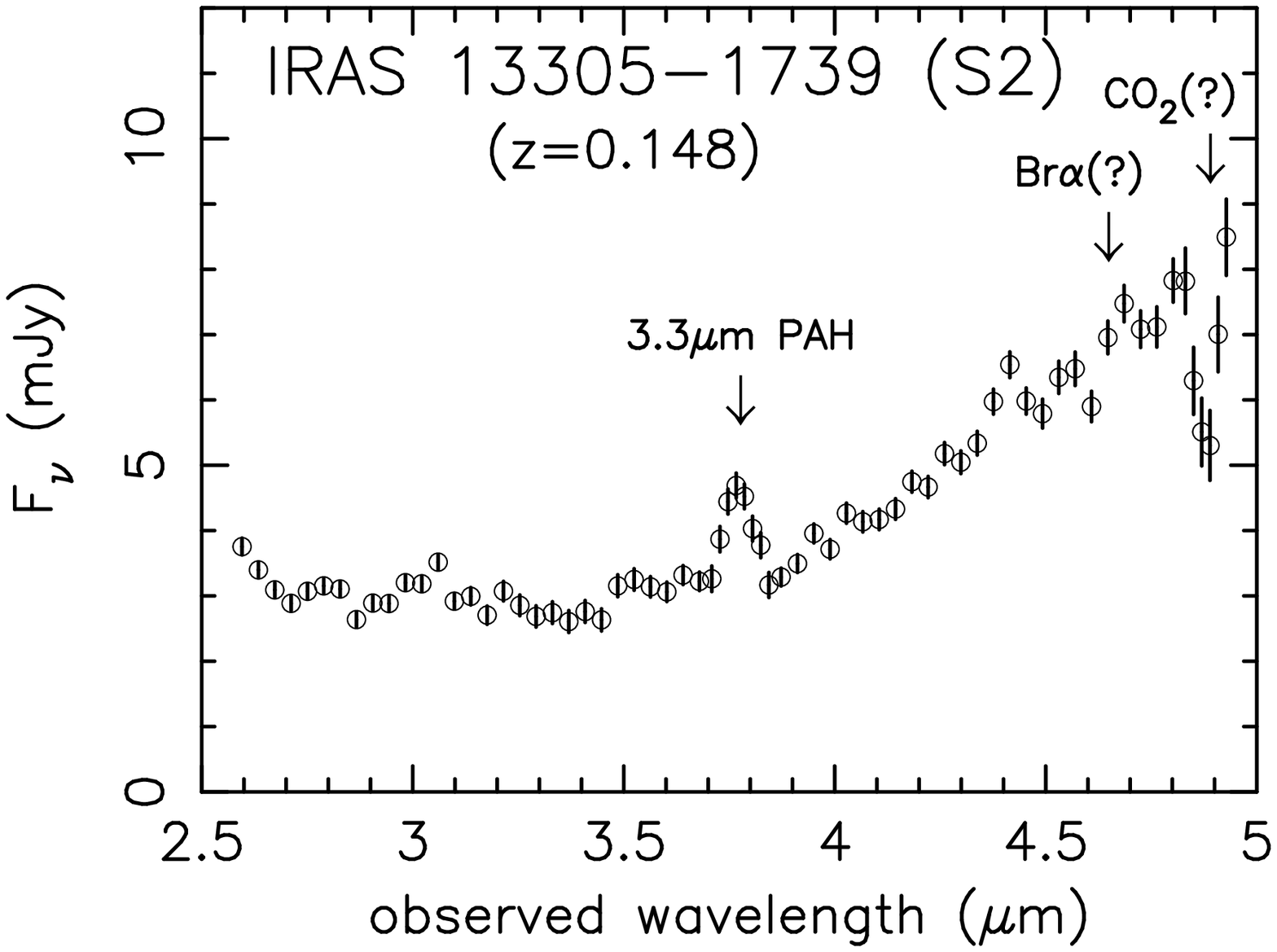}
\includegraphics[angle=0,scale=.27]{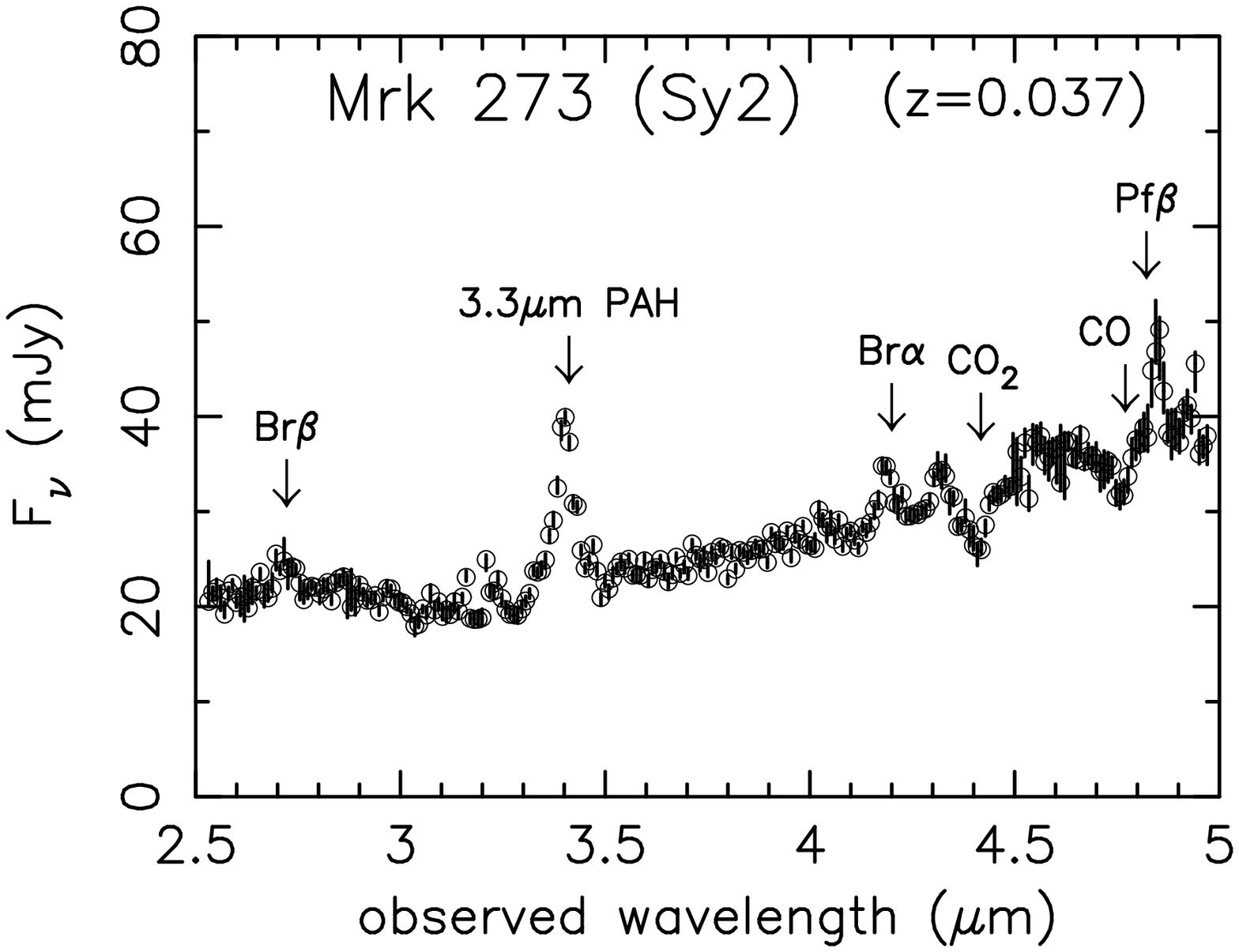} \\
\includegraphics[angle=0,scale=.27]{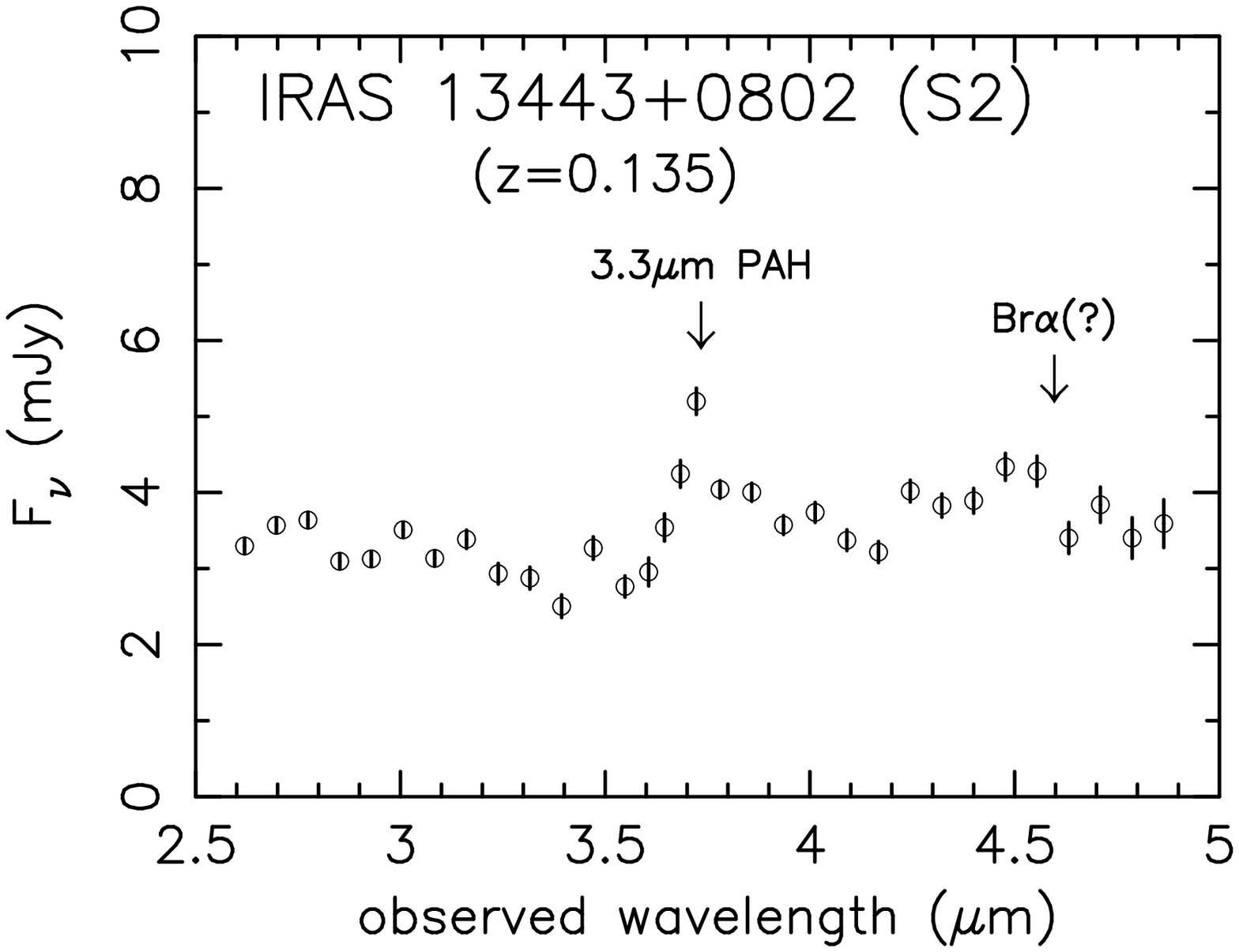}
\includegraphics[angle=0,scale=.27]{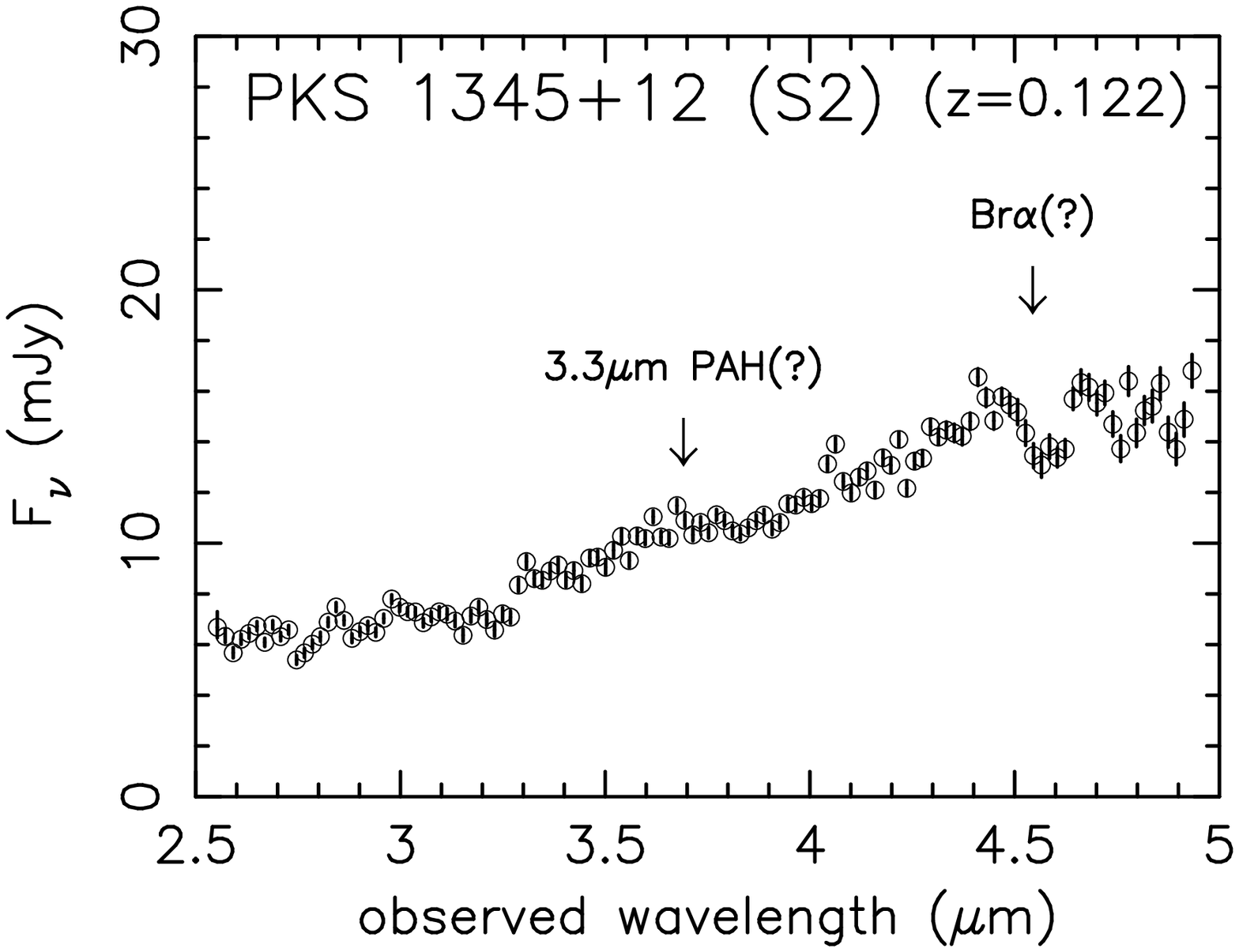}
\includegraphics[angle=0,scale=.27]{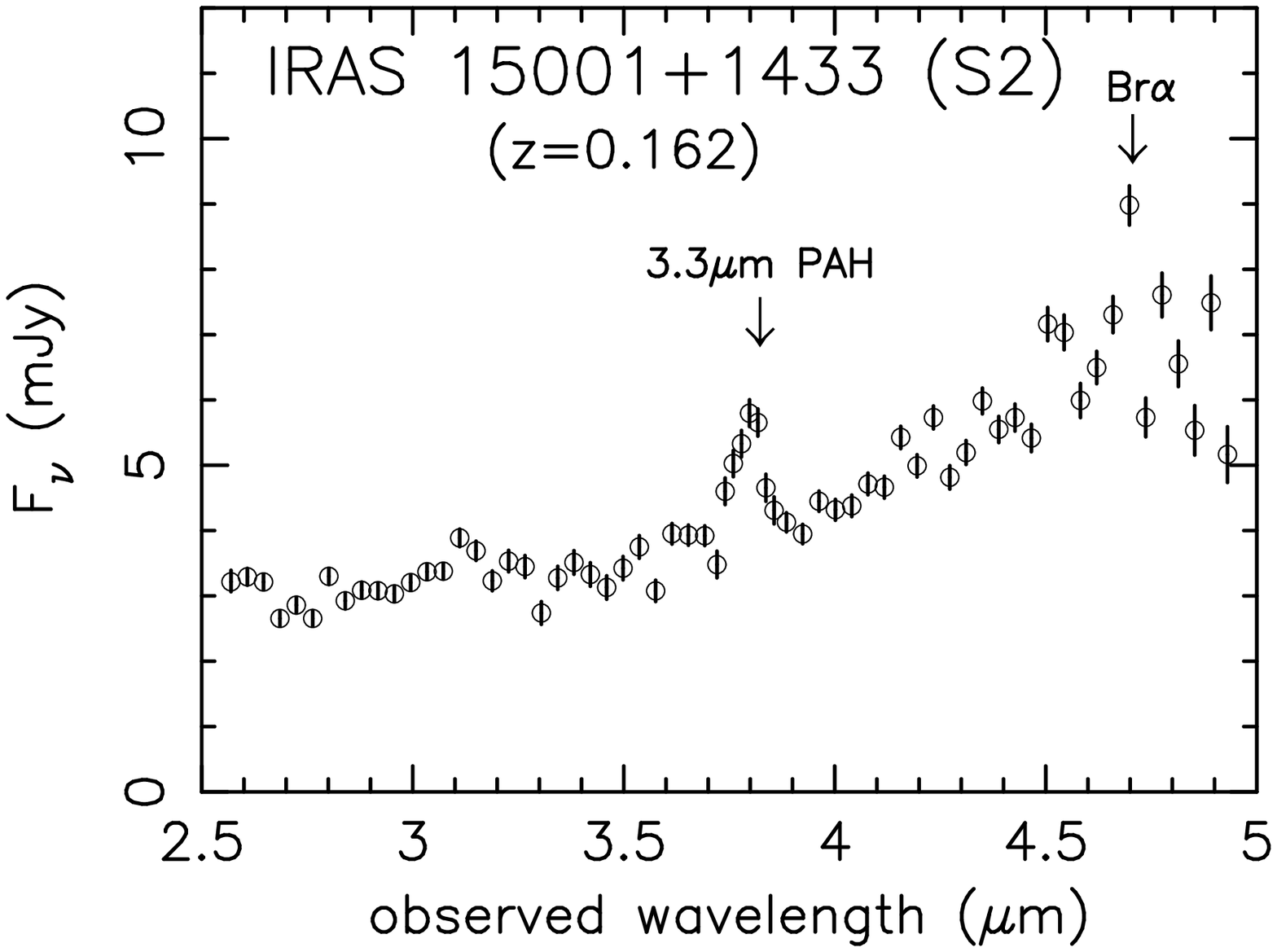} \\
\includegraphics[angle=0,scale=.27]{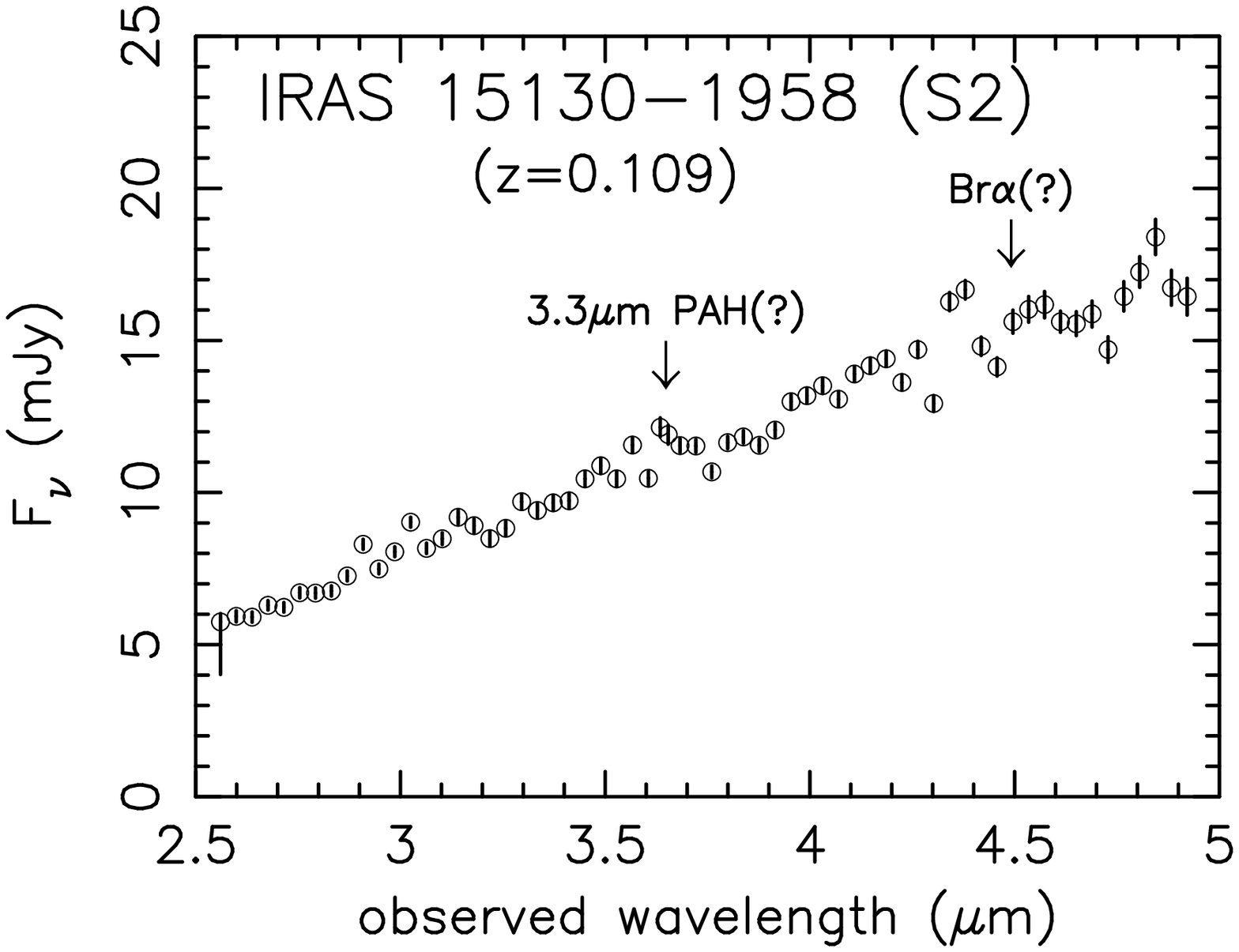}
\includegraphics[angle=0,scale=.27]{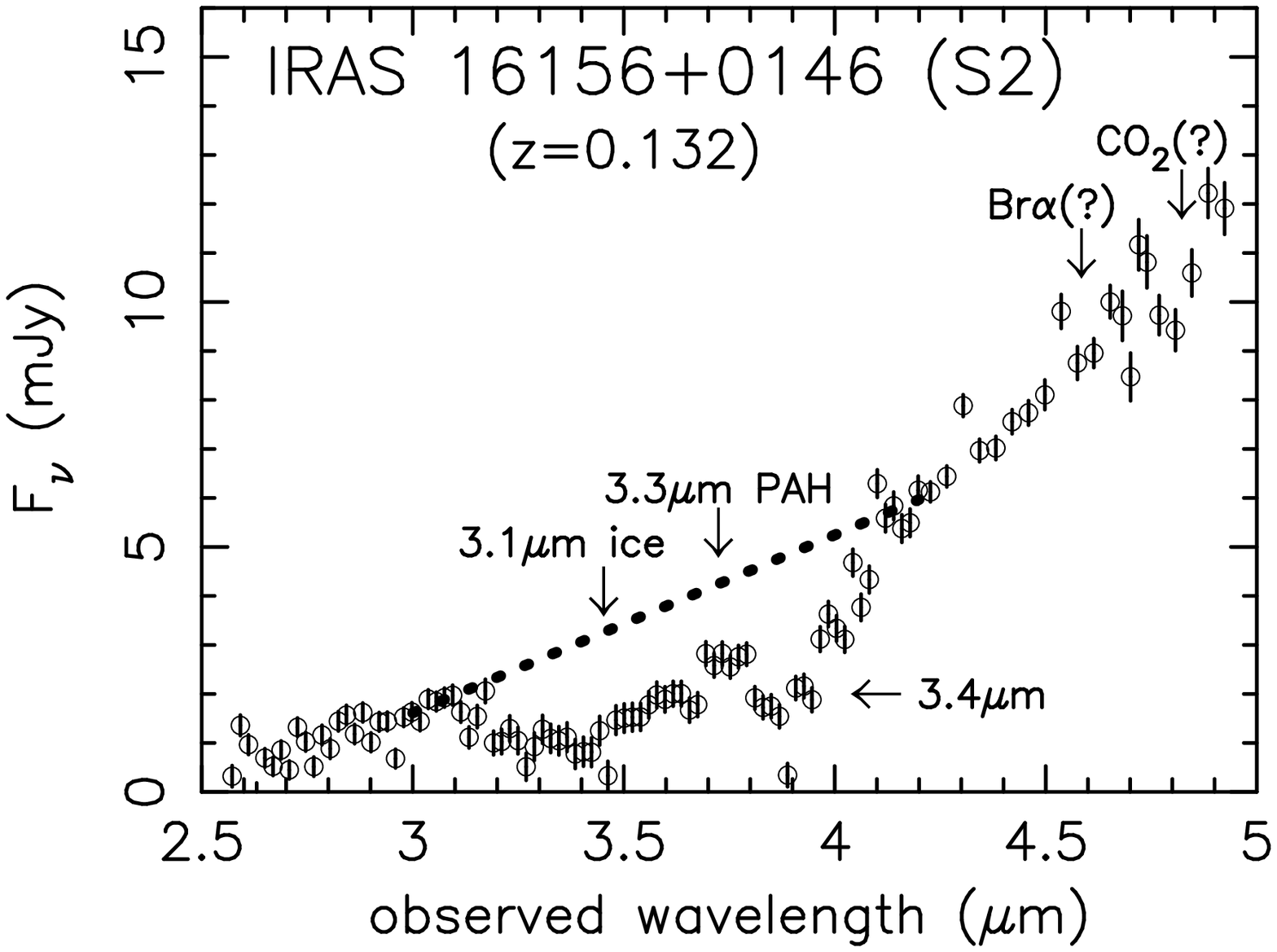}
\includegraphics[angle=0,scale=.27]{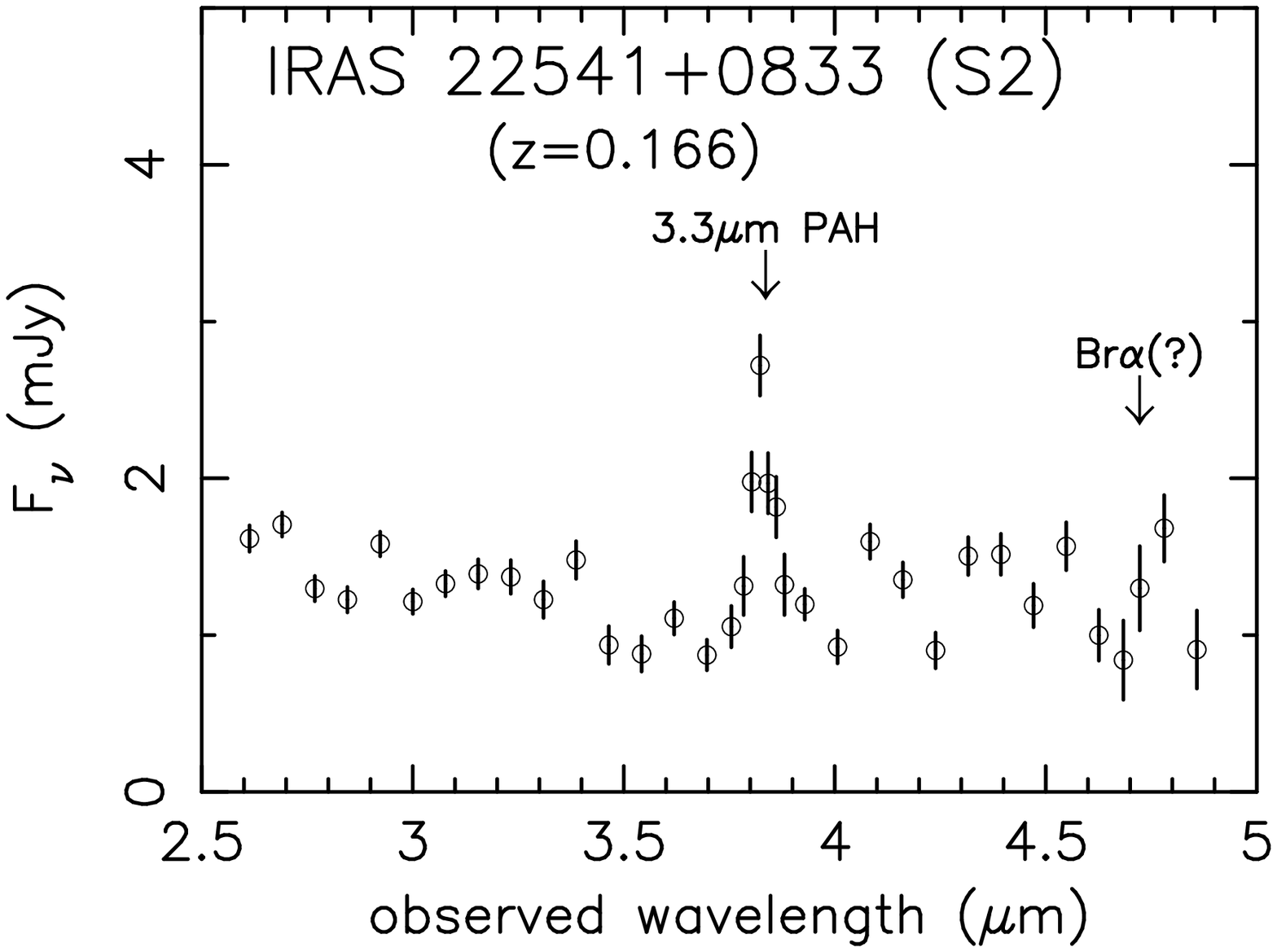} \\
\end{figure}

\clearpage

\begin{figure}
\includegraphics[angle=0,scale=.27]{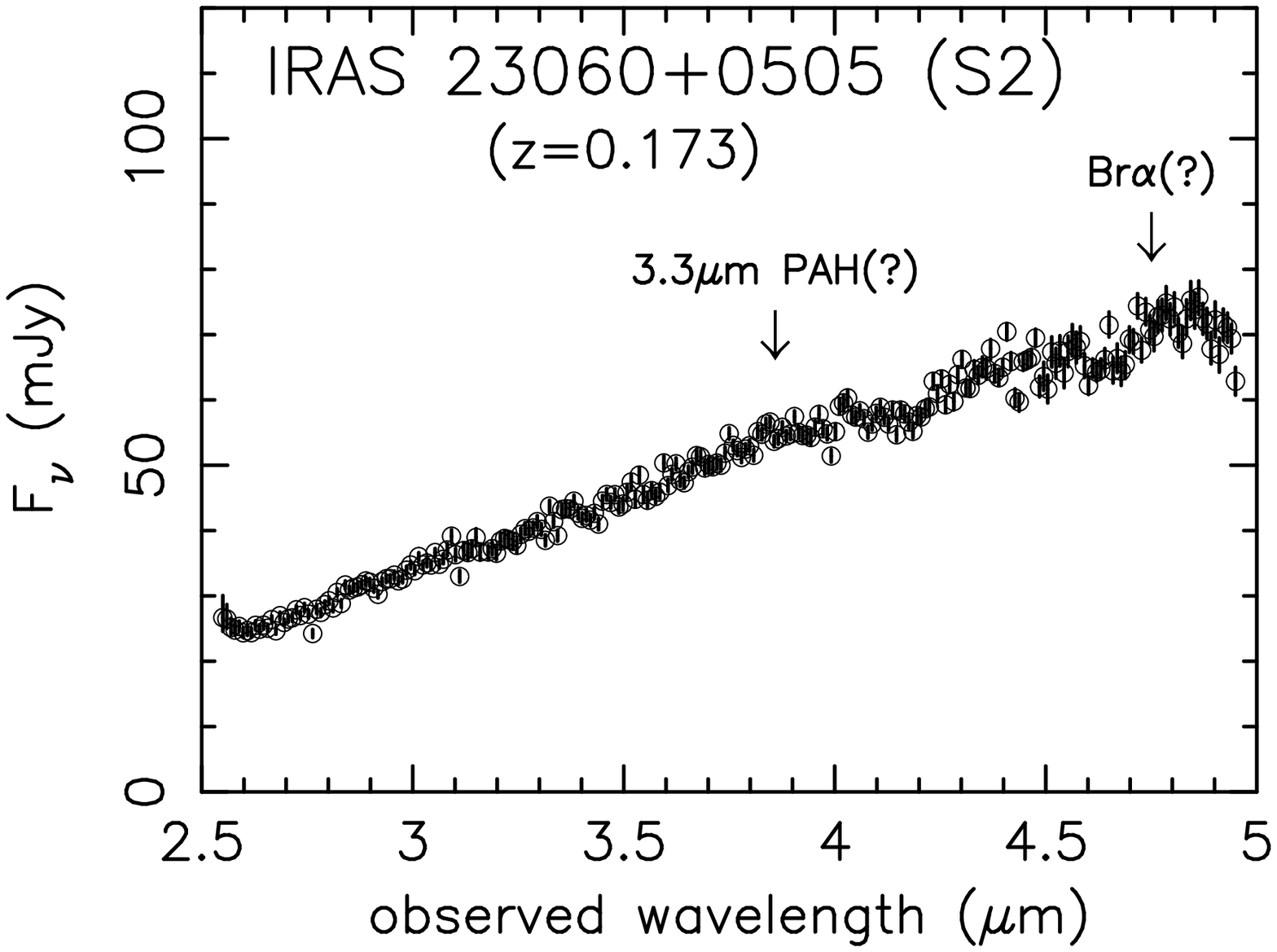}
\includegraphics[angle=0,scale=.27]{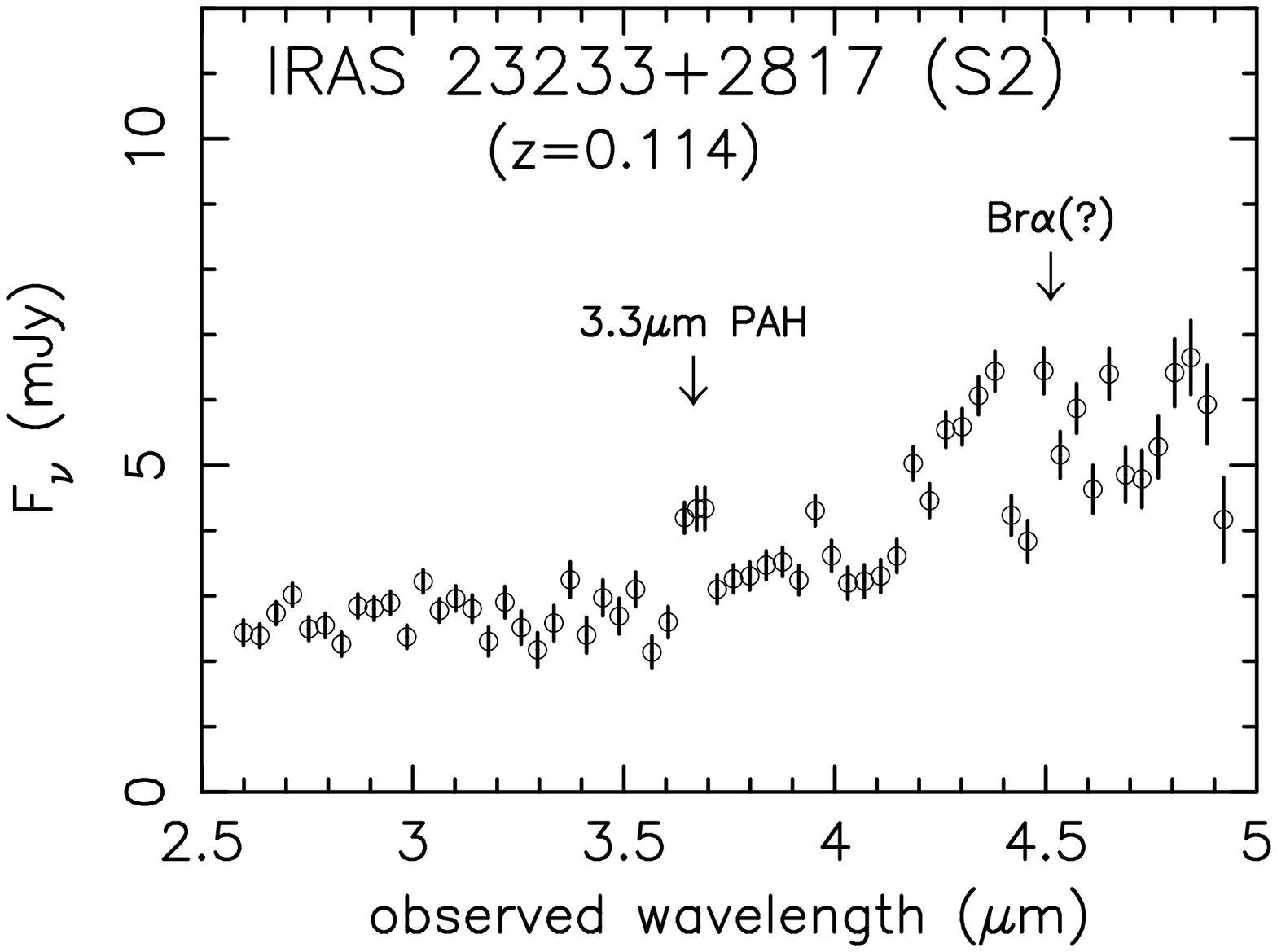}
\includegraphics[angle=0,scale=.27]{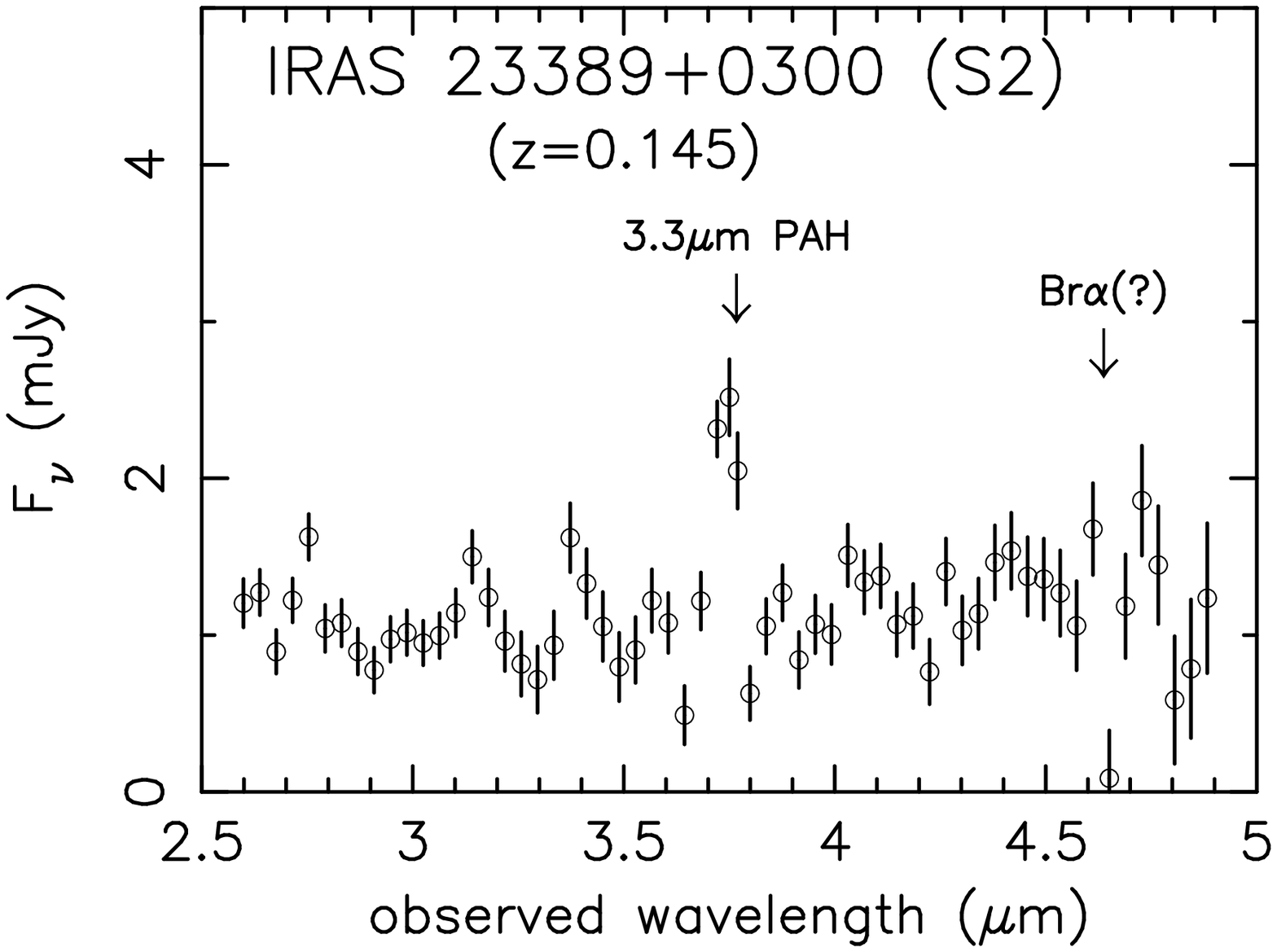} \\
\includegraphics[angle=0,scale=.27]{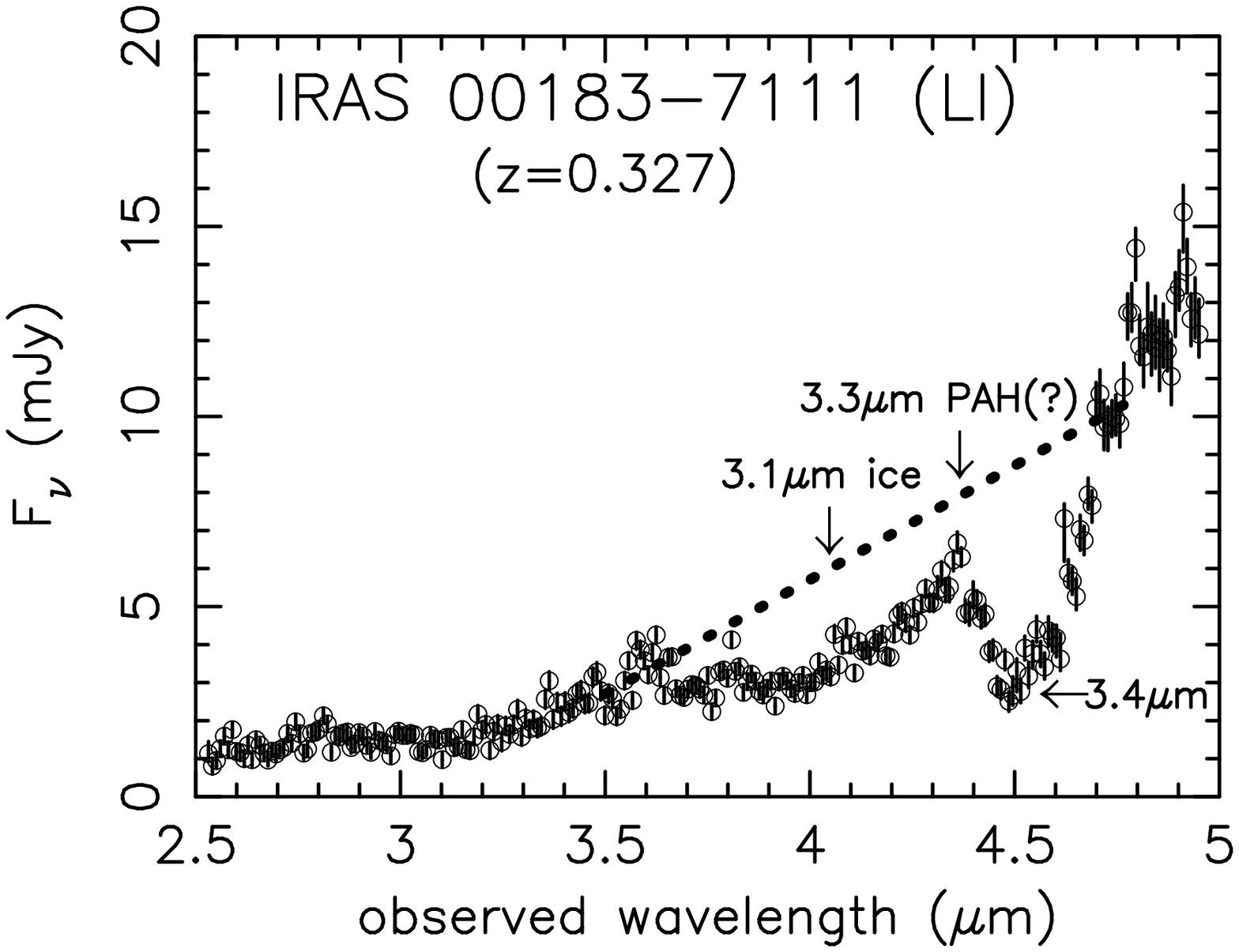}
\includegraphics[angle=0,scale=.27]{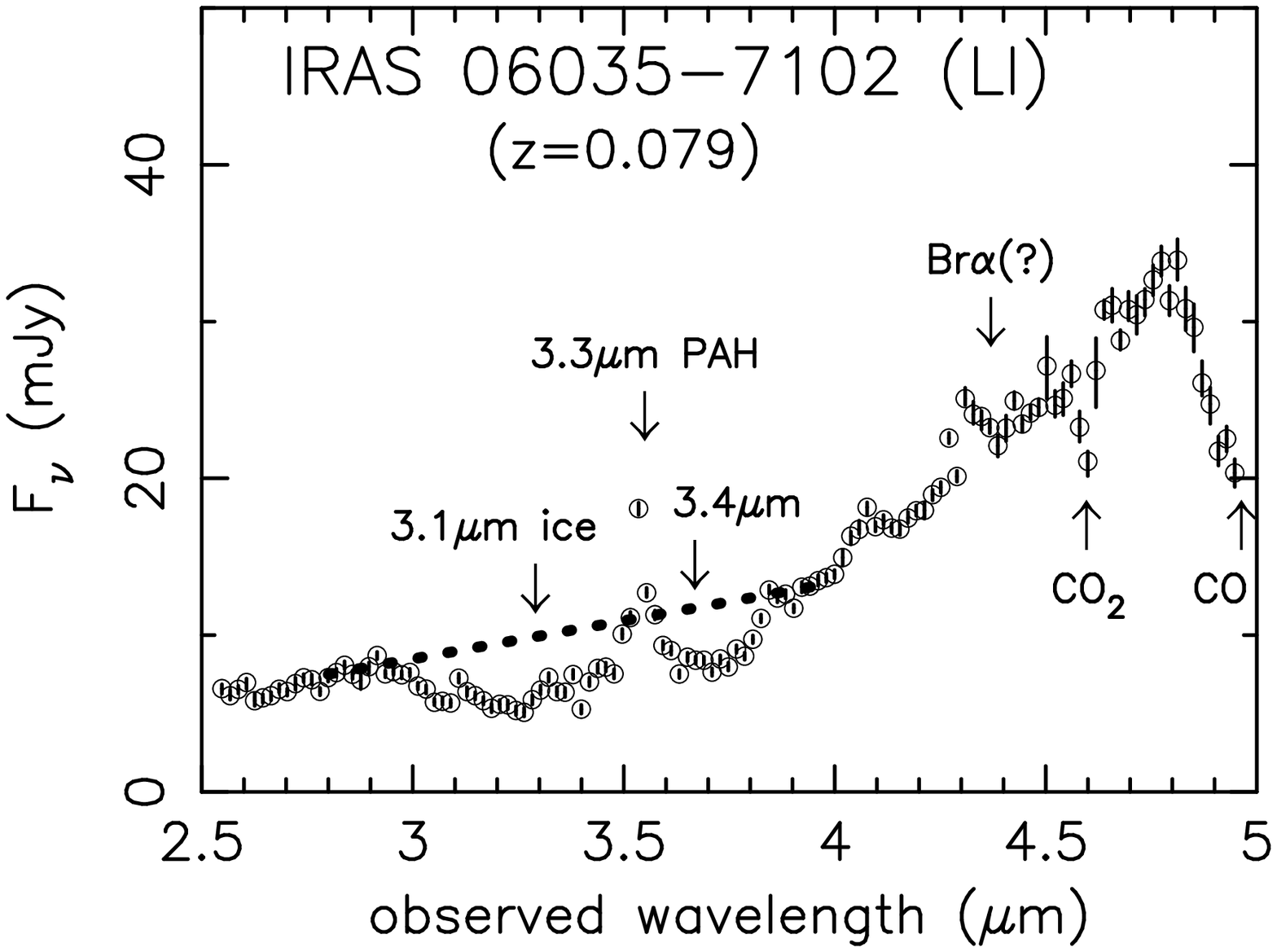}
\includegraphics[angle=0,scale=.27]{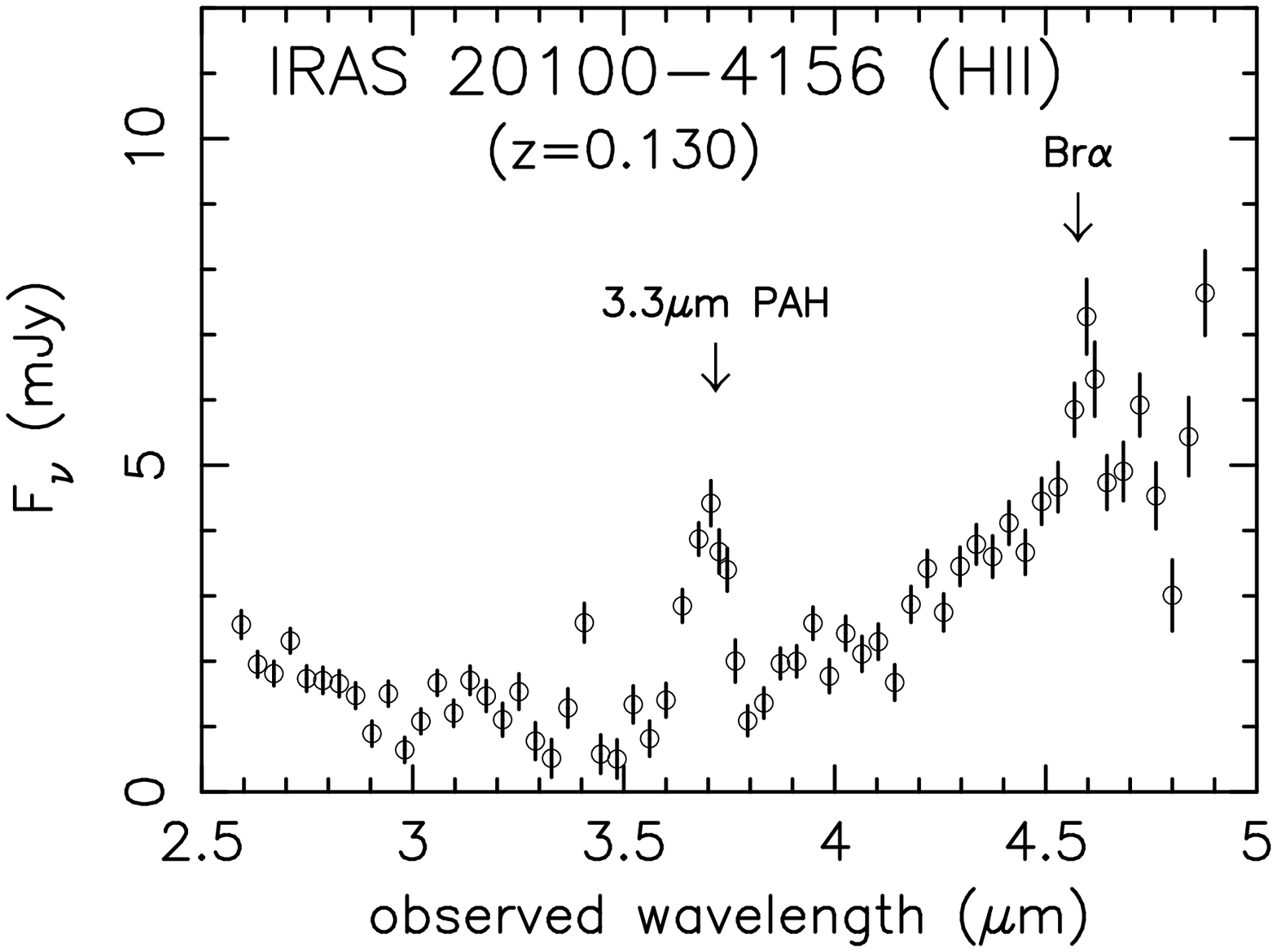} \\
\includegraphics[angle=0,scale=.27]{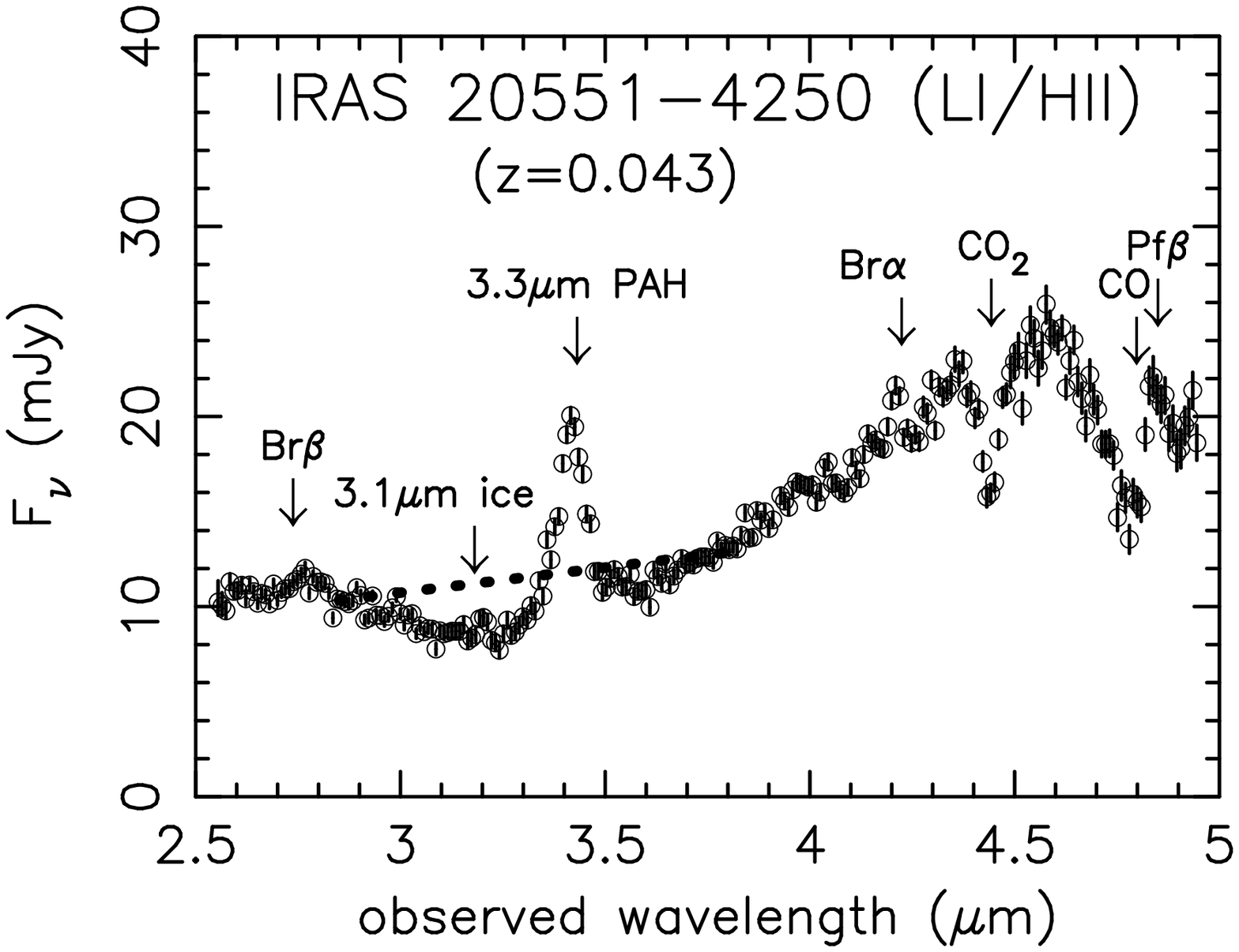}
\includegraphics[angle=0,scale=.27]{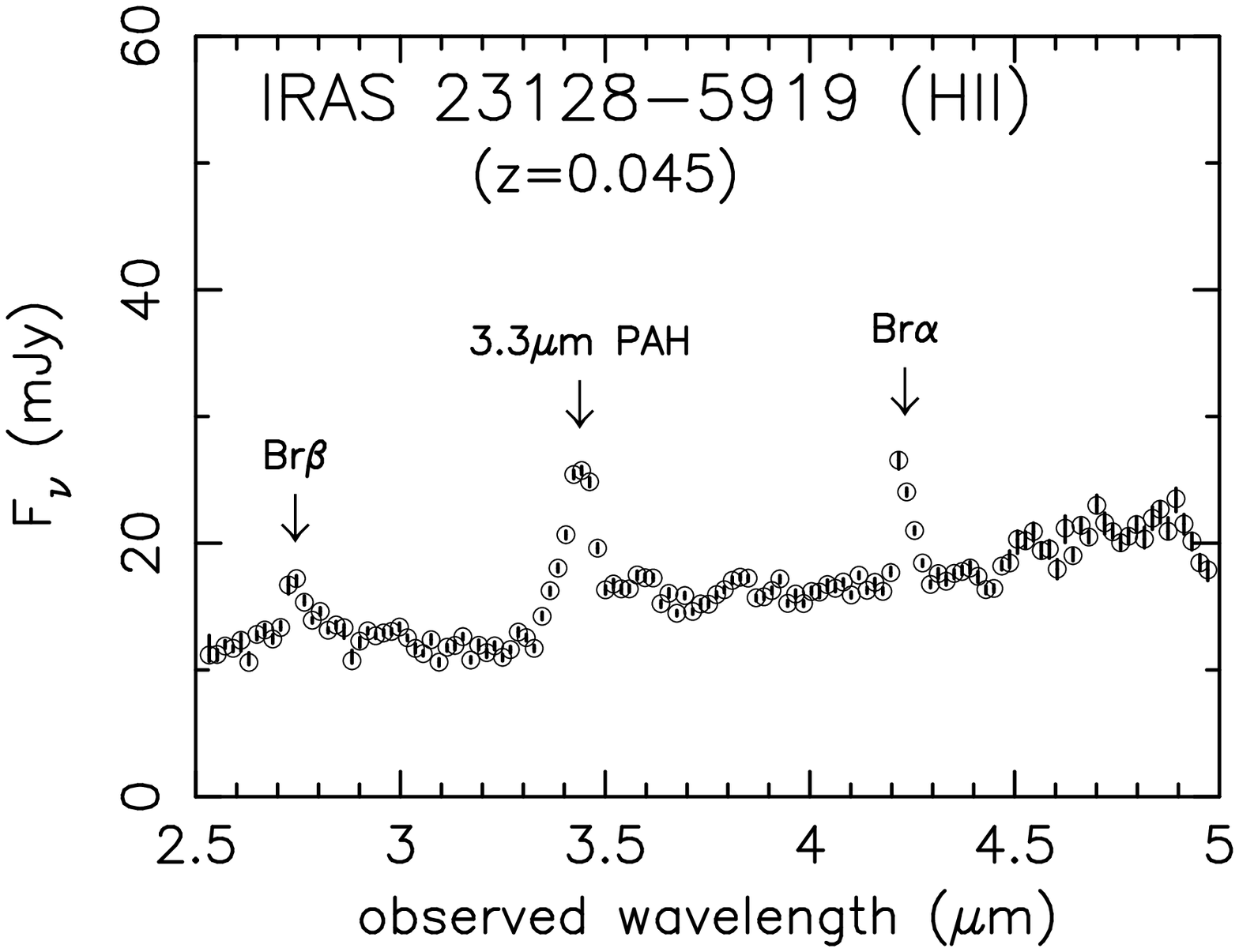} 
\includegraphics[angle=0,scale=.27]{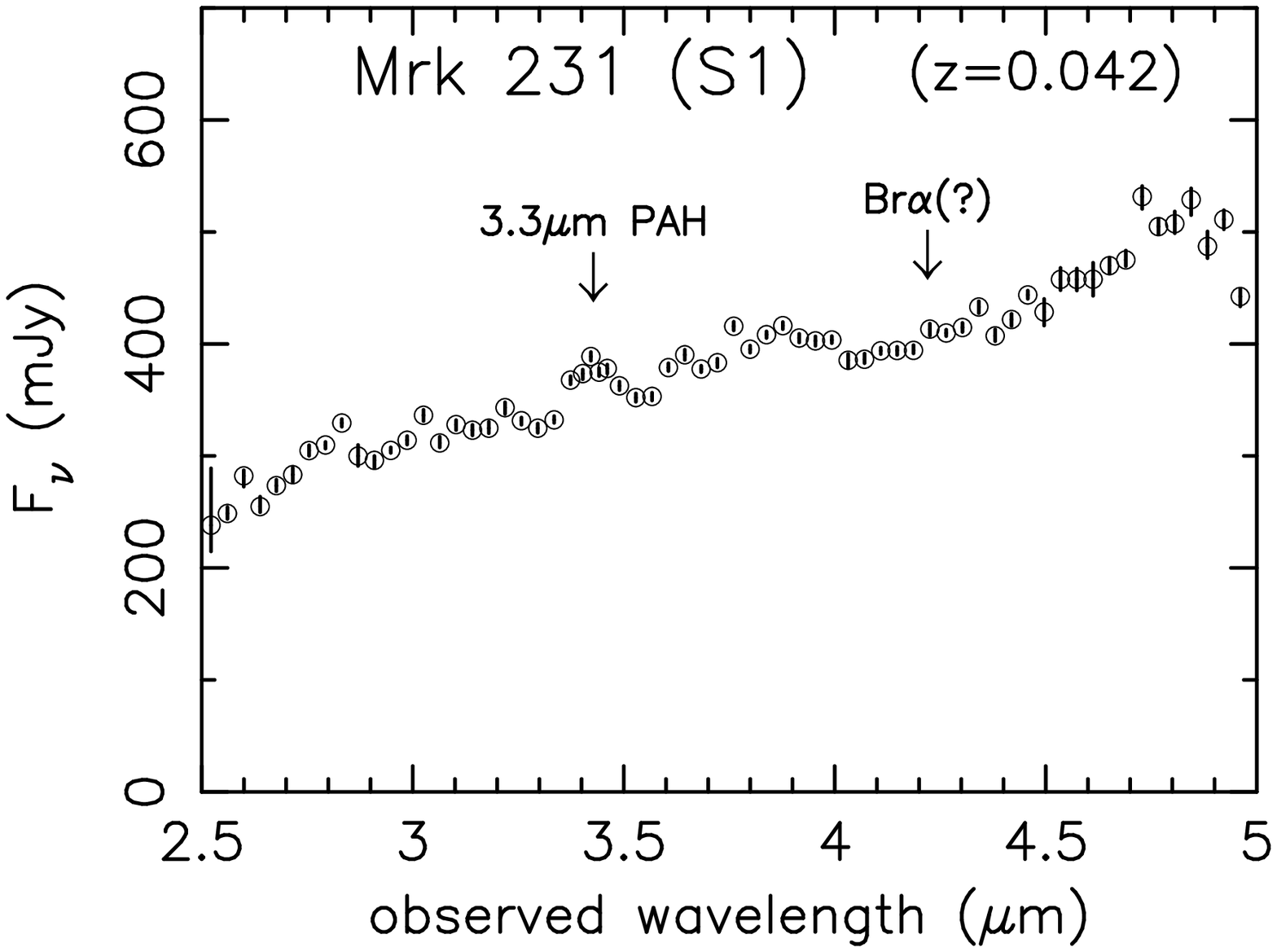} \\
\caption{AKARI IRC infrared 2.5--5 $\mu$m spectra of ULIRGs.
Symbols are the same as in Figure 1.
}
\end{figure}

\begin{figure}
\begin{center}
\includegraphics[angle=0,scale=.4]{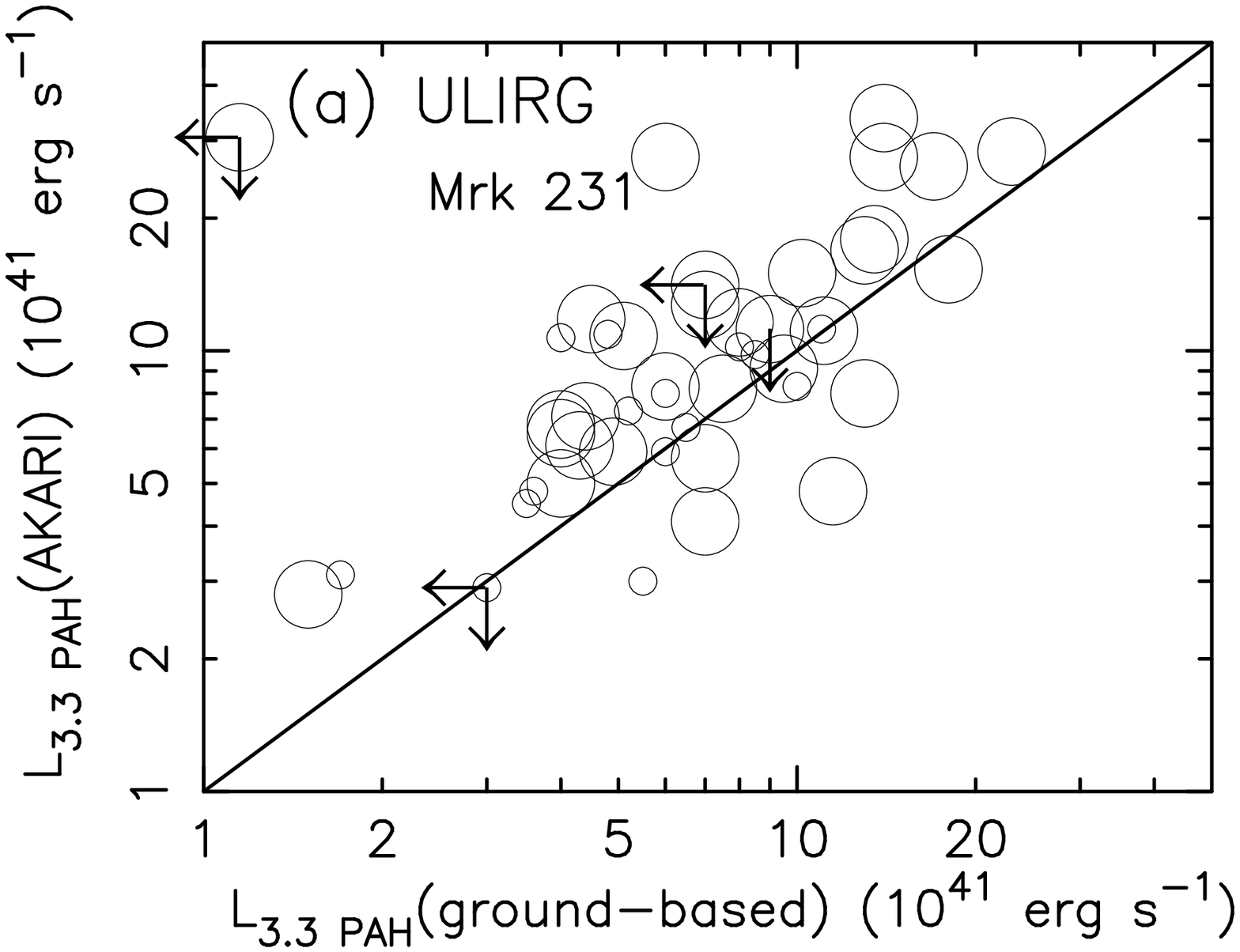}
\includegraphics[angle=0,scale=.4]{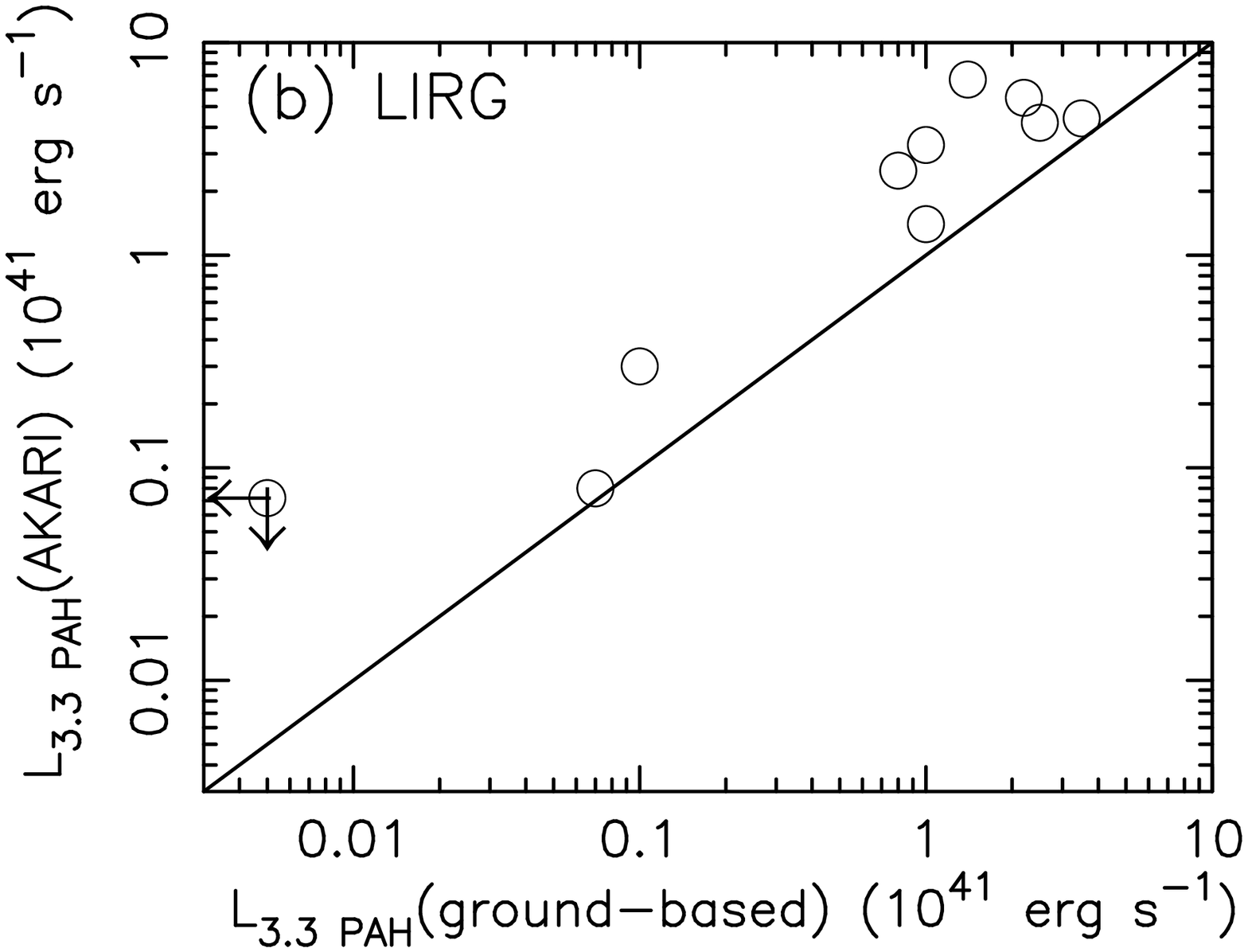}
\caption{
{\it (a)} : Comparison of the 3.3 $\mu$m PAH luminosities  
measured with ground-based slit spectra (abscissa) and AKARI IRC
slit-less spectra (ordinate) for ULIRGs for which both AKARI and
ground-based spectra at 3--4 $\mu$m are available
\citep{imd00,idm01,ris03,idm06,ima07b,ima08}.   
ULIRGs newly observed in this paper are represented by large open circles. 
ULIRGs previously observed with AKARI IRC \citep{ima08} are plotted
as small open circles.
For Mrk 231, Arp 220, and IRAS 08572+3915, more than one set of independent
ground-based 3--4 $\mu$m spectroscopic observations are available
\citep{idm06,ima07b}. 
Larger 3.3 $\mu$m PAH luminosities are adopted. 
The largest outlier is Mrk 231 (marked as Mrk 231), one of the closest
ULIRGs.
See $\S$5.1 for an explanation of the discrepancy.
{\it (b)}: Comparison of the 3.3 $\mu$m PAH luminosities  
measured with ground-based slit spectra (abscissa) and AKARI IRC
slit-less spectra (ordinate) for LIRGs for which both AKARI and
ground-based spectra at 3--4 $\mu$m are available.  
The plotted LIRGs with available ground-based 3--4 $\mu$m slit spectra
are VV 114E, NGC 2623, IC 694, NGC 3690, IC 860, Arp 193, NGC 5256 (Mrk
266) SW, IRAS 15250+3609,  NGC 4418, and NGC 1377
\citep{ima04,ima06,in06,ima07b,ima09b}.
}
\end{center}
\end{figure}

\begin{figure}
\begin{center}
\includegraphics[angle=0,scale=.4]{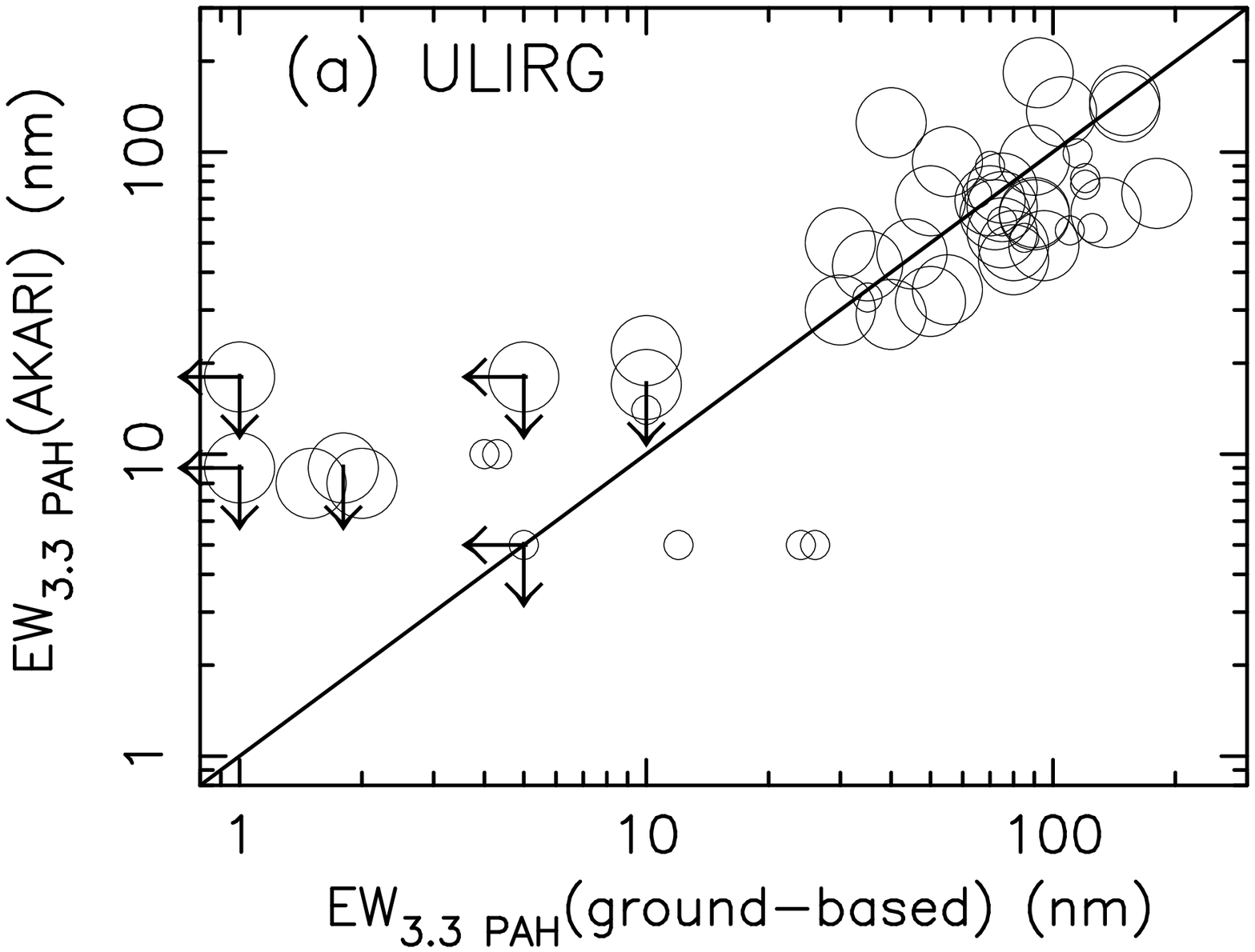}
\includegraphics[angle=0,scale=.4]{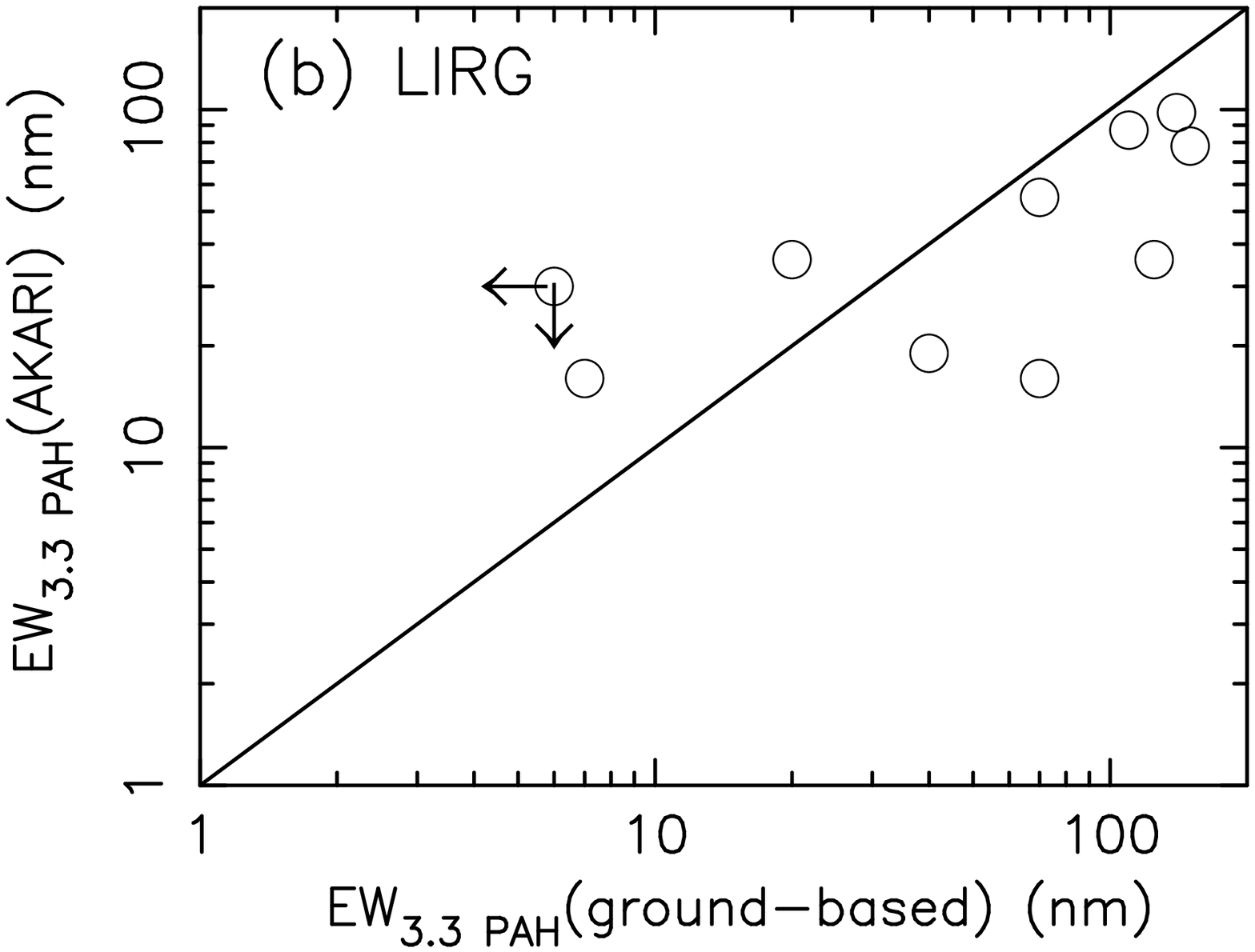}
\caption{
{\it (a)}: Comparison of rest-frame 3.3 $\mu$m PAH equivalent widths 
(EW$_{\rm 3.3PAH}$), measured with ground-based slit spectra (abscissa)
and AKARI IRC slit-less spectra (ordinate) for ULIRGs 
for which EW$_{\rm 3.3PAH}$ values have been estimated with both AKARI IRC and
ground-based spectra \citep{idm06,ris06,san08}. 
The additional interesting sources in the last rows of Tables 1 and 2
are also included, provided the EW$_{\rm 3.3PAH}$ values are estimated with
ground-based slit spectra.
For Arp 220, Mrk 231, IRAS 05189$-$2524, IRAS 23060+0505, IRAS
20551$-$4250, and Superantennae, EW$_{\rm 3.3PAH}$ values are estimated
with more than one independent reference
\citep{imd00,ris03,idm06,ris06,ima07b,san08}.  
Both values are plotted.
For IRAS 12112+0305, 14348$-$1447, and 12072$-$0444, 
EW$_{\rm 3.3PAH}$ values are separately estimated for individual nuclei 
using the ground-based slit spectra \citep{idm06,ris06}, but are estimated for the
combined emission of both nuclei with the AKARI IRC spectra.
These three sources are not plotted.
{\it (b)}: Comparison of EW$_{\rm 3.3PAH}$ values measured with
ground-based slit spectra (abscissa) and AKARI IRC slit-less
spectra (ordinate) for LIRGs for which EW$_{\rm 3.3PAH}$
values are estimated with both AKARI IRC and ground-based spectra 
\citep{ima04,ima06,in06,ima07b,ima09b}.
For NGC 3690, we assume that NGC 3690 B strongly dominates NGC 3690 C
in the AKARI IRC spectrum \citep{in06}. 
}
\end{center}
\end{figure}

\begin{figure}
\begin{center}
\includegraphics[angle=0,scale=.4]{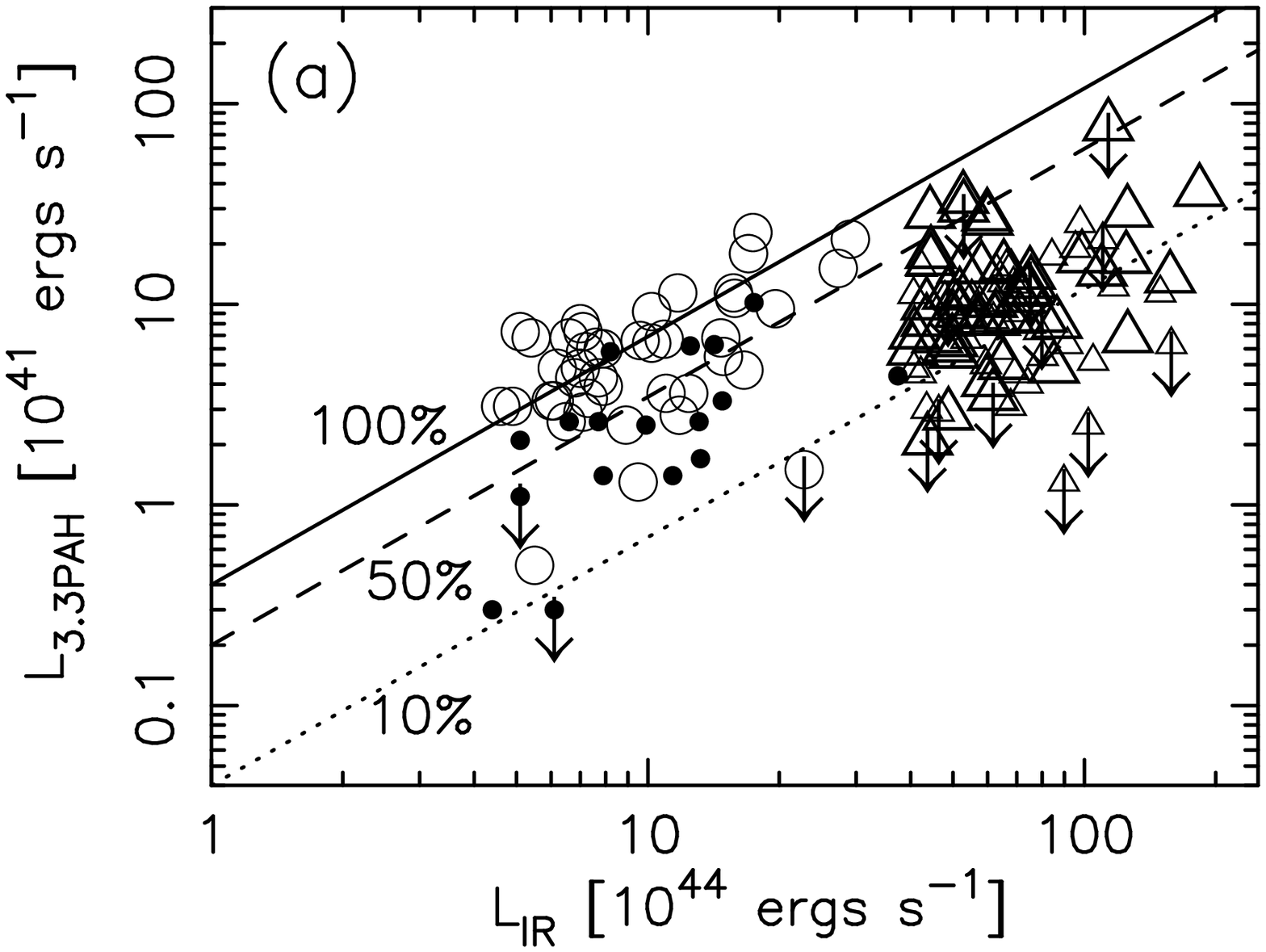}
\includegraphics[angle=0,scale=.4]{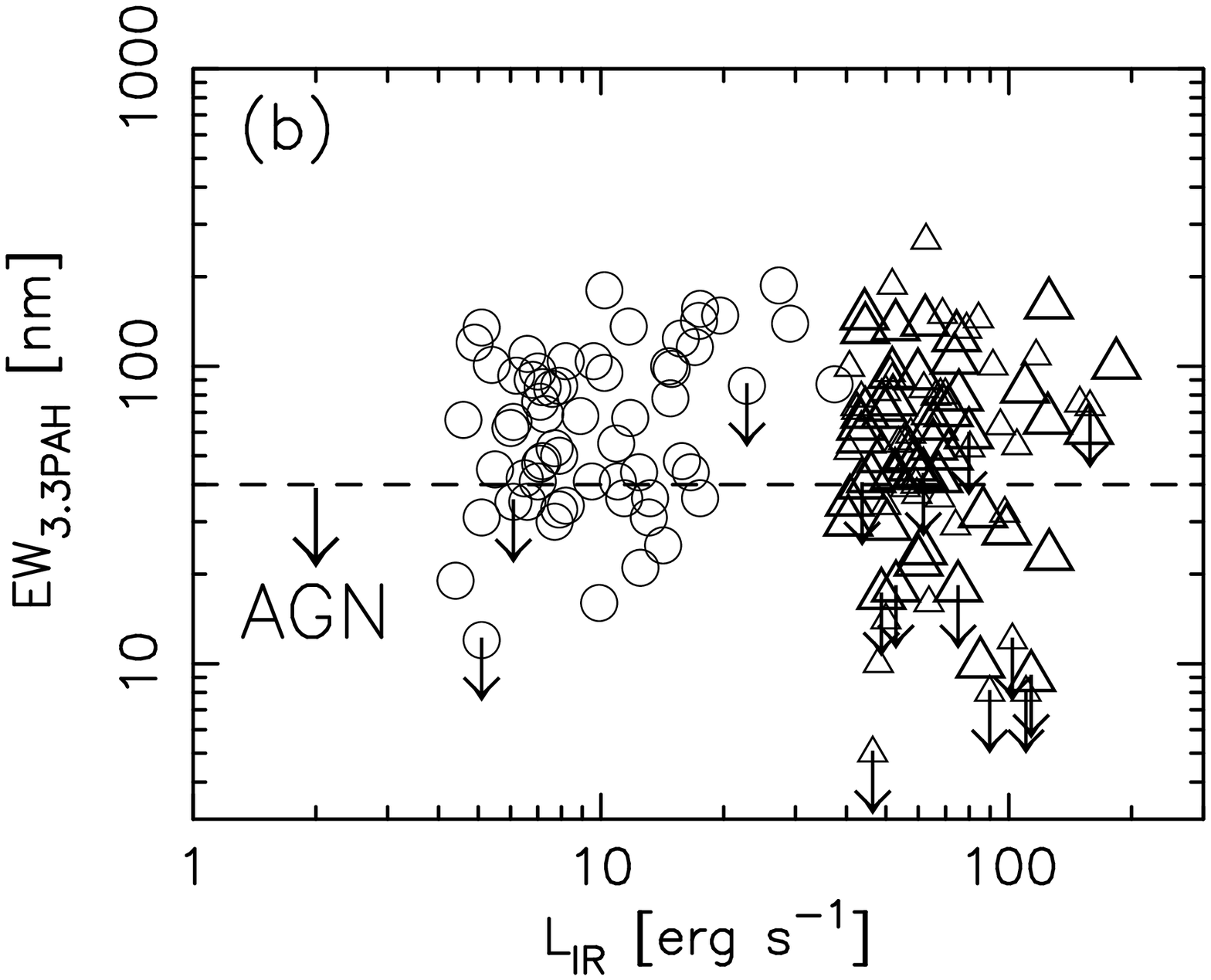}
\caption{
{\it (a) : } Comparison of the observed infrared luminosity (abscissa)
and 3.3 $\mu$m PAH luminosity (ordinate) in LIRGs and ULIRGs.
Open circles: LIRGs without obvious AGN signatures in the AKARI IRC
2.5--5 $\mu$m spectra. 
Filled circles: LIRGs with AGN signatures in the AKARI IRC spectra
(EW$_{\rm 3.3PAH}$ $<$ 40 nm).
Large triangles: ULIRGs observed in this paper. 
Small triangles: ULIRGs studied by \citet{ima08}.
The additional interesting sources in the last rows of Tables 1
and 2 are excluded from this plot, to observe the overall 
statistical trend in an unbiased manner.  
For LIRGs with multiple nuclei, the L$_{\rm 3.3PAH}$/L$_{\rm IR}$ ratios are
derived by combining the emission from all the nuclei. An exception is Arp 299,
for which the data from the IC 694 and NGC 3690 nuclei are plotted separately,
because the infrared luminosities from these individual nuclei are estimated
\citep{joy89,cha02}.  
The solid line represents L$_{\rm 3.3PAH}$/L$_{\rm IR}$ = 10$^{-3}$ (i.e.,
100\% of the infrared luminosities of galaxies could be accounted for by 
modestly obscured starburst activity probed by 3.3
$\mu$m PAH emission).
The dashed and dotted lines represent L$_{\rm 3.3PAH}$/L$_{\rm IR}$ =
0.5 $\times$ 10$^{-3}$ and 0.1 $\times$ 10$^{-3}$, respectively (in which 
50\% and 10\% of the infrared luminosities could be explained by 
3.3$\mu$m-PAH-probed starburst activity).
{\it (b) : } Comparison of infrared luminosity (abscissa) and the
rest-frame equivalent width of the 3.3 $\mu$m PAH emission 
(EW$_{\rm 3.3PAH}$) (ordinate). 
Open circles: LIRGs. 
Large triangles: ULIRGs observed in this paper. 
Small triangles: ULIRGs studied by \citet{ima08}.
The horizontal dashed line represents EW$_{\rm 3.3PAH}$ = 40 nm, below which 
strong contributions from the PAH-free continua of the AGNs to the observed AKARI
IRC 2.5--5 $\mu$m spectra are suggested. 
For sources with multiple nuclei, the EW$_{\rm 3.3PAH}$ values of individual
nuclei are plotted separately, but the L$_{\rm IR}$ values are the total
luminosities of the combined nuclei. Again, Arp 299 is the exception, for which 
the L$_{\rm IR}$ values for the individual IC 694 and NGC 3690 nuclei are derived
\citep{joy89,cha02}. 
The additional interesting sources are excluded.
}
\end{center}
\end{figure}

\begin{figure}
\begin{center}
\includegraphics[angle=0,scale=.4]{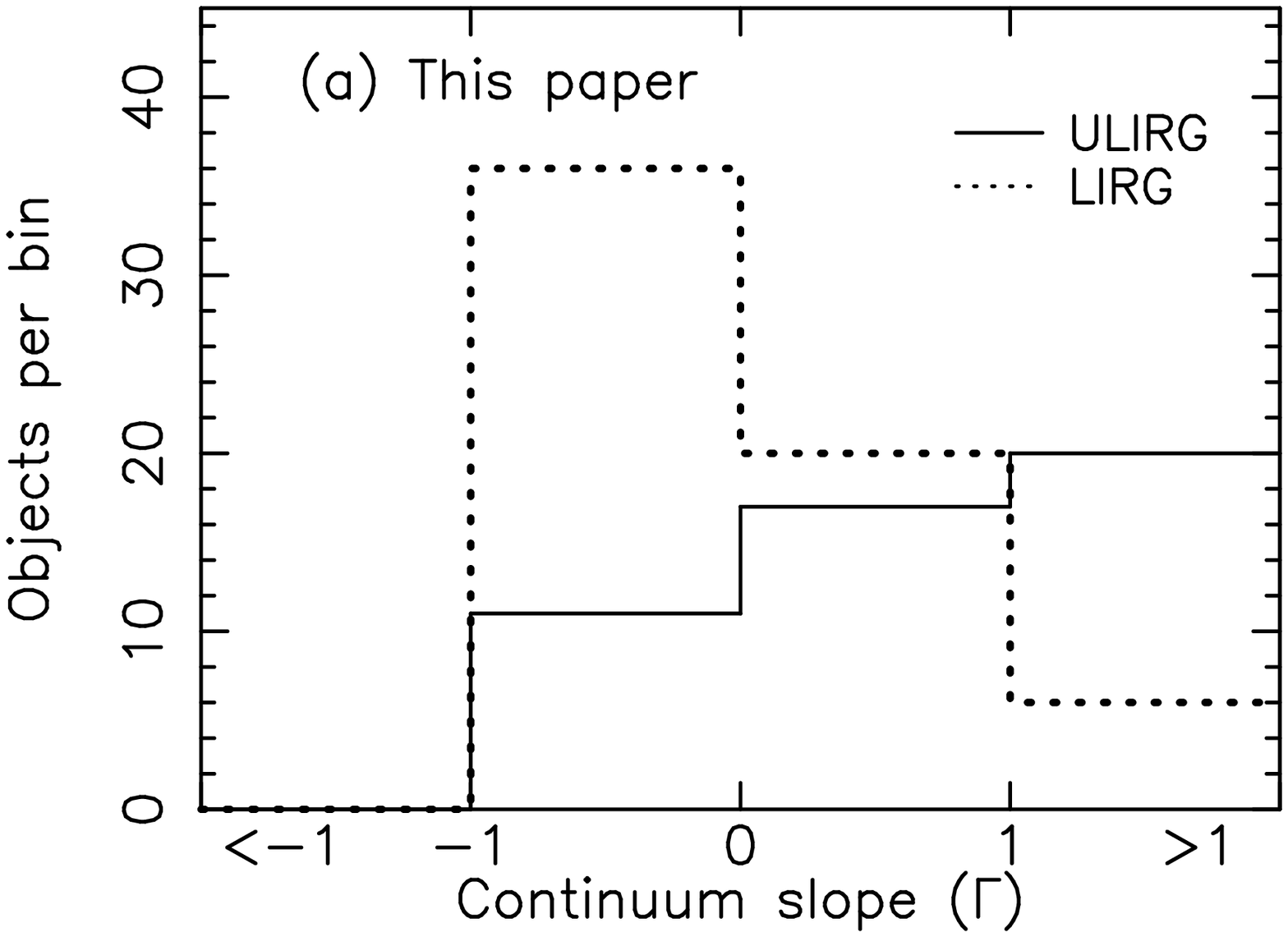}
\includegraphics[angle=0,scale=.4]{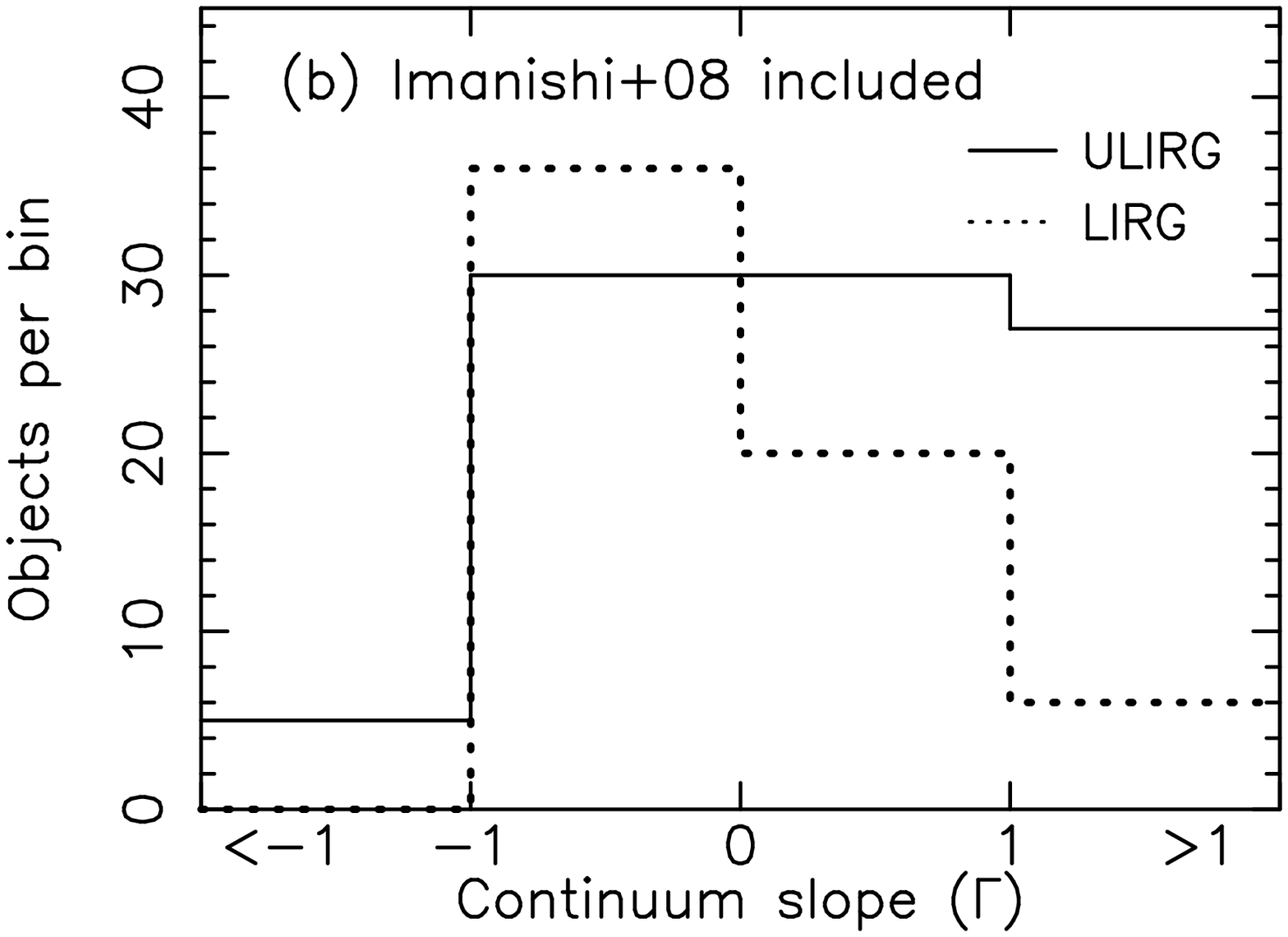} \\
\includegraphics[angle=0,scale=.4]{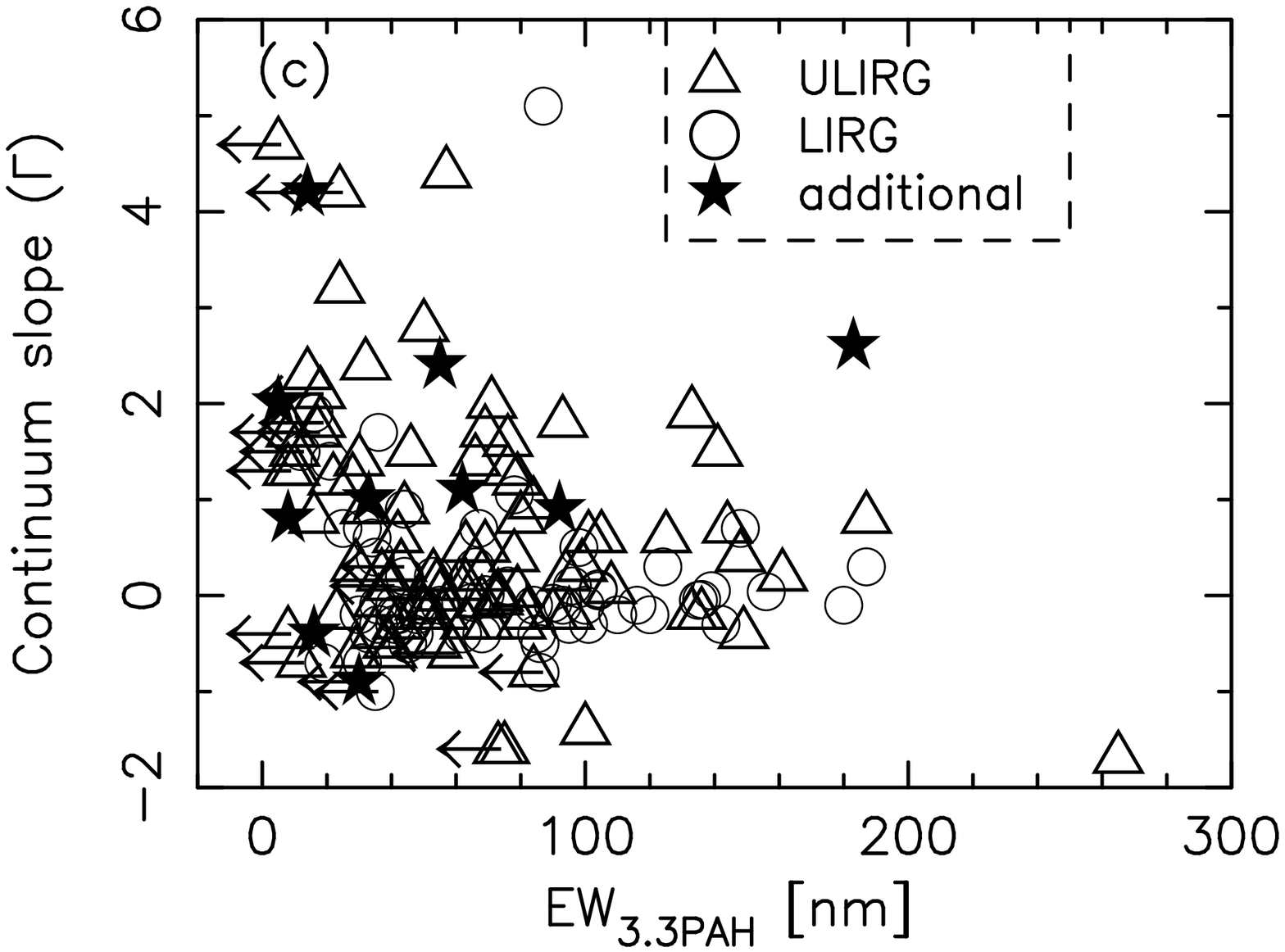}
\includegraphics[angle=0,scale=.4]{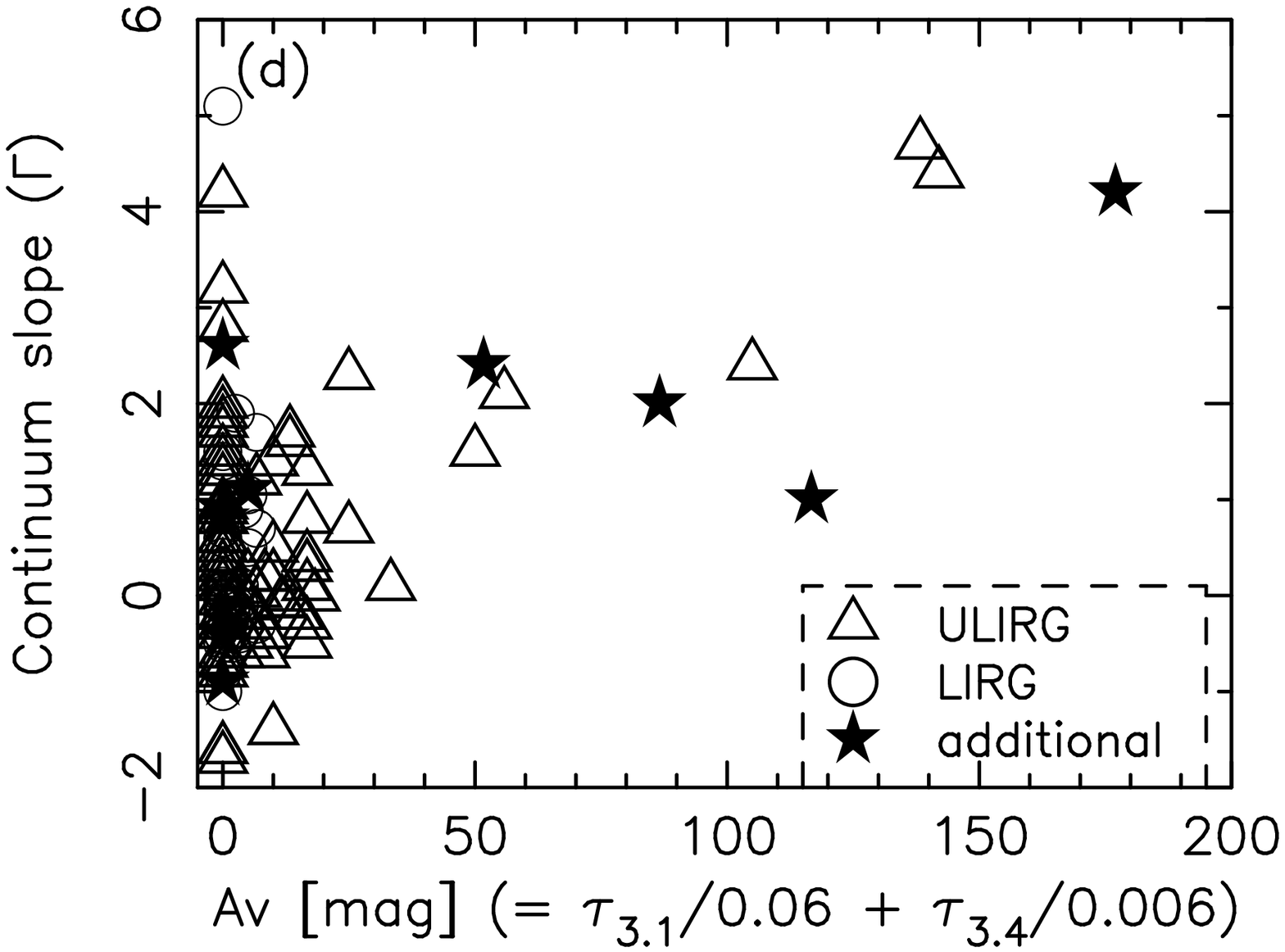}
\caption{
{\it (a) : } Histogram of the continuum slopes ($\Gamma$; F$_{\nu}$
$\propto$ $\lambda^{\Gamma}$) of LIRGs (62 nuclei) and ULIRGs (48
nuclei) studied in this paper.  
The additional interesting LIRGs (2 sources) and ULIRGs (6 sources) in
the last rows of Tables 1 and 2 are excluded, for the sake of an unbiased
statistical comparison. 
{\it (b) : } Histogram of the continuum slopes ($\Gamma$; F$_{\nu}$
$\propto$ $\lambda^{\Gamma}$) of ULIRGs and LIRGs. 
44 ULIRG nuclei studied by \citet{ima08} are added, after excluding 
the two additional interesting ULIRGs (UGC 5101 and
IRAS 19254$-$7245).
{\it (c) : } Relationship between EW$_{\rm 3.3PAH}$ (abscissa) and the
continuum slope ($\Gamma$) (ordinate) for all LIRGs and ULIRGs studied
in this paper and in \citet{ima08}. 
The additional interesting LIRGs and ULIRGs are included.
{\it (d) : } Relationship between dust extinction, estimated from the
formula A$_{\rm V}$ = $\tau_{\rm 3.1}$/0.06 + $\tau_{\rm 3.4}$/0.006
(abscissa), and the continuum slope ($\Gamma$) (ordinate) for all LIRGs
and ULIRGs studied in this paper and in \citet{ima08}.  
The additional interesting LIRGs and ULIRGs are included.
Because of the dilution of $\tau_{3.4}$ by the 3.4 $\mu$m PAH sub-peak,
the A$_{\rm V}$ values in the abscissa are only lower limits, and can be
substantially smaller than the actual dust extinction toward the 2.5--5
$\mu$m continuum emission regions, particularly for PAH-strong sources
with apparently small A$_{\rm V}$ values (see $\S$5.7). 
}
\end{center}
\end{figure}

\begin{figure}
\begin{center}
\includegraphics[angle=0,scale=.8]{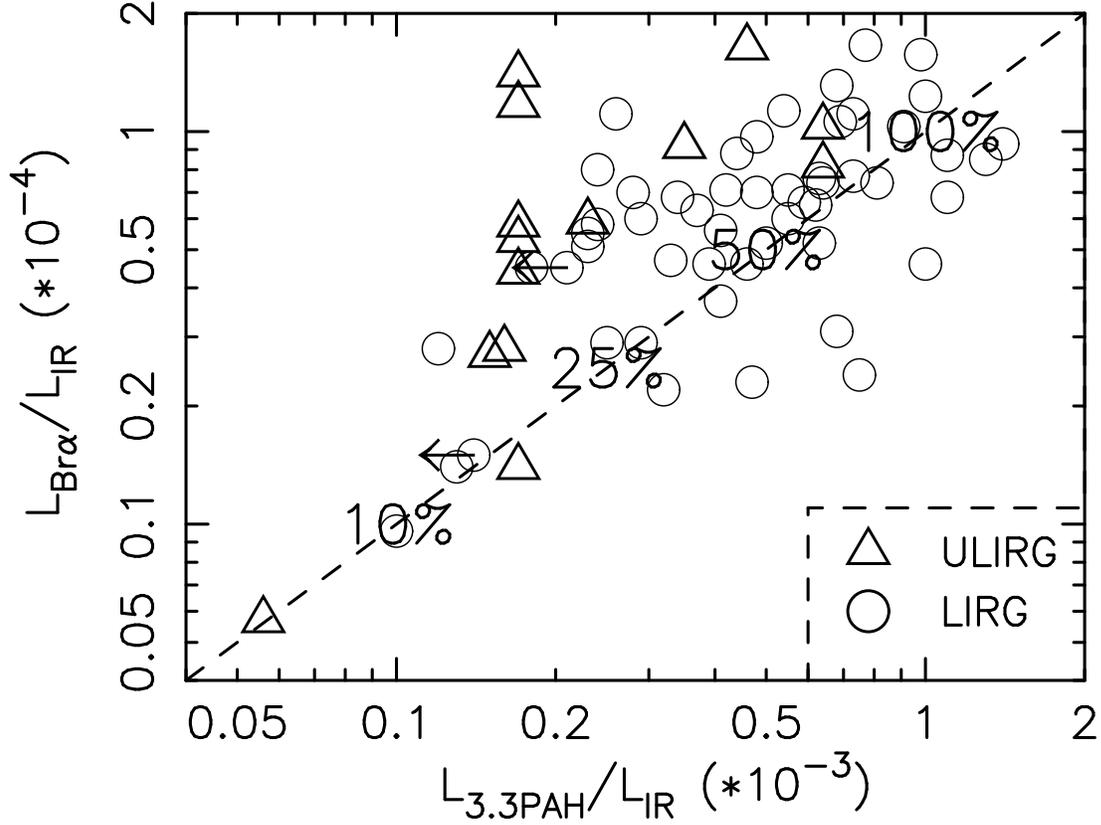}
\caption{
Comparison of the 3.3 $\mu$m PAH to infrared 
(L$_{\rm 3.3PAH}$/L$_{\rm IR}$) (abscissa) 
and Br$\alpha$ to infrared luminosity ratios 
(L$_{\rm Br\alpha}$/L$_{\rm IR}$) (ordinate) for sources with clearly 
detectable Br$\alpha$ emission.   
For starburst-dominated galaxies with modest dust obscuration, 
L$_{\rm 3.3PAH}$/L$_{\rm IR}$ $\sim$ 10$^{-3}$ and 
L$_{\rm Br\alpha}$/L$_{\rm IR}$ $\sim$ 10$^{-4}$ are the expected values. 
The symbol "100\%" indicates that all the infrared luminosity can be
accounted for by modestly obscured starburst emission probed by
3.3 $\mu$m PAH and Br$\alpha$ emission.
The symbols "50\%", "25\%", and "10\%" indicate, respectively, that 
50\%, 25\%, and 10\% of the observed infrared luminosities can be
explained by PAH- or Br$\alpha$-probed starburst activity.
For sources on the dashed line, both 3.3 $\mu$m PAH and Br$\alpha$
provide consistent starburst fractions of the observed infrared
luminosities.    
For Arp 299 (IC 694 + NGC 3690), NGC 3690 and IC 694 are plotted
separately, and we assume that 40\% and 60\% of the
infrared luminosity originates from NGC 3690 and IC 694, respectively 
\citep{joy89,cha02}.
}
\end{center}
\end{figure}

\begin{figure}
\begin{center}
\includegraphics[angle=0,scale=.8]{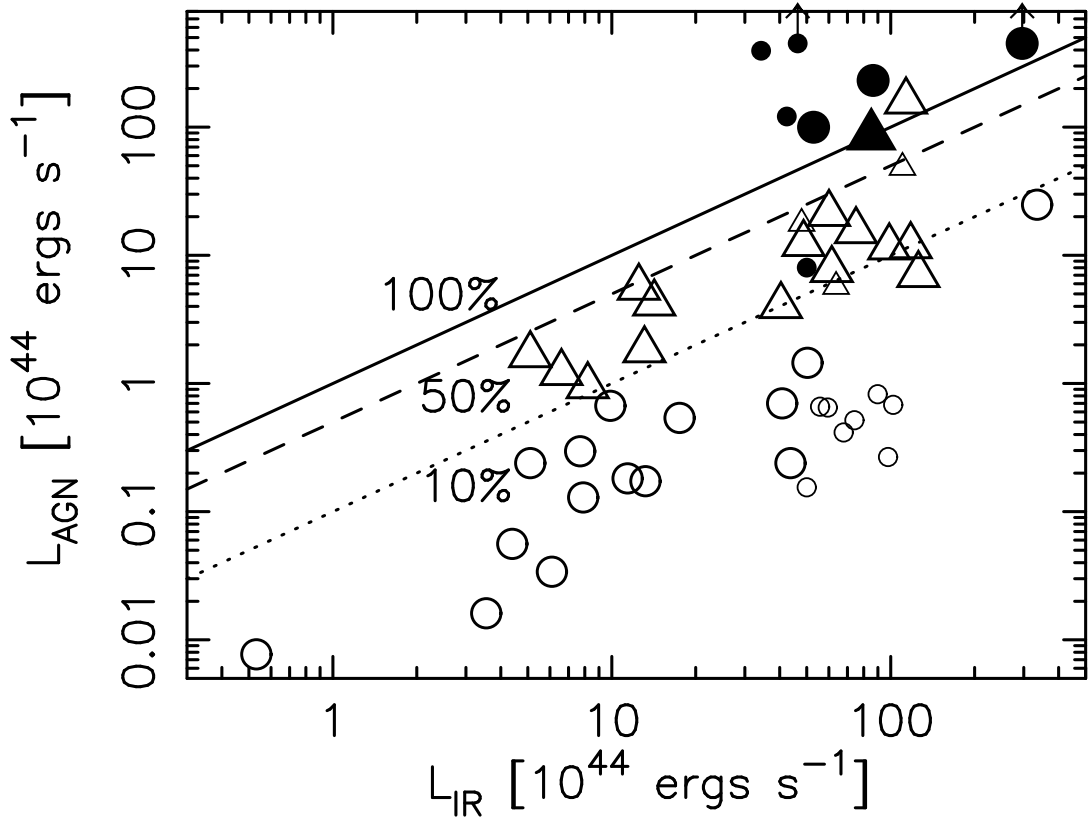}
\caption{
Comparison of the observed infrared luminosity (abscissa) and the
intrinsic, dust-extinction-corrected AGN luminosity, estimated from
AKARI IRC 2.5--5 $\mu$m spectra (ordinate) for sources with low 3.3
$\mu$m PAH equivalent widths (EW$_{\rm 3.3PAH}$ $<$ 40 nm). 
Circles: Optically non-Seyfert LIRGs and ULIRGs that display buried
AGN signatures. 
Triangles: LIRGs and ULIRGs classified optically as Seyferts.  
Filled and open symbols represent sources with detectable and
non-detectable 3.4 $\mu$m bare carbonaceous dust absorption features,
respectively.  
Large and small symbols represent sources studied in this paper and by
\citet{ima08}, respectively.
For Seyfert-type objects, the bolometric correction factor of 5 is
applied (see $\S$5.7). 
The solid line indicates that the estimated intrinsic AGN luminosity
equals the observed infrared luminosity. 
The dashed (dotted) line indicates that the intrinsic AGN
luminosity is 50\% (10\%) of the observed infrared luminosity. 
For Arp 299 (IC 694 + NGC 3690), NGC 3690 and IC 694 are plotted
separately, and we assume that 40\% of the infrared luminosity
originates from NGC 3690, and 60\% from IC 694 \citep{joy89,cha02}. 
}
\end{center}
\end{figure}

\begin{figure}
\begin{center}
\includegraphics[angle=0,scale=.8]{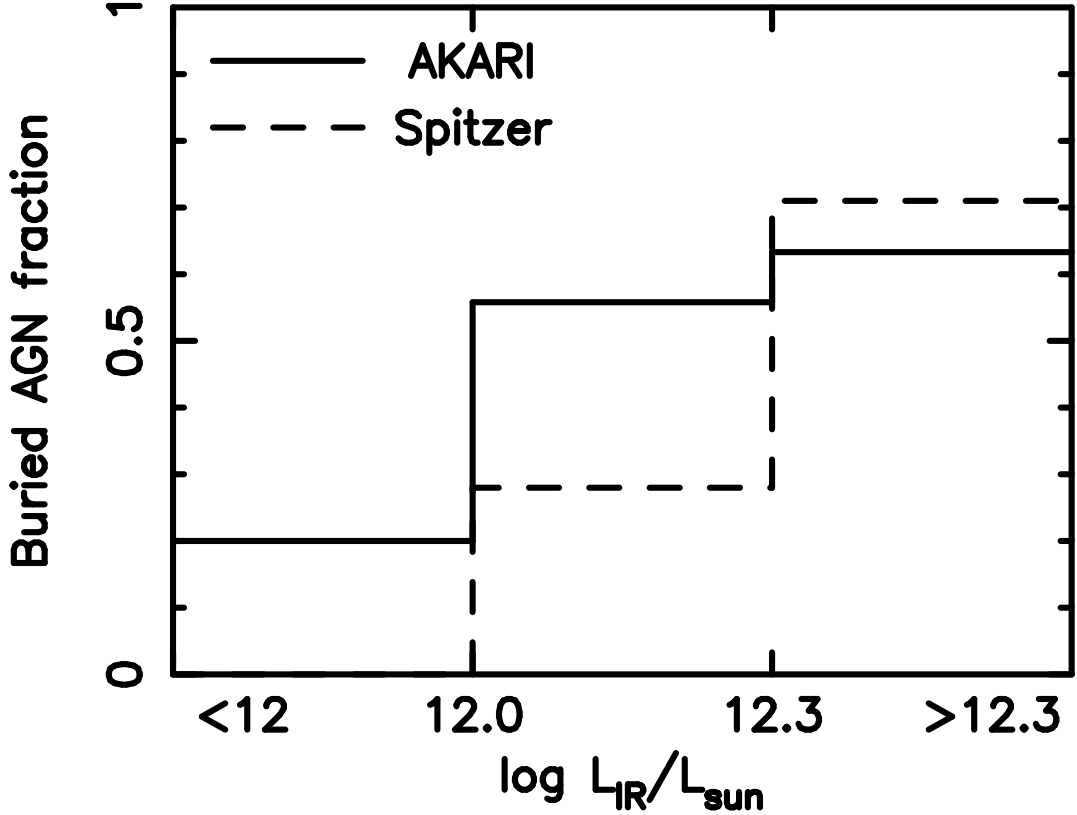}
\caption{
Detectable buried AGN fraction in optically non-Seyfert LIRGs and
ULIRGs as a function of galaxy infrared luminosity.
Sources with optical Seyfert signatures are excluded from this plot, because
our aim is to investigate the fraction of optically elusive buried AGNs. 
For sources with multiple nuclei, if optical Seyfert signatures are seen in
at least one nucleus, then these sources are excluded.
Solid line: buried AGN fraction in statistically significant number of
unbiased LIRGs and ULIRGs, as determined by AKARI IRC 2.5--5 $\mu$m
spectroscopy (this paper; Imanishi et al. 2008).  
The total number of sources is 50, 43, and 30 for 
LIRGs with L$_{\rm IR}$ $<$ 10$^{12}$L$_{\odot}$, 
ULIRGs with 10$^{12}$L$_{\odot}$ $\leq$ L$_{\rm IR}$ $<$
10$^{12.3}$L$_{\odot}$, and ULIRGs with 
L$_{\rm IR}$ $\geq$10$^{12.3}$L$_{\odot}$, respectively.
The number of sources is based on LIRGs or ULIRGs, and not on individual nuclei.
For sources with multiple nuclei, if buried AGN signatures are found in at
least one nucleus, then the sources are classified as buried AGNs.  
Dashed line: buried AGN fraction in LIRGs and ULIRGs, as determined by Spitzer IRS
5--35 $\mu$m slit spectroscopy \citep{ima07a,ima09,ima10a}.
For LIRGs, the fraction is derived only from a limited number of sources
\citep{bra06}. 
For ULIRGs, the sample is statistically complete because all ULIRGs
optically classified as non-Seyfert in the IRAS 1 Jy sample are covered.
The total number of sources is 18, 54, and 31 for 
LIRGs with L$_{\rm IR}$ $<$ 10$^{12}$L$_{\odot}$, 
ULIRGs with 10$^{12}$L$_{\odot}$ $\leq$ L$_{\rm IR}$ $<$
10$^{12.3}$L$_{\odot}$, and ULIRGs with 
L$_{\rm IR}$ $\geq$10$^{12.3}$L$_{\odot}$, respectively.
}
\end{center}
\end{figure}

\end{document}